\pdfoutput=1
\documentclass[11pt]{article}

\usepackage[english]{babel}
\usepackage{xcolor}
\usepackage{graphicx}

\usepackage[colorlinks,allcolors=blue,linktocpage=true]{hyperref}

\usepackage{cite}
\usepackage{subcaption}
\usepackage{amsmath}
\usepackage{amssymb}

\usepackage[top=15mm,bottom=12mm,left=30mm,right=30mm,foot=12mm,includefoot]{geo
metry}

\allowdisplaybreaks[3]

\numberwithin{equation}{section}

\newcommand{\sect}[1]{section~#1}
\newcommand{\sects}[1]{sections~#1}
\newcommand{\app}[1]{appendix~#1}

\newcommand{\fig}[1]{figure~#1}
\newcommand{\figs}[1]{figures~#1}

\newcommand{\tab}[1]{table~#1}

\newcommand{\eqn}[1]{equation~#1}
\newcommand{\eqs}[1]{equations~#1}

\newcommand{\re}{\operatorname{Re}}
\newcommand{\im}{\operatorname{Im}}

\newcommand{\tev}{\operatorname{TeV}}
\newcommand{\gev}{\operatorname{GeV}}

\newcommand{\fm}{\operatorname{fm}}

\newcommand{\ms}{\mskip 1.5mu}
\newcommand{\bs}{\mskip -1.5mu}

\newcommand{\tvec}[1]{\boldsymbol{#1}}

\newcommand{\chili}{\textsc{ChiliPDF}}

\newcommand{\SA}{S\bs A}
\newcommand{\AS}{A S}

\newcommand{\smalltenbar}{\scalebox{0.7}{$\overline{10}$}}
\newcommand{\smallten}{\scalebox{0.7}{$10\phantom{\overline{1}}\hspace{-1ex}$}}
\newcommand{\smalltentenbar}{\smallten\ms\smalltenbar}
\newcommand{\smalltenbarten}{\smalltenbar\ms\smallten}

\newcommand{\tenbar}{\scalebox{0.95}{$\overline{10}$}}
\newcommand{\ten}{\scalebox{0.95}{$10\phantom{\overline{1}}\hspace{-1ex}$}}
\newcommand{\tentenbar}{\ten\ms\tenbar}
\newcommand{\tenbarten}{\tenbar\ms\ten}

\newcommand{\twensev}{27\,27}

\newcommand{\Rbar}{\overline{R}\ms}

\newcommand{\Rp}[1]{R{}^{\ms \prime}_{#1}}
\newcommand{\Rpbar}[1]{\overline{R}{}^{\ms \prime}_{#1}}

\newcommand{\pr}[2]{{}^{#1}\bs #2}      
\newcommand{\prb}[2]{{}^{#1}\! #2}      
\newcommand{\prn}[2]{{}^{#1} #2}        

\newcommand{\fsqrt}[1]{\sqrt{#1\rule{0pt}{1.7ex}}}

\newcommand{\xav}[1]{x_{#1, \text{avg}}}

\newcommand{\NLLp}{NLL${}^\prime$}
\newcommand{\NNLLp}{NNLL${}^\prime$}

\newcommand{\mustar}{\mu_{y^*}}

\graphicspath{{graphs/},{plots/}}

\newcommand{\rev}[1]{#1}

\usepackage{placeins}

\begin{document}

\begin{flushright}
DESY-23-157 \\
\href{https://arxiv.org/abs/2310.16432}{arXiv:2310.16432 [hep-ph]}
\end{flushright}

\begin{center}
\vspace{4\baselineskip}
\textbf{\Large Evolution of colour correlated double parton distributions:
\\[0.1em]
a quantitative study
} \\
\vspace{3\baselineskip}
Markus~Diehl, Florian Fabry and Peter Pl{\"o}{\ss}l
\end{center}

\vspace{2\baselineskip}

\noindent
Deutsches Elektronen-Synchrotron DESY, Notkestr.~85, 22607 Hamburg, Germany

\vspace{3\baselineskip}

\parbox{0.9\textwidth}{
}

Double parton distributions satisfy the same evolution equations as ordinary
single-parton densities, \rev{provided that the colours of the two partons are
uncorrelated.  The situation is different for colour correlated parton pairs,
where evolution to higher scales} results in a suppression by Sudakov double
logarithms.  We perform a detailed study of evolution \rev{for colour correlated
double parton distributions}, both analytically and numerically, at lowest order
and beyond. When the two observed partons originate from the perturbative
splitting of a single one, the Sudakov suppression of colour correlations at the
cross section level is not as strong as one might expect.

\vfill

\newpage

\tableofcontents

\begin{center}
\rule{0.6\textwidth}{0.3pt}
\end{center}


\section{Introduction}
\label{sec:introduction}

An intriguing mechanism in proton-proton collisions is double parton scattering
(DPS), where two partons in each proton participate in two separate
hard-scattering processes.  A wealth of experimental evidence for this mechanism
has been obtained at the Tevatron and the LHC (see \cite{Abe:1997xk,
Abazov:2015nnn, Aaij:2016bqq, Aaboud:2018tiq, CMS:2022pio} and references
therein), and more detailed studies from the LHC are eagerly awaited.
Building on pioneering work in the 1980s \cite{Paver:1982yp, Mekhfi:1983az,
Sjostrand:1986ep}, a detailed theoretical understanding of double parton
scattering in QCD has been achieved in the last decade \cite{Blok:2010ge,
Gaunt:2011xd, Ryskin:2011kk, Blok:2011bu, Diehl:2011yj, Manohar:2012jr,
Manohar:2012pe, Ryskin:2012qx, Gaunt:2012dd, Blok:2013bpa, Diehl:2017kgu,
Cabouat:2019gtm, Cabouat:2020ssr}.  Recent work on theory and phenomenology can
for instance be found in \cite{Fedkevych:2020cmd, Ceccopieri:2021luf,
Blok:2022mtv, Golec-Biernat:2022wkx}.  An comprehensive overview of the subject
is given in \cite{Bartalini:2017jkk}.

The non-perturbative quantities needed to compute DPS cross sections are double
parton distributions (DPDs), which generalise the familiar concept of
single-parton densities (PDFs) to the case of two partons.  Most theoretical
work so far has focused on the case in which the colour state of each parton is
summed over separately.  In general, however, two partons inside a proton are
correlated in their colour \cite{Mekhfi:1985dv}.  It has been realised long ago
that such colour correlations are suppressed by Sudakov logarithms in processes
at high momentum scales \cite{Mekhfi:1988kj}, and a simple quantitative estimate
of this effect was given in \cite{Manohar:2012jr}. However, there are several
reasons why one should not discard colour correlations in DPDs altogether.  One
is that the Sudakov suppression becomes weaker in the case where at least one of
the hard scatters takes place at a scale of a few $\gev$; this concerns in
particular the production of additional jets (or mini-jets) in what is commonly
called the underlying event.  A second reason is that the two partons described
by a DPD may originate from the perturbative splitting of a single parton.  As
was observed out in \cite{Blok:2022mtv}, the Sudakov logarithms in this case are
only built up in the kinematic region between the scale of this splitting and
the scale of the hard scatter, which significantly reduces their size.

The study in \cite{Blok:2022mtv} was performed in the leading logarithmic (LL)
approximation, using the approach to DPS developed in \cite{Blok:2010ge,
Blok:2011bu, Blok:2013bpa}.  In the present work, we use instead the formalism
described in \cite{Diehl:2011yj, Diehl:2017kgu, Buffing:2017mqm}, which allows
for a systematic inclusion of higher-order corrections.  In this formalism, the
scale evolution for DPDs without colour correlations is described by the DGLAP
equations familiar from the evolution of PDFs.  DPDs with colour correlations
additionally depend on a rapidity parameter $\zeta$, in the same way as
transverse-momentum dependent distributions (TMDs) for a single parton.  The
dependence on $\zeta$ is described by a Collins-Soper equation
\cite{Collins:1981uk, Collins:2011zzd}, the solution of which exponentiates the
Sudakov logarithms mentioned above.  The evolution kernels and anomalous
dimensions needed for the evolution of colour correlated DPDs were calculated at
leading order (LO) in \cite{Diehl:2011yj, Buffing:2017mqm} and at
next-to-leading order (NLO) in the recent work \cite{Diehl:2022rxb}.  In the
present paper, we use these ingredients to perform a detailed study of DPD
evolution in the presence of colour correlations, both analytically and
numerically.  This will allow us to assess in which kinematic situations these
correlations may be neglected and in which ones they should be taken into
account.

This paper is organised as follows.  In \sect{\ref{sec:basics}} we recall the
basics of the formalism used in our work, with a focus on the colour structure
of DPDs and on their scale dependence.  The initial conditions we use when
solving the evolution equations are described in \sect{\ref{sec:input}}.  In
\sect{\ref{sec:solutions}} we discuss two semi-analytic solutions of these
equations, where the Sudakov logarithms are exponentiated in closed form.
Numerical results are presented in \sect{\ref{sec:dpds}} for DPDs and in
\sect{\ref{sec:lumis}} double parton luminosities, which directly enter in the
cross section formula for DPS.  We conclude in \sect{\ref{sec:summary}}.  A
number of technical aspects is presented in \app{\ref{sec:colour-proj}} to
\app{\ref{sec:chili}}, and additional numerical results are shown in
\app{\ref{sec:more-lumis}}.

\section{Basics of colour non-singlet DPDs}
\label{sec:basics}

In this section we recall the formalism to compute double parton scattering, as
it was set up in our previous work \cite{Diehl:2011yj, Diehl:2017kgu,
Buffing:2017mqm}.


\subsection{Distributions and cross section}
\label{sec:cross-sect-basics}

\begin{figure}
\centering
\includegraphics[width=0.7\textwidth]{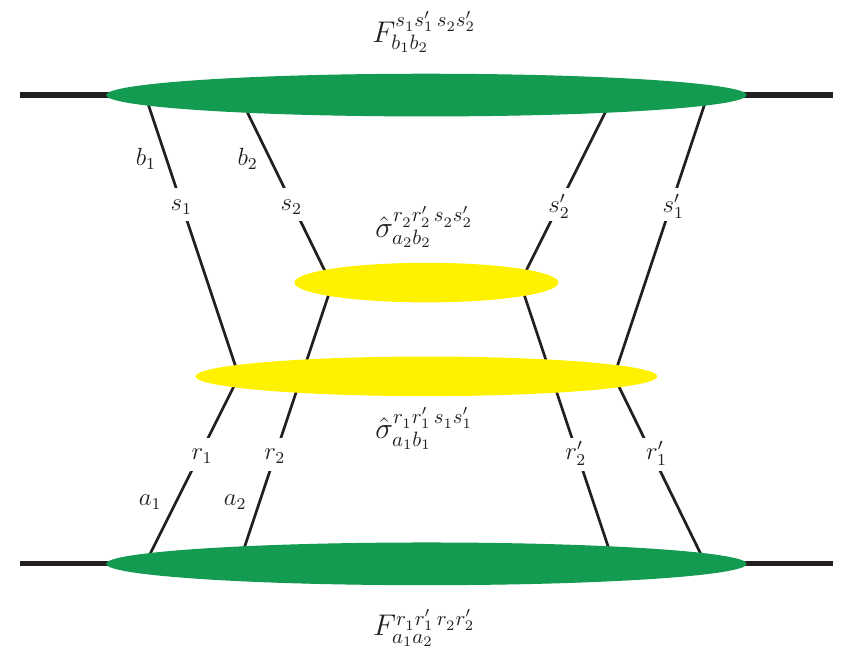}
\caption{\label{fig:xsect-graph} Factorised representation of the DPS cross
section in terms of parton-level cross sections $\hat{\sigma}$ and DPDs $F$.
The labels $a_i$ and $b_i$ denote the parton type and polarisation, whereas
$r_i^{}, r_i', s_i^{}$, and $s_i'$ are colour indices in the fundamental or
adjoint representation.}
\end{figure}

To begin with, we present the factorisation formula for DPS in terms of
parton-level cross sections $\hat{\sigma}$ and double parton distributions $F$,
as graphically represented in \fig{\ref{fig:xsect-graph}}.
Consider the cross section for $p + p \to A_1 + A_2 + X$, where $A_i$ denotes
the system of observed particles produced in the hard subprocess number $i$, and
$X$ denotes the unobserved part of the final state.   If the hard subprocesses
are computed at tree level, then the DPS cross section is given by
\begin{align}
   \label{dps-Xsect}
   &
   \frac{d \sigma_{\text{DPS}}}{d x_1\, d x_2\, d\bar{x}_1\, d\bar{x}_2}
   =
   \frac{1}{1 + \delta_{A_1 A_2}} \,
   \sum_{R_1 R_2 R_3 R_4} \, \sum_{a_1 a_2\, b_1 b_2}
   \prn{\Rbar_{1} \Rbar_{3} \ms}{\hat{\sigma}}_{a_1 b_1}(x_1, \bar{x}_1,
      s; \mu_1) \;
   \prn{\Rbar_{2} \Rbar_{4} \ms}{\hat{\sigma}}_{a_2 b_2}(x_2, \bar{x}_2,
      s; \mu_2) \;
   \notag\\
   & \qquad \times
   2\pi \int_{b_0 /\nu}^{\infty} d y\, y\;
   \prn{R_1 R_2}{F}_{a_1 a_2}(x_1, x_2, y; \mu_1, \mu_2, \zeta) \;
   \prn{R_3 R_4}{F}_{b_1 b_2}(\bar{x}_1, \bar{x}_2, y;
      \mu_1, \mu_2, \bar{\zeta})
\end{align}
Here the DPDs are evaluated for parton momentum fractions
\begin{align}
   \label{x-fractions}
   x_i
   &=
   \frac{M_i}{\sqrt{s}}\; e^{Y_i} \,,
   &
   \bar{x}_i
   &=
   \frac{M_i}{\sqrt{s}}\; e^{- Y_i} \,,
   &
   i=1,2,
\end{align}
where $\sqrt{s}$ is the c.m.\ energy of the proton-proton collision, $M_i$ is
the invariant mass and $Y_i$ the rapidity of the observed system $A_i$.  The
differentials in the cross section \eqref{dps-Xsect} can alternatively be
written as
\begin{align}
   d x_i \, d\bar{x}_i
   &=
   s^{-1}\, d M_i^2\, d Y_i
   \,.
\end{align}
If the hard-scattering cross sections are evaluated beyond tree level, a
four-fold integral over parton momentum fractions appears on the r.h.s.\ of
\eqref{dps-Xsect}.

The function $\prn{R_1 R_2}{F}_{a_1 a_2}(x_1, x_2, y; \mu_1, \mu_2,
\zeta)$ in \eqref{dps-Xsect} depends on the longitudinal momentum fractions
$x_1$ and $x_2$ of the two partons, on their distance $y$ in the plane
transverse to the proton momentum, and on the factorisation scales $\mu_1$ and
$\mu_2$ associated with the two hard-scattering processes.  The labels $a_1$ and
$a_2$ specify the flavour and polarisation of each parton.  We note that DPDs
involving transverse quark or linear gluon polarisation have open Lorenz indices
and depend on the difference $\tvec{y}$ of the transverse parton positions
rather than on the length $y$ of this vector.  In this case one should replace
$2 \pi\ms d y\, y$ in \eqref{dps-Xsect} with $d^2 \tvec{y}$ and restrict the
integral to the region $y > b_0/\nu$.

The transverse distance $y$ between the partons is integrated with a lower
cutoff $b_0 / \nu$, where
\begin{align}
   \label{b0-def}
   b_0 &= 2 e^{-\gamma_E} \approx 1.12
\end{align}
with the Euler-Mascheroni constant $\gamma_E$, and the scale $\nu$ should be
taken as
\begin{align}
   \label{nu-choice}
   \nu \sim  \min(\mu_1, \mu_2)
\end{align}
as explained in \sect{6} of \cite{Diehl:2017kgu}.  In physical terms this scale
is associated with the separation between DPS and single parton scattering, as
the factorisation scales $\mu_i$ are associated with the separation between
parton distributions and hard-scattering cross
sections.

Implicit in the factorisation formula \eqref{dps-Xsect} is a soft factor that
describes gluon exchange between the \rev{partons in one and the partons in the
other proton}.  This factor has been split into two parts, \rev{each of which is
absorbed into} one of the two DPDs.
The technical procedure is given in \sects{3} to 5 of \cite{Buffing:2017mqm} and
closely resembles the method developed for TMDs in \cite{Collins:2011zzd}.  The
manner in which the soft factor is split is described by the rapidity parameters
$\zeta$ and $\bar{\zeta}$, which satisfy the kinematic constraint
\begin{align}
   \label{zeta-basic-constraint}
   \bigl(\ms \zeta \bar{\zeta} \,\bigr)^{1/2}
   &= s
   \,.
\end{align}
Their specific definition depends on the regulator used to handle rapidity
divergences in intermediate steps; expressions for the Collins and the $\delta$
regulator are given in \sect{2.2} of \cite{Diehl:2021wpp}.  An important feature
is that $\fsqrt{\zeta}$ is proportional to the momentum of the proton, or more
precisely to the proton plus-momentum if the proton moves into the positive
$z$ direction.

\paragraph{Colour structure.}  The factorisation formula can be written with
explicit colour indices for each parton line that leaves a DPD and enters a
hard-scattering graph, as shown in \fig{\ref{fig:xsect-graph}}.  In the final
form \eqref{dps-Xsect} we have projected these indices on irreducible
representations of the colour group for each pair of partons in the scattering
amplitude and its conjugate.  The representations are indicated by the labels
\begin{equation}
   R_1 \text{ for } (r_1^{}, r_1') \,,
   \quad
   R_2 \text{ for } (r_2^{}, r_2') \,,
   \quad
   R_3 \text{ for } (s_1^{}, s_1') \,,
   \quad
   R_4 \text{ for } (s_2^{}, s_2')
   \,.
\end{equation}
The relevant representations are
\begin{itemize}
\item for quarks and antiquarks: the singlet ($R=1$) and the octet ($R=8$),
\item for gluons: the singlet ($R=1$), a symmetric octet ($R=S$), an
antisymmetric octet ($R=A$), a decuplet and an antidecuplet ($R=10$ and
$R=\tenbar$), and an icosaheptaplet ($R=27$).
\end{itemize}
By $\Rbar$ we denote the conjugate of the representation of $R$, with the
convention that $R=\Rbar$ for the singlet, all octets and the icosaheptaplet.

Since colour is conserved, the four parton lines of each DPD must couple to an
overall colour singlet, as well as the four parton lines of each hard-scattering
cross section.  As a result, the sum over colour representations $R_i$ in the
factorisation formula \eqref{dps-Xsect} is restricted to combinations in which
all four representations have the same dimension.

Our normalisation conventions for colour projected DPDs and parton-level cross
sections are given in \app{\ref{sec:colour-proj}}.  The detailed algebraic
manipulations (including the treatment of the soft factor) leading to the cross
section formula \eqref{dps-Xsect} can be found in \sect{4} and \app{I.1} of
\cite{Buffing:2017mqm}.

In colour singlet DPDs $\prn{11}{F}_{a_1 a_2}$, the colour states of each parton
$a_1$ and $a_2$ are summed over separately, as is done in single parton
distributions. The colour singlet combination $\prn{11}{\hat{\sigma}}_{a b}$ is
the familiar parton-level cross section averaged over the colours of the
incoming partons $a$ and $b$, as is done for single parton scattering.  Most
theoretical studies of DPS consider only the terms with colour singlet labels in
the factorisation formula.

We note that the combinations $\prn{\SA}{\hat{\sigma}}_{g g}$ and
$\prn{\AS}{\hat{\sigma}}_{g g}$ are zero if the observed final state is
invariant under charge conjugation. This is because the two octets $S$ and $A$
have opposite charge conjugation parity.

For the production of a colour-singlet final state, one obtains the simple
colour dependence
\begin{align}
   \label{Xsect-for-singlet}
   \prn{\Rbar \Rpbar{} \ms}{\hat{\sigma}}_{a b}
   &=
   \delta_{R \Rpbar{}} \; \eta(R) \,
   \prn{11}{\hat{\sigma}}_{a b}
   \,,
\end{align}
where $\eta(A) = -1$ and $\eta(R) = 1$ for $R \neq A$.  The colour sum in the
factorisation formula them simplifies to
\begin{align}
   \label{fact-for-singlet}
   \sum_{R_1 R_2 R_3 R_4}
   &
   \prn{\Rbar_{1} \Rbar_{3} \ms}{\hat{\sigma}}_{a_1 b_1} \,
   \prn{\Rbar_{2} \Rbar_{4} \ms}{\hat{\sigma}}_{a_2 b_2} \,
   \prn{R_1 R_2}{F}_{a_1 a_2} \,
   \prn{R_3 R_4}{F}_{b_1 b_2}
   \nonumber \\
   & =
   \prn{11}{\hat{\sigma}}_{a_1 b_1} \,
   \prn{11}{\hat{\sigma}}_{a_2 b_2}
   \sum_{R_1 R_2}
   \eta(R_1) \, \eta(R_2) \;
   \prn{R_1 R_2}{F}_{a_1 a_2} \;
   \prn{\Rbar_{1} \Rbar_{2}}{F}_{b_1 b_2}
   \,.
\end{align}


\subsection{Evolution equations}
\label{sec:evolution-eqs}

The scale dependence of DPDs is given by a DGLAP-type equation
\begin{align}
   \label{DGLAP-zeta}
   &
   \frac{d}{d \ln\mu_1}\,
   \prn{R_1 R_2}{F}_{a_1 a_2}(x_1, x_2, {y}; \mu_1, \mu_2, \zeta)
   \nonumber \\
   &\quad =
   - \;\prn{R_1}{\gamma_J}(\mu_1) \;
   \ln \biggl( \frac{x_1 \fsqrt{\zeta}}{\mu_1} \biggr) \;\,
   \prn{R_1 R_2}{F}_{a_1 a_2}(x_1, x_2, {y}; \mu_1, \mu_2, \zeta)
   \nonumber \\
   & \qquad
   + 2 \sum_{b_1, \Rp{1}}
   \int_{x_1}^1 \frac{d z}{z} \;
   \pr{R_1^{} \Rpbar{1}}{P}_{a_1 b_1}\biggl(
      \frac{x_1}{z};\ms \mu_1^{} \biggr) \,
   \prn{\Rp{1} R_2^{}}{F}_{b_1 a_2}(z, x_2^{}, {y}; \mu_1, \mu_2, \zeta)
\end{align}
and its analogue for the dependence on $\mu_2$.
We note that the evolution equations can be written in more compact form by
using kernels $P(z; \mu, \zeta)$ with a rapidity parameter dependence.  Making
this dependence explicit, as in \eqn{(2.10)} of \cite{Diehl:2022rxb}, one
obtains the rapidity independent kernels $P(z; \mu)$ that we will use throughout
this work.

The combination $x_1 \fsqrt{\zeta}$ in the second line of \eqref{DGLAP-zeta} can
be defined in terms of the momentum of the first parton, without reference to
the proton momentum.  This must be the case, since the renormalisation procedure
resulting in the evolution equation does not make reference to the proton
momentum.  On the other hand, it is the rapidity parameter $\zeta$ that remains
constant under the convolution integral in \eqref{DGLAP-zeta}, where the
momentum of the first parton varies but the momentum of the proton does not.

The only cases in which the sum over $\Rp{1}$ in \eqref{DGLAP-zeta} contains
more than one term is for the quark-gluon mixing kernels \rev{$\pr{8S}{P}_{q
g}$} and
$\pr{8A}{P}_{q g}$ (and their antiquark equivalents).  The kernels
$\pr{\AS}{P}_{g g}$ and $\pr{\SA}{P}_{g g}$ are zero at all orders in $\alpha_s$
because of charge conjugation invariance.

The dependence on the rapidity parameter is given by a Collins-Soper equation
\begin{align}
   \label{CS-equation}
   & \frac{d}{d \ln\fsqrt{\zeta}}\,
   \prn{R_1 R_2}{F}_{a_1 a_2}(x_1, x_2, {y}; \mu_1, \mu_2, \zeta)
   \nonumber \\[0.5em]
   &\quad
   = \prb{R_1}{J}({y};\mu_1,\mu_2) \,
   \prn{R_1 R_2}{F}_{a_1 a_2}(x_1, x_2, {y}; \mu_1, \mu_2, \zeta)
   \,.
\end{align}
The Collins-Soper kernel has a scale dependence
\begin{align}
   \label{CS-RGE}
   \frac{d}{d \ln\mu_1}\, \prb{R}{J}({y};\mu_1,\mu_2)
   &=
   - \;\prn{R}{\gamma_J}(\mu_1)
\end{align}
and an analogous dependence on $\mu_2$, so that
\begin{align}
   \label{CS-solved}
   \prb{R}{J}({y};\mu_1,\mu_2)
   &= \prb{R}{J}({y};\mu_{01},\mu_{02})
   - \int_{\mu_{01}}^{\mu_1} \frac{d\mu}{\mu}\; \prn{R}{\gamma}_J(\mu)
   - \int_{\mu_{02}}^{\mu_2} \frac{d\mu}{\mu}\; \prn{R}{\gamma}_J(\mu)
   \,.
\end{align}
The kernel $\prb{R}{J}$ depends only on the dimension of $R$, i.e.\ $\pr{8}{J} =
\pr{S}{J} = \pr{A}{J}$ and $\pr{\smallten}{J} = \pr{\smalltenbar}{J}$.  The same
holds of course for the cusp anomalous dimensions $\prn{R}{\gamma}_J$. In
the colour singlet sector, one has
\begin{align}
   \label{J-singlet}
   \pr{1}{J} &= 0
   \,,
   &
   \prn{1}{\gamma}_J &= 0
   \,,
\end{align}
i.e.\ colour-singlet DPDs are $\zeta$ independent.  The corresponding DGLAP
kernels $\prn{11}{P}_{a b}$ are identical to the DGLAP kernels for ordinary
PDFs.

The system of the evolution equations just presented has a unique solution for
given initial conditions at $\mu_{01}$, $\mu_{02}$, and $\zeta_0$, i.e.\ one
obtains the same result at any $\mu_1$, $\mu_2$, and $\zeta$ regardless of the
path of evolution in the space of these three variables.  This is readily shown
by verifying that the second derivatives commute for any pair of these variables
(see \app{A} in \cite{Diehl:2023cth} for the analogous argument for colour
singlet DPDs).
In practice, the evolution kernels and anomalous dimensions are truncated at
some order in $\alpha_s$.  To preserve the path independence of evolution, we
solve \eqref{DGLAP-zeta} and \eqref{CS-RGE} exactly for given perturbative
orders, i.e.\ we do not omit subleading orders in the solutions of the
differential equations.  In particular, we use the form \eqref{CS-solved} of the
Collins-Soper kernel for a fixed choice of $\mu_{01}$ and $\mu_{02}$, and we
take the same order of $\gamma_J$ in this expression and in \eqref{DGLAP-zeta}.


\paragraph{Comparison with TMD evolution.}  It is instructive to compare the
evolution equations for collinear DPDs with the ones for single-parton TMDs,
which can for instance be found in \cite{Collins:2011zzd, Collins:2014jpa}.
They read
\begin{align}
   \label{TMD-RGE}
   &
   \frac{d}{d \ln\mu}\, f_a(x, b; \mu, \zeta)
   =
   {}- \gamma_{K, a}(\mu) \;
     \ln\biggl( \frac{x \fsqrt{\zeta}}{\mu} \biggr) \;
     f_a(x, b; \mu, \zeta)
   + \gamma_a(\mu) \, f_a(x, b; \mu, \zeta)
\end{align}
and
\begin{align}
   \label{TMD-CS}
   & \frac{d}{d \ln\fsqrt{\zeta}}\, f_a(x, b; \mu, \zeta)
   =
   K_a(b ;\mu) \, f_a(x, b; \mu, \zeta)
   \,,
\end{align}
where the scale dependence of the Collins-Soper kernel is given by
\begin{align}
   \label{TMD-CS-RGE}
   \frac{d}{d \ln\mu}\, K_a(b ;\mu)
   &=
   {}- \gamma_{K, a}(\mu)
   \,.
\end{align}
The Collins-Soper equations \eqref{CS-equation} and \eqref{TMD-CS} for DPDs and
TMDs have the same structure, and the same holds for the scale dependence
\eqref{CS-RGE} and \eqref{TMD-CS-RGE} of the respective Collins-Soper kernels.
Correspondingly, the evolution equations \eqref{DGLAP-zeta} and \eqref{TMD-RGE}
for the distributions both have a term depending on their respective cusp
anomalous dimensions times a rapidity logarithm.  However, the evolution of DPDs
involves a Mellin convolution in the parton momentum fraction, whereas for TMDs
it is local in $x$.  We will return to this point in
\sects{\ref{sec:second-solution}} and~\ref{sec:exponentiation}.

Note that we use a rapidity parameter $\zeta$ defined with reference to the
proton momentum, so that the evolution equation \eqref{TMD-RGE} involves the
logarithm of $x \fsqrt{\zeta} / \mu$ rather than $\fsqrt{\zeta} / \mu$, in
analogy to the expression \eqref{DGLAP-zeta} for DPDs.  For TMDs, which evolve
without any cross talk between different $x$ values, one can easily trade
$\zeta$ for the rescaled variable $\fsqrt{\zeta}_{\text{resc}} = x\ms
\fsqrt{\zeta}$, which
refers directly to the parton momentum. This has become common in the TMD
literature since the book \cite{Collins:2011zzd} by Collins.\footnote{It is
amusing to note that $\zeta$ in the original work \cite{Collins:1981uk} of
Collins and Soper was defined with respect to the hadron momentum, as it is done
here.}
As discussed after \eqn{\eqref{DGLAP-zeta}}, the evolution equations for DPDs
have a simpler form if we do not use a rescaling of this kind.


\paragraph{Perturbative results.}  The colour dependent DGLAP kernels and the
anomalous dimensions $\gamma_J$  in \eqref{DGLAP-zeta} are known up to NLO.
In the following, we recall some basic features of the results, referring the
reader to \cite{Diehl:2022rxb} for the full expressions of the NLO results.

The cusp anomalous dimension has a perturbative expansion
\begin{align}
   \label{gamma-J-as}
   \prn{R}{\gamma}_J(\mu)
   &=
   \sum_{n=0}^{\infty} \,
   \biggl[ \frac{\alpha_s(\mu)}{2\pi} \biggr]^{n+1} \,
   \prn{R}{\gamma}_J^{(n)}
\end{align}
with lowest-order coefficients
\begin{align}
   \label{gamma-J-0}
   \prn{8}{\gamma}_J^{(0)} &= 2 C_A = 6
   \,,
   &
   \prn{10}{\gamma}_J^{(0)} &= 12
   \,,
   &
   \prn{27}{\gamma}_J^{(0)} &= 16
   \,.
\end{align}
Throughout this work, all colour factors are given for $N_c = 3$ colours;
expressions for generic $N_c$ can be found in \cite{Diehl:2022rxb}.
The perturbative expansion of the DGLAP kernels has the form
\begin{align}
   \label{P-as}
   \prb{R \Rp{}}{P}_{a b}(z; \mu)
   &=
   \sum_{n=0}^{\infty} \,
   \biggl[ \frac{\alpha_s(\mu)}{2\pi} \biggr]^{n+1} \,
   \prb{R \Rp{}}{P}_{a b}^{(n)}(z)
   \,.
\end{align}
At LO and NLO ($n=0$ and $n=1$), the kernels have the structure
\begin{align}
   \label{P-structure}
   \prb{R \Rp{}}{P}_{a b}^{(n)}(z)
   &=
   \frac{1}{2} \ms \delta_{R \Rpbar{}}\, \delta_{a b}
   \biggl[
      \prn{R}{d}_a^{\ms (n)}\, \delta(1-z)
      +
      \frac{\pr{R}{s}_a^{(n)}}{(1-z)_+}
   \biggr]
   + \prb{R \Rp{}}{P}_{a b,\ms \text{reg}}^{(n)}(z)
   \,,
\end{align}
where $1/(1-z)_+$ is the plus-distribution familiar from the ordinary DGLAP
kernels, and $P_{\text{reg}}$ are ordinary functions of $z$.  The coefficients
of the $\delta(1-z)$ terms can be written as
\begin{align}
   \label{delta-coeffs}
   \prn{R}{d}_a^{\ms (0)} &= \prn{1}{d}_a^{\ms (0)}
   \,,
   &
   \prn{R}{d}_a^{\ms (1)} &= \prn{1}{d}_a^{\ms (1)} + \prn{R}{c}^{(1)}
\end{align}
with $d_q^{\ms (0)} = 3 C_F$ and $d_g^{\ms (0)} = \beta_0$, where
\begin{align}
   \label{beta-0}
   \beta_0 &= 11 \ms C_A / 3 - 2 \ms n_f / 3
\end{align}
is the leading-order coefficient of the QCD $\beta$ function for $n_f$ active
quark flavours.  The colour singlet coefficients $\smash{\prn{1}{d}_a^{\ms
(n)}}$ are independent of the parton polarisation, and the coefficients
$\prn{R}{c}^{(1)}$ depend only on the dimension of $R$.
One finds that the cusp anomalous dimensions as well as $\prn{R}{c}^{(1)}$ obey
Casimir scaling
\begin{align}
   \label{gamma-J-Casimir}
   \frac{\prn{R}{\gamma}_J^{(0)}}{\prn{8}{\gamma}_J^{(0)}}
   &=
   \frac{\prn{R}{\gamma}_J^{(1)}}{\prn{8}{\gamma}_J^{(1)}}
   =
   \frac{\prn{R}{c}^{(1)}_{\phantom{J}}}{\prn{8}{c}^{(1)}_{\phantom{J}}}
   =
   \frac{C_R}{C_A}
   &
   \text{ for } R = 10, 27,
\end{align}
where
\begin{align}
   \label{higher-Casimirs}
   C_{10} &= 2 C_A = 6
   \,,
   &
   C_{27} &= 8
\end{align}
are the eigenvalues of the quadratic Casimir operators for the decuplet and
icosaheptaplet of $\text{SU}(3)$ \cite{Bali:2000un, Dokshitzer:2005ig}.

The one- and two-loop coefficients of the plus-distribution and the one-loop
part of the regular kernels have a simple multiplicative colour dependence
\begin{align}
   \label{LO-scaling}
   \prn{R}{s}_a^{(n)}
   &=
   c_{a a}^{}(R \Rbar)\; \prn{1}{s}_a^{(n)}
   \,,
   &
   \prn{1}{s}_q^{(n)} / C_F
   &=
   \prn{1}{s}_g^{(n)} / C_A
   &
   \text{ for } n=0,1
   \,,
   \notag \\
   \prb{R \Rp{}}{P}_{a b,\ms \text{reg}}^{(0)}(z)
   &=
   c_{a b}(R \Rp{})\; \pr{11}{P}_{a b,\ms \text{reg}}^{(0)}(z)
\end{align}
with proportionality factors
\begin{align}
   \label{P-factors-q}
   c_{q q}(88) = c_{\bar{q} \bar{q}}(88) &= - 1/8
   \,,
   \notag \\
   c_{q g}(8S) = c_{\bar{q} g}(8S) &= \sqrt{5}/4
   \,,
   &
   c_{q g}(8A) = {}- c_{\bar{q} g}(8A) &= 3/4
   \,,
   \notag \\
   c_{g q}(S8) = c_{g \bar{q}}(S8) &= \sqrt{5}/4
   \,,
   &
   c_{g q}(A8) = {}- c_{g \bar{q}}(A8) &= 3/4
\end{align}
and
\begin{align}
   \label{P-factors-g}
   c_{g g}(S S)
   &=
   c_{g g}(A A)
   = 1/2
   \,,
   &
   c_{g g}(\tentenbar)
   &=
   c_{g g}(\tenbarten)
   = 0
   \,,
   &
   c_{g g}(\twensev) &= -1/3
   \,.
\end{align}
Notice that $|c_{a b}(R \Rp{})| < 1$ for all colour non-singlet channels.

The two-loop functions $\prb{R \Rp{}}{P}_{a b,\ms \text{reg}}^{(1)}$ have a more
complicated structure, except in the decuplet sector, where they are zero.  One
therefore has
\begin{align}
   \label{decuplet-kernels}
   \pr{\smalltentenbar}{P}_{g g}^{(n)}(z)
   &=
   \pr{\smalltenbarten}{P}_{g g}^{(n)}(z)
   =
   \frac{1}{2} \, \prn{10}{d}_{g}^{(n)} \, \delta(1-z)
   &
   \text{ for } n=0,1
\end{align}
and identical expressions for polarised gluons.


\subsection{Heavy-flavour matching}
\label{sec:flavour-matching}

Our presentation so far was for a fixed number $n_f$ of active quark flavours.
For DPDs, one can have different active flavour numbers $n_{f 1}$ and $n_{f 2}$
for the two partons (just as one can have different scales $\mu_1$ and
$\mu_2$).  This can be implemented by specifying the active number of flavours
via the renormalisation prescription of the twist-two operators that define DPDs
and PDFs.

DPDs with different flavour numbers are related by matching equations, which
read
\begin{align}
   \label{DPD-matching}
   &
   \prn{R_1 R_2}{F}_{a_1 a_2}^{(n_{f 1} + 1, n_{f 2})}(x_1, x_2, {y};
      \mu_1, \mu_2, \zeta)
   \notag \\
   &\qquad =
   \sum_{b_1, \Rp{1}} \int_{x_1}^1 \frac{d z}{z} \,\,
   \prn{R_1^{} \Rpbar{1}}{A}_{a_1 b_1}^{n_{f 1}}\biggl(
      \frac{x_1}{z}, m_Q; \mu_1 \bigg) \,
   \prn{\Rp{1} R_2^{}}{F}_{b_1 a_2}^{(n_{f 1}, n_{f 2})}(
      z, x_2^{}, {y}; \mu_1, \mu_2, \zeta)
\end{align}
for the first parton and likewise for the second one.  The matching kernel has a
perturbative expansion
\begin{align}
   \label{A-as}
   \prb{R \Rp{}}{A}_{a b}^{n_f}(z, m_Q; \mu)
   &=
   \sum_{k=0}^{\infty} \,
   \Biggl[ \frac{\alpha_s^{(n_f + 1)}(\mu)}{2\pi} \Biggr]^{k} \;
      \prb{R \Rp{}}{A}_{a b}^{n_f (k)}(z, m_Q/\mu)
   \,,
\end{align}
where it is conventional to use the coupling $\alpha_s$ for $n_f + 1$ rather
than $n_f$ flavours on the r.h.s.  The tree-level matching coefficients are
\begin{align}
   \label{A-LO}
   \prb{R \Rp{}}{A}_{a b}^{n_f (0)}(z, m_Q/\mu)
   &=
   \begin{cases}
      \delta{}_{R \Rpbar{}} \, \delta_{a b}  \, \delta(1-z)
         & \text{if $a$ is one of the $n_f$ light flavours or a gluon,} \\
      0  & \text{otherwise.}
   \end{cases}
\end{align}
The one-loop coefficients for flavour matching of PDFs are well known, and it is
easy to generalise them to colour non-singlet channels in DPDs.  For the
transition from a gluon to a heavy quark, one has
\begin{align}
   \label{AQg1}
   \prb{R \Rp{}}{A}^{n_f (1)}_{Q g}(z, m_Q/\mu)
   &=
   \prb{R \Rp{}}{P}^{(0)}_{q g}(z) \, \log \frac{\mu^2}{m_Q^2}
   \hspace{1.05em}
   = \ms
   c_{q g}(R \Rp{}) \, \prb{11}{P}^{(0)}_{q g}(z) \, \log \frac{\mu^2}{m_Q^2}
   \,,
   \notag \\[0.2em]
   \prb{R \Rp{}}{A}^{n_f (1)}_{\Delta Q \ms \Delta g}(z, m_Q/\mu)
   &=
   \prb{R \Rp{}}{P}^{(0)}_{\Delta q \ms \Delta g}(z) \,
      \log \frac{\mu^2}{m_Q^2}
   = \ms
   c_{q g}(R \Rp{}) \, \prb{11}{P}^{(0)}_{\Delta q \ms \Delta g}(z) \,
      \log \frac{\mu^2}{m_Q^2}
\end{align}
for the combinations $(R R') = (1 1), (8 S)$ and $(8 A)$.  Corresponding
relations hold for antiquarks.  The one-loop matching coefficients for
gluon-to-gluon transitions originate from a heavy-quark loop on one gluon leg
and are therefore independent of how the colour of the two gluons is coupled:
\begin{align}
   \label{Agg1}
   \pr{R \Rbar{}}{A}^{n_f (1)}_{g g \phantom{\delta}}(z, m_Q/\mu)
   &=
   \pr{R \Rbar{}}{A}^{n_f (1)}_{\Delta g \ms \Delta g}(z, m_Q/\mu)
   =
   \pr{R \Rbar{}}{A}^{n_f (1)}_{\delta g \ms \delta g}(z, m_Q/\mu)
   \notag \\
   &=
   - \frac{1}{3} \ms \delta(1-z) \, \log \frac{\mu^2}{m_Q^2}
   \,.
\end{align}
In the above equations, $\Delta q$ and $\Delta g$ denotes longitudinal quark and
gluon polarisation, whereas $\delta g$ indicates linearly polarised gluons.  As
in the case of the DGLAP kernels, the coefficients $\prb{\SA}{A}_{g g}$ and
$\prb{\AS}{A}_{g g}$ are zero at all orders due to charge conjugation
invariance.

\section{Perturbative and non-perturbative input}
\label{sec:input}

At small values of the distance $y$, both DPDs and their Collins-Soper kernel
$J$ are constrained by perturbative dynamics, whereas at large $y$ they are
purely non-perturbative quantities.  In this section, we review the constraints
at small $y$ and present the model assumptions we make for large $y$ in the
numerical studies of this work.


\subsection{The Collins-Soper kernel}
\label{sec:cs-kernel}

At small $y$, the Collins-Soper kernel can be expanded in $\alpha_s$ as
\begin{align}
   \label{J-as}
   \pr{R}{J}(y; \mu, \mu)
   &=
   \sum_{n=0}^{\infty} \,
   \biggl[ \frac{\alpha_s(\mu)}{2\pi} \biggr]^{n+1} \,\,
   \sum_{k=0}^{n+1} \biggl[ \ln \frac{y \mu}{b_0} \biggr]^{k} \,
      \pr{R}{J}^{(n,k)}
   \,,
\end{align}
where the highest power of the logarithm is dictated by the renormalisation
group equation \eqref{CS-RGE} and its analogue for the $\mu_2$ dependence.  The
LO coefficients read \cite{Buffing:2017mqm}
\begin{align}
   \label{CS-LO-coeffs}
   \pr{R}{J}^{(0, 0)}
   &= 0
   \,,
   &
   \pr{R}{J}^{(0, 1)}
   &=
   - 2\; \prn{R}{\gamma}_J^{(0)}
   \,.
\end{align}
The kernel for the colour octet channel is actually known up to order
$\alpha_s^3$, thanks to the relation \cite{Vladimirov:2016qkd}
\begin{align}
   \label{J-K-equal}
   \prn{8}{J}(y; \mu, \mu) &= K_g(y; \mu)
   \,,
\end{align}
which connects the colour-octet kernel for DPDs with the Collins-Soper kernel
$K_g$ for gluon TMDs.  This relation follows from an identity between the
Wilson-line operators defining the corresponding soft factors and is therefore
exact.  The two- and three-loop expressions for $K_g$ can for instance be found
in \app{D} of reference \cite{Echevarria:2016scs}, using the relation $K_g(b_T;
\mu) = - 2 \ms \mathcal{D}^g(\mu, b_T)$ between the notation here and there.

At large $y$, the kernel is a non-perturbative quantity, and to obtain a smooth
transition between small and large $y$, we adapt the so-called $b^*$
prescription, which has been devised by Collins and Soper for
transverse-momentum dependent distributions \cite{Collins:1981va}:
\begin{align}
   \label{J-star-prescription}
   \prb{R}{J}(y; \mustar,\mustar)
   &=
   \prb{R}{J}(y^*; \mustar,\mustar)
   + \prn{R}{\Delta J}(y)
\end{align}
with a function $y^*(y)$ that tends to $y$ at small values of $y$, whereas at
large $y$ it saturates at a value that is small enough for the associated scale
\begin{align}
   \label{mu-star-def}
   \mustar &= b_0 / y^*(y)
\end{align}
to remain in the region where perturbation theory can be used.  In our study, we
choose
\begin{align}
   \label{star-choice}
   y^*(y) &= \frac{y}{\sqrt[4]{1 + y^4 / y_{\text{max}}^4}}
   \,,
\end{align}
which leads to a minimal value $\mu_{\text{min}} = b_0 / y_{\text{max}}$ of the
scale $\mustar$.  We take
\begin{align}
   \label{mu-min}
   \mu_{\text{min}} &= 2 \gev
   \,.
\end{align}
Compared with the more common form that has a square root and second instead of
fourth powers, the function \eqref{star-choice} tends more rapidly to $y$ at
small distances.

The first term on the r.h.s.\ of \eqref{J-star-prescription} can be evaluated
reliably in fixed-order perturbation theory, as it contains no large logarithms.
The function $\Delta J(y)$ tends to zero for small $y$ and controls the
behaviour of $J$ at large distances.  As follows from the renormalisation group
equation \eqref{CS-RGE}, it is scale independent (but it depends of course on
the choice of the function $y^*(y)$).

To evaluate for arbitrary scales $\mu_1, \mu_2$, we use the solution of the
renormalisation group equation
\begin{align}
   \label{J-evolved-form}
   \prb{R}{J}(y; \mu_1, \mu_2)
   &=
   \prb{R}{J}(y; \mustar,\mustar)
   - \int_{\mustar}^{\mu_1} \frac{d\mu}{\mu}\, \gamma_J(\mu)
   - \int_{\mustar}^{\mu_2} \frac{d\mu}{\mu}\, \gamma_J(\mu)
   \,.
\end{align}

The above construction is valid for given numbers $n_{f 1}$, $n_{f 2}$ of active
quark flavours.  To switch between different values of these numbers, one can
use a matching equation, which has the generic form
\begin{align}
   \label{J-matching}
   \prb{R}{J}^{(n_{f 1} + 1, n_{f 2})}(y; \mu_1, \mu_2)
   &=
   \prb{R}{J}^{(n_{f 1}, n_{f 2})}(y; \mu_1, \mu_2)
   +
   \mathcal{O}(\alpha_s^2)
\end{align}
and likewise for a change in $n_{f 2}$.  In analogy to the DPD case discussed
in \sect{\ref{sec:flavour-matching}}, separate values of $n_{f 1}$ and $n_{f 2}$
can
be introduced by the renormalisation prescription for the Wilson lines operators
defining the soft factor and its associated Collins-Soper kernel (see \sect{3.2}
of \cite{Buffing:2017mqm} for the explicit form of these operators).
The first graphs that contribute to a non-trivial matching relation appear at
order $\alpha_s^2$, where one of the gluons connecting different Wilson lines in
the soft factor can have a heavy-quark loop insertion.\footnote{For a discussion
of mass effects in the two-loop soft factor for TMDs, we refer to \sect{5} in
\cite{Pietrulewicz:2017gxc}.}
Since heavy-quark effects are not at the centre of our present investigations,
and since most of our study uses $J$ computed at order $\alpha_s$, we will not
further investigate the $\alpha_s^2$ terms in \eqref{J-matching} here.


\paragraph{Model for the behaviour at large distances.}
Starting point of our model for large $y$ is the exact relation
\eqref{J-K-equal} between $\pr{8}{J}$ and the Collins-Soper kernel for gluon
TMDs.
Motivated by the observation that Casimir scaling holds for $\pr{R}{J}$ at order
$\alpha_s$ and for the anomalous dimension $\pr{R}{\gamma}_{J}$ up to order
$\alpha_s^2$, see \eqs{\eqref{CS-LO-coeffs}} and \eqref{gamma-J-Casimir},
we \emph{assume} Casimir scaling also in the non-perturbative region, i.e.\
\begin{align}
   \label{J-Casimir}
   \prn{R}{\Delta J}(y) \, \big/ \, \prn{8}{\Delta J}(y)
   &=
   C_{R} \, \big/ \, C_{A}
   &
   \text{ for } R = 10, 27,
\end{align}
with $C_R$ given in \eqref{higher-Casimirs}.

Since the Collins-Soper kernel for gluon TMDs is not well known at large
distances, we connect it to the one for quarks by \emph{assuming} Casimir
scaling
\begin{align}
   \label{K-Casimir}
   K_g(b, \mu) &= \frac{C_A}{C_F}\, K_q(b, \mu)
   \,,
\end{align}
which holds in perturbation theory up to $\mathcal{O}(\alpha_s^3)$ (see e.g.\
\cite{Li:2016ctv} and \app{D} in \cite{Echevarria:2016scs}).\footnote{Starting
at $\mathcal{O}(\alpha_s^4)$ Casimir scaling is broken for the cusp anomalous
dimension \cite{Moch:2017uml} and therefore also for the Collins-Soper kernel.}

The Collins-Soper kernels for quark TMDs has been extensively studied in recent
years, both in fits to experimental data and in lattice computations.  The fits
we consider in the following all use the $b^*$ prescription and write
\begin{align}
   \label{K-star-prescription}
   K_q(b, \mu) &= K_q\bigl( b^*(b), \mu \bigr) + \Delta K_q(b)
\end{align}
where the first term on the r.h.s.\ is evaluated in perturbation theory and the
second one is parameterised and determined by a fit.

Combining \eqref{J-star-prescription} with \eqref{J-K-equal}, \eqref{K-Casimir},
and \eqref{K-star-prescription}, we obtain
\begin{align}
   \label{Delta-J-model}
   \pr{8}{\Delta J}(y)
   &=
   \biggl[\ms
      \frac{C_A}{C_F} \, K_q\bigl( b^*(y), \mustar \bigr)
      - \prb{8}{J}\bigl( y^*(y); \mustar, \mustar \bigr)
   \ms\biggr]
   + \frac{C_A}{C_F} \, \Delta K_q(b)
   \,,
\end{align}
where $y^*(y)$ is our choice \eqref{star-choice} and $b^*(y)$ the function used
in the different determinations of $K_q$.  For simplicity, we evaluate both
terms in square brackets at the same order in $\alpha_s$, using the strong
coupling from \eqn{\eqref{as-lo}} or \eqref{as-nlo}.\footnote{For a more precise
determination of $K_q$, one should take the perturbative order and the
$\alpha_s$ value used in the respective fits for the first term in
\protect\eqref{Delta-J-model}.  Given the inherent uncertainties due to the
Casimir scaling assumption in our construction of $\Delta J$, we consider our
simplified procedure to be sufficient here.}

Whilst the behaviour of $K_q(b, \mu)$ is relatively well known at intermediate
values of $b$, its limiting behaviour for $b\to \infty$ cannot be determined
from phenomenology or the lattice.  We have therefore selected a number of fits
from recent phenomenological determinations \cite{Scimemi:2019cmh,
Hautmann:2020cyp, Bacchetta:2022awv, Moos:2023yfa} that together span a wide
range in this limiting behaviour, from a constant limit that was argued for in
\cite{Collins:2014jpa} to a quadratic growth in $b$ that has been widely used in
the older literature.
The fits of SV19 \cite{Scimemi:2019cmh}, MAP22 \cite{Bacchetta:2022awv}, and
ART23 \cite{Moos:2023yfa} agree reasonably well with recent lattice
determinations, which cover a range of $b$ up to $0.84 \fm$, see \fig{3} in
\cite{Shu:2023cot}, \fig{13} in \cite{LatticePartonLPC:2023pdv}, and \fig{15} in
\cite{Avkhadiev:2023poz}.

\begin{table}
\centering
\renewcommand{\arraystretch}{1.3}
\begin{tabular}{l l l l l l}
\hline
key & $\Delta K_q(b)$ & $c_0$      & $b_{\text{max}}$ & reference \\
    &                 & $[\gev^2]$ & $[\gev^{-1}]$    &           \\
\hline
SV19 (N$^3$LO) & $- 2 c_0 \ms b \, b^*\bs(b)$ & $0.0427$ & $1.93$
               & \cite[eqs.~(2.88), (2.89) and tab.~9]{Scimemi:2019cmh} \\
HSV20 case 2 & $- 2 c_0 \ms b^2$ & $0.032$ & $2.2$
             & \cite[eqs.~(5), (6), (7) and tab.~3]{Hautmann:2020cyp} \\
HSV20 case 4 & $- 2 c_0 \ms b \, b^*\bs(b)$ & $0.04$ & $3.0$
             & \cite[eqs.~(5), (6), (8) and tab.~3]{Hautmann:2020cyp} \\
HSV20 case 6 & $- 2 c_0 \ms \bigl[ b^*\bs(b) \bigr]^2$ & $0.05$ & $2.43$
             & \cite[eqs.~(5), (6), (9) and tab.~3]{Hautmann:2020cyp} \\
MAP22 & $- 2 c_0 \ms b^2$ & $0.0154$ & $1.123$
           & \cite[eqs.~(33), (34), (36) and tab.~V]{Bacchetta:2022awv} \\
ART23 & \multicolumn{2}{c}{see eq.~\protect\eqref{msvz-form} here} & $1.56$
       & \cite[eqs.~(2.33), (2.34), (5.1)]{Moos:2023yfa} \\
\hline
\end{tabular}
\caption{\label{tab:Delta_Kq} Functional forms and parameters for the
non-perturbative part $\Delta K_q(b)$ of the Collins-Soper kernel for quark TMDs
in recent phenomenological extractions.  The associated functions $b^*(b)$ are
given in \protect\eqref{std-bstar} and \protect\eqref{bstar-pavia}.  The value
of $c_0$ for MAP22 has been rounded in the table; we use its exact form $c_0 =
(0.124 \gev)^2$ in our numerical calculation.}
\end{table}

In \tab{\ref{tab:Delta_Kq}} we list the functional form and the parameters of
$\Delta K_q(b)$ for our selected fits, except for the more lengthy form used by
ART23, which reads
\begin{align}
   \label{msvz-form}
      \Delta K_q(b )
   &=
   - 2 \ms b \, b^*\bs(b)
   \biggl[ c_0 + c_1 \ln\dfrac{b^*\bs(b)}{b_{\text{max}}} \ms\biggr]
   &
   \text{ with $c_0 = 0.0369 \gev^2$, $c_1 = 0.0582 \gev^2$.}
\end{align}
The associated form of $b^*(b)$ is
\begin{align}
   \label{std-bstar}
   b^*(b) &= b \, / \sqrt{1 + b^2 / b_{\text{max}}^2}
\end{align}
for SV19, HSV20, and ART23, whilst MAP22 use
\begin{align}
   \label{bstar-pavia}
   b^*(b)
   &=
   b_{\text{max}} \,
   \sqrt[4]{1 - \exp\bigl(- b^4 / b_{\text{max}}^4 \bigr)}
   \,,
\end{align}
where we have set $b_{\text{min}} = 0$ in their \eqn{(33)}.

\begin{figure}
\vspace{1em}
\centering
\subfloat[$\mu = 2 \gev$]{
   \includegraphics[width=0.48\textwidth,trim=0 0 30
34,clip]{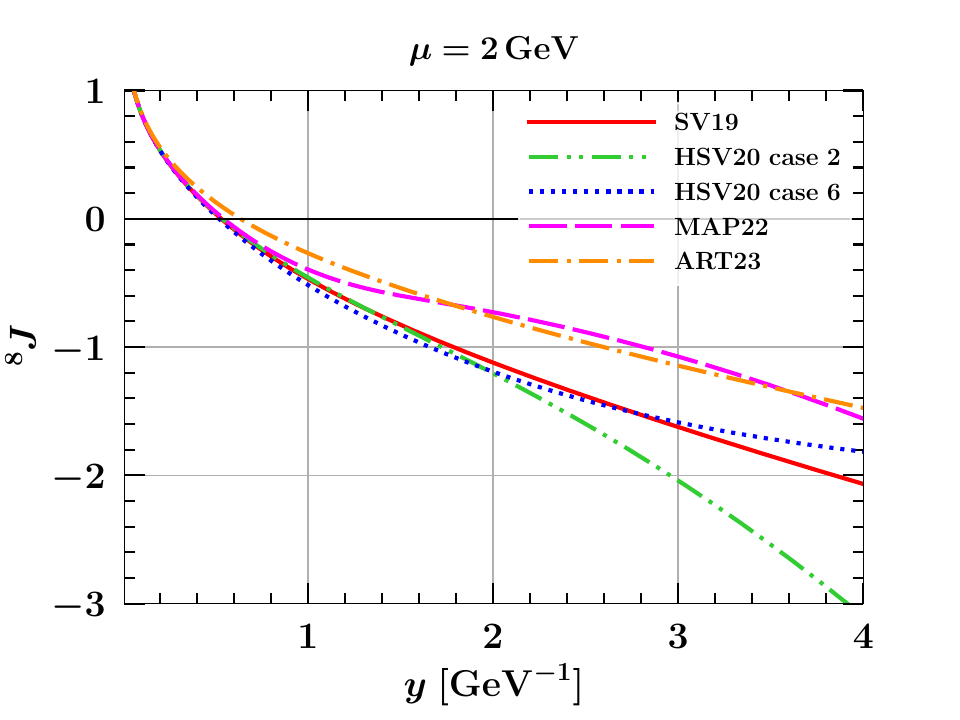}
}
\subfloat[$\mu = \mustar$]{
   \includegraphics[width=0.48\textwidth,trim=0 0 30
34,clip]{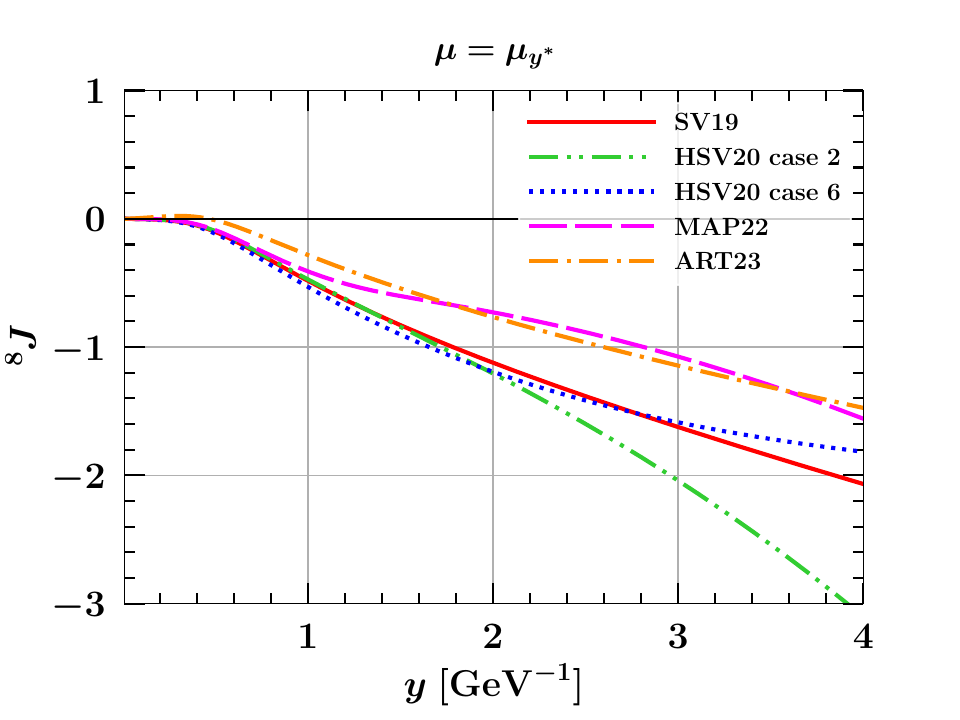}
}
\caption{\label{fig:CS-kernel} The Collins Soper kernel $\pr{8}{J}(y; \mu, \mu)$
for colour octet DPDs, obtained with our model prescription
\protect\eqref{Delta-J-model} from different fits of $K_q$.  The anomalous
dimension $\gamma_J$ and the perturbative part of $J$ have been evaluated at order $\alpha_s$.  With the assumption \protect\eqref{J-Casimir} of Casimir scaling, the kernels for $R=10$ and $R=27$ are obtained by multiplication with $2$ and $8/3$, respectively.}
\end{figure}

The result of our procedure for the Collins-Soper kernel $\pr{8}{J}(y; \mu,
\mu)$ is shown in \fig{\ref{fig:CS-kernel}}, with the exception of the curves
for HSV20 case 4, which are barely distinguishable from the ones for SV19.  In
panel (b) we see that at the $y$ dependent scale $\mu_{y*}$ entering in
\eqref{J-evolved-form}, the kernel is negative in most of the $y$ range, except
for an excursion to very small positive values at low $y$ in the case of ART23.

\subsection{Initial conditions for DPD evolution}
\label{sec:init-cond}

In the limit of small $y$, the dominant contribution to a DPD comes from the
perturbative splitting of a single parton $a_0$ into the two observed ones,
$a_1$ and $a_2$, plus possibly more partons at higher orders.  At lowest order
this gives \cite{Diehl:2011yj}
{\allowdisplaybreaks[0]
\begin{align}
   \label{splitting-DPD-pt}
   &
   \prn{R_1 R_2}{F}_{a_1 a_2}^{\text{spl, pt}}(x_1, x_2, {y};
      \mu, \mu, \zeta)
   \notag \\[0.2em]
   & \qquad =
   k_{a_1 a_2}(R_1 R_2) \,
   \frac{1}{\pi y^2} \;
   \frac{\alpha_s(\mu)}{2\pi} \,
   V_{a_1 a_2, a_0}\biggl( \frac{x_1}{x_1 + x_2} \biggr)
   \frac{f_{a_0}(x_1 + x_2; \mu)}{x_1 + x_2}
   \,,
\end{align}
}
where $f_{a_0}(x; \mu)$ denotes the PDF for parton $a_0$.  The DPD splitting
kernel $V_{a_1 a_2, a_0}(z)$ is obtained from the LO DGLAP kernel
$\prn{11}{P}_{a_1 a_0}^{(0)}(z)$ by removing the $\delta(1-z)$ terms and the
plus-prescription for $1/(1-z)$.

The colour factors $k$ are by definition unity for the singlet,
\begin{align}
   \label{splitting-singlet-factor}
   k_{a_1 a_2}(11) = 1
   \,,
\end{align}
and for colour non-singlet combinations they read
\begin{align}
   \label{splitting-colour-factors}
   k_{q \bar{q}}(88) &= - 1 / \sqrt{8}
   \,,
   \notag \\
   k_{q g}(8S) = k_{\bar{q} g}(8S) &= \sqrt{5/2}
   \,,
   &
   k_{q g}(8A) = - k_{\bar{q} g}(8A) &= - 3/\sqrt{2}
\end{align}
and
\begin{align}
   \label{splitting-colour-factors-gg}
   k_{g g}(S S) &= \sqrt{2}
   \,,
   &
   k_{g g}(A A) &= - \sqrt{2}
   \,,
   &
   k_{g g}(S A)
   =
   k_{g g}(A S) &= 0
   \,,
   \notag \\
   k_{g g}(\twensev) &= - \sqrt{3}
   \,,
   &
   &
   &
   k_{g g}(\tentenbar)
   =
   k_{g g}(\tenbarten)
   &= 0
   \,.
\end{align}
Notice that for most combinations one has $|k_{a_1 a_2}| > 1$, in contrast to
the colour factors $c_{a b}$ for the LO DGLAP kernels in \eqref{P-factors-q} and
\eqref{P-factors-g}.
The $\alpha_s^2$ corrections to \eqref{splitting-DPD-pt} have been computed in
\cite{Diehl:2019rdh, Diehl:2021wpp} for unpolarised partons.  We do not use
these in the present work, where our focus is the behaviour of DPDs under scale
evolution rather than their initial conditions.

In addition to the splitting contribution, there is an ``intrinsic''
contribution to the DPD, which lacks the $1/y^2$ enhancement at small $y$ but
starts at order $\alpha_s^0$ rather than $\alpha_s^{}$ \cite{Diehl:2017kgu}.  As
a model ansatz, we write the full DPD as
\begin{align}
   \label{DPD-sum}
   F &= F^{\text{spl}} + F^{\text{intr}}
\end{align}
in the full $y$ range.
The splitting part $F^{\text{spl}}$ is an extension of the perturbative form
\eqref{splitting-DPD-pt} to all $y$:
\begin{align}
   \label{splitting-DPD}
   &
   \prn{R_1 R_2}{F}_{a_1 a_2}^{\text{spl}}(x_1, x_2, {y};
      \mustar, \mustar, \zeta_{y^*})
   =
   \exp \biggl[ \frac{-y^2}{4 h_{a_1 a_2}} \biggr] \;
   \prn{R_1 R_2}{F}_{a_1 a_2}^{\text{spl, pt}}(x_1, x_2, {y};
      \mustar, \mustar, \zeta_{y^*})
\end{align}
with a Gaussian factor that should give a more realistic behaviour at large
$y$ while not significantly affecting the perturbative region.
To avoid large logarithmic corrections from higher-order corrections, the
perturbative expression ${F}^{\text{spl, pt}}$ is evaluated at the scale
$\mustar$ defined in \eqref{mu-star-def} and at the associated rapidity
parameter
\begin{align}
   \label{zeta-star-def}
   \zeta_{y^*} &= \mustar^2 / (x_1 x_2)
   \,.
\end{align}
Higher-order corrections contain logarithms of $\zeta$ in the combination $(x_1
x_2 \ms \zeta / \mu^2)$, which can be defined in terms of the parton momenta
rather that the proton momentum (on which the DPD splitting kernels do not
depend).  This explains the factor $1/(x_1 x_2)$ in our choice
\eqref{zeta-star-def}.


\paragraph{Intrinsic part.}  For the colour singlet channel, the intrinsic
contribution has long been studied in the literature.  We take an ansatz that
was developed in \cite{Diehl:2020xyg} in an effort to approximately satisfy the
number and momentum sum rules for DPDs \cite{Gaunt:2009re, Diehl:2018kgr}.  As
is common, it is constructed from the product of two PDFs with additional
factors, namely
\begin{align}
   \label{int-DPD-singlet}
   &
   \prn{11}{F}_{a_1 a_2}^{\text{intr}}(x_1, x_2, {y};
      \mustar, \mustar)
   \notag \\
   & \qquad =
   n_{a_1 a_2} \;
   \frac{1}{4 \pi h_{a_1 a_2}}
   \exp \biggl[ \frac{-y^2}{4 h_{a_1 a_2}} \biggr] \;
   \frac{(1 - x_1 - x_2)^2}{(1 - x_1)^2 \ms (1 - x_2)^2} \;
   f_{a_1}(x_1; \mustar) \, f_{a_2}(x_2; \mustar)
\end{align}
with
\begin{align}
   n_{a_1 a_2}
   &=
   \begin{cases}
      0   & \text{ if } a_1 = a_2 = d - \bar{d}, \\
      1/2 & \text{ if } a_1 = a_2 = u - \bar{u}, \\
      1   & \text{ otherwise. }
   \end{cases}
\end{align}
Here the parton labels $a_1$ and $a_2$ have been transformed to the valence-sea
basis, i.e.\ to the linear combinations $q - \bar{q}$ and $\bar{q}$ for each
quark flavour.  The Gaussian parameters are taken as
\begin{align}
   \label{damping-parameters}
   h_{g g} = 4.66 \gev^{-2}
   \,,
   && h_{g q} = h_{q g} = 5.86 \gev^{-2}
   \,,
   && h_{q q} = 7.06 \gev^{-2}
\end{align}
with equal values for quarks and antiquarks (and hence for valence and sea
quarks).  A motivation of these values is given in \sect{9.2.1} of
\cite{Diehl:2017kgu}.  For simplicity and lack of better guidance, the same
parameters are used in the splitting part \eqref{splitting-DPD}.

In contrast to \cite{Diehl:2018kgr} and other studies, we make the product
ansatz \eqref{int-DPD-singlet} at the natural scale $\mustar$ associated with
the distance $y$, rather than at a fixed scale $\mu_0$.  This makes no
difference at large $y$, where $\mustar$ approaches a constant value. Even for
small $y$ there is little difference between the two choices unless the momentum
fractions are close to the kinematic limit $x_1 + x_2 = 1$.  Away from this
limit, a colour singlet DPD given by the product of two PDFs at an input scale
is approximately equal to the product of the evolved PDFs at any other scale
\cite{Diehl:2014vaa}.  The reason is of course that the DGLAP equations for
colour singlet DPDs and for PDFs have the same form.  This is however not the
case if one extends the product ansatz to colour non-singlet DPDs, as we will do
now.

Unfortunately, next to nothing is known about the intrinsic part of colour
non-singlet DPDs.  As a guide for constructing a model ansatz, we will use the
positivity bounds in colour space, which have been derived in
\cite{Kasemets:2014yna} and which may be regarded as an extension to colour
degrees of freedom of the Soffer bound \cite{Soffer:1994ww} for polarised parton
densities.

To obtain the bounds in colour space, one couples the colour indices
of partons $1$ and $2$ to irreducible representations $R$ in the amplitude
and $\Rp{}$ in its complex conjugate.  This corresponds to coupling the index
pairs $(r_1^{} r_2^{})$ and $(r_1', r_2')$ for the DPD shown at the bottom of
\fig{\ref{fig:xsect-graph}}.  We refer to this as the $s$-channel
basis and denote the resulting distributions by $F^{R \Rp{}\!}$, as opposed to
the DPDs $\pr{R_1 R_2}{F}$ in the $t$-channel basis discussed so far.  The
distributions $F^{R \Rbar{}}$ can be interpreted as number densities, which must
be non-negative.  Bounds on $\pr{R_1 R_2}{F}$ are then obtained by a linear
transformation between the two colour bases (see \app{\ref{sec:colour-proj}}).

As in the familiar case of PDFs, the density interpretation of $s$-channel DPDs
is invalidated by the renormalisation of the twist-two operators in their
definition.  Moreover, LO evolution of the DPD to higher scales does \emph{not}
conserve positivity \cite{Diehl:2021wvd}, in contrast to the Soffer bound for
PDFs.  The colour space bounds therefore do not provide strict constraints on
DPDs.  For lack of better guidance, we nevertheless construct an ansatz in which
these bounds are saturated, with the understanding that this corresponds to the
case where colour correlation effects are large (although not maximal, as would
be the case if the bounds were rigorous).

The positivity bounds are saturated \rev{if at least one $s$-channel DPD is
zero.  In the following, we consider a simple subset of these configurations,
namely the cases where exactly one $s$-channel DPD is nonzero.  One such case is
in fact realised in} the LO splitting contribution \eqref{splitting-DPD-pt},
where the parton pair $(a_1 a_2)$ is in the same
colour representation as the parton $a_0$ from which it originates.  Assuming
the same colour structure for the intrinsic part of the DPD, we obtain our
default ansatz
\begin{align}
   \label{int-DPD-colour}
   \prn{R_1 R_2}{F}_{a_1 a_2}^{\text{intr}}(x_1, x_2, {y};
      \mustar, \mustar, \zeta_{y^*})
   &=
   k_{a_1 a_2}(R_1 R_2) \,
   \prn{11}{F}_{a_1 a_2}^{\text{intr}}(x_1, x_2, {y};
      \mustar, \mustar)
\end{align}
with the colour factors given in \eqref{splitting-colour-factors} and
\eqref{splitting-colour-factors-gg}.  To complete the model, we must specify
the corresponding factors for channels with two quarks or two antiquarks, which
are absent in the LO splitting formula.  We take
\begin{align}
   \label{extra-colour-factors}
   k_{q q}(88)
   &=
   k_{\bar{q} \bar{q}}(88)
   =
   - \sqrt{2}
   \,,
\end{align}
which corresponds to a $q q$ pair coupled to an anti-triplet or a $\bar{q}
\bar{q}$ pair coupled to a triplet in the $s$-channel.  Note that for two
quarks this corresponds to the colour coupling in a three-quark wave
function, where the only colour structure is $\epsilon_{i j k}\, u_i u_j d_k$.
We use the colour factors $k_{q \bar{q}}$ and $k_{q q}$ for channels with equal
or different flavours of the two partons.

Since this ansatz yields vanishing decuplet distributions, we consider as an
alternative the case in which two gluons are coupled to a colour singlet in the
$s$ channel.  The corresponding colour factors are
\begin{align}
   \label{alt-colour-factors}
   k_{g g}(S S) &= \sqrt{8}
   \,,
   &
   k_{g g}(A A) &= - \sqrt{8}
   \,,
   &
   k_{g g}(A S)
   =
   k_{g g}(S A)
   &=
   0
   \,,
   \notag \\
   k_{g g}(\twensev) &= \sqrt{27}
   \,,
   &
   &
   &
   k_{g g}(\tentenbar)
   =
   k_{g g}(\tenbarten)
   &=
   \sqrt{10}
\end{align}
instead of those in \eqref{splitting-colour-factors-gg}.

\begin{figure}
\centering
\subfloat[]{
   \includegraphics[width=0.48\textwidth]{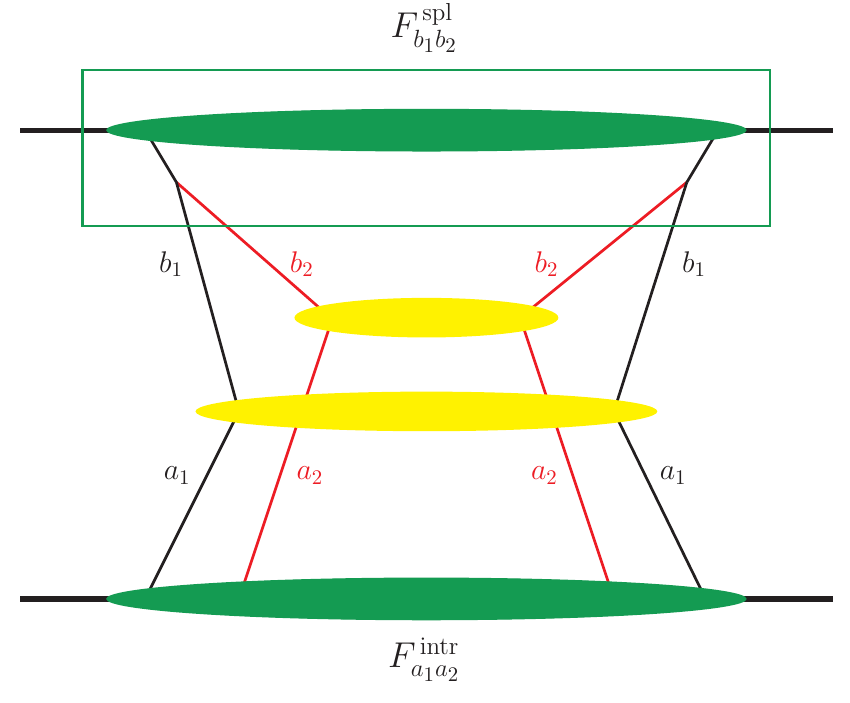}
}
\subfloat[]{
   \includegraphics[width=0.48\textwidth]{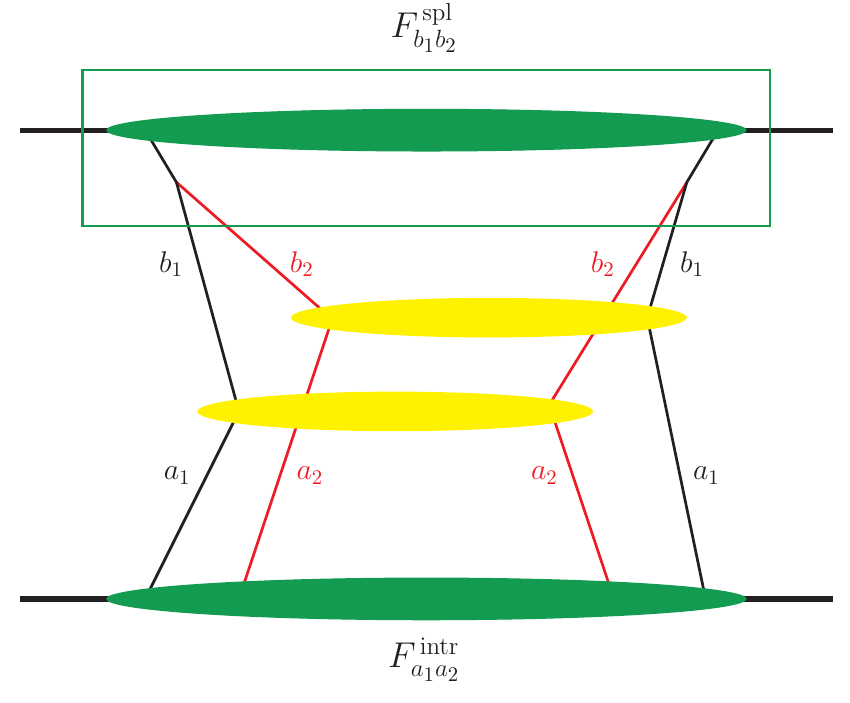}
}
\caption{\label{fig:std-int} (a) A graph for the DPDs cross section.  (b) An
``interference graph'' as invoked in reference \protect\cite{Blok:2022mtv}.  The
upper box in each graph denotes the splitting part of the DPD at order
$\alpha_s$, given by the expression \protect\eqref{splitting-DPD-pt}.}
\end{figure}

The necessity to model the intrinsic part of colour non-singlet DPDs presents a
significant challenge for phenomenological predictions.  The  work in
\cite{Blok:2022mtv} avoids this problems by invoking an ``interference term'' in
which the connection of parton legs with the hard-scattering processes is
swapped on the r.h.s.\ of the final state cut, as shown in
\fig{\ref{fig:std-int}}b (see \figs{2} and 3 of reference \cite{Blok:2022mtv}).
In such a term one can indeed have the partons ($b_1$ and $b_2$) in the
splitting DPD coupled to colour octets whilst the partons ($a_1$ and $a_2$) in
the intrinsic DPD are coupled to colour singlets.
However, the assignment of parton legs to DPDs in \fig{\ref{fig:std-int}}b
is incorrect, even if one takes identical partons $a_1=a_2$ with identical
momentum fractions $x_1=x_2$ and scales $\mu_1=\mu_2$ as done in
\cite{Blok:2022mtv}.  This assignment must match the flow of transverse
parton momenta in the corresponding graphs, as can be seen in the derivation of
the DPS factorisation formula (see for instance \sect{2.1.2} of
\cite{Diehl:2011yj}).  In particular, the DGLAP equations for DPDs apply to the
pairs of parton legs that enter one of the two hard scatters, and not to the
pairs with equal parton labels in \fig{\ref{fig:std-int}}b.  The correct
assignment of parton legs to DPDs can always be represented as in
\fig{\ref{fig:std-int}}a.


\paragraph{PDF input.}  To evaluate the splitting and product forms
\eqref{splitting-DPD-pt} and \eqref{int-DPD-singlet}, one needs PDFs as input.
We use the default LO or NLO sets of the MSHT20 fit \cite{Bailey:2020ooq},
depending on the order at which we evolve the DPDs.  The strong coupling used in
these sets is
\begin{align}
   \label{as-lo}
   \smash{\alpha_s^{(n_f=5)}(m_Z)} &= 0.13
   &&
   \text { at LO}
   \intertext{and}
   \label{as-nlo}
   \smash{\alpha_s^{(n_f=5)}(m_Z)} &= 0.118
   &&
   \text{ at NLO.}
\end{align}
The PDF values are obtained with the LHAPDF interface \cite{Buckley:2014ana},
where the respective sets have the keys \texttt{MSHT20lo\_as130} and
\texttt{MSHT20nlo\_as118}.


\paragraph{Number of active quark flavours.}  The initial conditions for
$F^{\text{spl}}$ and $F^{\text{intr}}$ should be formulated with an active
number $n_f$ of flavours that is appropriate for the initial scale $\mustar$.
In our present work, we use the simplified setting that was discussed in
\cite{Diehl:2022dia} under the name of ``massless scheme''.  For $b_0 / y <
m_b$, we initialise the DPDs with $n_f = 4$ flavours and match them to $n_f = 5$
flavours for each parton at the scale $\mu_i = m_b$.  For $b_0 / y \ge m_b$, the
DPDs are initialised with $n_f = 5$ flavours.  The $b$ quark mass is taken as
$m_b = 4.75 \gev$ in accordance with the above PDF sets.  Note that we do not
invoke distributions with $n_f = 3$ flavours because the smallest scale
$\mu_{\text{min}}$ in our construction is above the charm quark mass.

\section{Semi-analytic solutions of the evolution equations}
\label{sec:solutions}

The system \eqref{TMD-RGE} to \eqref{TMD-CS-RGE} of equations for the evolution
of TMDs in $\mu$ and $\zeta$ has a simple and well-known analytic solution.  In
particular, the term proportional to $\gamma_{K, a}$ in \eqref{TMD-RGE} gives
rise to Sudakov double logarithms that are exponentiated in the solution of the
equations.

The evolution equations for DPDs cannot be solved in the same way, because they
contain Mellin convolutions in $x_1$ and $x_2$.  In this section we present two
semi-analytic solutions, in which the Sudakov double logarithms are
exponentiated explicitly and a ``reduced'' system of integro-differential
equations remains to be solved.  The first version is suitable for a numerical
implementation, whereas the second one allows us to understand qualitative
features of evolution for colour non-singlet DPDs.
In subsection \ref{sec:orders} we discuss how to combine the perturbative orders
of the different ingredients in the DPS cross section.
An approximate analytic solution to evolution can be obtained near the
kinematic limit $x_1 + x_2 = 1$ of the momentum fractions.  We consider this to
be of more theoretical than phenomenological interest and hence relegate its
discussion to \app{\ref{sec:large-x}}.

We argued in \sect{\ref{sec:init-cond}} that initial conditions for DPD
evolution are more naturally taken at fixed $x_1 x_2 \ms \zeta$ rather than
fixed $\zeta$.  In the course of this section, we will see that this rescaled
rapidity parameter is also advantageous for discussing evolved DPDs.  We
therefore introduce a ``reduced'' rapidity parameter $\xi$ as
\begin{align}
  \label{xi-def}
  \zeta &= \xi /(x_1 x_2)
  \,,
\end{align}
where it is understood that $\xi$ is independent of $x_1$ and $x_2$.

For the sake of legibility, representation labels $R_1$, $R_2$, \ldots and the
corresponding sums are suppressed throughout this section.


\subsection{First representation}
\label{sec:first-solution}

In \cite{Diehl:2023cth}, a numerical method for the evolution of colour singlet
DPDs was presented, together with its implementation in the C++ library \chili.
A crucial part of the method is the use of Chebyshev interpolation on grids in
$x_1$, $x_2$, and $y$.  The Mellin convolution in the DGLAP equations is turned
into the multiplication of a matrix constructed from the splitting function
$P(x/z)$ with the vector of the discretised distribution $F(z,\ldots)$. The
resulting coupled system of linear differential equations in $\mu_1$ or $\mu_2$
is solved by a higher-order Runge-Kutta method.

For the present work, we generalised this implementation to the evolution of
colour non-singlet DPDs.  To this end, we write the DPDs in the form
\begin{align}
   \label{solution-1}
   &
   F_{a_1 a_2}\bigl( x_1,x_2,{y};\mu_1,\mu_2, \xi /(x_1 x_2) \bigr)
   \nonumber \\[0.5em]
   &\quad
   = \exp\,\biggl[- \frac{1}{2}\ms J({y}; \mu_1, \mu_2) \ln (x_1 x_2)
      \,\biggr]\,
   F_{a_1 a_2}\bigl( x_1,x_2,{y};\mu_1,\mu_2, \xi \bigr)
   \nonumber \\[0.5em]
   &\quad
   =
   \exp\,\biggl[\ms
   - \int_{\mu_{01}}^{\mu_1} \frac{d\mu}{\mu}\, \gamma_J(\mu)\,
     \ln\frac{\xav{1} \fsqrt{\xi}}{\mu}
   - \int_{\mu_{02}}^{\mu_2} \frac{d\mu}{\mu}\, \gamma_J(\mu)\,
     \ln\frac{\xav{2} \fsqrt{\xi}}{\mu}
   + \frac{1}{2}\ms J({y};{\mu_{01},\mu_{02}}) \,
      \ln\frac{\xi}{\xi_0}
   \ms\biggr]
   \nonumber \\
   &\qquad \times
   \exp\,\biggl[- \frac{1}{2}\ms J({y}; \mu_1, \mu_2) \ln (x_1 x_2)
      \,\biggr]\,
   \widehat{F}_{a_1 a_2\ms (\mu_{01},\mu_{02},\xi_0)}(x_1,x_2, {y};
      \mu_1,\mu_2)
   \,,
   \phantom{\frac{1}{1}}
\end{align}
where the distribution $\widehat{F}$ evolves as
\begin{align}
   \label{DGLAP-red-1}
   &
   \frac{d}{d \ln\mu_1}\,
   \widehat{F}_{a_1 a_2\ms (\mu_{01},\mu_{02},\xi_0)}(x_1,x_2,
   {y};\mu_1,\mu_2)
   \nonumber \\[0.2em]
   & \qquad =
   - \;\gamma_J(\mu_1) \, \ln \biggl( \frac{x_1}{\xav{1}} \bigg) \;
   \widehat{F}_{a_1 a_2\ms (\mu_{01},\mu_{02},\xi_0)}(x_1,x_2,{y};
     \mu_1,\mu_2)
   \nonumber \\[0.3em]
   & \qquad\quad
   + 2 \sum_{b_1}
   \int_{x_1}^1 \frac{d z}{z} \;
   P_{a_1 b_1}\biggl( \frac{x_1}{z};\ms \mu_1^{} \biggr) \,
   \widehat{F}_{b_1 a_2\ms (\mu_{01},\mu_{02},\xi_0)}(z,x_2^{},{y};
      \mu_1,\mu_2)
\end{align}
in $\mu_1$ and correspondingly in $\mu_2$.  A rapidity parameter dependence only
appears in the initial condition
\begin{align}
   \label{DGLAP-red-1-init}
   &
   \widehat{F}_{a_1 a_2\ms (\mu_{01},\mu_{02},\xi_0)}(x_1,x_2,
   {y};\mu_{01},\mu_{02})
   = F_{a_1 a_2}(x_1,x_2,{y};\mu_{01},\mu_{02},\xi_0)
   \nonumber \\[0.4em]
   &\qquad
   = \exp\,\biggl[\frac{1}{2}\ms J({y}; \mu_{01}, \mu_{02}) \,
      \ln (x_1 x_2)  \,\biggr] \,
   F_{a_1 a_2}\bigl( x_1,x_2,{y};\mu_{01},\mu_{02},
      \xi_0 / (x_1 x_2) \bigr)
\end{align}
but not in the evolution equation \eqref{DGLAP-red-1}.  For the above
construction it is essential that $\xi_0$ in the initial conditions is
\emph{independent} of the momentum fractions.

\rev{The term with $\gamma_J$ on the r.h.s.\ of \eqref{DGLAP-red-1} is
multiplied with a logarithm and can become quite large, in particular for higher
colour representations. This requires a finer step size of the Runge-Kutta
algorithm, which increases computing time.  To alleviate this problem, we
introduced the parameters $\xav{1}$ and $\xav{2}$ in \eqref{solution-1}.  We
take $\xav{1} = \fsqrt{x_{1, \text{min}}}$, where $x_{1, \text{min}}$ is the
smallest value on the discretisation grid in $x_1$, which minimises the
magnitude of the $\gamma_J$ term in~\eqref{DGLAP-red-1} over the $x_1$ range of
interest.}  An analogous choice is made for $\xav{2}$.

The solution of \eqref{DGLAP-red-1} can be written as
\begin{align}
   &
   \widehat{F}_{a_1 a_2\ms (\mu_{01},\mu_{02},\xi_0)}(x_1,x_2,{y};
      \mu_1,\mu_2)
   \nonumber \\
   &\qquad
   =
   \sum_{b_1}
   \int d x_1' \, U_{a_1 b_1}(x_1^{}, x_1'; \mu_{1}^{}, \mu_{01}^{}) \,
   \widehat{F}_{b_1 a_2\ms (\mu_{01},\mu_{02},\xi_0)}(x_1', x_2^{},
      {y};\mu_{01}^{},\mu_2^{})
   \,,
\end{align}
where $U$ is the Green function of the differential equation
\eqref{DGLAP-red-1}, i.e.\ it satisfies \eqref{DGLAP-red-1} for the active
variables $x_1$ and $\mu_1$, with the initial condition $U_{a_1 b_1}(x_1^{},
x_1'; \mu_{01}^{}, \mu_{01}^{}) = \delta_{a_1 b_1} \, \delta(x_1^{} - x_1')$.
Rather than solving \eqref{DGLAP-red-1} directly, we use the Runge-Kutta
algorithm to compute the matrix that discretises the Green function $U_{a b}(x,
x')$ in the momentum fractions for given initial and final scales.

The overall solution of the evolution equations can then be obtained as
\begin{align}
   \label{solution-1-final}
   &
   F_{a_1 a_2}\bigl( x_1,x_2,{y};\mu_1,\mu_2, \xi /(x_1 x_2) \bigr)
   \nonumber \\[0.5em]
   &\quad
   = \exp\,\biggl[\ms
   - \int_{\mu_{01}}^{\mu_1} \frac{d\mu}{\mu}\, \gamma_J(\mu)\,
     \ln\frac{\xav{1} \fsqrt{\xi}}{\mu}
   - \int_{\mu_{02}}^{\mu_2} \frac{d\mu}{\mu}\, \gamma_J(\mu)\,
     \ln\frac{\xav{2} \fsqrt{\xi}}{\mu}
   + \frac{1}{2}\ms J({y};{\mu_{01},\mu_{02}}) \,
      \ln \frac{\xi}{\xi_0}
   \ms\biggr]
   \nonumber \\[0.5em]
   &\qquad \times
   \sum_{b_1 b_2}
   \exp\,\biggl[- \frac{1}{2}\ms J({y}; \mu_1, \mu_2) \ln (x_1 x_2)
      \,\biggr]\,
   \int d x_1'\, d x_2'\,
      U_{a_1 b_1}(x_1^{}, x_1'; \mu_{1}^{}, \mu_{01}^{}) \,
      U_{a_2 b_2}(x_2^{}, x_2'; \mu_{2}^{}, \mu_{02}^{}) \,
   \nonumber \\[0.3em]
   &\qquad\quad \times
   \exp\,\biggl[\frac{1}{2}\ms J({y}; \mu_{01}, \mu_{02}) \ln (x_1' x_2')
      \,\biggr]\,
   F_{b_1 b_2}\bigl(x_1', x_2',{y};\mu_{01}^{},\mu_{02}^{},
      \xi_0 /(x_1' x_2') \bigr)
   \,.
\end{align}
Discretised in the momentum fractions and in $y$, this expression is computed in
\chili\ for given $\mu_1$, $\mu_2$ and $\xi$.  The initial scales $\mu_{01},
\mu_{02}$ and $\xi_0$ are allowed to depend on~$y$.  The integrals over $\mu$ in
the first line of \eqref{solution-1-final}, as well as the ones in the evolved
expression \eqref{CS-solved} of the Collins-Soper kernel, are evaluated
analytically.  We use the exact expressions from \cite{Ebert:2021aoo}, which do
not involve any $\alpha_s$ expansion beyond the perturbative truncation of the
anomalous dimensions and the QCD $\beta$ function.  At lowest order, one has the
simple forms
\begin{align}
   \label{LL-RGE-ints}
   \int_{\mu_a}^{\mu_b} \frac{d\mu}{\mu}\, \gamma_J(\mu)\,
   \ln \frac{\mu_b}{\mu}
   &=
   \gamma_{J}^{(0)}\, \frac{2 \pi}{\beta_0^2} \,
   \biggl[ \frac{1}{\alpha_s(\mu_b)}
      \ln \frac{\alpha_s(\mu_a)}{\alpha_s(\mu_b)}
      - \frac{1}{\alpha_s(\mu_b)} + \frac{1}{\alpha_s(\mu_a)}
   \ms \biggr]
   \,,
   \notag \\[0.2em]
   \int_{\mu_a}^{\mu_b} \frac{d\mu}{\mu}\, \gamma_J(\mu)\,
   &=
   \gamma_{J}^{(0)}\, \frac{1}{\beta_0} \,
   \ln \frac{\alpha_s(\mu_a)}{\alpha_s(\mu_b)}
\end{align}
with $\gamma_J^{(0)}$ given in \eqref{gamma-J-0} and $\beta_0$ in
\eqref{beta-0}.


\subsection{Second representation}
\label{sec:second-solution}

Defining distributions $\widetilde{F}$ via
\begin{align}
   \label{Ftilde}
   &
   F_{a_1 a_2}\bigl( x_1,x_2,{y};\mu_1,\mu_2, \xi /(x_1 x_2) \bigr)
   \nonumber \\[0.5em]
   &\quad
   =
   \exp\,\biggl[- \frac{1}{2}\ms J({y}; \mu_1, \mu_2) \ln (x_1 x_2)
     \,\biggr]\,
   F_{a_1 a_2}\bigl( x_1,x_2,{y};\mu_1,\mu_2, \xi \bigr)
   \nonumber \\[0.5em]
   &\quad
   =
   \exp\,\biggl[\ms
   - \int_{\mu_{01}}^{\mu_1} \frac{d\mu}{\mu}\, \gamma_J(\mu)\,
     \ln\frac{x_{1} \fsqrt{\xi}}{\mu}
   - \int_{\mu_{02}}^{\mu_2} \frac{d\mu}{\mu}\, \gamma_J(\mu)\,
     \ln\frac{x_{2} \fsqrt{\xi}}{\mu}
   + \frac{1}{2}\ms J({y};{\mu_{01},\mu_{02}}) \,
     \ln \frac{x_1 x_2\, \xi}{\xi_0}
   \ms\biggr]
   \nonumber \\
   &\qquad \times
   \exp\,\biggl[- \frac{1}{2}\ms J({y}; \mu_1, \mu_2) \ln (x_1 x_2)
      \,\biggr]\,
   \widetilde{F}_{a_1 a_2\ms (\mu_{01},\mu_{02},\xi_0)}(x_1,x_2, {y};
      \mu_1,\mu_2)
   \,,
   \phantom{\frac{1}{1}}
\end{align}
we find that the evolution equation \eqref{DGLAP-zeta} turns into
\begin{align}
   \label{DGLAP-red-2}
   &
   \frac{d}{d \ln\mu_1}\,
   \widetilde{F}_{a_1 a_2\ms (\mu_{01},\mu_{02},\xi_0)}(x_1,x_2,{y};
      \mu_1,\mu_2)
   \nonumber \\[0.2em]
   & \qquad =
   2 \sum_{b_1}
   \int_{x_1}^1 \frac{d z}{z} \;
   \widetilde{P}_{a_1 b_1\ms (\mu_{01})}\biggl(
      \frac{x_1}{z};\ms \mu_1^{}, J({y}; \mu_{01}, \mu_{02}) \biggr) \,
   \widetilde{F}_{b_1 a_2\ms (\mu_{01},\mu_{02},\xi_0)}(z,x_2^{},{y};
      \mu_1,\mu_2)
\end{align}
with a kernel
\begin{align}
   \label{eff-kernel}
   \widetilde{P}_{a b\ms (\mu_{0})}(z; \mu, J)
   &=
   \exp\, \biggl[\ms
   \biggl(\ms \int_{\mu_{0}}^{\mu} \frac{d\mu}{\mu}\, \gamma_J(\mu)
            - \frac{1}{2}\ms J
   \biggr) \ms
   \ln z
  \,\biggr] \,
  P_{a b}(z; \mu)
\end{align}
and initial conditions
\begin{align}
   \label{DGLAP-red-2-init}
   \widetilde{F}_{a_1 a_2\ms (\mu_{01},\mu_{02},\xi_0)}(x_1,x_2,
   {y};\mu_{01},\mu_{02})
   &= F_{a_1 a_2}\bigl( x_1,x_2,{y};\mu_{01},\mu_{02},
      \xi_0 / (x_1 x_2) \bigr)
   \,.
\end{align}
Note that the kernel on the r.h.s.\ of \eqref{DGLAP-red-2} depends explicitly on
$\mu_{01}$, $\mu_{02}$, and ${y}$, in contrast to the one in
\eqref{DGLAP-red-1}. The formulation in \eqref{Ftilde} to
\eqref{DGLAP-red-2-init} is hence not well suited for a numerical implementation
in \chili, since one would need to compute discretised evolution kernels
$\widetilde{P}(x_1/z)$ and Green functions $\widetilde{U}(x_1^{}, x_1')$ for
each of the variables just mentioned.  However, \rev{it} provides insight into
how
evolution changes the $x_1$ and $x_2$ dependence of colour non-singlet DPDs.

\begin{figure}
\centering
\subfloat[$\mu = 10 \gev$]{
   \includegraphics[width=0.48\textwidth,trim=0 25 35 47,clip]{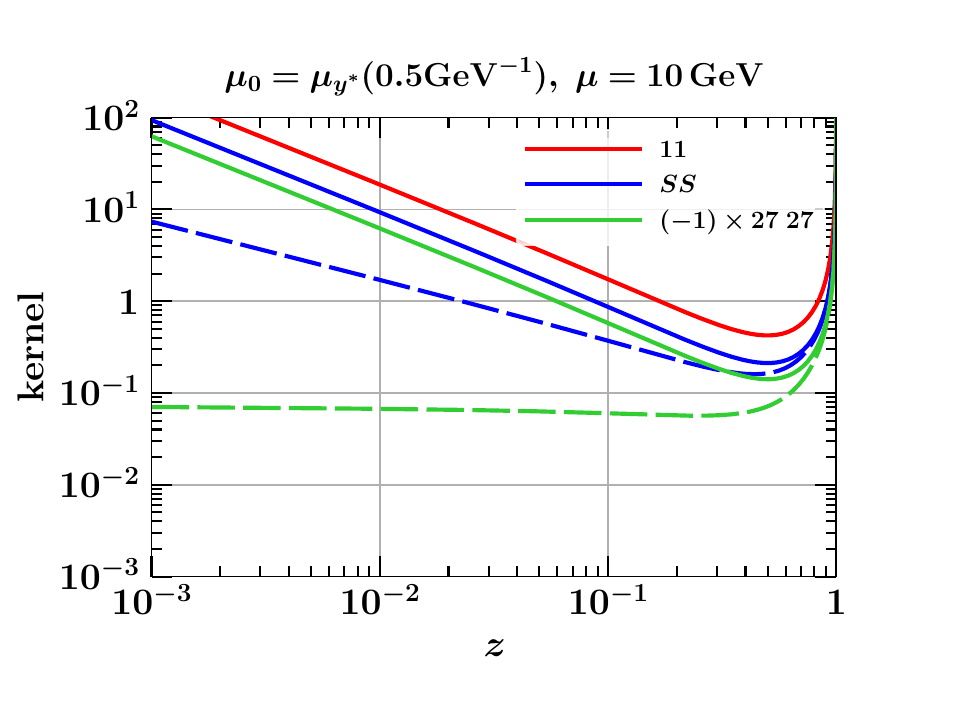}
}
\subfloat[$\mu = 80 \gev$]{
   \includegraphics[width=0.48\textwidth,trim=0 25 35 47,clip]{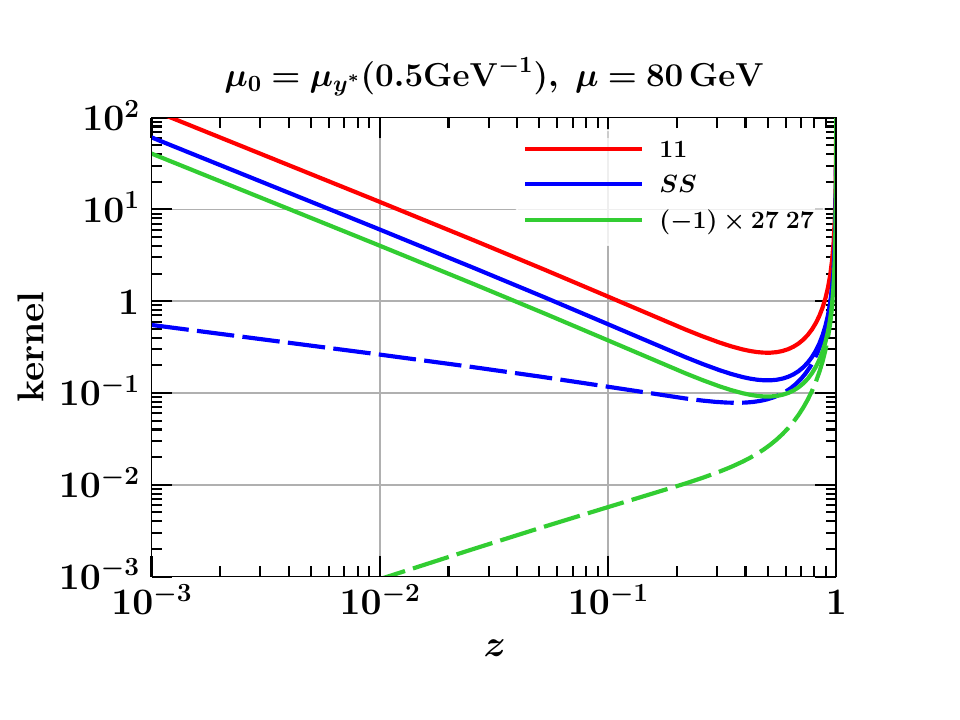}
}
\caption{\label{fig:eff-DGLAP-kernel} Dashed lines: the effective kernel
$\widetilde{P}_{g g\ms (\mustar)}\bigl(z; \mu, J(y; \mustar, \mustar)
\bigr)$ at $y = 0.5 \gev^{-1}$.
Solid lines: the original kernel $P_{\smash[v]{g g}}(z, \mu)$. Here $P_{g g}$,
$\gamma_J$, and the perturbative part of $J$ have been evaluated at
order~$\alpha_s$.
The shorthand notation $(-1) \times \twensev$ indicates that the curves for
$(R_1 R_2) = (\twensev)$ have been multiplied with $-1$.  The same convention is
used in all figures of this work.}
\end{figure}

Consider forward evolution ($\mu_1 > \mu_{01}$ and $\mu_2 > \mu_{02}$) and
assume that ${y}$ is such that $J(y; \mu_{01}, \mu_{02}) < 0$.  As we see in
\fig{\ref{fig:CS-kernel}}b, the latter condition is in particular fulfilled for
initial scales $\mu_{01} = \mu_{02} = \mustar$ and $y$ not too small.  The
prefactor of $\ln z$ in the exponential of \eqref{eff-kernel} is then positive,
which results in a damping of $\widetilde{P}(z)$ compared with $P(z)$ that
becomes increasingly strong with decreasing $z$.  This is illustrated in
\fig{\ref{fig:eff-DGLAP-kernel}}, where we compare the original and the
effective gluon splitting kernel for $\mu_{01} = \mu_{02} = \mustar$ at
\begin{align}
   \label{std-y-choice}
   y &= 0.5 \gev^{-1}
   \,,
   &
   \mustar &\approx 2.54 \gev
   \,,
\end{align}
where $\mustar$ is defined in \eqref{mu-star-def} and \eqref{star-choice}.
With increasing distance between initial and final scales, the evolution of
$\widetilde{F}$ thus becomes increasingly ``local'' in the momentum fractions,
since the exponential damping in $\widetilde{P}$ limits the effective range of
integration in \eqref{DGLAP-red-2} to a region of $z$ increasingly close to
$x_1$.  As seen in the figure, the damping also becomes stronger with increasing
dimension of the colour representation, as this increases $\gamma_J$ and $-J$.
According to \fig{\ref{fig:CS-kernel}}b, the damping also becomes stronger with
increasing $y$.

\begin{figure}
\centering
\subfloat[$S_{\text{RGE}}$]{
   \includegraphics[width=0.48\textwidth,trim=0 15 35 47,clip]{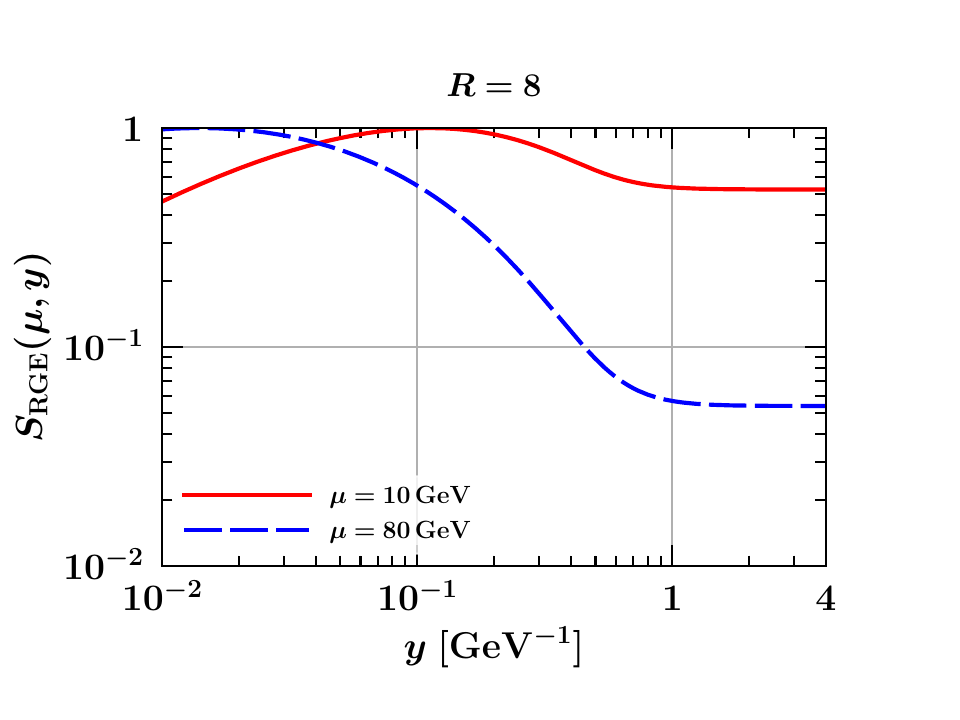}
}
\subfloat[\label{fig:sudavok-exp-J} $S_{J}$]{
   \includegraphics[width=0.48\textwidth,trim=0 15 35 47,clip]{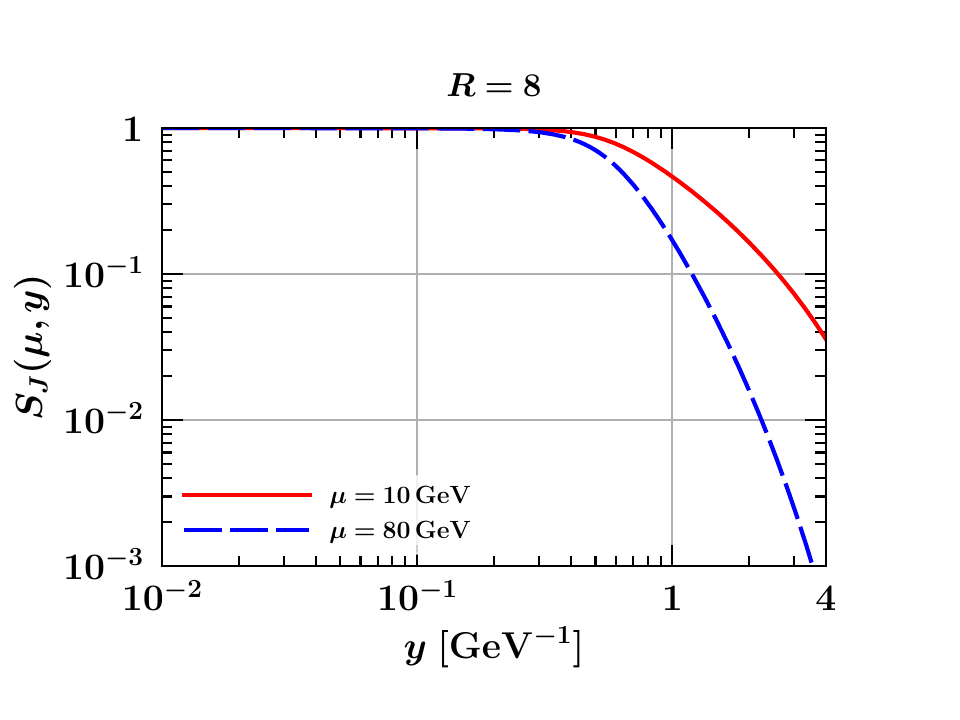}
}
\caption{\label{fig:sudakov-exp} The exponential factors defined in
\protect\eqs{\eqref{sudakov-rge}} and \protect\eqref{sudakov-J} for two choices
of $\mu$.  The curves are for the colour octet ($R=8$) and have been computed
with $J$ obtained from the SV19 fit as described in
\sect{\protect\ref{sec:cs-kernel}}.}
\end{figure}

Using the relation \eqref{CS-solved}, we can simplify \eqref{Ftilde} to
\begin{align}
   \label{solution-2}
   &
   F_{a_1 a_2}\bigl( x_1,x_2,{y};\mu_1,\mu_2, \xi /(x_1 x_2) \bigr)
   \nonumber \\[0.5em]
   &\qquad
   =
   \exp\,\biggl[\ms {}
   - \int_{\mu_{01}}^{\mu_1} \frac{d\mu}{\mu}\, \gamma_J(\mu)\,
     \ln\frac{\fsqrt{x_1 \xi / x_2}}{\mu}
   - \int_{\mu_{02}}^{\mu_2} \frac{d\mu}{\mu}\, \gamma_J(\mu)\,
     \ln\frac{\fsqrt{x_2\ms \xi / x_1}}{\mu} \ms\biggr]
   \nonumber \\
   &\qquad\quad \times
   \exp\,\biggl[\frac{1}{2}\ms J({y}; \mu_{01}, \mu_{02}) \,
      \ln\frac{\xi}{\xi_0}
      \,\biggr] \,
   \widetilde{F}_{a_1 a_2\ms (\mu_{01},\mu_{02},\xi_0)}(x_1,x_2,{y};
      \mu_1,\mu_2)
   \,.
\end{align}
For $\mu_{01} = \mu_{02}$ and $\mu_1 = \mu_2$, the factors $x_1/x_2$ and
$x_2/x_1$ in the logarithms of the second line in \eqref{solution-2} cancel
against each other.  Otherwise, the exponential in the second line gives rise to
powers of $x_1$ and $x_2$ that can be either positive or negative.
In \fig{\ref{fig:sudakov-exp}} we show
\begin{align}
   \label{sudakov-rge}
   S_{\text{RGE}}(\mu, y)
   &=
   \exp\,\biggl[\ms
   - 2 \int_{\mustar}^{\mu} \frac{d\mu'}{\mu'}\, \gamma_J(\mu')\,
     \ln\frac{\mu}{\mu'}
   \ms\biggr]
   \intertext{and}
   \label{sudakov-J}
   S_{J}(\mu, y)
   &=
   \exp\, \biggl[\ms
      J(y; \mustar, \mustar) \, \ln \frac{\mu}{\mustar}
   \biggr]
   \,,
\end{align}
which correspond to the exponential factors in \eqref{solution-2} with scales
$\mu_{01}^{} = \mu_{02}^{} = \mustar$ and $\mu_1^{} = \mu_2^{} = \mu$ and with
reduced rapidity parameters $\xi_0^{} = \mustar^2$ and $\xi = \mu_{}^2$.  The
perturbative factor $S_{\text{RGE}}$ depends on $y$ only via $\mustar$ and
therefore saturates at large $y$.  We note that the exponent of $S_{\text{RGE}}$
corresponds to double logarithms whereas the one of $S_{J}$ has only a single
logarithm.  For $y$ in the perturbative region, this counting of large
logarithms correctly reflects the relative size of the two factors, but we see
in the figure that this is no longer the case at large $y$, where $S_{J}$
provides a much stronger suppression than $S_{\text{RGE}}$.

Trading the combination $\xi / (x_1 x_2)$ in \eqref{solution-2} for the original
variable $\zeta$, we get
\begin{align}
   \label{solution-3}
   &
   F_{a_1 a_2}\bigl( x_1,x_2,{y};\mu_1,\mu_2, \zeta \bigr)
   \nonumber \\[0.5em]
   &\qquad
   =
   \exp\,\biggl[\ms {}
   - \int_{\mu_{01}}^{\mu_1} \frac{d\mu}{\mu}\, \gamma_J(\mu)\,
     \ln\frac{x_1 \fsqrt{\zeta}}{\mu}
   - \int_{\mu_{02}}^{\mu_2} \frac{d\mu}{\mu}\, \gamma_J(\mu)\,
     \ln\frac{x_2 \fsqrt{\zeta}}{\mu} \ms\biggr]
   \nonumber \\
   &\qquad\quad \times
   \exp\,\biggl[\frac{1}{2}\ms J({y}; \mu_{01}, \mu_{02}) \,
      \ln\frac{x_1 x_2\ms \zeta}{\xi_0}
      \,\biggr] \,
   \widetilde{F}_{a_1 a_2\ms (\mu_{01},\mu_{02},\xi_0)}(x_1,x_2,{y};
      \mu_1,\mu_2)
   \,.
\end{align}
For forward evolution, the exponential in the second line results in a power of
$x_1$ that is negative and always smaller than the power of $x_1$ resulting from
the second line of \eqref{solution-2}.  The exponential in the third line of
\eqref{solution-3} also results in a negative power of $x_1$.  After evolution,
colour non-singlet DPDs hence decrease more steeply with $x_1$ if one keeps
$\zeta$ fixed rather than taking $\zeta = \xi / (x_1 x_2)$.  This decrease
becomes stronger with $\mu_1$, with ${y}$, and with the dimension of the colour
representation. An analogous statement holds for the behaviour on $x_2$.

Since accurate Chebyshev interpolation requires denser grids for steeper
functions, DPDs in \chili\ are discretised in $x_1$ and $x_2$ for fixed
$\xi = x_1 x_2 \ms \zeta$, as is reflected in the equations of
\sect{\ref{sec:first-solution}}.  However, we see in \eqref{solution-1-final}
that convolutions in $x_1'$ and $x_2'$ with the Green functions $U$ are computed
for
\begin{align}
   \label{internal-convol}
   \exp\,\biggl[\frac{1}{2}\ms J({y}; \mu_{01}, \mu_{02}) \ln (x_1' x_2')
     \,\biggr] \,
   F(x_1', x_2', \ldots)
   \,,
\end{align}
which for large initial scales or large $y$ has a significantly steeper
behaviour than $F(x_1', x_2', \ldots)$. Discretisation errors tend to partially
cancel under integrals, because the interpolant oscillates around the exact
value of a function.  Nevertheless, we use finer discretisation grids for colour
non-singlet DPDs than for their colour singlet counterparts.  Some details about
the discretisation grids and the resulting numerical accuracy are given in
\app{\ref{sec:chili}}.


\paragraph{Products of DPDs in the cross section.}

The cross section formula \eqref{dps-Xsect} involves the product of DPDs at two
rapidity parameters, whose product is fixed by the process kinematics as
\begin{align}
   \label{zeta-constraints}
   \bigl(\ms \zeta \bar{\zeta} \,\bigr)^{1/2}
   &= s
   = \frac{M_1^2}{x_1 \bar{x}_1}
   = \frac{M_2^2}{x_2 \ms \bar{x}_2}
   = \frac{M_1 M_2}{\fsqrt{x_1 \bar{x}_1\, x_2 \ms \bar{x}_2}}
\end{align}
according to the relations \eqref{x-fractions} and
\eqref{zeta-basic-constraint}.  These constraints are fulfilled by the choice
\begin{align}
   \label{zeta-choice}
   \zeta &= \xi \ms/\ms (x_1 x_2)
   \,,
   &
   \bar{\zeta} &= \xi \ms/\ms (\bar{x}_1 \ms \bar{x}_2)
   \,,
   &
   \xi &= M_1 M_2
   \,.
\end{align}
Inserting this into the representation \eqref{solution-2} of evolved DPDs, we
obtain
\begin{align}
   \label{global-Sudakov}
   &
   F_{a_1 a_2}\bigl( x_1,x_2,{y};
      \mu_1,\mu_2, M_1 M_2/(x_1 x_2) \bigr) \,
   F_{b_1 b_2}\bigl( \bar{x}_1,\bar{x}_2,{y};
     \mu_1,\mu_2, M_1 M_2/(\bar{x}_1 \ms \bar{x}_2) \bigr)
   \nonumber \\[0.5em]
   &\qquad
   = \exp\,\biggl[
      - 2 \int_{\mu_{01}}^{\mu_1} \frac{d\mu}{\mu}\, \gamma_J(\mu)\,
         \ln\frac{M_1}{\mu}
      - 2 \int_{\mu_{02}}^{\mu_2} \frac{d\mu}{\mu}\, \gamma_J(\mu)\,
         \ln\frac{M_2}{\mu}
      + J({y}; \mu_{01}, \mu_{02}) \, \ln\frac{M_1 M_2}{\xi_0}
   \ms\biggr]
   \nonumber \\[0.5em]
   &\qquad\quad \times
   \widetilde{F}_{a_1 a_2\ms (\mu_{01},\mu_{02},\xi_0)}(x_1,x_2,
      {y};\mu_1,\mu_2) \,
   \widetilde{F}_{b_1 b_2\ms (\mu_{01},\mu_{02},\xi_0)}(
      \bar{x}_1,\bar{x}_2,{y};\mu_1,\mu_2)
   \,,
\end{align}
where we used that the colour representations in both DPDs must have the same
dimension, so that the associated Collins-Soper kernels and cusp anomalous
dimensions are the same.  Note that on the r.h.s.\ all reference to the
particular choice of $\zeta$ and $\bar{\zeta}$ has disappeared, whilst on the
l.h.s.\ the DPDs are evaluated at rapidity parameters of the form
\eqref{xi-def}.


\subsection{Combining perturbative orders}
\label{sec:orders}

The computation of evolved DPDs and of DPS cross sections involves a number of
perturbative ingredients, of which some are associated with the resummation of
double or single logarithms, whilst others appear as fixed-order ingredients.
The orders of their expansion in $\alpha_s$ should be coordinated in a suitable
manner.  In table~\ref{tab:orders} we propose a scheme for DPS cross sections in
collinear factorisation, obtained by adapting the schemes discussed in
\cite{Abbate:2010xh, Berger:2010xi, Stewart:2013faa} and
\cite{Bacchetta:2019sam} for other processes involving Sudakov logarithms.

\begin{table}
  \centering
  \setlength\tabcolsep{1em}
  \renewcommand{\arraystretch}{1.25}
\begin{tabular}{c c c c c c c} \hline
  accuracy
  & ${\gamma_J}$ & $\beta$ & ${J}(y^*; \mustar)$
  & ${F}^{\text{spl}}(\ldots; \mustar,\ms \zeta_{y^*})$
  & ${P}(z; \mu)$ & ${\hat{\sigma}}$
  \\ \hline
  LL  & $\alpha_s$ & $\alpha_s^2$ & $-$ &
        $\alpha_s$ & $\alpha_s$ & LO
  \\
  NLL & $\alpha_s^2$ & $\alpha_s^3$ & $\alpha_s$ &
        $\alpha_s$ & $\alpha_s$ & LO
  \\
  \NLLp & $\alpha_s^2$ & $\alpha_s^3$ & $\alpha_s$ &
          $\alpha_s^2$ & $\alpha_s^2$ & NLO
  \\
  NNLL & $\alpha_s^3$ & $\alpha_s^4$ & $\alpha_s^2$ &
         $\alpha_s^2$ & $\alpha_s^2$ & NLO
  \\
  \NNLLp & $\alpha_s^3$ & $\alpha_s^4$ & $\alpha_s^2$ &
           $\alpha_s^3$ & $\alpha_s^3$ & NNLO
  \\ \hline
\end{tabular}
\caption{\label{tab:orders} Different levels of accuracy of the DPS cross
section and the associated orders of $\alpha_s$ for its perturbative
ingredients.}
\end{table}

Let us explain and comment on the different entries in the table.
\begin{itemize}
\item The cusp anomalous dimension $\gamma_J$ and the QCD $\beta$ function
appear in the resummation of double logarithms, as in the second line of
\eqref{solution-2}.  They are evaluated at the same order,\footnote{Here and in
the following, ``at the same order'' means ``both at LO'', ``both at NLO'', etc., which can result in different powers of $\alpha_s$.}
which determines the number of ``N''s before the ``LL'' in the nomenclature.

We evaluate $\gamma_J$ at the same order in \emph{all} terms that influence the
scale dependence, including single logarithmic ones as in \eqref{CS-solved} and
\eqref{eff-kernel} and including the evolution equation \eqref{DGLAP-red-1}.
This ensures the internal consistency of the coupled set of evolution equations
in \eqn{\eqref{DGLAP-zeta}} to \eqref{CS-RGE}, and in particular the path
independence of evolution discussed below \eqn{\eqref{J-singlet}}.
\item The Collins-Soper kernel at the scale $\mustar$ appears as the
perturbative starting condition for the kernel at general scales according to
\eqref{J-star-prescription} and \eqref{J-evolved-form}.  It is associated with
resummed single logarithms and can hence be evaluated one order lower than
$\gamma_J$.
We note that setting $J(y; \mustar, \mustar)$ to zero (in the LL entry of
the table) is tantamount to evaluating it at LO, thanks to the vanishing
coefficient $\pr{R}{J}^{(0, 0)}$ in \eqref{J-as}.

The non-perturbative part $\Delta J$ of the kernel is not easily classified in a
perturbative power-counting scheme.  We argue that it should always be included,
including at LL, because of its numerical importance seen in
\fig{\ref{fig:sudakov-exp}}b.
\item The fifth column in the table refers to the perturbative splitting form
\eqref{splitting-DPD-pt} evaluated at the initial scales of evolution.  This is
a term evaluated in fixed-order perturbation theory, as are the hard-scattering
cross sections in the last column.
In the studies at order \NLLp\ and NNLL of the present work, we depart from the
scheme of the table by using only the $\alpha_s$ approximation of
$F^{\text{spl}}$ for the reason given below
\eqn{\eqref{splitting-colour-factors-gg}}.
\item As is customary in single parton scattering, we use the same perturbative
order for the DGLAP evolution kernels $P(z; \mu)$ as for the hard-scattering
cross sections.  For consistency, the same order is also used for the evolution
of the PDFs that appear in the initial conditions of DPD evolution (see
\eqs{\eqref{splitting-DPD-pt}} and \eqref{int-DPD-singlet}).
Notice that at NLL and NNLL accuracy, $\beta$ and hence the evolution of
$\alpha_s$ is taken at one order higher than the DGLAP kernels.  This presents a
somewhat awkward situation, because almost all PDF fits take the same order for
the DGLAP kernels and the evolution of~$\alpha_s$.  Different solutions to this
dilemma have been adopted in the TMD literature, where PDFs appear as an input
to computing TMDs at small impact parameter $b$, see e.g.\ \sect{4} in
\cite{Stewart:2010qs} and \sect{2.4} in \cite{Bacchetta:2019sam}.  The bulk of
studies in the present work uses LL or \NLLp, where this problem does not
appear.

We note that in TMD factorisation one takes the non-cusp anomalous dimension
$\gamma_a$ in \eqref{TMD-RGE} at one order less than the cusp anomalous
dimension $\gamma_{K, a}$ (see for instance \tab{1} in
\cite{Bacchetta:2019sam}).  This is the same order as for $J(y^*, \mustar)$ in
our table, with the same justification that these quantities accompany single
logarithms in the resummed expressions.  As a consequence, the orders of
$\gamma_a$ in TMD factorisation differ from the ones of the DGLAP kernels in DPD
factorisation for LL, \NLLp, and \NNLLp, despite the formal similarity between
these quantities in the respective evolution equations.
\item Not included in the table is the order of heavy-flavour matching for DPDs
and PDFs.  As is customary, we take this at the same order as the DGLAP kernels.
Note that, if one performs flavour matching at scales equal to the quark masses,
there is actually no difference between matching at tree level (LO) and at order
$\alpha_s$ (NLO), due to the form of the matching kernels in \eqs{\eqref{AQg1}}
and \eqref{Agg1}.
\end{itemize}
Let us finally see which are the highest orders presently known for the
different ingredients.  The octet Collins-Soper kernel $\pr{8}{J}$ and its
anomalous dimension would allow an analysis up to \NNLLp\ (and even higher)
thanks to their relation \eqref{J-K-equal} with the kernel $K_{g}$ for gluon
TMDs.  The available results for $\pr{R}{J}$ and $\prn{R}{\gamma}_J$ with $R=10$
and $R=27$ are sufficient for \NLLp\ accuracy.  The splitting formula for DPDs
is known at order $\alpha_s^2$ for unpolarised partons \cite{Diehl:2019rdh,
Diehl:2021wpp} and at order $\alpha_s$ for polarised ones
\cite{Buffing:2017mqm}. The DGLAP kernels for colour non-singlet DPDs have been
computed up to NLO \cite{Diehl:2022rxb}, except for linearly polarised gluons.
The current bottleneck are the hard-scattering cross sections in colour
non-singlet channels.  They are reasonably simple to obtain at LO (see e.g.\
\cite{Dokshitzer:2005ig} for $g g \to g g$ and \cite{Blok:2022mtv} for $g g \to
q \bar{q}$), but their computation at higher orders will require non-trivial
adjustments of the available calculations for the colour singlet case.

\section{Results: evolved DPDs}
\label{sec:dpds}

In this section, we study numerically the evolution of DPDs with the initial
conditions described in \sect{\ref{sec:input}}.  In the next two subsections, we
consider evolution at LL accuracy as defined in \tab{\ref{tab:orders}}, and in
\sect{\ref{sec:dpds-higher-orders}} we investigate how results change when we
evolve at \NLLp\ or at NNLL.
Throughout this section, DPDs at scales $\mu_1$ and $\mu_2$ are evaluated at the
rapidity parameter
\begin{align}
   \zeta &= \mu_1 \mu_2 / (x_1 x_2)
   \,.
\end{align}
Unless specified otherwise, we show DPDs at $y = 0.5 \gev^{-1}$, corresponding
to an initial scale $\mustar \approx 2.54 \gev$, and with the non-perturbative
part of the Collins-Soper kernel obtained from the SV19 fit.

\rev{For this and the next section, we produced plots for different flavour
combinations, with final scales of $10 \gev$ or $80 \gev$ for the two partons.
We will not show all of these combinations, but instead present a selection of
plots that we found to be indicative of the general situation.  We have chosen
$10 \gev$ as an example for a rather low hard scale and $80 \gev$ as a typical
scale for jet production or electroweak processes.}


\subsection{Exponentiation versus full evolution}
\label{sec:exponentiation}

To begin with, let us investigate how strongly the evolution of DPDs is
influenced by the $z$ dependent part of the DGLAP kernels, including the
subtraction term included in the plus-distribution $1/(1-z)_+$.  To this end,
we compare the solution of the full evolution equations with the result of
replacing the full DGLAP kernels in \eqref{DGLAP-zeta} with the terms
proportional to $\delta(1-z)$:
\begin{align}
   \label{simple-kernels}
   \prb{R \Rp{}}{P}_{a b}(z; \mu)
   & \to
   \frac{1}{2} \, \delta_{R \Rpbar{}}\, \delta_{a b} \,
   \delta(1-z) \, \prn{R}{d}_a(\mu)
   \,.
\end{align}
The resulting evolution equations can be solved by explicit exponentiation:
\begin{align}
   \label{simple-solution}
   &
   \prn{R_1 R_2}{F}_{a_1 a_2}\bigl(
      x_1,x_2,{y};\mu_1,\mu_2, \xi /(x_1 x_2) \bigr)
   \nonumber \\[0.5em]
   &\qquad
   =
   \exp\,\biggl[\ms {}
   - \int_{\mu_{01}}^{\mu_1} \frac{d\mu}{\mu}\, \prn{R_1}{\gamma}_J(\mu)\,
     \ln\frac{\fsqrt{x_1 \xi / x_2}}{\mu}
   - \int_{\mu_{02}}^{\mu_2} \frac{d\mu}{\mu}\, \prn{R_2}{\gamma}_J(\mu)\,
     \ln\frac{\fsqrt{x_2\ms \xi / x_1}}{\mu} \ms\biggr]
   \nonumber \\[0.2em]
   &\qquad\quad \times
   \exp\,\biggl[\frac{1}{2}\, \pr{R_1}{J}({y}; \mu_{01}, \mu_{02}) \,
      \ln\frac{\xi}{\xi_0}
      \,\biggr] \,
   \exp\,\biggl[\ms
     \int_{\mu_{01}}^{\mu_1} \frac{d\mu}{\mu}\; \prn{R_1}{d}_{a_1}(\mu)\,
   + \int_{\mu_{02}}^{\mu_2} \frac{d\mu}{\mu}\; \prn{R_2}{d}_{a_2}(\mu)\,
   \ms\biggr] \,
   \nonumber \\
   &\qquad\quad \times
   \prn{R_1 R_2}{F}_{a_1 a_2}\bigl(
      x_1,x_2,{y}; \mu_{01},\mu_{02}, \xi_0 /(x_1 x_2)  \bigr)
   \,.
   \phantom{\biggl[ \biggr]}
\end{align}
We refer to this as the ``exponentiated result''.  The first two exponentials
are the same as in \eqref{solution-2}, and they reduce to unity for colour
singlet DPDs.  The third exponential is colour independent at LO according to
\eqref{delta-coeffs}.
As follows from \eqref{decuplet-kernels}, the solution of the full evolution
equations for decuplet gluon distributions is actually given by
\eqref{simple-solution}, both at LO and at NLO accuracy.

In the following study, evolution is performed at LL accuracy.  For simplicity,
we omit flavour matching at the $b$ quark mass, since this would require
separation of the exponential factors in \eqref{simple-solution} into terms for
$n_f = 4$ and $n_f = 5$.  As initial conditions we take the intrinsic part of
the DPDs, with our default choice specified in \eqref{splitting-colour-factors},
\eqref{splitting-colour-factors-gg}, and \eqref{extra-colour-factors}  for the
colour factors $k_{a_1 a_2}(R_1 R_2)$ in \eqref{int-DPD-colour}.

\begin{figure}
\centering
\subfloat[$(\mu_1, \mu_2) = (\mustar,\mustar)$]{
   \includegraphics[width=0.47\textwidth,trim=0 25 35
45,clip]{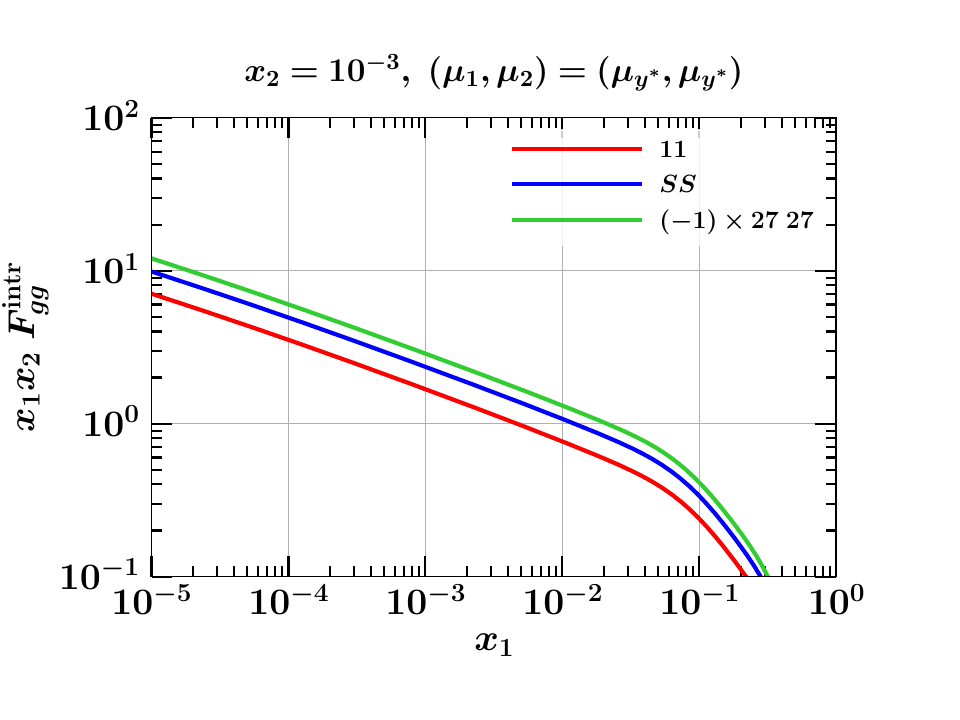}
}
\subfloat[$(\mu_1, \mu_2) = (10 \gev, 10 \gev)$]{
   \includegraphics[width=0.47\textwidth,trim=0 25 35
45,clip]{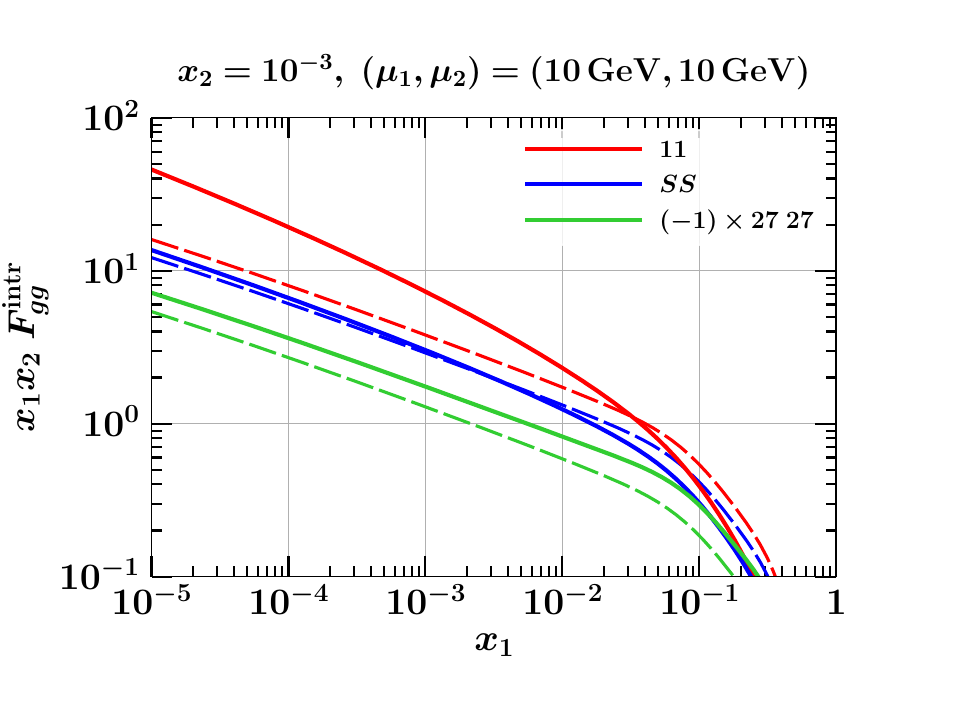}
}
\\[1.5em]
\subfloat[$(\mu_1, \mu_2) = (10 \gev, 80 \gev)$]{
   \includegraphics[width=0.47\textwidth,trim=0 25 35
45,clip]{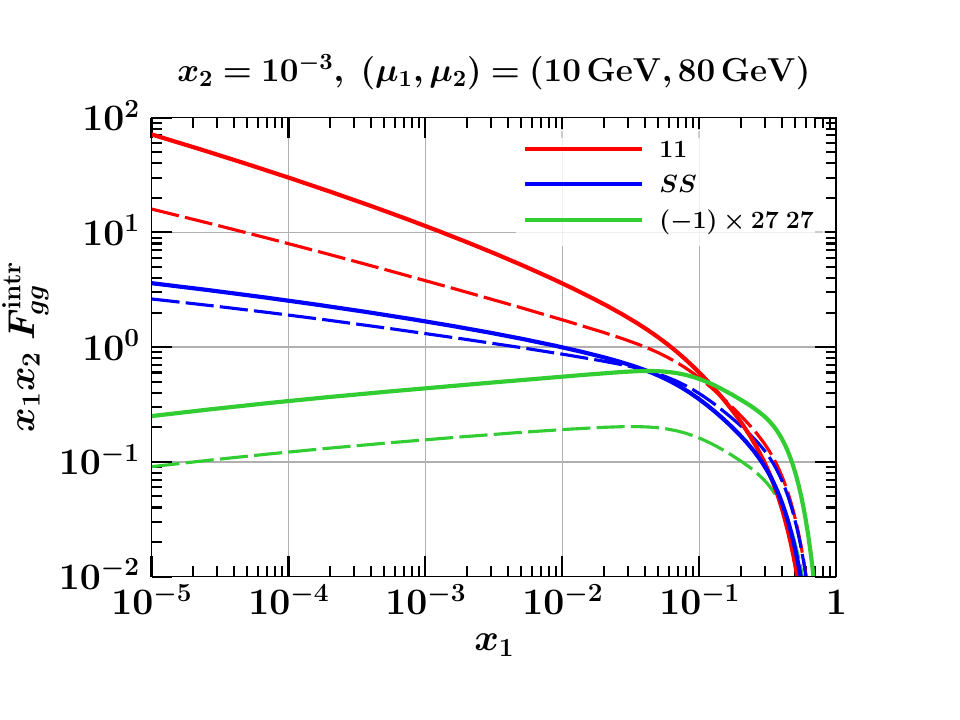}
}
\subfloat[$(\mu_1, \mu_2) = (80 \gev, 80 \gev)$]{
   \includegraphics[width=0.47\textwidth,trim=0 25 35
45,clip]{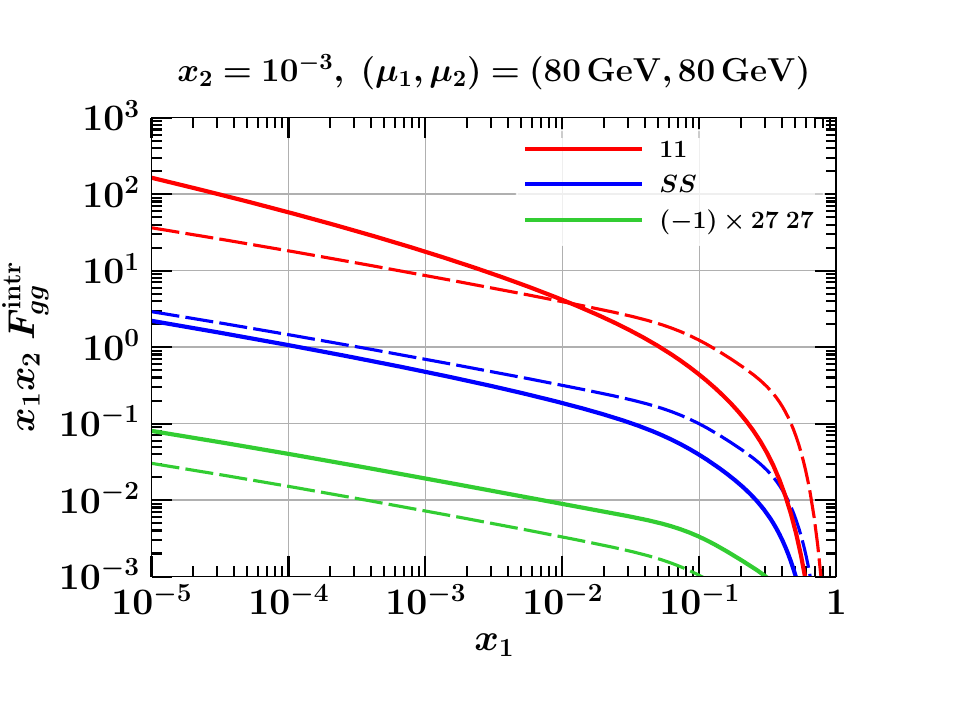}
}
\caption{\label{fig:F_gg-sudakov} The intrinsic part $F_{g g}^{\text{intr}}$ of
the two-gluon distribution at $x_2 = 10^{-3}$ and $y = 0.5 \gev^{-1}$.  Full
lines show the solution of the full evolution equations, and dashed lines the
exponentiated result \protect\eqref{simple-solution}. Evolution is performed at
LL accuracy and with $n_f = 4$ active flavours.  The shorthand notation $(-1)
\times \twensev$  is explained in \fig{\protect\ref{fig:eff-DGLAP-kernel}}.
DPDs are shown in units of $\gev^2$ in this and all following figures.
}
\end{figure}

The comparison between full evolution and the exponentiated result is shown in
\fig{\ref{fig:F_gg-sudakov}} for the distribution $F_{g g}^{\text{intr}}$.  We
see that for the colour singlet, the $z$ dependent part of $P(z)$ has a strong
effect: compared with the exponentiated result it leads to larger distributions
at small $x_1$ and to smaller ones at large $x_1$.  The sign of this shift is
readily understood.  At small $x_1$ the convolution integral in the evolution
equation receives a large positive contribution from $P(z)$ at small $z$,
whereas at large $x_1$ the convolution has a limited integration range and is
dominated by the negative contribution from the subtraction term in $1 /
(1-z)_+$.
The situation for $R_1 = R_2 = 27$ is very different: the result of full
evolution is always larger than the exponentiated result and has a very similar
shape in $x_1$.  To understand this, we recall from
\fig{\ref{fig:eff-DGLAP-kernel}} that the corresponding effective kernel
$\widetilde{P}(z)$ is strongly suppressed for $z<1$.  The convolution in the
evolution equation \eqref{DGLAP-red-2} is hence dominated by the subtraction
term in $1 / (1-z)_+$, which yields a positive result because the coefficient
$c_{g g}(\twensev)$ in \eqref{LO-scaling} is negative.
For $R_1 = R_2 = S$, the suppression of the effective kernels for $z<1$ is
weaker, and the competition between positive and negative contributions results
in a small convolution integral at low $x_1$.  The curves for full evolution are
then close to the exponentiated result.  In \fig{\ref{fig:F_gg-sudakov}}c, we
also see the impact of the $x_1$ dependence in the first exponential of
\eqn{\eqref{simple-solution}}, which leads to a modified small $x_1$ behaviour
of colour non-singlet distributions.  The curves for $R_1 = R_2 = A$ are
identical or very close to the ones for $R_1 = R_2 = S$ and not shown in the
figure.

\begin{figure}
\centering
\subfloat[$(\mu_1, \mu_2) = (\mustar,\mustar)$]{
   \includegraphics[width=0.49\textwidth,trim=0 30 35
52,clip]{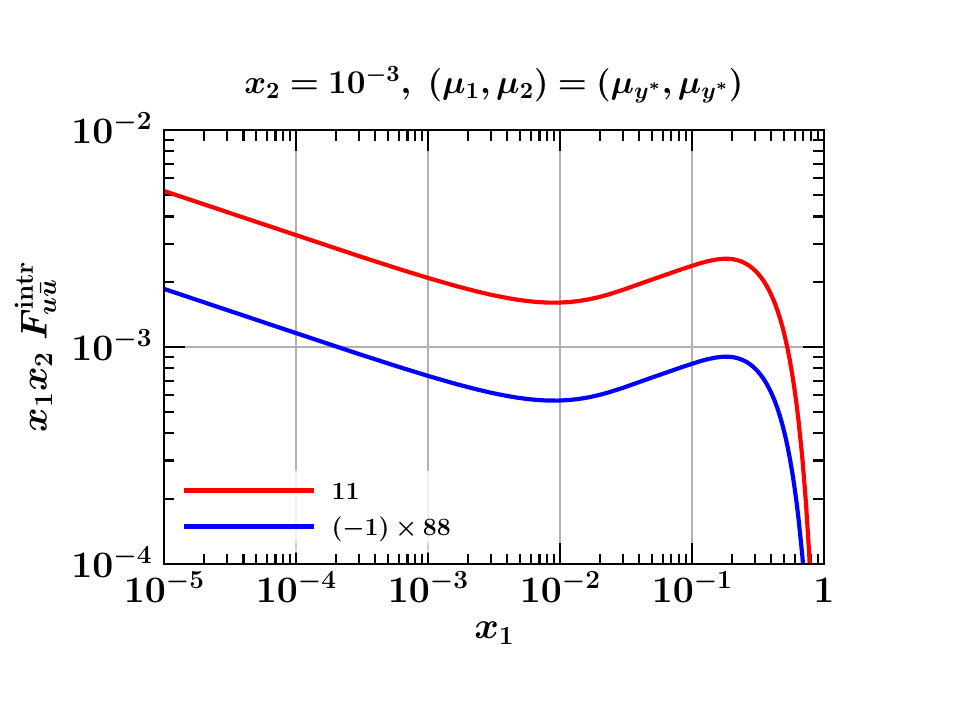}
}
\subfloat[$(\mu_1, \mu_2) = (10 \gev, 10 \gev)$]{
   \includegraphics[width=0.47\textwidth,trim=0 25 35
45,clip]{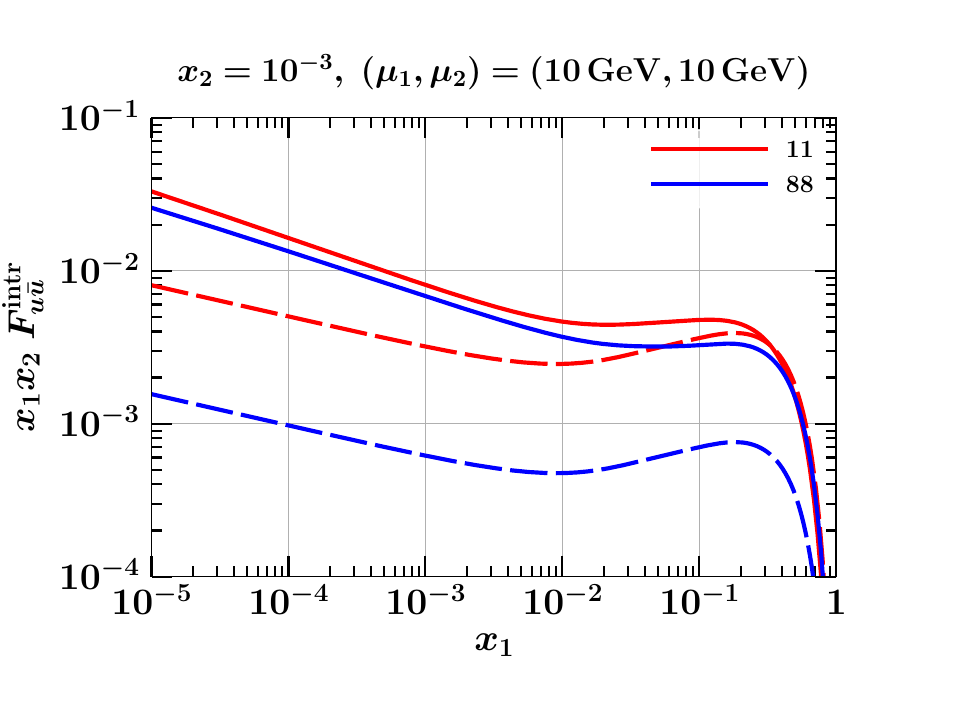}
}
\\[1.5em]
\subfloat[$(\mu_1, \mu_2) = (10 \gev, 80 \gev)$]{
   \includegraphics[width=0.47\textwidth,trim=0 25 35
45,clip]{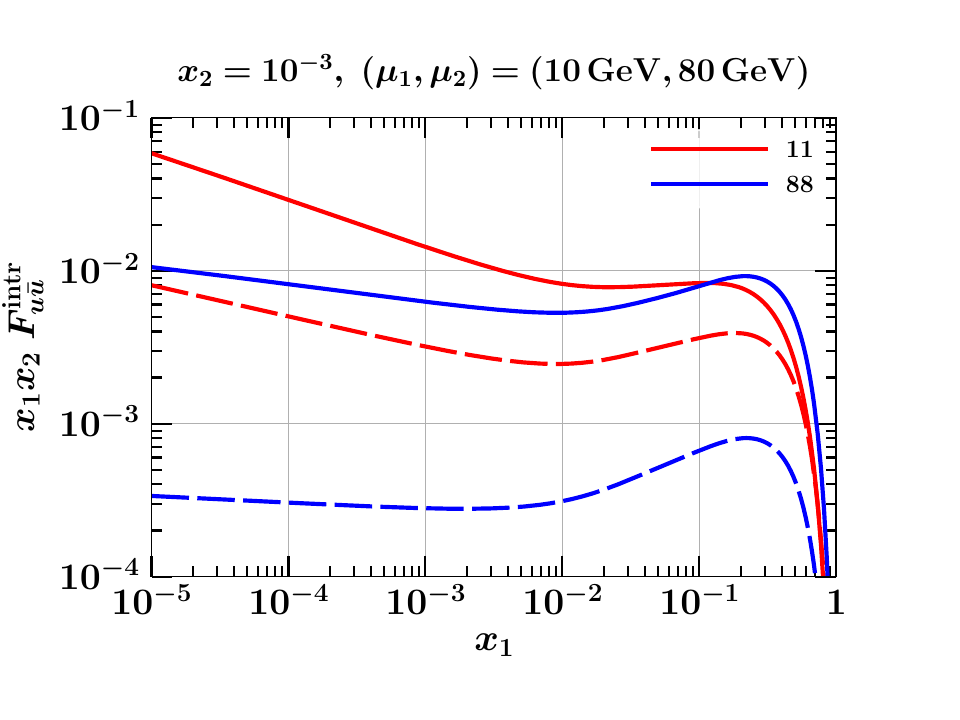}
}
\subfloat[$(\mu_1, \mu_2) = (80 \gev, 80 \gev)$]{
   \includegraphics[width=0.47\textwidth,trim=0 25 35
45,clip]{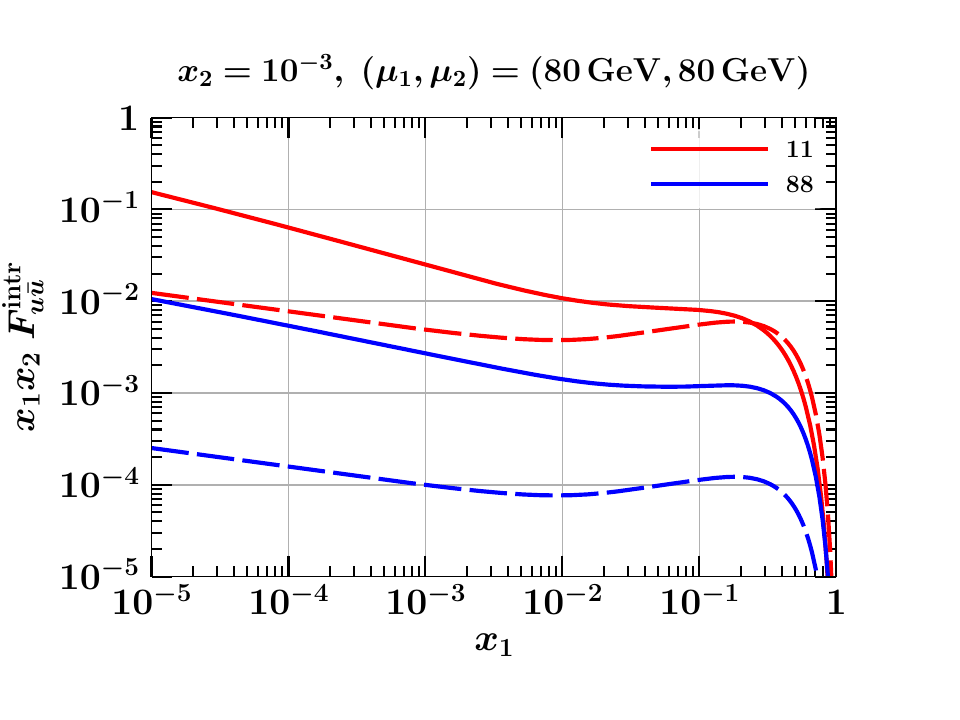}
}
\caption{\label{fig:F_uubar-sudakov} As \fig{\protect\ref{fig:F_gg-sudakov}} but
for the distribution $F_{u\bar{u}}^{\text{intr}}$.}
\end{figure}

A very different picture is obtained for the distribution
$F_{u\bar{u}}^{\text{intr}}$, as shown in \fig{\ref{fig:F_uubar-sudakov}}.  In a
wide range of $x_1$, there is a huge difference between full evolution and the
exponentiated result. This is because the evolution of this distribution is
dominated by mixing with gluons, both in the colour singlet and the colour octet
channels.  The same holds for other distributions, such as $F_{u
g}^{\text{intr}}$ and $F_{\bar{u} g}^{\text{intr}}$, which are not shown here.

Globally, we find that colour non-singlet distributions evolved to higher scales
tend to be smaller than their colour singlet counterparts for the same parton
combination.  This is to be expected from the exponential factors in
\eqref{solution-2}.  However, the actual suppression may be rather weak as in
\fig{\ref{fig:F_uubar-sudakov}}b, and at large $x_1$ there are cases where
evolved colour non-singlet distributions are larger, as seen
\figs{\ref{fig:F_gg-sudakov}}c and \ref{fig:F_uubar-sudakov}c.


\subsection{DPDs evolved at LL accuracy}
\label{sec:dpds-at-ll}

\begin{figure}[p]
\centering
\subfloat[$x_2 = x_1$ at $\mu_1 = \mu_2 = \mustar$]{
   \includegraphics[width=0.48\textwidth,trim=0 25 35
50,clip]{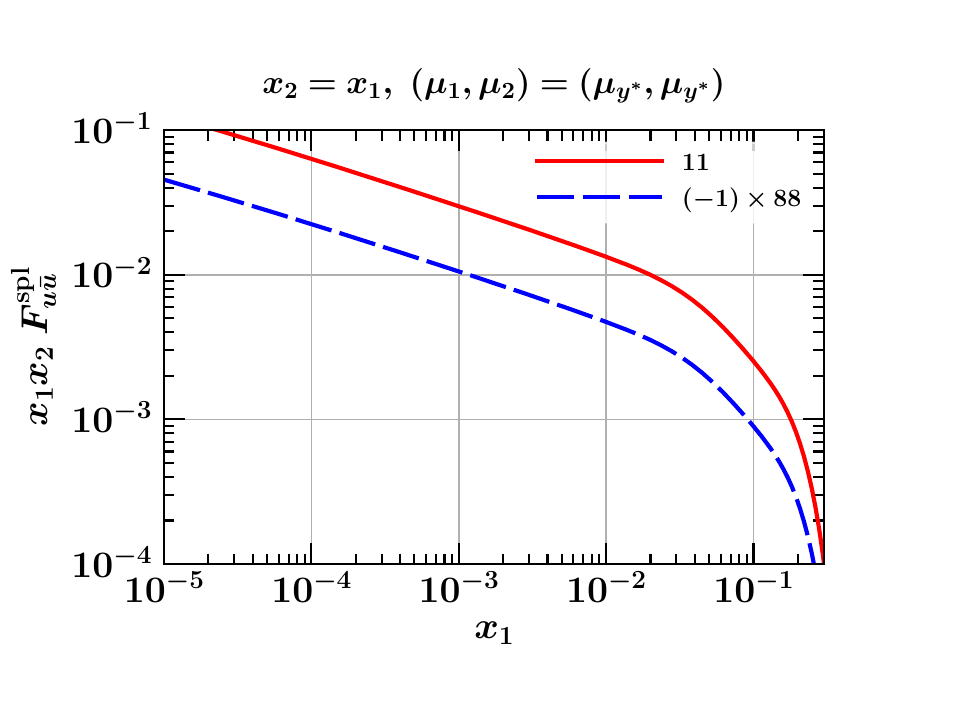}
}
\subfloat[$x_2 = x_1$ at $\mu_1 = \mu_2 = 10 \gev$]{
   \includegraphics[width=0.48\textwidth,trim=0 25 35
50,clip]{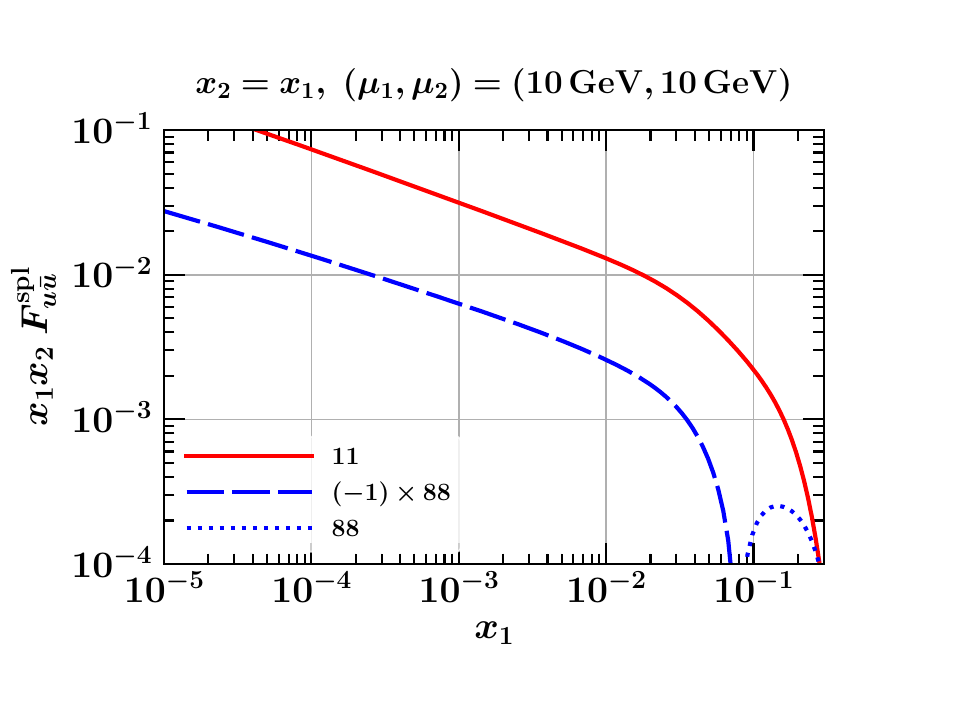}
}
\\[1.5em]
\subfloat[$x_2 = 0.3$ at $\mu_1 = \mu_2 = \mustar$]{
   \includegraphics[width=0.48\textwidth,trim=0 25 35
50,clip]{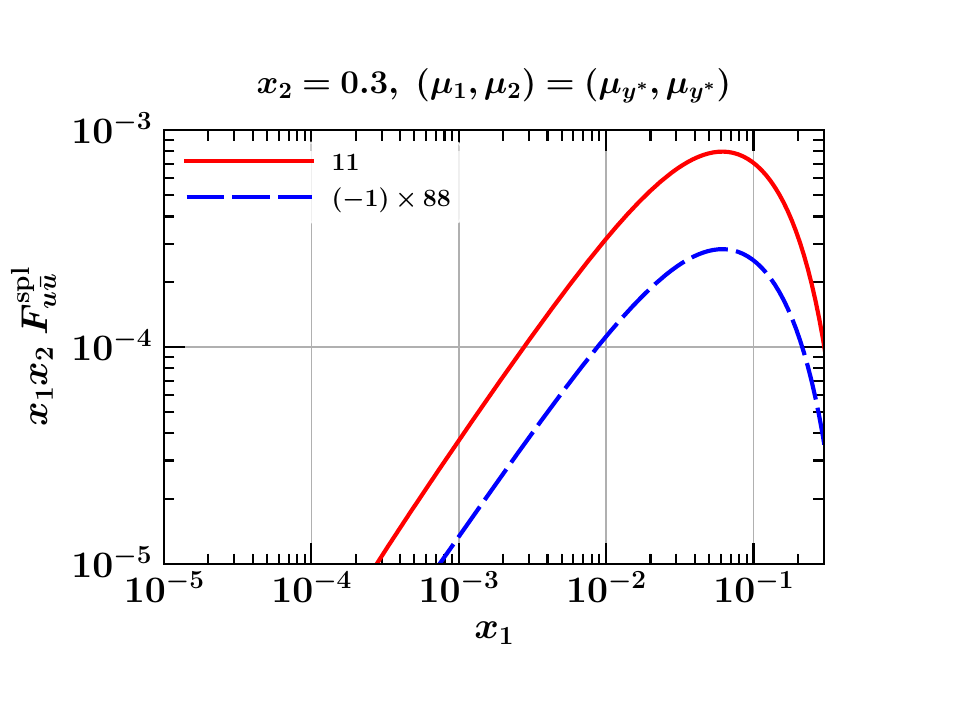}
}
\subfloat[$x_2 = 0.3$ at $\mu_1 = \mu_2 = 10 \gev$]{
   \includegraphics[width=0.48\textwidth,trim=0 25 35
50,clip]{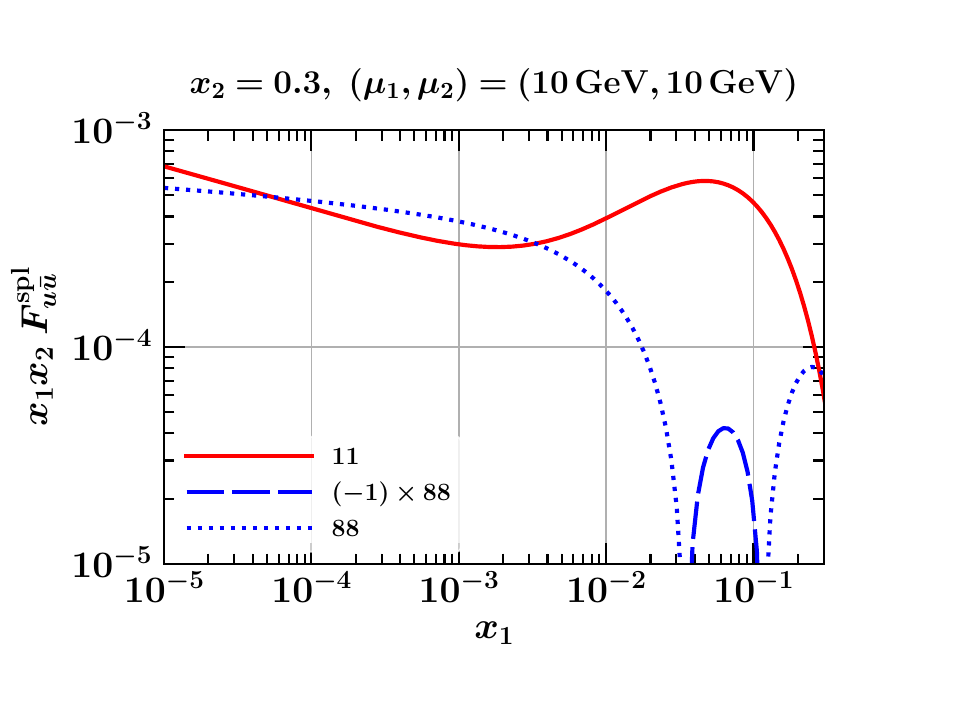}
}
\\[1.5em]
\subfloat[$x_2 = 10^{-3}$ at $\mu_1 = \mu_2 = \mustar$]{
   \includegraphics[width=0.48\textwidth,trim=0 25 35
50,clip]{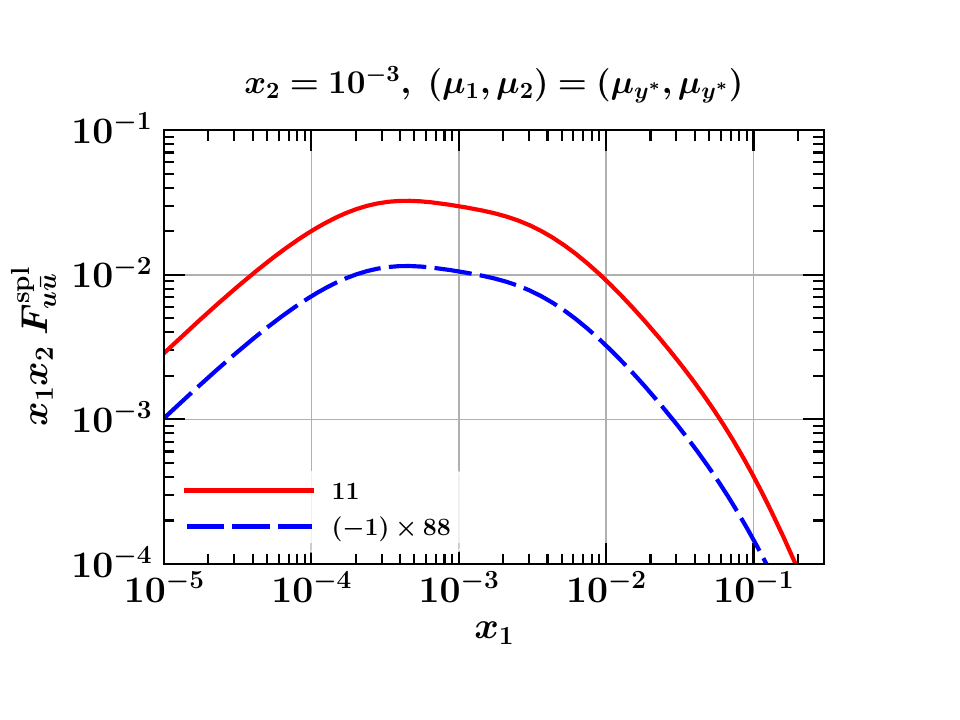}
}
\subfloat[$x_2 = 10^{-3}$ at $\mu_1 = \mu_2 = 10 \gev$]{
   \includegraphics[width=0.48\textwidth,trim=0 25 35
50,clip]{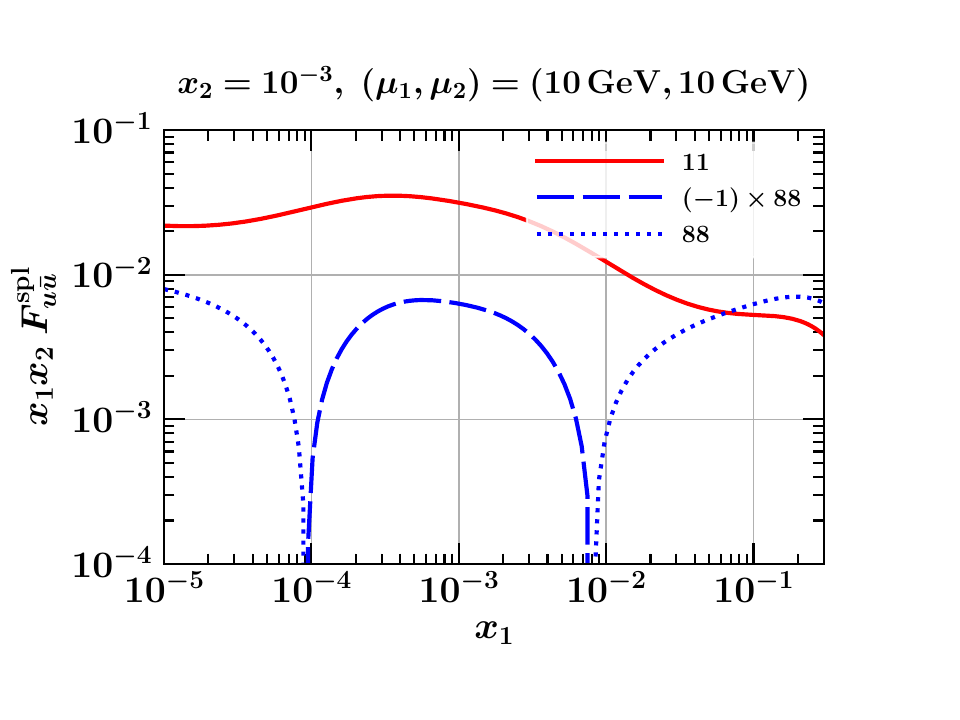}
}
\caption{\label{fig:F_uubar-x2} $F_{u\bar{u}}^{\text{spl}}$ as a function of
$x_1$ for different values of $x_2$.  The values of $y$ and $\mustar$ for this
and the following figures are given in \protect\eqref{std-y-choice}.}
\end{figure}

Let us now take a closer look at DPDs evolved at LL accuracy from different
initial conditions.  From now on, we always show evolved DPDs for $n_f = 5$
flavours, with flavour matching carried out as described at the end of
\sect{\ref{sec:init-cond}}.
In \fig{\ref{fig:F_uubar-x2}} we compare the splitting part
$F_{u\bar{u}}^{\text{spl}}$ at its initial scale and at $\mu_1 = \mu_2 = 10
\gev$.  We observe an intricate pattern of sign changes in the evolved octet
distributions; the relative importance of $q \to q$ and $g \to q$ transitions
(and their analogues for antiquarks) can obviously change significantly in
different regions of $x_1$ and $x_2$.

%

%
\begin{figure}[p]
\centering
\subfloat[$u g$ at $\mu_1 = \mu_2 = \mustar$]{
   \includegraphics[width=0.48\textwidth,trim=0 25 35 50,clip]{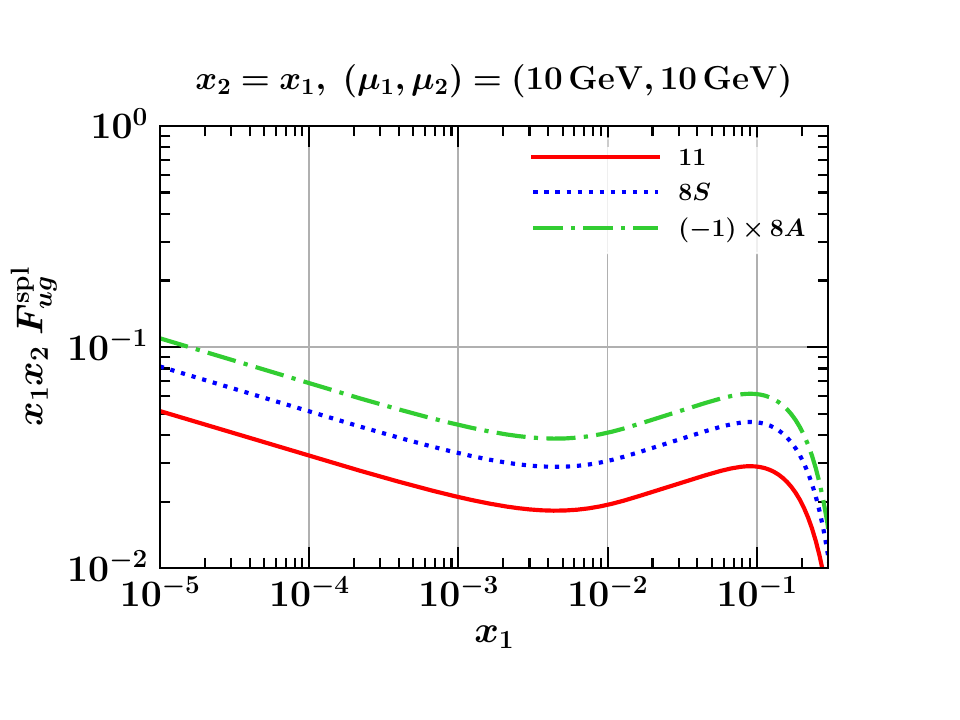}
}
\subfloat[$u g$ at $\mu_1 = \mu_2 = 10 \gev$]{
   \includegraphics[width=0.48\textwidth,trim=0 25 35
50,clip]{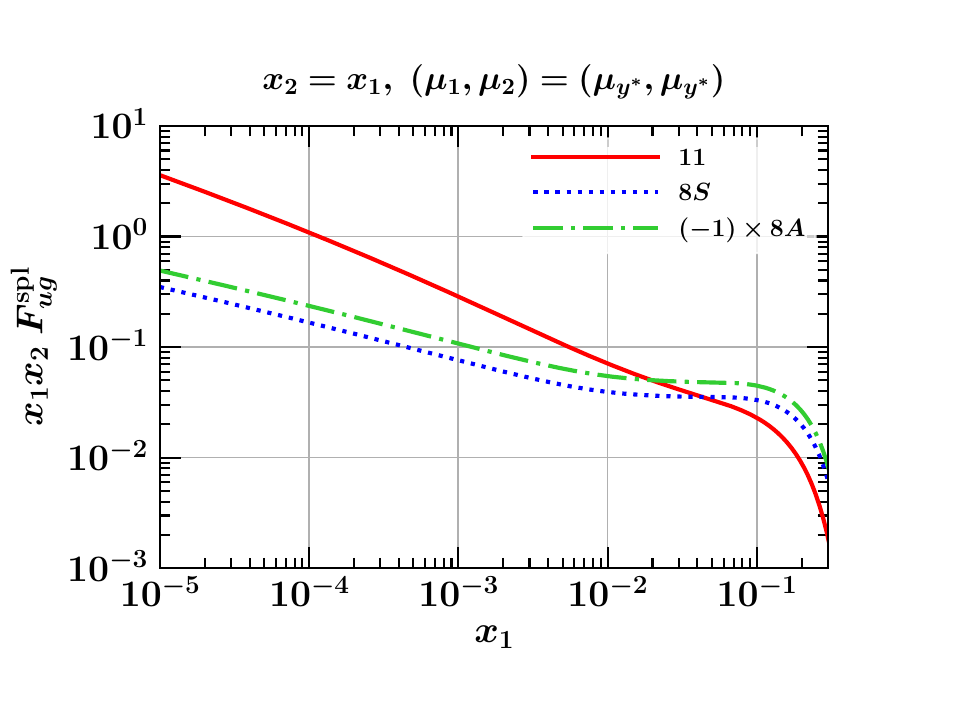}
}
\\[1.5em]
\subfloat[$\bar{u} g$ at $\mu_1 = \mu_2 = \mustar$]{
   \includegraphics[width=0.48\textwidth,trim=0 25 35
50,clip]{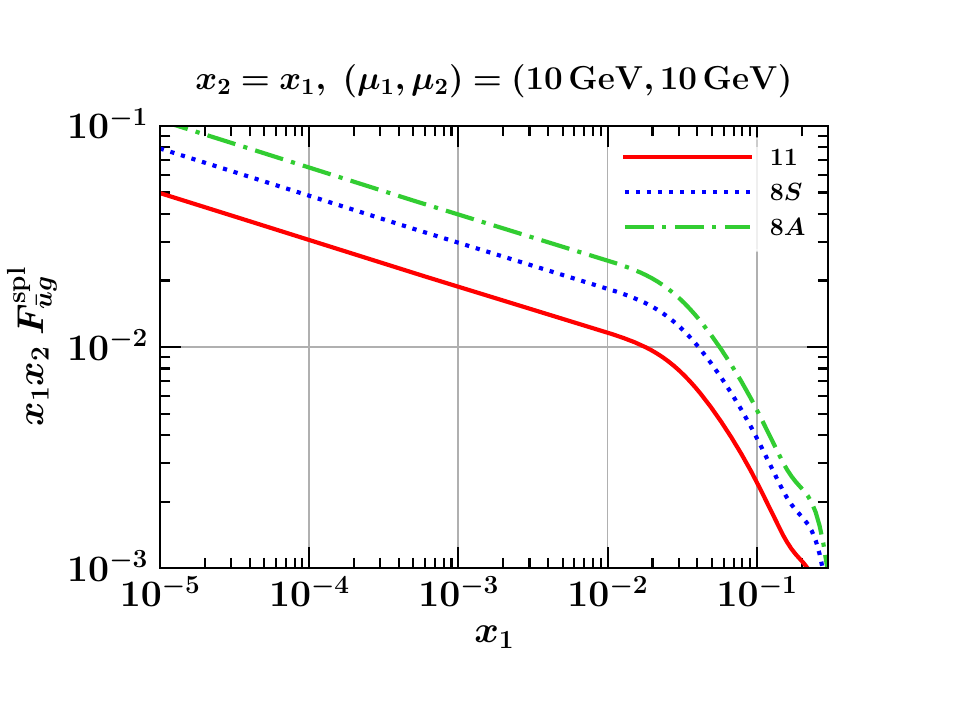}
}
\subfloat[$\bar{u} g$ at $\mu_1 = \mu_2 = 10 \gev$]{
   \includegraphics[width=0.48\textwidth,trim=0 25 35
50,clip]{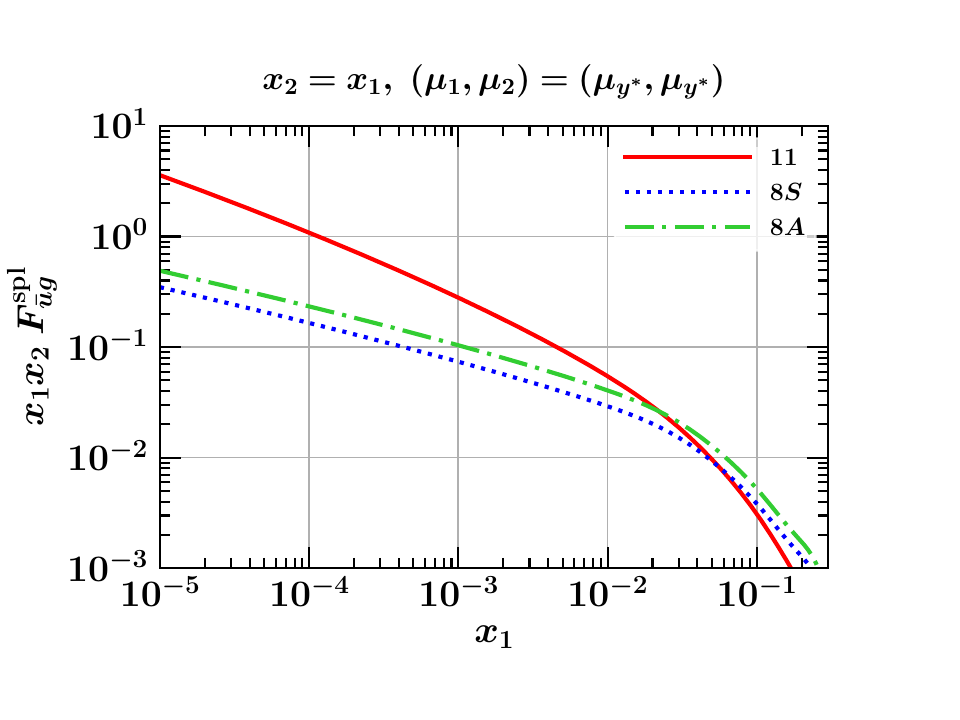}
}
\caption{\label{fig:F_ug} $F_{u g}^{\text{spl}}$ and $F_{\bar{u}
g}^{\text{spl}}$ at $x_1 = x_2$.}
\end{figure}
\begin{figure}[p]
\centering
\subfloat[$u \bar{d}$ at $\mu_1 = \mu_2 = 10 \gev$]{
   \includegraphics[width=0.48\textwidth,trim=0 25 35
50,clip]{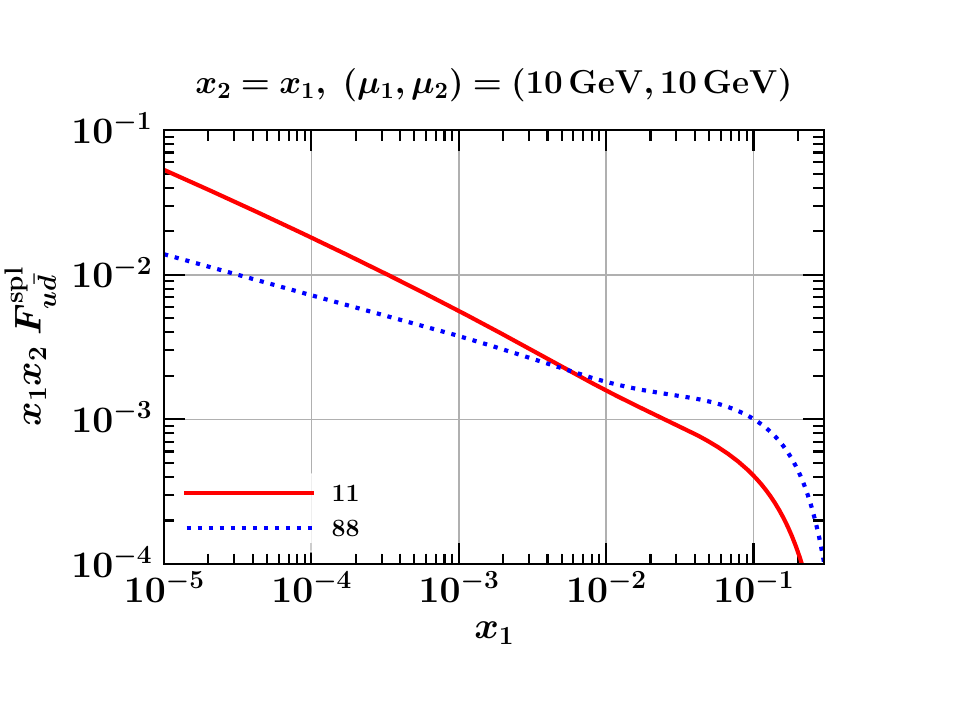}
}
\subfloat[$b \bar{b}$ at $\mu_1 = \mu_2 = 10 \gev$]{
   \includegraphics[width=0.48\textwidth,trim=0 25 35
50,clip]{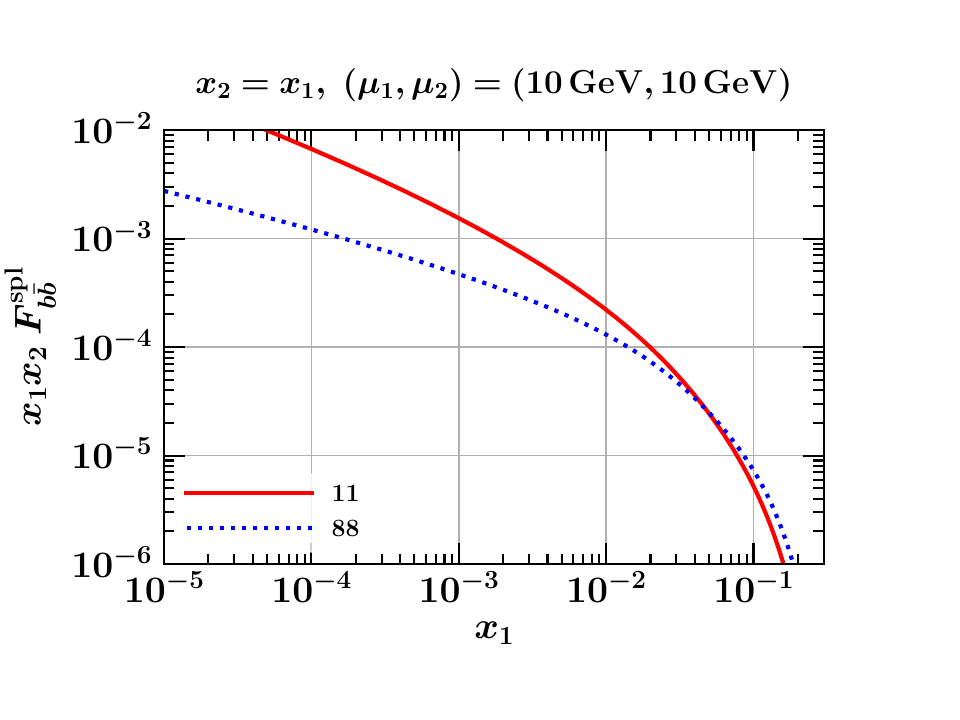}
}
\caption{\label{fig:F_flavours} Splitting DPDs at $x_1 = x_2$ for flavour
combinations that are zero at the initial scale $\mustar$.}
\end{figure}

%
\begin{figure}
\centering
\subfloat[$\Delta g \Delta g$ at $\mu_1 = \mu_2 = \mustar$]{
   \includegraphics[width=0.48\textwidth,trim=0 25 35
50,clip]{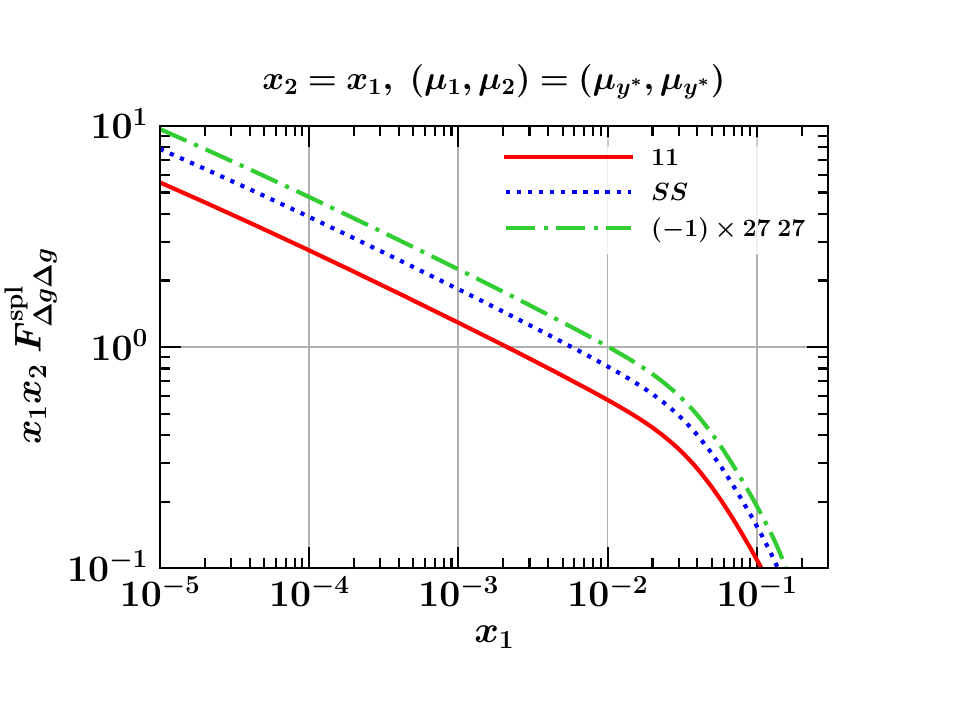}
}
\subfloat[$\Delta g \Delta g$ at $\mu_1 = \mu_2 = 10 \gev$]{
   \includegraphics[width=0.48\textwidth,trim=0 25 35
50,clip]{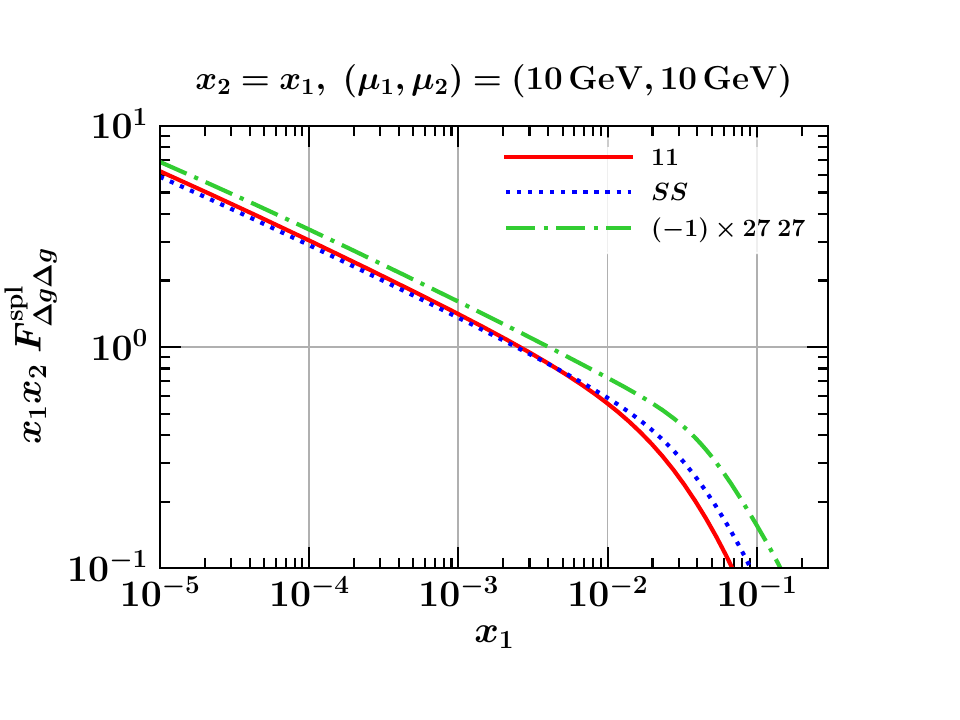}
}
\\[1.5em]
\subfloat[$\Delta u \Delta g$ at $\mu_1 = \mu_2 = \mustar$]{
   \includegraphics[width=0.48\textwidth,trim=0 25 35
50,clip]{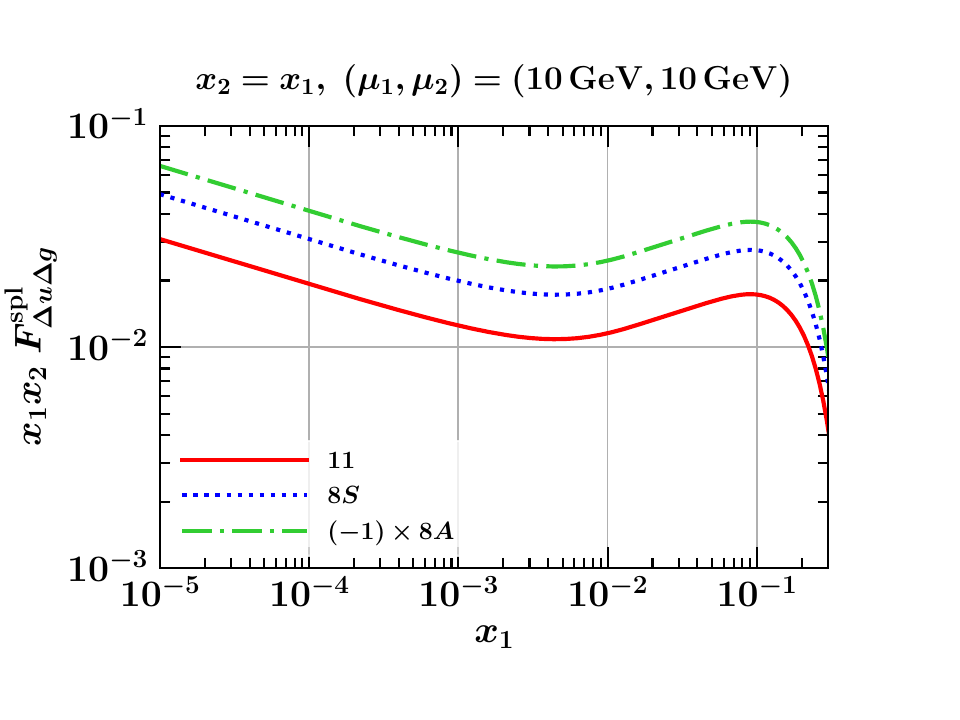}
}
\subfloat[$\Delta u \Delta g$ at $\mu_1 = \mu_2 = 10 \gev$]{
   \includegraphics[width=0.48\textwidth,trim=0 25 35
50,clip]{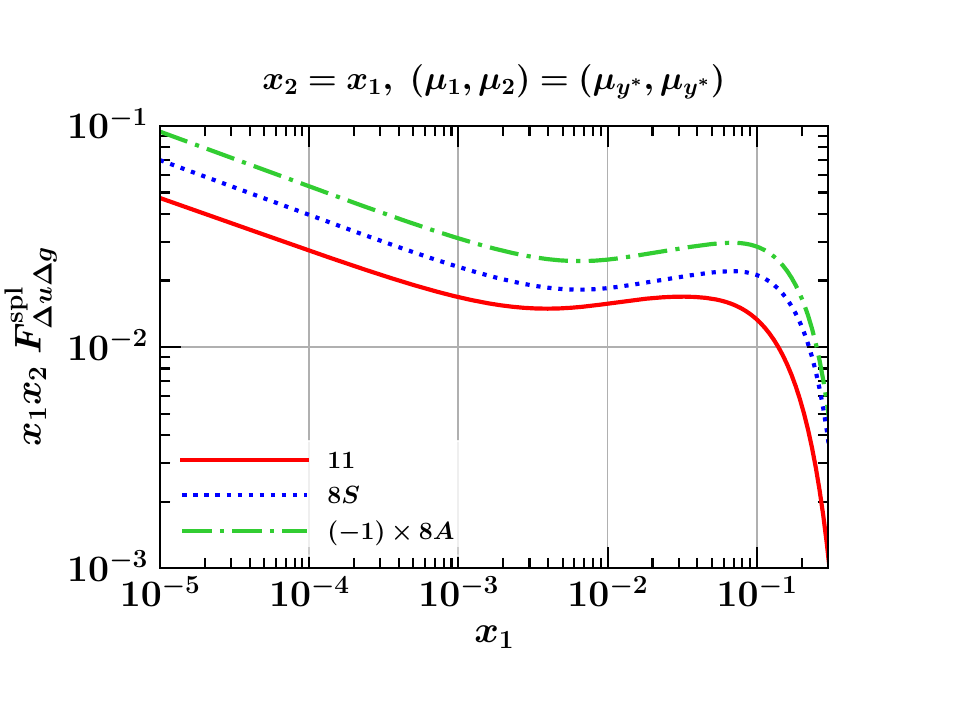}
}
\\[1.5em]
\subfloat[$\Delta u \Delta\bar{u}$ at $\mu_1 = \mu_2 = \mustar$]{
   \includegraphics[width=0.48\textwidth,trim=0 25 35
50,clip]{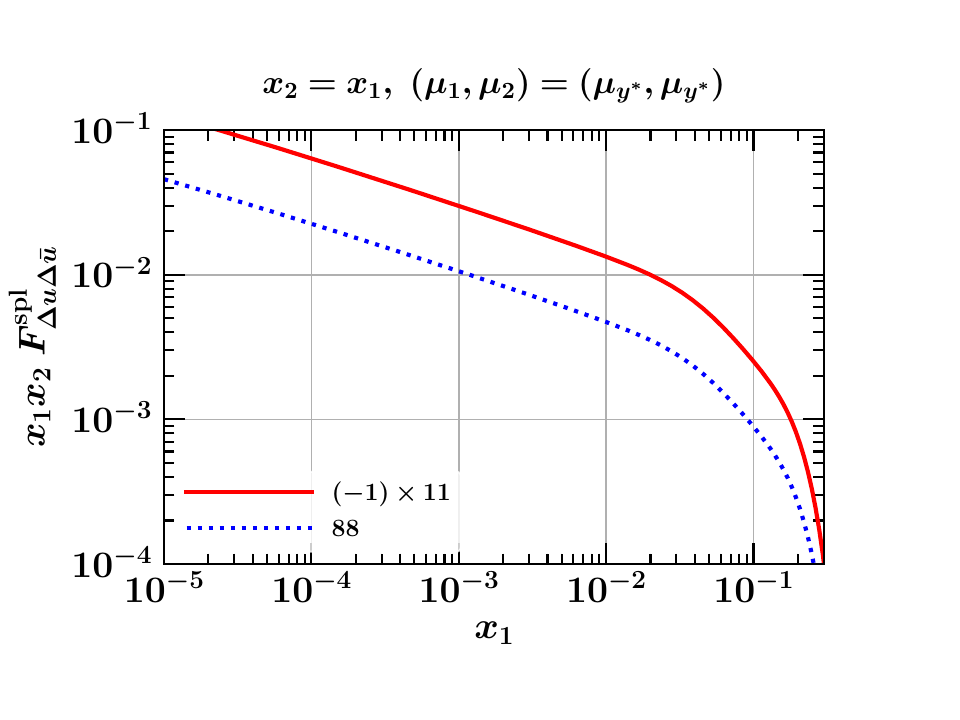}
}
\subfloat[$\Delta u \Delta\bar{u}$ at $\mu_1 = \mu_2 = 10 \gev$]{
   \includegraphics[width=0.48\textwidth,trim=0 25 35
50,clip]{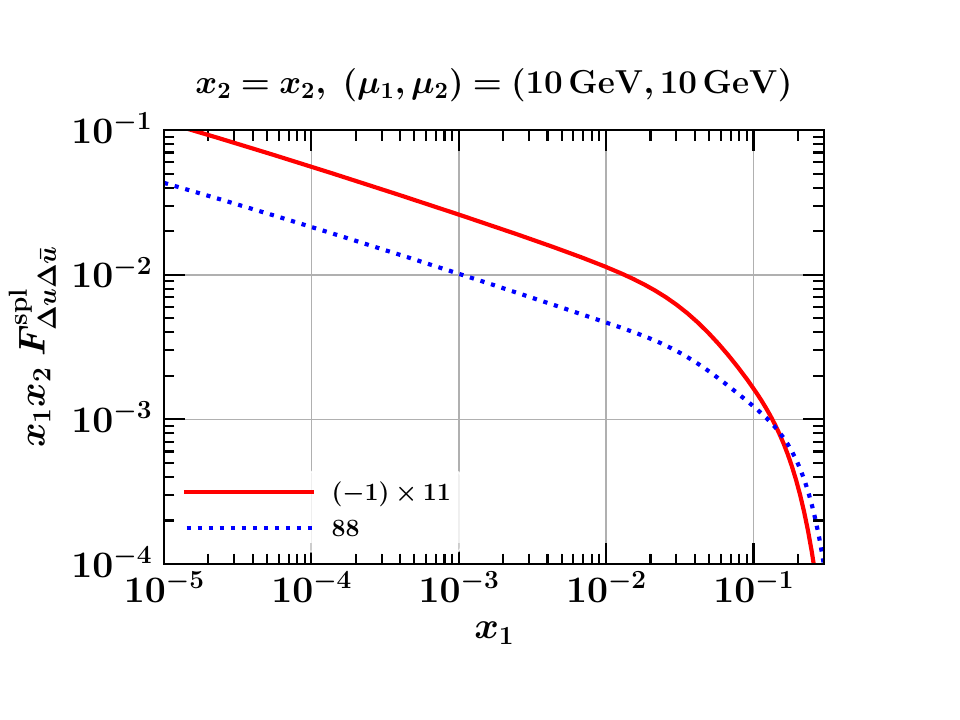}
}
\caption{\label{fig:F_longit} Splitting DPDs at $x_1 = x_2$ for longitudinally
polarised partons.}
\end{figure}

In \fig{\ref{fig:F_ug}}, we show mixed quark-gluon and antiquark-gluon
distributions.  At $\mu_1 = \mu_2 = 10 \gev$ the two octet combinations are
smaller than the singlet for small $x_1 = x_2$, whereas at large momentum
fractions they are larger.   The relative hierarchy between the two octet
channels does not change between the starting scale and $10 \gev$.

Examples of evolved splitting DPDs that are zero at the starting scale are shown
in \fig{\ref{fig:F_flavours}}: the combination $u \bar{d}$ is not generated by
LO splitting, whilst $b \bar{b}$ is not an active flavour pair at the starting
scale and only generated by radiation at scales above $m_b$.  After evolution,
the relative size between colour octet and colour singlet channels for
these distributions is qualitatively similar to the $u g$ and $\bar{u} g$
distributions in \fig{\ref{fig:F_ug}}.

DPDs for longitudinally polarised partons are shown in \fig{\ref{fig:F_longit}}.
At small momentum fractions, they evolve much less rapidly than unpolarised
distributions.  The reason is that the splitting kernel $P_{g g}(z)$ grows like
$1/z$ at low $z$ whilst $P_{\Delta g \Delta g}(z)$ tends to a constant in
that limit.  This directly impacts $F_{\Delta g \Delta g}$ and indirectly the
distributions whose evolution is gluon driven.  These observations apply to all
colour channels.

\begin{figure}[p]
\centering
\subfloat[splitting form, $\mu_1 = \mu_2 = \mustar$]{
   \includegraphics[width=0.48\textwidth,trim=0 25 35
48,clip]{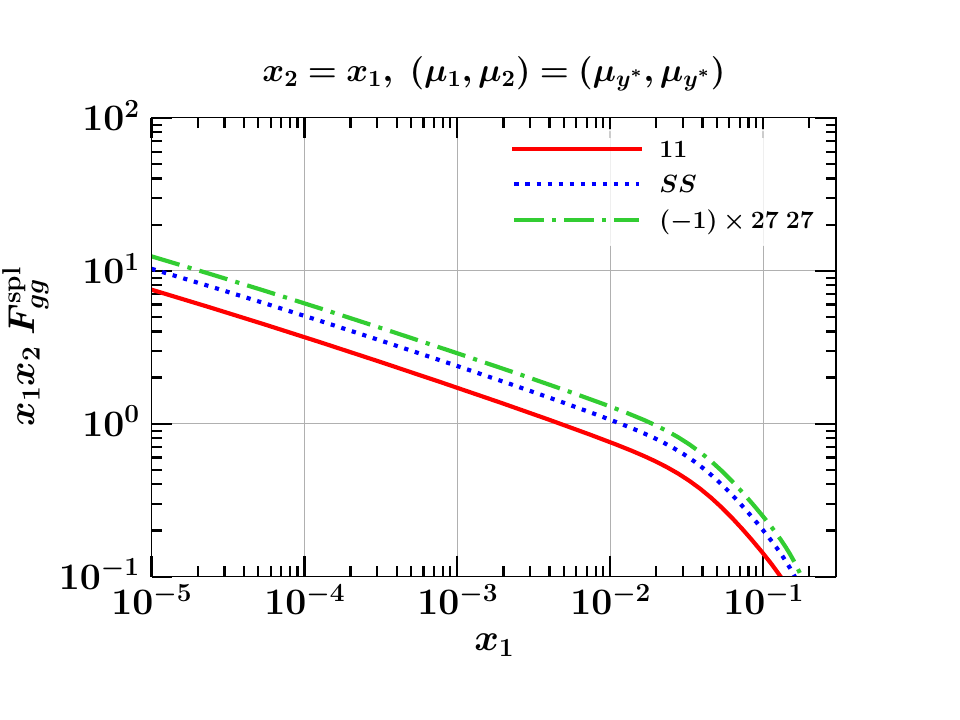}
}
\subfloat[splitting form, $\mu_1 = \mu_2 = 10 \gev$]{
   \includegraphics[width=0.48\textwidth,trim=0 25 35
48,clip]{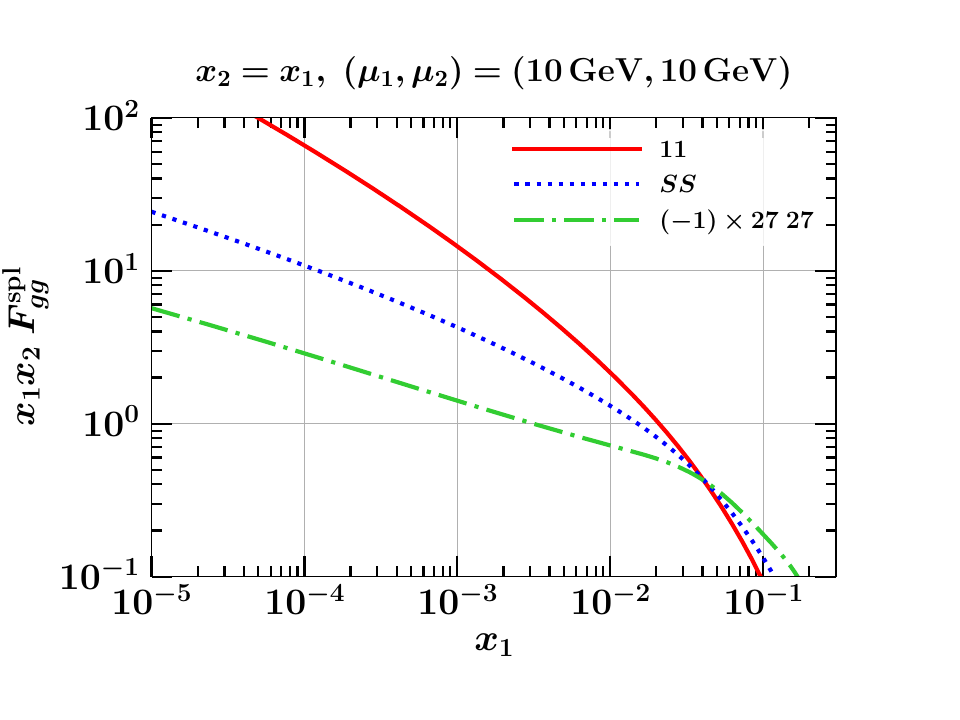}
}
\\[1.5em]
\subfloat[product form, $\mu_1 = \mu_2 = \mustar$]{
   \includegraphics[width=0.48\textwidth,trim=0 25 35
48,clip]{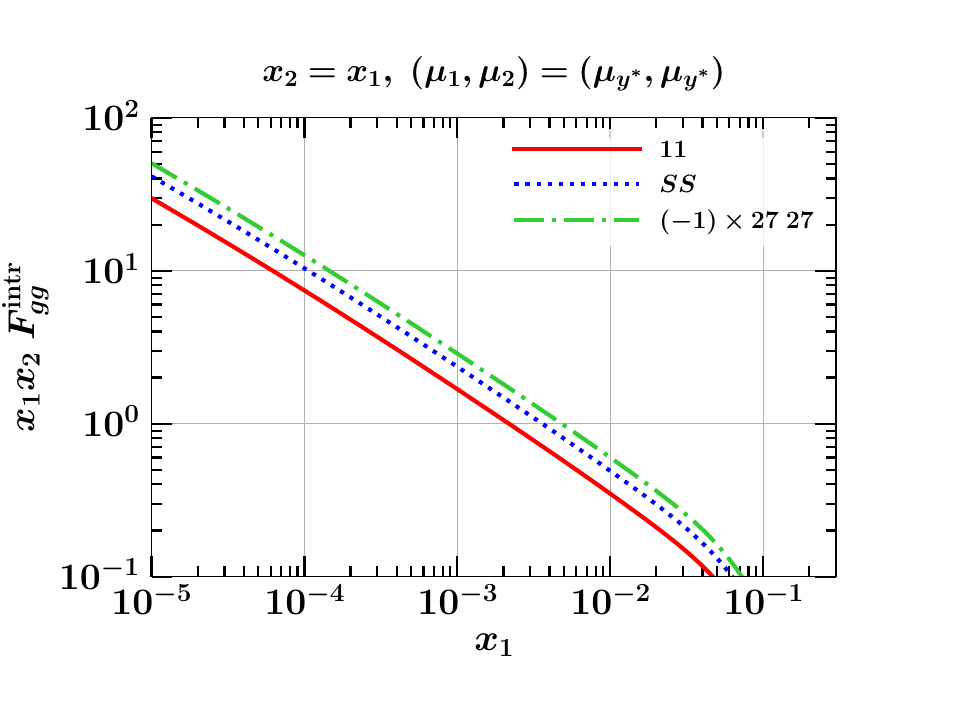}
}
\subfloat[product form, $\mu_1 = \mu_2 = 10 \gev$]{
   \includegraphics[width=0.48\textwidth,trim=0 25 35
48,clip]{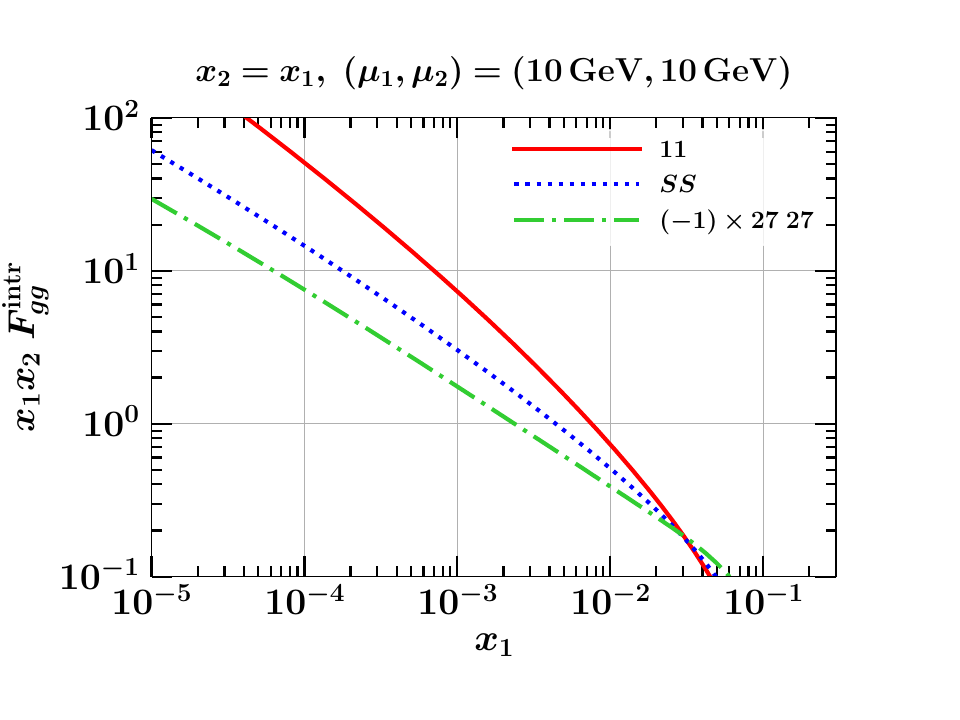}
}
\\[1.5em]
\subfloat[alternative product form, $\mu_1 = \mu_2 = \mustar$]{
   \includegraphics[width=0.48\textwidth,trim=0 25 35
50,clip]{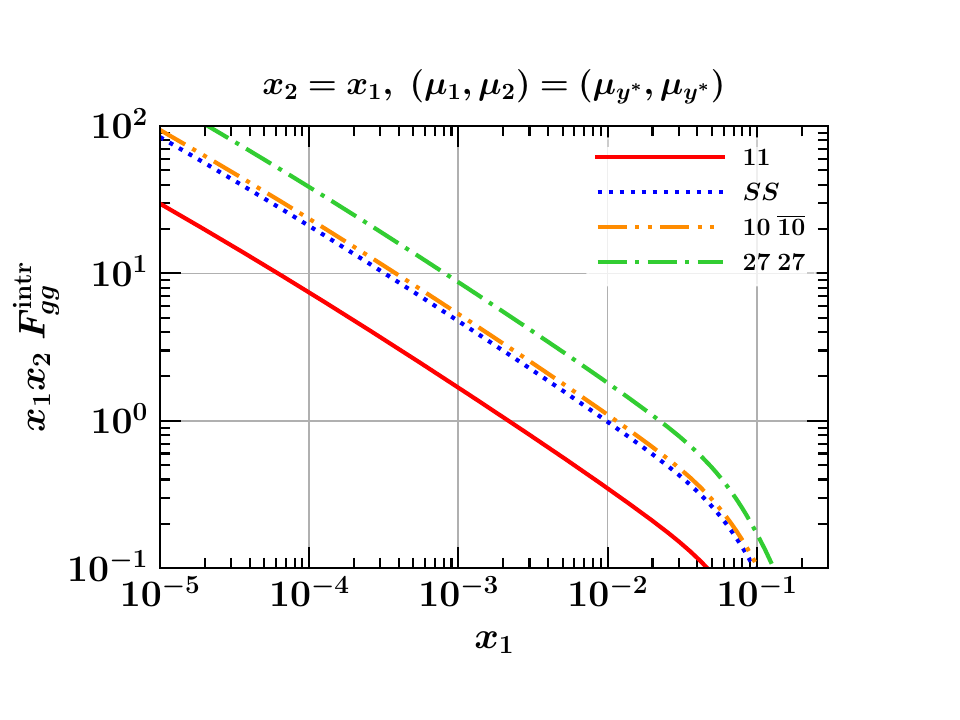}
}
\subfloat[alternative product form, $\mu_1 = \mu_2 = 10 \gev$]{
   \includegraphics[width=0.48\textwidth,trim=0 25 35
50,clip]{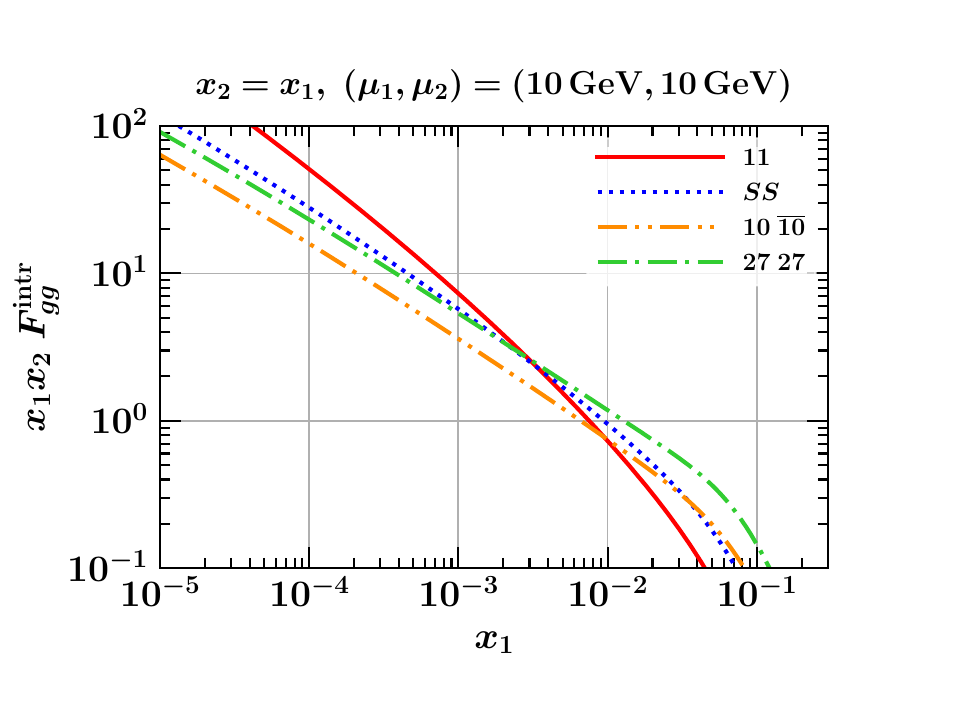}
}
\caption{\label{fig:F_gg_models} $F_{g g}$ at $x_1 = x_2$ for different initial
conditions.}
\end{figure}

The impact of the initial conditions on the evolved result is illustrated in
\fig{\ref{fig:F_gg_models}}, where we show $F_{g g}$ for the splitting form (top
row), for the product ansatz with our default choice of colour factors in
\eqref{splitting-colour-factors}, \eqref{splitting-colour-factors-gg}, and
\eqref{extra-colour-factors} (middle row), and for the product ansatz with the
alternative choice of colour factors in \eqref{splitting-colour-factors},
\eqref{extra-colour-factors}, and \eqref{alt-colour-factors}.  In the latter
case, we have nonzero decuplet distributions.  Notice the change in relative
size of the distributions with $R=10$ and $R=27$ under evolution, which is a
consequence of the stronger Sudakov suppression for higher dimensional
colour representations.

%

\begin{figure}[p]
\centering
\subfloat[splitting form, $\mu_1 = \mu_2 = \mustar$]{
   \includegraphics[width=0.48\textwidth,trim=0 22 35
48,clip]{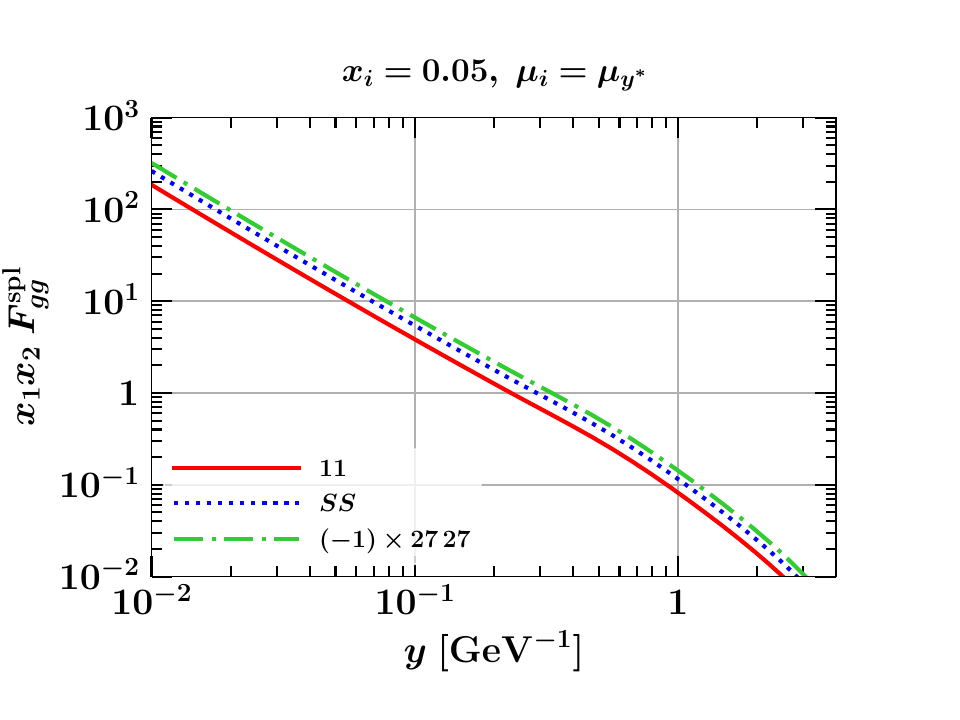}
}
\subfloat[splitting form, $\mu_1 = \mu_2 = 80 \gev$]{
   \includegraphics[width=0.48\textwidth,trim=0 22 35
48,clip]{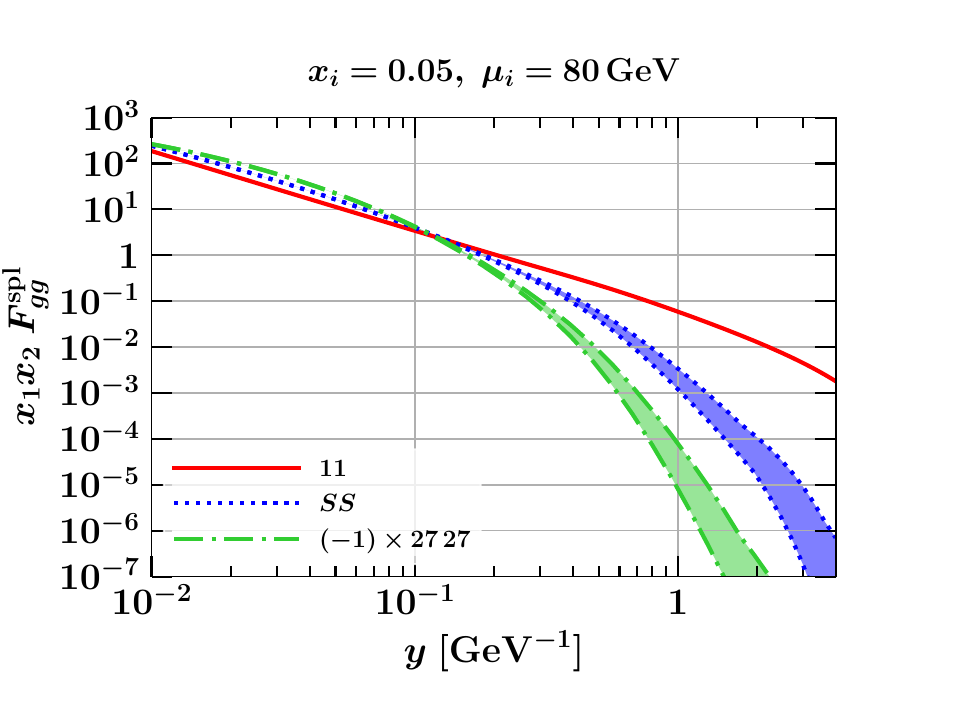}
}
\\[1.5em]
\subfloat[product form, $\mu_1 = \mu_2 = \mustar$]{
   \includegraphics[width=0.48\textwidth,trim=0 22 35
48,clip]{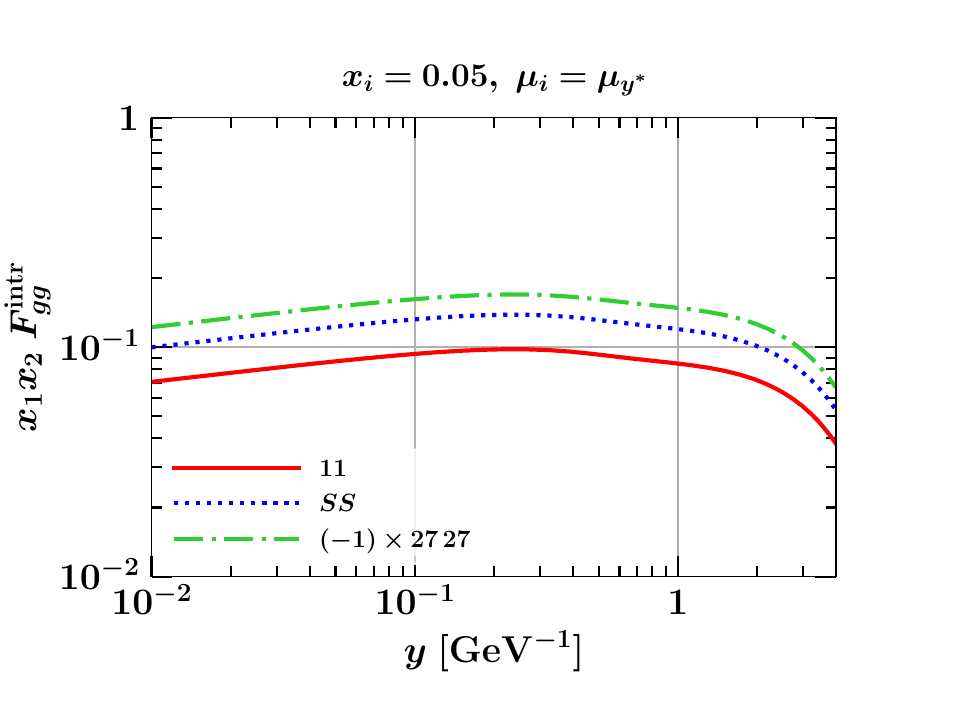}
}
\subfloat[product form, $\mu_1 = \mu_2 = 80 \gev$]{
   \includegraphics[width=0.48\textwidth,trim=0 22 35
48,clip]{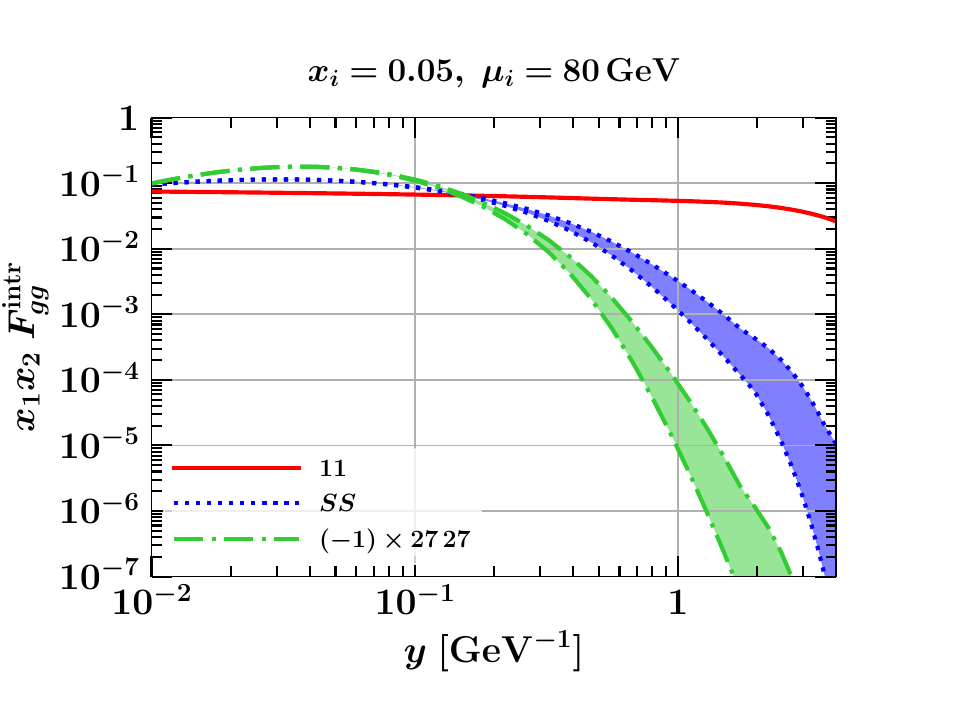}
}
\\[1.5em]
\subfloat[alternative product form, $\mu_1 = \mu_2 = \mustar$]{
   \includegraphics[width=0.48\textwidth,trim=0 22 35
48,clip]{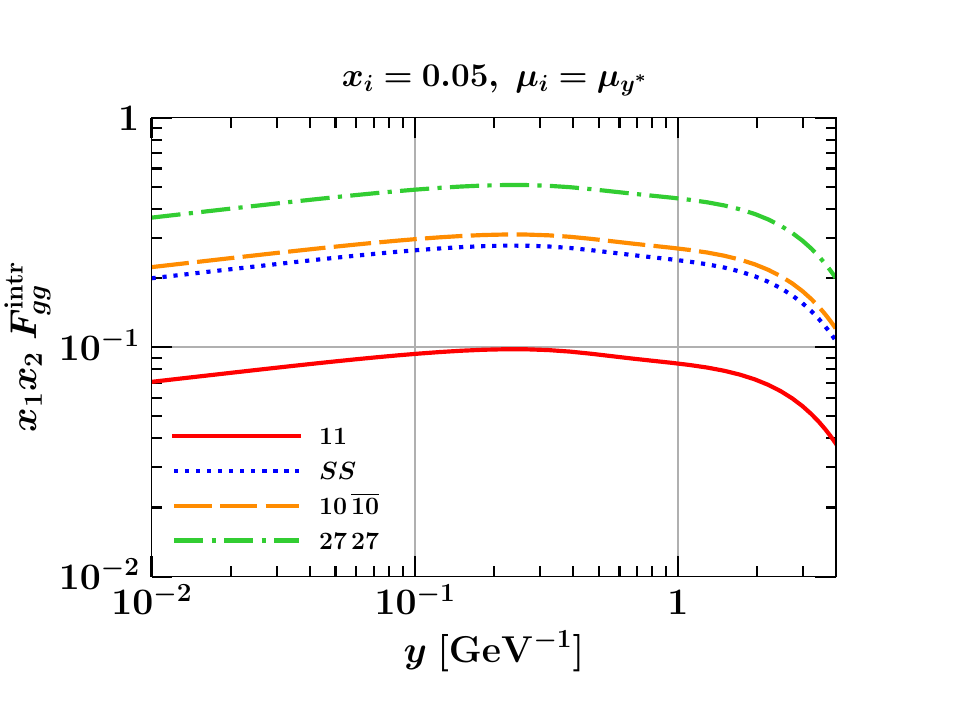}
}
\subfloat[alternative product form, $\mu_1 = \mu_2 = 80 \gev$]{
   \includegraphics[width=0.48\textwidth,trim=0 22 35
48,clip]{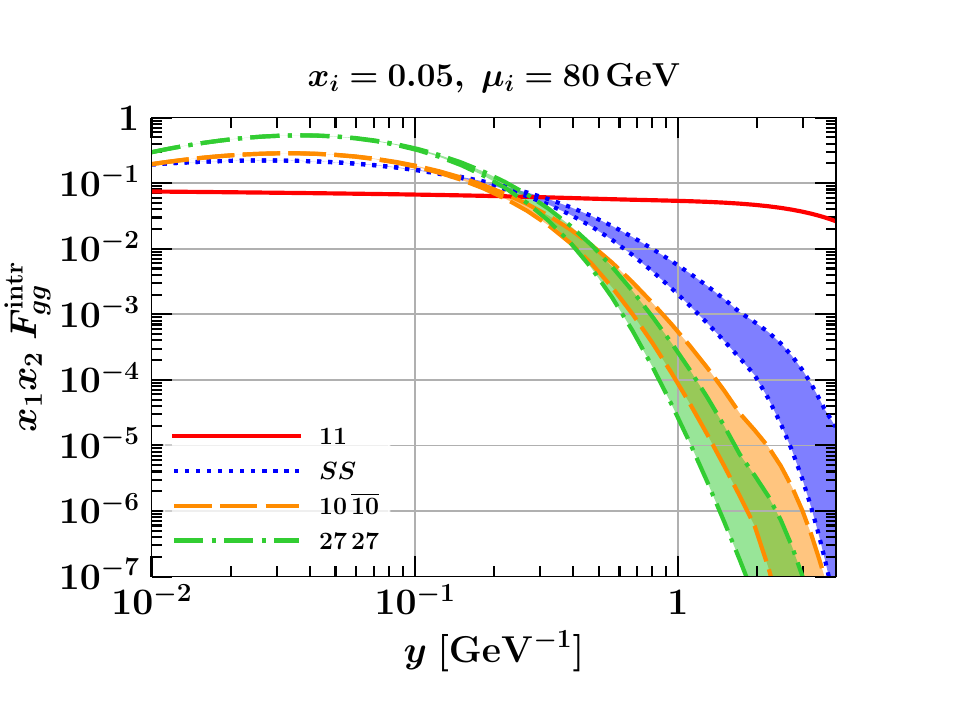}
}
\caption{\label{fig:F-vs-y} $F_{g g}$ at $x_1 = x_2 = 0.05$ for the same set of
initial conditions as in \fig{\protect\ref{fig:F_gg_models}}.  The bands are the
envelopes of results obtained with the different fits in
\tab{\protect\ref{tab:Delta_Kq}} when modelling the non-perturbative part of the
Collins-Soper kernel.  \rev{Notice that the final scale here is $80 \gev$ rather
than $10 \gev$, which enhances the effect of Collins-Soper evolution at large
$y$ (see \fig{\protect\ref{fig:sudavok-exp-J}}).}}
\end{figure}

So far we have focused on DPDs at $y = 0.5 \gev^{-1}$, where the
non-perturbative part $\Delta J$ of the Collins-Soper kernel is small according
to \fig{\ref{fig:CS-kernel}}b.  This changes rapidly with increasing $y$, where
colour non-singlet distributions receive an increasingly strong suppression from
the second exponential in \eqref{solution-2}, as is evident in
\fig{\ref{fig:F-vs-y}}.  To assess the model uncertainty of Collins-Soper
evolution in this region, we show bands that cover the smallest and largest
results obtained with the different TMD fits we use to construct $\Delta J$.  We
find that this uncertainty does not affect the overall picture.  Of course, the
bands do not reflect the model dependence due to our assumption of Casimir
scaling in \eqref{J-Casimir} and \eqref{K-Casimir}.

As one goes to small $y$, the initial scale $\mustar$ grows, and evolved DPDs
differ less and less from the initial conditions.  Note that the gluon DPDs
obtained with the product ansatz slightly decrease at the initial scale when
going to smaller $y$, because the gluon PDF at $x_1 + x_2 = 0.1$ decreases with
growing scale.

%


\subsection{Impact of higher orders}
\label{sec:dpds-higher-orders}

\paragraph{\NLLp\ accuracy.}  Let us now compare the DPDs evolved at LL, as done
so far, with the result of evolving at \NLLp, as defined in
\tab{\ref{tab:orders}}.  Going to \NLLp\ accuracy implies several changes:
\begin{itemize}
\item PDFs are evolved at NLO instead of LO.  This changes the initial
conditions of DPDs, both in the splitting part \eqref{splitting-DPD-pt} and in
the ansatz \eqref{int-DPD-singlet} for the intrinsic part.  As discussed in
\sect{\ref{sec:orders}}, we stay at LO in the evaluation of the splitting
formula for practical reasons.
\item DPDs are evolved with NLO kernels $P$ and with the two-loop expressions
of $\gamma_J$ and the $\beta$ function.  The Collins-Soper kernel at the input
scale $\mustar$ does not change, as mentioned in \sect{\ref{sec:orders}}.
\item $\alpha_s$ is taken from the PDF fit, which changes the order of its
running and its value at the $Z$ mass (see \eqref{as-lo} and \eqref{as-nlo}).
This affects both the evolution of the DPDs and the initial conditions for
$F^{\text{spl}}$.
\end{itemize}
In the following, we present the comparison for the splitting DPDs
$F^{\text{spl}}$.  We find effects of similar size for the intrinsic part with
our ansatz for the initial conditions.

In order to distinguish the effects of changed initial conditions and changed
evolution, we introduce the single ratio
\begin{align}
   \label{r-ratio-NLLp}
   \prn{R_1 R_2}{r}_{a_1 a_2,\ms\text{\NLLp}\ms}(\ldots; \mu_1, \mu_2)
   &=
   \frac{\prn{R_1 R_2}{F}_{a_1 a_2,\ms\text{\NLLp}\ms}\bigl(
      \ldots; \mu_1, \mu_2, \mu_1 \mu_2/(x_1 x_2) \bigr)}{
      \prn{R_1 R_2}{F}_{a_1 a_2,\ms\text{LL}}\bigl(
      \ldots; \mu_1, \mu_2, \mu_1 \mu_2/(x_1 x_2) \bigr)}
\end{align}
and the double ratio
\begin{align}
   \label{rho-ratio-NLLp}
   \pr{R_1 R_2}{\rho}_{a_1 a_2,\ms\text{\NLLp}\ms}^{}(\ldots; \mu_1, \mu_2)
   &=
   \frac{\prn{R_1 R_2}{r}_{a_1 a_2,\ms\text{\NLLp}\ms}(\ldots; \mu_1, \mu_2)}{
      \prn{R_1 R_2}{r}_{a_1 a_2,\ms\text{\NLLp}\ms}(\ldots; \mustar, \mustar)}
   \;,
\end{align}
where the ellipsis denotes the arguments $x_1, x_2$, and $y$.  The single ratio
\eqref{r-ratio-NLLp} at scale $\mustar$ probes only the initial conditions,
whereas in the double ratio \eqref{rho-ratio-NLLp} the change in initial
conditions partially (but not fully) cancels out.

\begin{figure}
\centering
\includegraphics[width=0.5\textwidth,trim=0 20 45
48,clip]{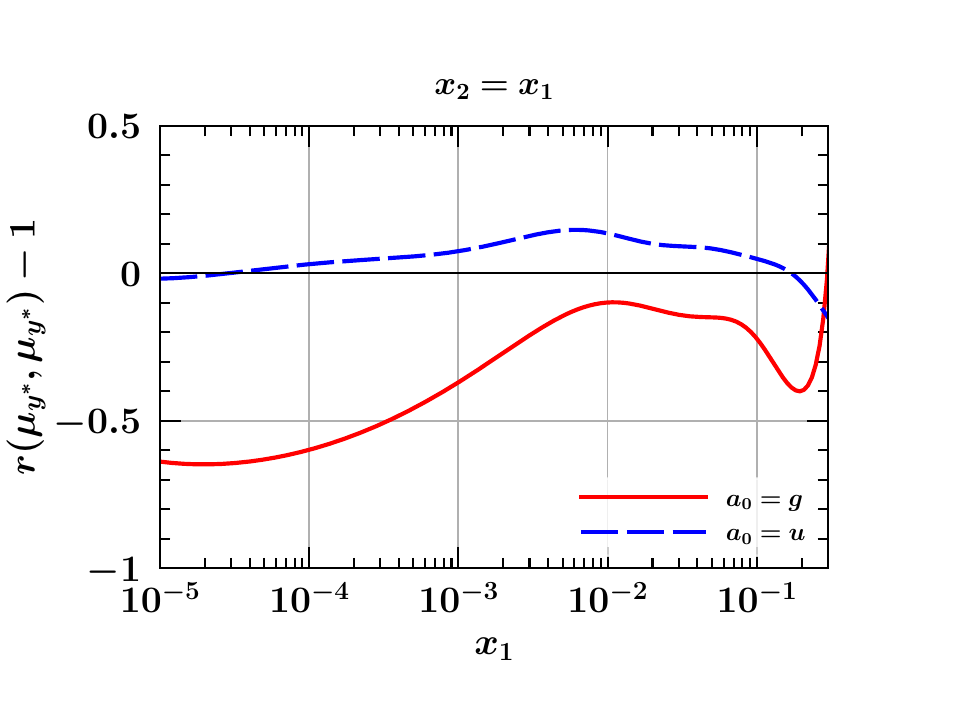}
\caption{\label{fig:r-ratio} The ratio \protect\eqref{r-ratio-initial} evaluated
for $x_1 = x_2$ and $y = 0.5 \gev^{-1}$. Shown is $r - 1$, such that the zero
line is where the DPDs are equal at LL and \NLLp.}
\end{figure}

In \fig{\ref{fig:r-ratio}} we show $r_{\text{\NLLp}}$ at the initial scale
$\mustar$.  For the LO splitting formula \eqref{splitting-DPD-pt}, this ratio is
given by
\begin{align}
   \label{r-ratio-initial}
   \prn{R_1 R_2}{r}_{a_1 a_2,\ms\text{\NLLp}\ms}(x_1, x_2, y; \mustar, \mustar)
   &=
   \frac{\bigl[ \alpha_s(\mustar) \, f_{a_0}(x_1 + x_2; \mustar)
            \ms\bigr]_{\text{NLO}}}{
         \bigl[ \alpha_s(\mustar) \, f_{a_0}(x_1 + x_2; \mustar)
            \ms\bigr]_{\text{LO}}}
\end{align}
and thus colour independent.  The strongest change when going from LO to NLO
PDFs is seen for the gluon distribution at small $x$, which is well known.

\begin{figure}[p]
\centering
\subfloat[$g g$]{
   \includegraphics[width=0.46\textwidth,trim=0 20 45
50,clip]{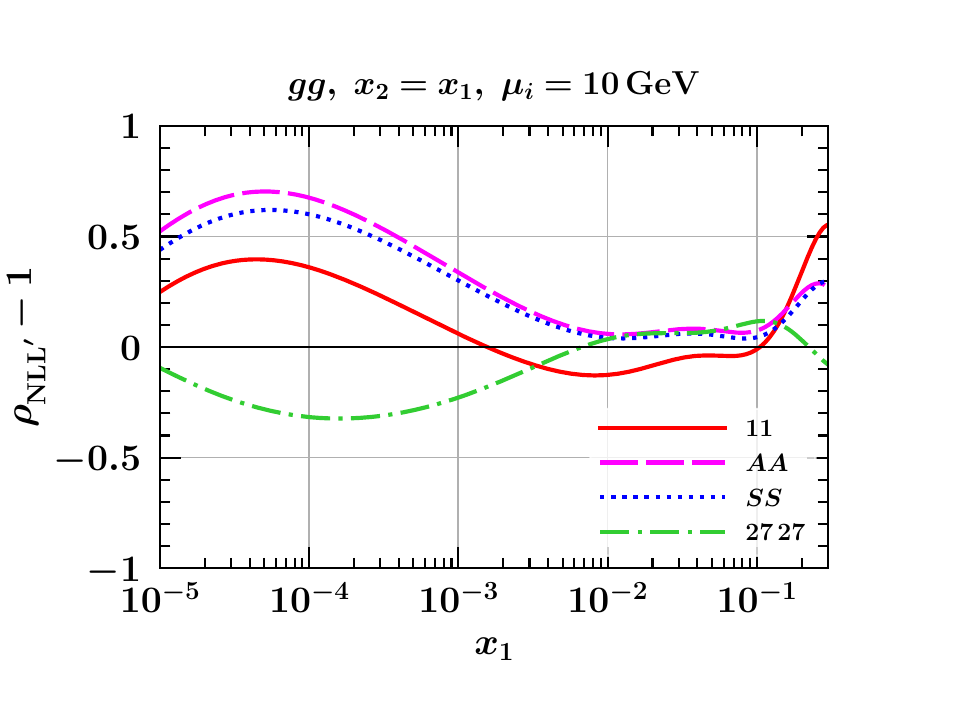}
}
\hspace{0.02\textwidth}
\subfloat[$\Delta g \Delta g$]{
   \includegraphics[width=0.46\textwidth,trim=0 20 45
50,clip]{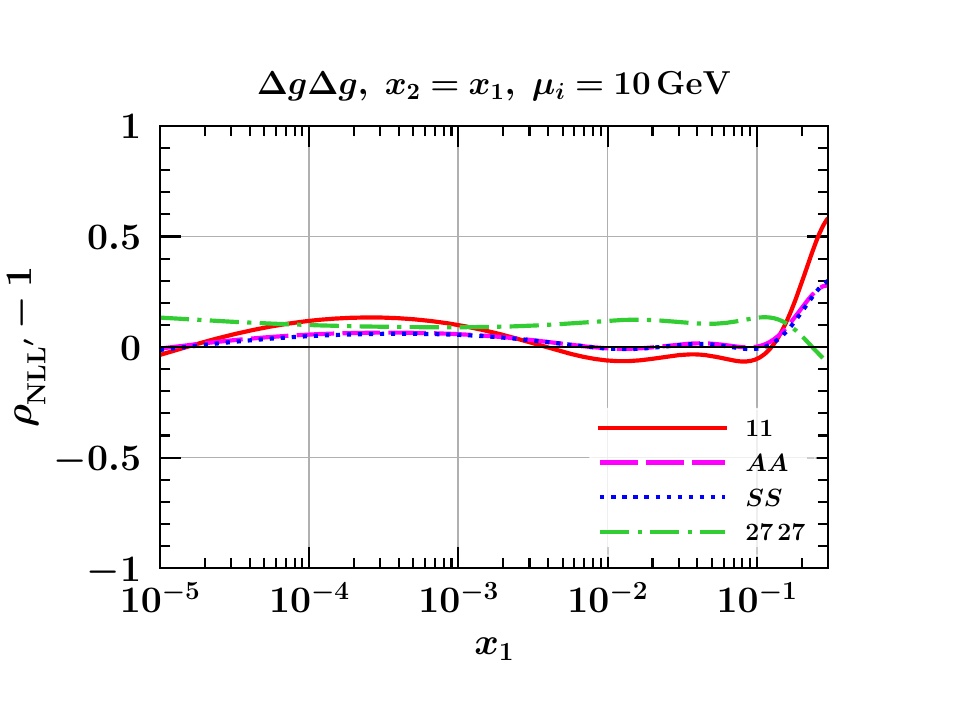}
}
\\[1.5em]
\subfloat[$u \bar{u}$]{
   \includegraphics[width=0.49\textwidth,trim=0 35 45
60,clip]{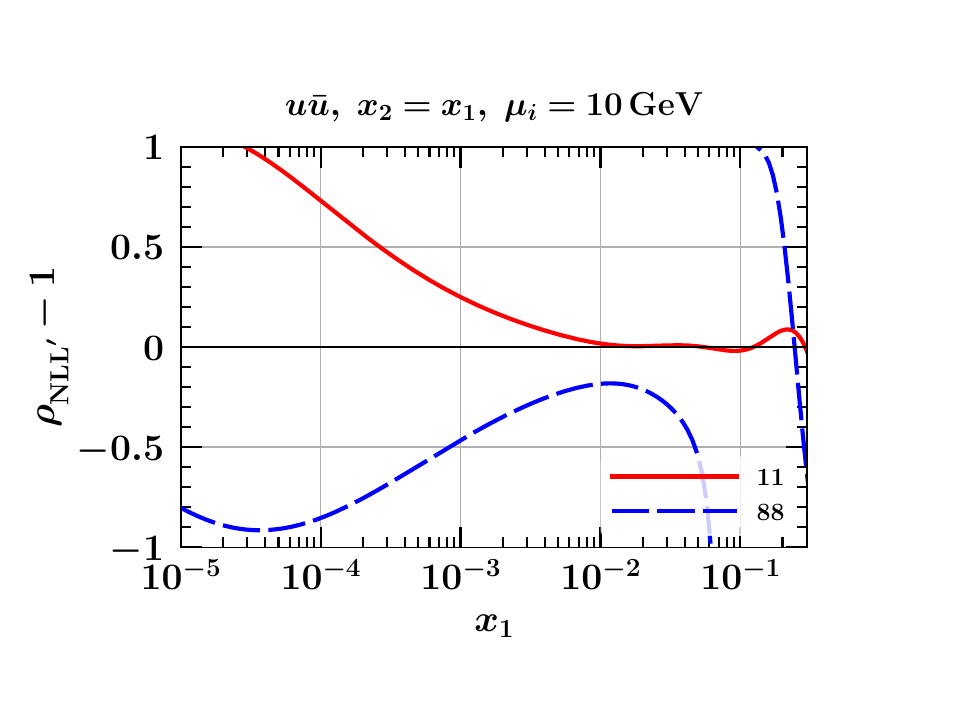}
}
\subfloat[$\Delta u \Delta\bar{u}$]{
   \includegraphics[width=0.49\textwidth,trim=0 35 45
60,clip]{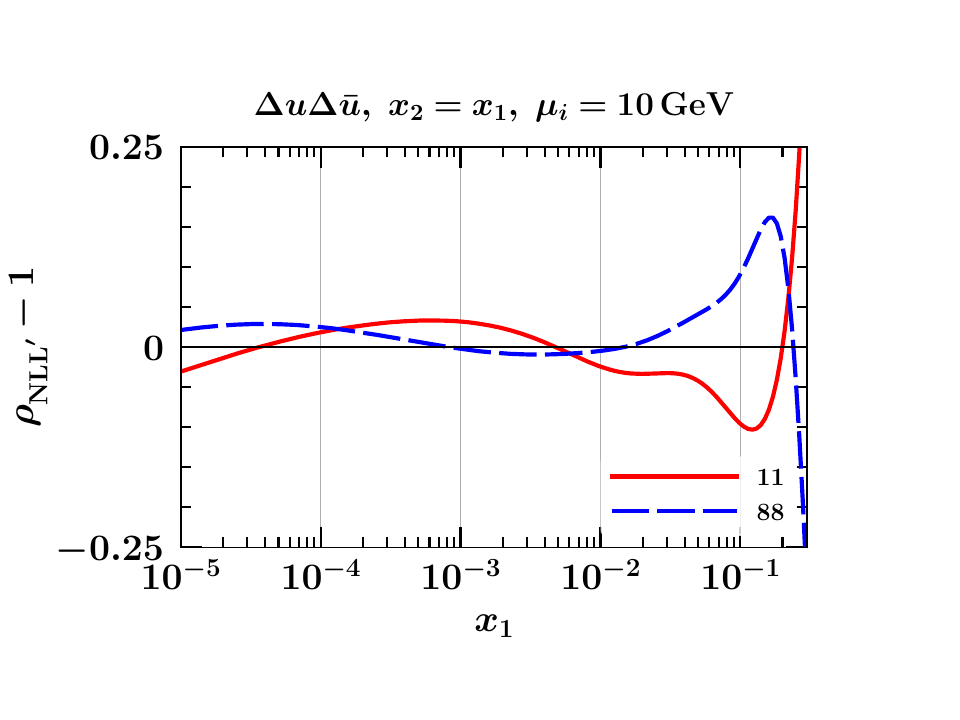}
}
\\[1.5em]
\subfloat[$u g$]{
   \includegraphics[width=0.49\textwidth,trim=0 35 45
60,clip]{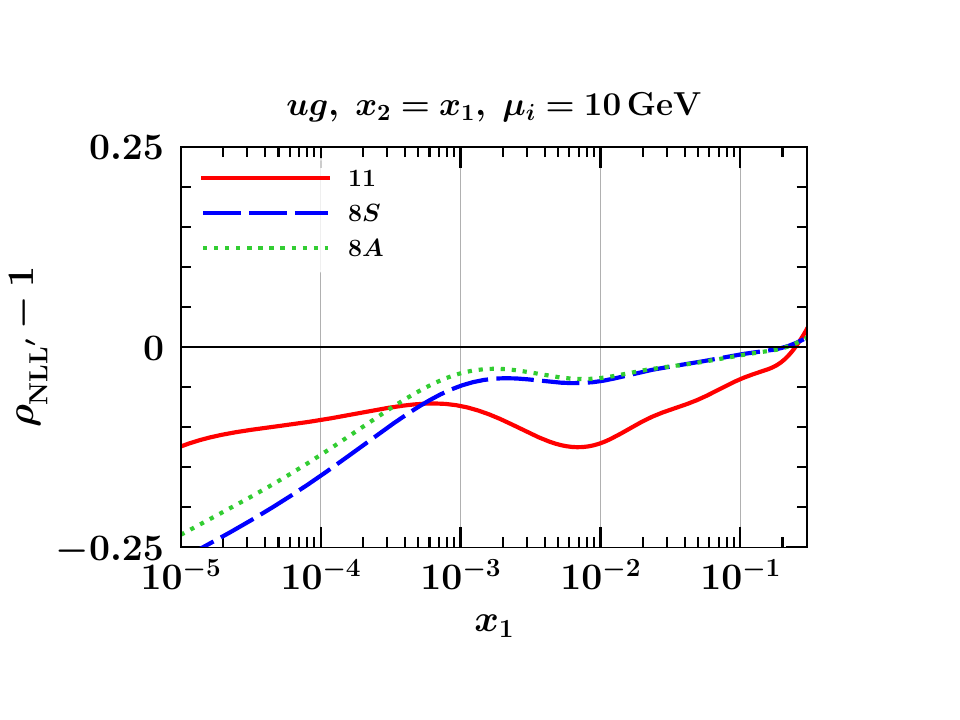}
}
\subfloat[$\delta u \delta\bar{u}$]{
   \includegraphics[width=0.49\textwidth,trim=0 35 45
60,clip]{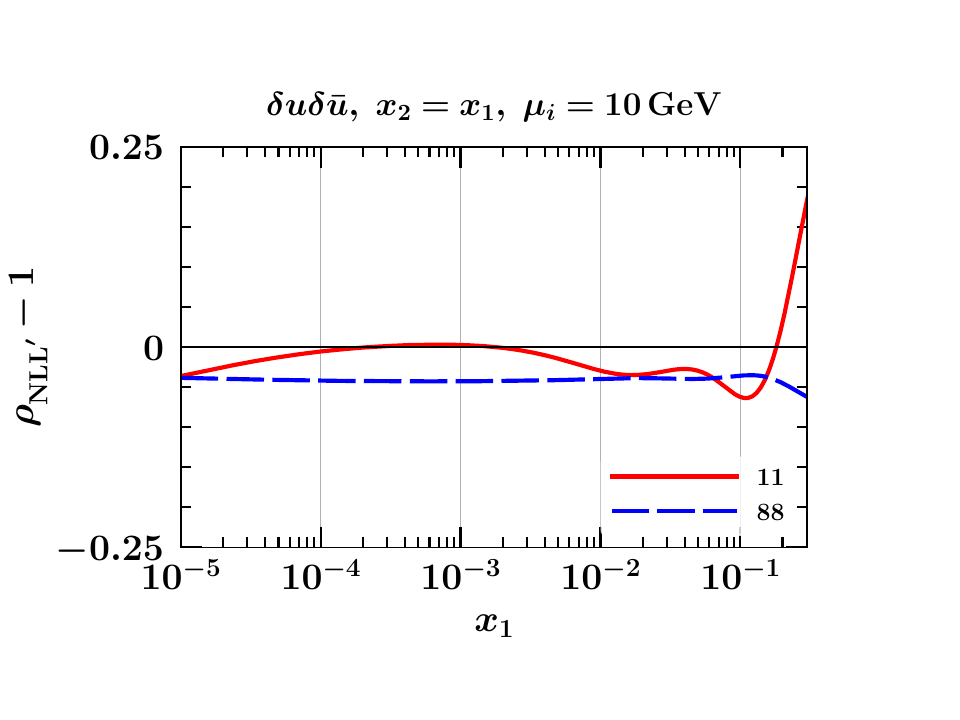}
}
\caption{\label{fig:rho-ratio} The double ratio \protect\eqref{rho-ratio-NLLp},
evaluated at $\mu_1 = \mu_2 = 10 \gev$ for splitting DPDs with $x_1 = x_2$ and
$y = 0.5 \gev^{-1}$.  Notice the different range of $\rho - 1$ in the first
three and last three panels.  The divergent behaviour of the ratio for $u
\bar{u}$ reflects the zero of the LL distribution around $x_1 = x_2 = 0.1$,
which is seen in \fig{\protect\ref{fig:F_uubar-x2}}b.}
\end{figure}

The double ratio $\rho_{\text{\NLLp}}$ is shown in \fig{\ref{fig:rho-ratio}} for
different parton combinations at $\mu_1 = \mu_2 = 10 \gev$.  The results for
other parton combinations are similar.  We see substantial effects of the
higher-order terms in the evolution, which reach $50$ to $100 \%$ in some cases,
especially at low or very high momentum fractions.  The corrections for
polarised partons tend to be lower than for unpolarised ones. Since the
Collins-Soper kernel and $\gamma_J$ are polarisation independent, this points to
the importance of the NLO corrections to the splitting kernels $P$.
We see no universal pattern for the relative importance of \NLLp\ corrections in
the different colour channels.

\paragraph{NNLL accuracy.}  We have also computed splitting DPDs at NNLL
accuracy, as specified in \tab{\ref{tab:orders}} and the associated text.
Compared with \NLLp, this means that the orders of $\gamma_J$ and the $\beta$
function are increased, as well as the order of the perturbative input to the
Collins-Soper kernel.  As explained in \sect{\ref{sec:cs-kernel}}, the NLO
expression for $\pr{8}{J}(y^*; \mustar, \mustar)$ is known.  For the higher
representations $R=10$ and $R=27$ it is not, and we assume Casimir scaling for
these cases, as we did for the non-perturbative part in \eqref{J-Casimir}.

For the problem of the $\alpha_s$ order discussed in \sect{\ref{sec:orders}}, we
adopt the prescription described in \sect{2.4} in \cite{Bacchetta:2019sam},
taking the $\beta$ function at the same order as the running coupling in the
NLO PDF set.

\begin{figure}
\centering
\subfloat[$\mu_1 = \mu_2 = 10 \gev$]{
   \includegraphics[width=0.49\textwidth,trim=0 35 45
60,clip]{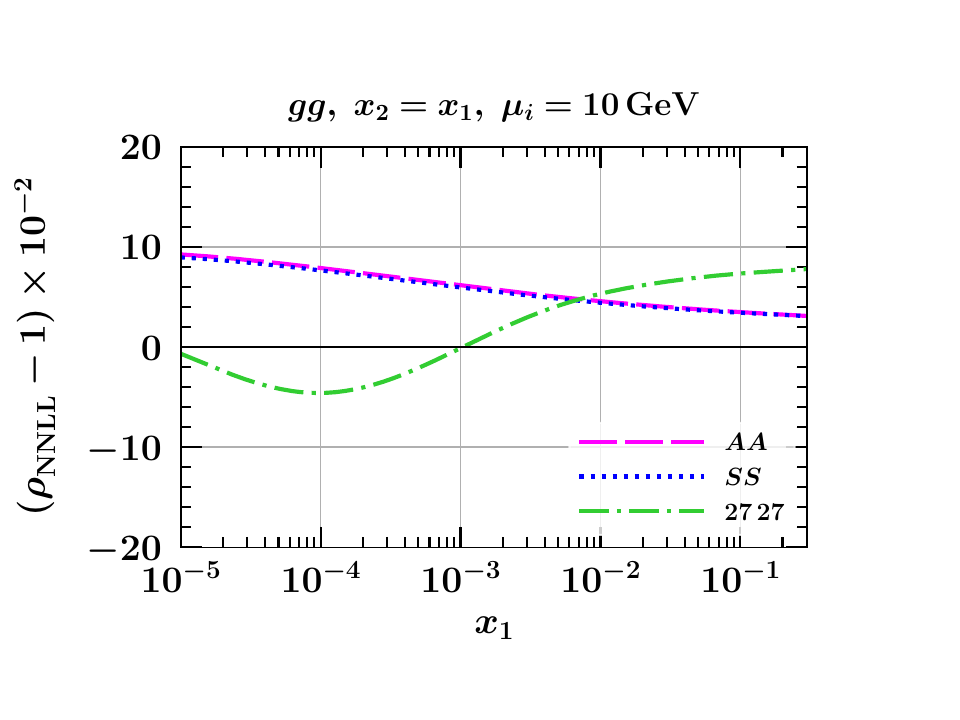}
}
\subfloat[$\mu_1 = \mu_2 = 80 \gev$]{
   \includegraphics[width=0.49\textwidth,trim=0 35 45
60,clip]{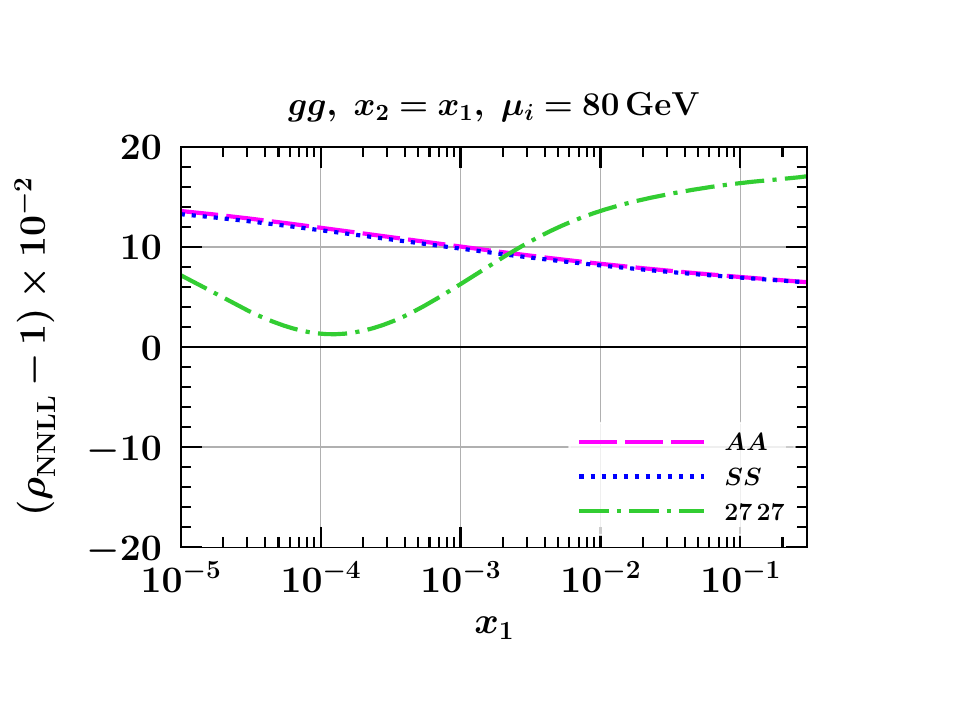}
}
\caption{\label{fig:F-NNLL-ratio} The ratio \protect\eqref{rho-ratio-NNLL},
evaluated for splitting DPDs with $x_1 = x_2$ and $y = 0.5 \gev^{-1}$ for two
different final scales.  Notice that $\rho - 1$ is given in units of percent.}
\end{figure}

The initial conditions of the DPDs do not change between NNLL and \NLLp, so that
the ratio
\begin{align}
   \label{rho-ratio-NNLL}
   \prn{R_1 R_2}{\rho}_{\ms\text{NNLL}}^{}(\ldots; \mu_1, \mu_2)
   &=
   \frac{\prn{R_1 R_2}{F}_{\ms\text{NNLL}}\bigl(
      \ldots; \mu_1, \mu_2, \mu_1 \mu_2/(x_1 x_2) \bigr)}{
      \prn{R_1 R_2}{F}_{\ms\text{\NLLp}}\bigl(
      \ldots; \mu_1, \mu_2, \mu_1 \mu_2/(x_1 x_2) \bigr)}
\end{align}
directly reflects the impact of the corrections to the Collin-Soper kernel and
its anomalous dimension.  It is equal to $1$ for the colour singlet.  Results
for colour non-singlet two-gluon distributions are shown in
\fig{\ref{fig:F-NNLL-ratio}}.  We find corrections up to 10\% at $\mu_1 = \mu_2
= 10 \gev$ and up to 20\% at $\mu_1 = \mu_2 = 80 \gev$.

\section{Results: double parton luminosities}
\label{sec:lumis}

In the cross section formula \eqref{dps-Xsect}, DPDs enter via the combinations
\begin{align}
   \label{dpd-lumi-def}
   &
   \prn{R_1 R_2,\ms R_3 R_4 \ms}{\mathcal{L}}_{a_1 a_2,\ms b_1 b_2}(
      x_1, x_2, \bar{x}_1, \bar{x}_2; \mu_1, \mu_2, \nu, s)
   \notag \\
   &\qquad
   =
   2\pi \int_{b_0 /\nu}^{\infty} d y\, y\;
   \prn{R_1 R_2}{F}_{a_1 a_2}(x_1, x_2, y; \mu_1, \mu_2, \zeta) \;
   \prn{R_3 R_4}{F}_{b_1 b_2}(\bar{x}_1, \bar{x}_2, y;
      \mu_1, \mu_2, \bar{\zeta})
   \,,
\end{align}
which we call ``double parton luminosities'' in analogy to the familiar case of
single hard scattering \cite[\sect{6.2}]{Campbell:2006wx}.  The $s$ dependence
arises from the constraint \eqref{zeta-basic-constraint} on the rapidity
parameters.  In the present section, we investigate the colour dependence of these quantities, which indicates how important colour non-singlet DPDs are in physical cross sections. Note that for the production of colour singlet particles, such as $W$ or $Z$ bosons, the luminosities $\mathcal{L}$ for different colour combinations enter the cross section with relative weights $\pm 1$, as seen in~\eqref{fact-for-singlet}.

Throughout this section, we choose scales as\,\footnote{This choice of $\mu_1$
and $\mu_2$ is quite natural for the production of single particles such as a
$W$, $Z$, or Higgs boson.  For the production of jets, a scale that depends on
their transverse momentum appears to be more adequate.  To study this at the
level of parton luminosities is more involved since it introduces a dependence
on an additional variable.}
\begin{align}
   \mu_1 &= M_1 \,,
   &
   \mu_2 &= M_2 \,,
   &
   \nu &= \min(M_1, M_2)
   \,,
\end{align}
where we recall that $M_i$ is the invariant mass of the observed particle or
system of particles produced by the hard scatter number $i$.  As seen in
\eqref{global-Sudakov}, the specific choice of $\zeta$ and $\bar{\zeta}$ is not
relevant in luminosities as long as \eqref{zeta-basic-constraint} is fulfilled.

\begin{figure}[ht]
\vspace{1em}
\centering
\subfloat[1v1]{
   \includegraphics[width=0.32\textwidth]{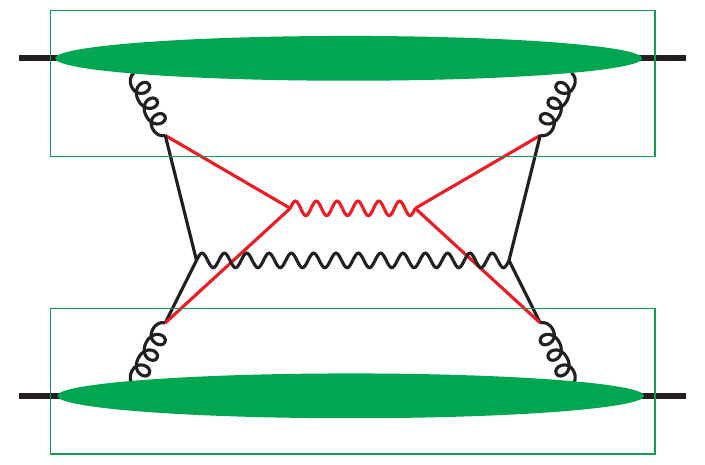}
}
\subfloat[2v1]{
   \includegraphics[width=0.32\textwidth]{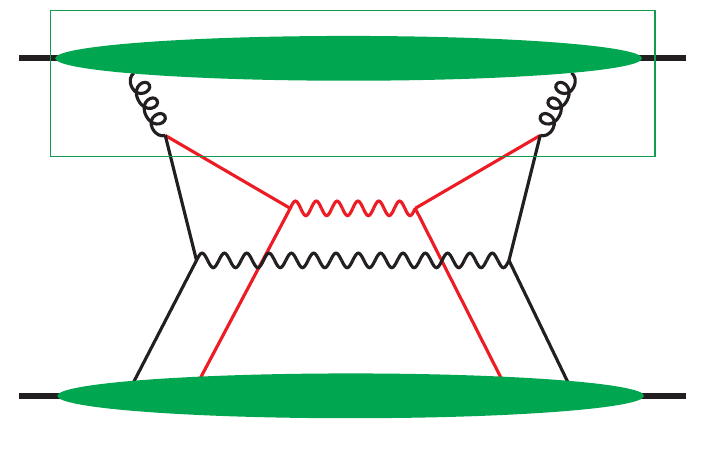}
}
\subfloat[2v2]{
   \includegraphics[width=0.32\textwidth]{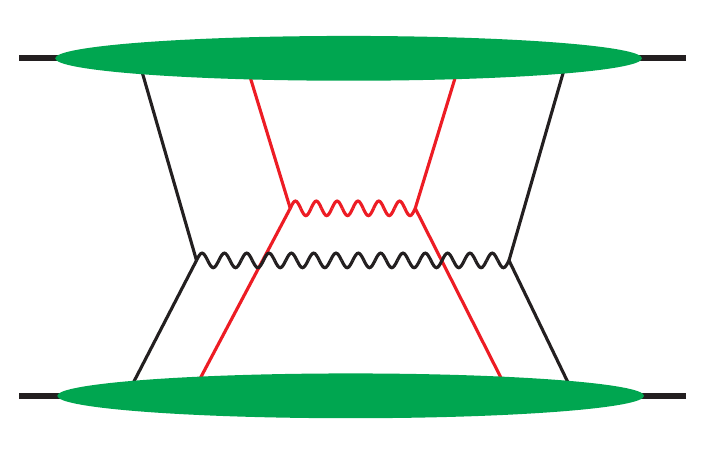}
}
\caption{\label{fig:n-vs-m} Different contributions to the DPS cross section for
the production of two gauge bosons.  The boxes indicate the splitting part of
DPDs.  The 1v2 contribution is obtained by interchanging the upper and lower
protons in graph b.}
\end{figure}

Let us first recall a number of general results, which were obtained in
\cite{Diehl:2017kgu}.
The decomposition \eqref{DPD-sum} of a DPD into splitting and intrinsic parts
induces a decomposition of $\mathcal{L}$ into the four components
\begin{align}
   \mathcal{L}_{\text{1v1}}
   & \leftrightarrow
   F^{\text{spl}} \, F^{\text{spl}}
   \,,
   &
   \mathcal{L}_{\text{1v2}}
   & \leftrightarrow
   F^{\text{spl}} \, F^{\text{intr}}
   \,,
   &
   \mathcal{L}_{\text{2v1}}
   & \leftrightarrow
   F^{\text{intr}} \, F^{\text{spl}}
   \,,
   &
   \mathcal{L}_{\text{2v2}}
   & \leftrightarrow
   F^{\text{intr}} \, F^{\text{intr}}
   \,,
\end{align}
which are shown in \fig{\ref{fig:n-vs-m}} at the level of graphs.  The notation ``$m$v$n$'' follows the parlance of \cite{Gaunt:2012dd} and indicates that $m$
partons in one proton scatter on $n$ partons in the other if one resolves the
splitting part of DPDs, as indicated by the boxes in the figure.

In general, the 1v1 contribution is strongly dominated by small values of $y$
because of the power-law rise of $F^{\text{spl}}$ with decreasing $y$.  In the
region $y \sim 1/\nu$, the graph in \fig{\ref{fig:n-vs-m}}a should not be
computed in terms of double parton distributions: the characteristic distance
scale of at least one hard scatter is then comparable to the transverse distance
between the partons in each DPD, so that the two hard scatters overlap and are
no longer independent of each other.  Instead, one should remove the boxes in
the graph and compute it as a higher-order contribution to single parton
scattering (SPS).

The scheme developed in \cite{Diehl:2017kgu} avoids double counting between
single and double parton scattering by writing the full cross section as
\begin{align}
   \label{full-Xsect}
   \sigma
   &=
   \sigma_{\text{DPS}}(\nu) - \sigma_{\text{1v1, sub}}(\nu)
   + \sigma_{\text{SPS}}
   \,,
\end{align}
with $\sigma_{\text{DPS}}$ given by the factorisation formula \eqref{dps-Xsect}.
The double counting subtraction term $\sigma_{\text{1v1, sub}}$ is easy to
construct in the case of equal scales $M_1 = M_2 = M$.  It is obtained by
replacing the evolved full DPDs in \eqref{dps-Xsect} with the purely
perturbative expression $F^{\text{spl,pt}}$ given in \eqref{splitting-DPD-pt},
evaluated at $\mu = M$ and $\zeta = M^2 / (x_1 x_2)$. The dependence on the
cutoff scale $\nu$ for the $y$ integration largely cancels in
\eqref{full-Xsect}, up to higher-order contributions beyond the accuracy of the
calculation and up to a logarithmic dependence due to the 1v2 and 2v1 contributions.
The subtraction term for unequal scales requires a more elaborate construction
that involves some kind of interpolation, as discussed in \sect{6.3} of
\cite{Diehl:2017kgu}.  We postpone this to future work.

At the level of double parton luminosities, we can form the combination
$\mathcal{L} - \mathcal{L}_{\text{1v1, sub}}$ in analogy to \eqref{full-Xsect}.
In the following plots, we show the partial combination
$\mathcal{L}_{\text{1v1}} - \mathcal{L}_{\text{1v1, sub}}$ along with
$\mathcal{L}_{\text{1v1}}$ for the cases where $\mu_1 = \mu_2$.


\subsection{Double parton luminosities at LL accuracy}
\label{sec:lumis-LL}

In this subsection, we present luminosities computed at LL accuracy.  We focus
on parton combinations that appear in four-jet production or in the production
of $W$ pairs.  Plots for additional combinations are shown in
\app{\ref{sec:more-lumis}}.

We always plot luminosities for specified invariant masses $M_1$ and $M_2$ as a
function of $Y = Y_1 = - Y_2$, i.e.\ for opposite rapidities of the two hard
systems.  The collision energy is set to $\sqrt{s} = 14 \tev$.  According to
\eqref{x-fractions}, this corresponds to momentum fractions
\begin{align}
   x_1
   =
   \bar{x}_1
   &\approx
   \begin{cases}
      7.1 \times 10^{-4}  & \text{ for $M_1 = 10 \gev$} \\
      5.7 \times 10^{-3}  & \text{ for $M_1 = 80 \gev$}
   \end{cases}
\end{align}
if $Y_1 = 0$ and
\begin{align}
   x_1
   &\approx
   \begin{cases}
      3.9 \times 10^{-2} \\
      3.1 \times 10^{-1}
   \end{cases}
   \qquad
   \bar{x}_1
   \approx
   \begin{cases}
      1.3 \times 10^{-5}     & \text{ for $M_1 = 10 \gev$} \\
      1.0 \times 10^{-4}     & \text{ for $M_1 = 80 \gev$}
   \end{cases}
\end{align}
if $Y_1 = 4$, and to corresponding values for $x_2$ and $\bar{x}_2$.


\subsubsection{Double dijets}
\label{sec:double-dijets}

\begin{figure}[p]
\centering
\subfloat[2v2 \, ($g g, g g$)]{
   \includegraphics[width=0.48\textwidth,trim=0 15 25 38,clip]{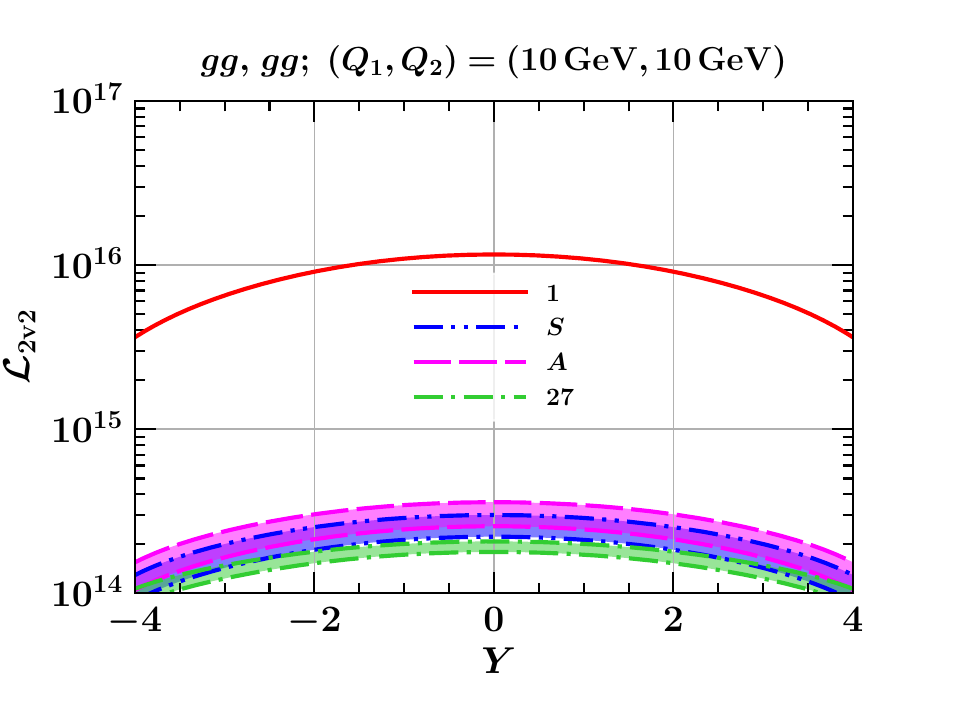}
}
\subfloat[1v2 + 2v1 \, ($g g, g g$)]{
   \includegraphics[width=0.48\textwidth,trim=0 15 25 38,clip]{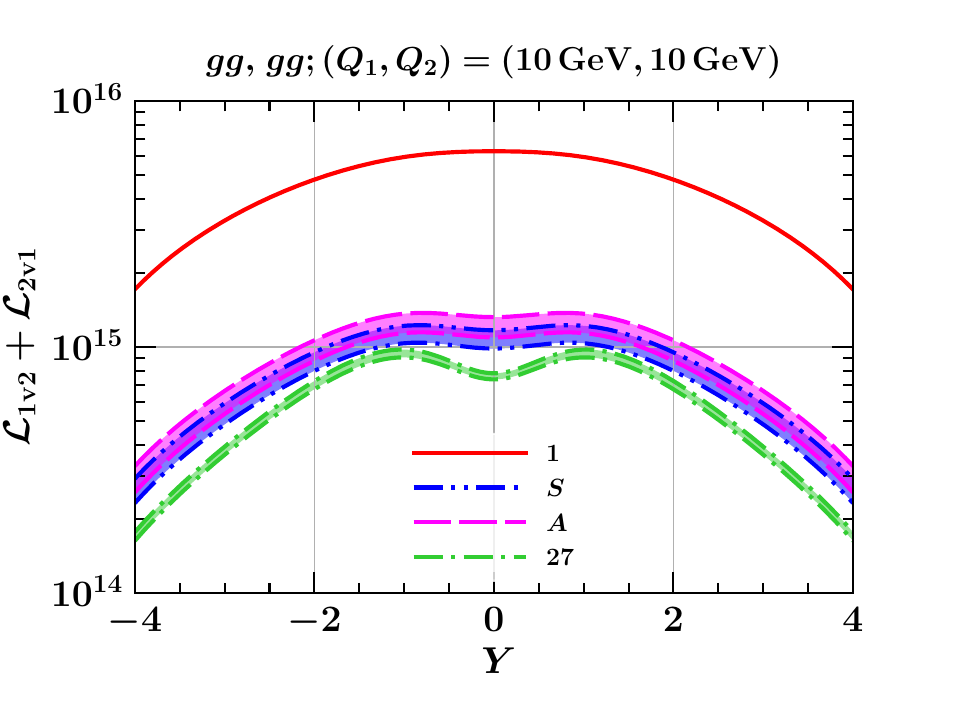}
}
\\[1.5em]
\subfloat[1v1 \, ($g g, g g$)]{
   \includegraphics[width=0.48\textwidth,trim=0 15 25 38,clip]{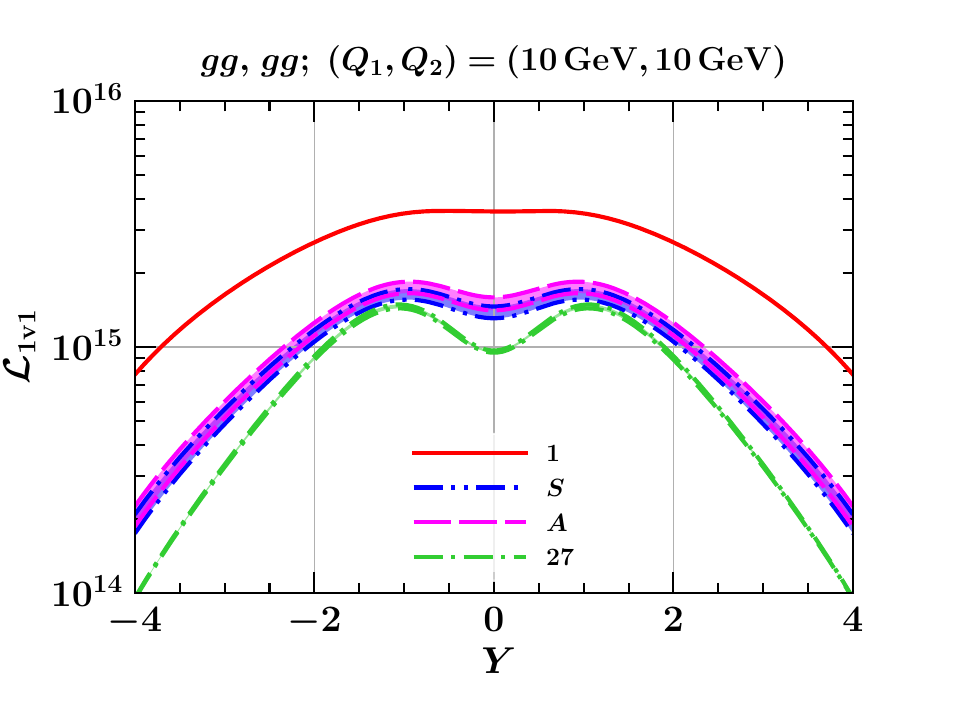}
}
\subfloat[1v1 subtracted \, ($g g, g g$)]{
   \includegraphics[width=0.48\textwidth,trim=0 15 25 38,clip]{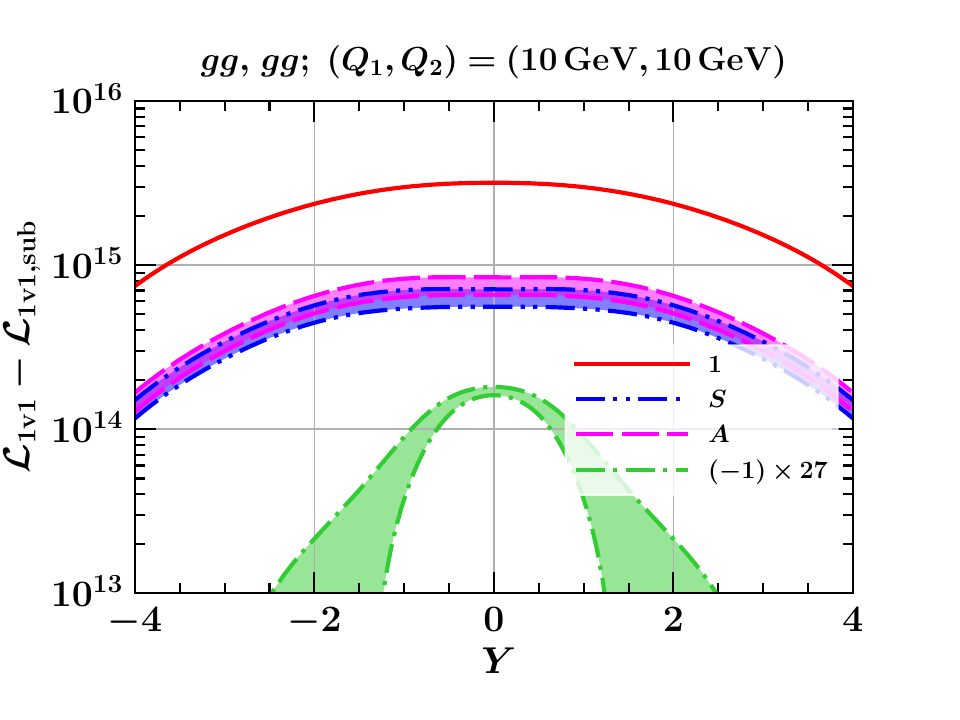}
}
\caption{\label{fig:lumis-4g-10-10} Double parton luminosities $\mathcal{L}_{g
g, g g}$ for four gluons with $(M_1, M_2) = (10 \gev, 10 \gev)$.  The two hard
systems have opposite rapidities $Y_1 = Y$ and $Y_2 = -Y$.
Here and in the following plots, bands reflect the uncertainty due to modelling
the Collins-Soper kernel at large $y$, as explained in the caption of
\fig{\protect\ref{fig:F-vs-y}}.  We write $1$, $S$, $A$, etc.\ in the figure
keys if all four colour representation labels are identical.  In this and all
following figures, double parton luminosities are given in units of $\gev^2$.}
\end{figure}

\begin{figure}[p]
\centering
\subfloat[2v2 \, ($g g, g g$)]{
   \includegraphics[width=0.48\textwidth,trim=0 15 25 38,clip]{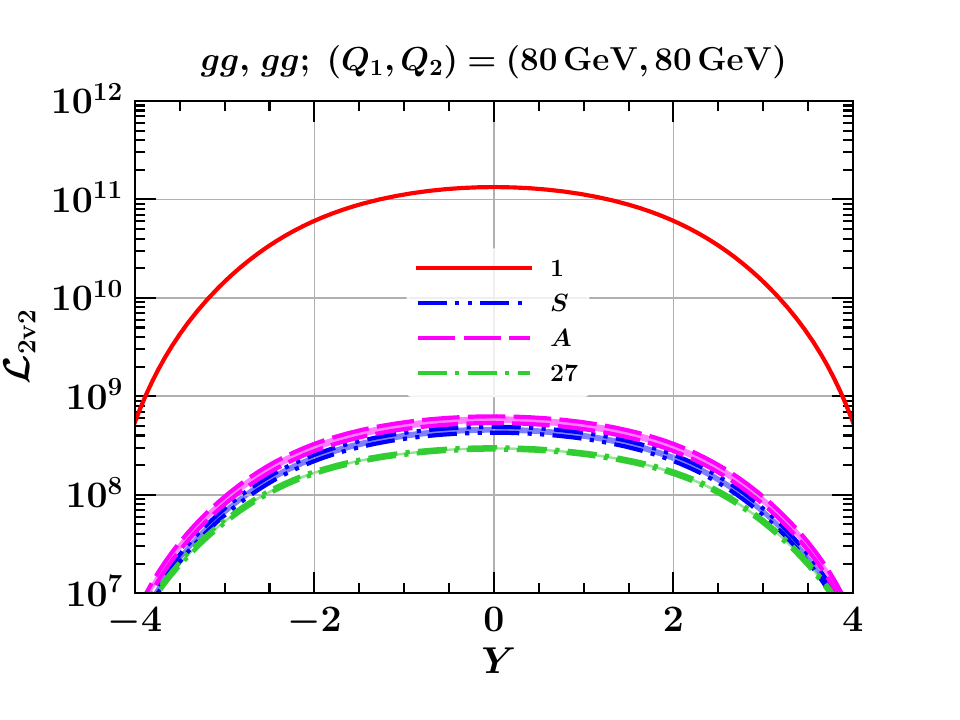}
}
\subfloat[1v2 + 2v1 \, ($g g, g g$)]{
   \includegraphics[width=0.48\textwidth,trim=0 15 25 38,clip]{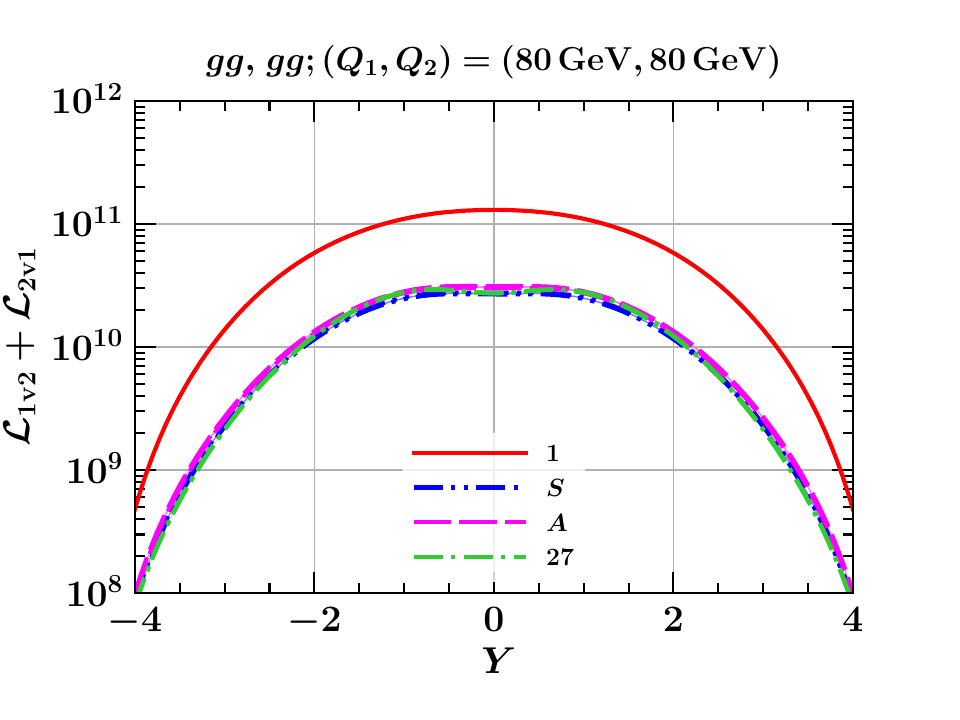}
}
\\[1.5em]
\subfloat[1v1 \, ($g g, g g$)]{
   \includegraphics[width=0.48\textwidth,trim=0 15 25 38,clip]{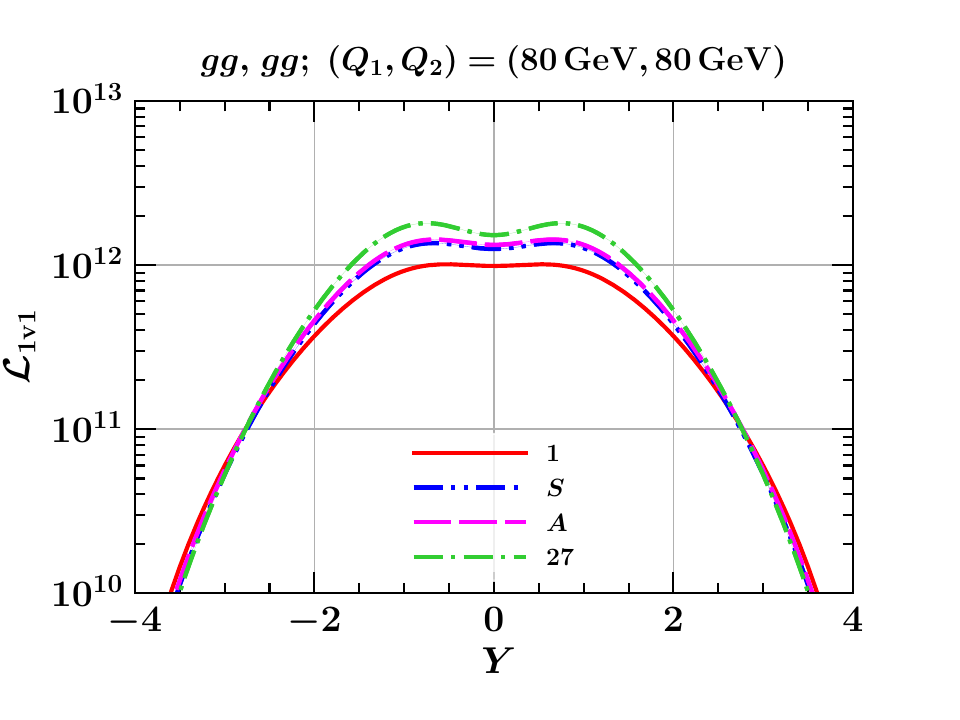}
}
\subfloat[1v1 subtracted \, ($g g, g g$)]{
   \includegraphics[width=0.48\textwidth,trim=0 15 25 38,clip]{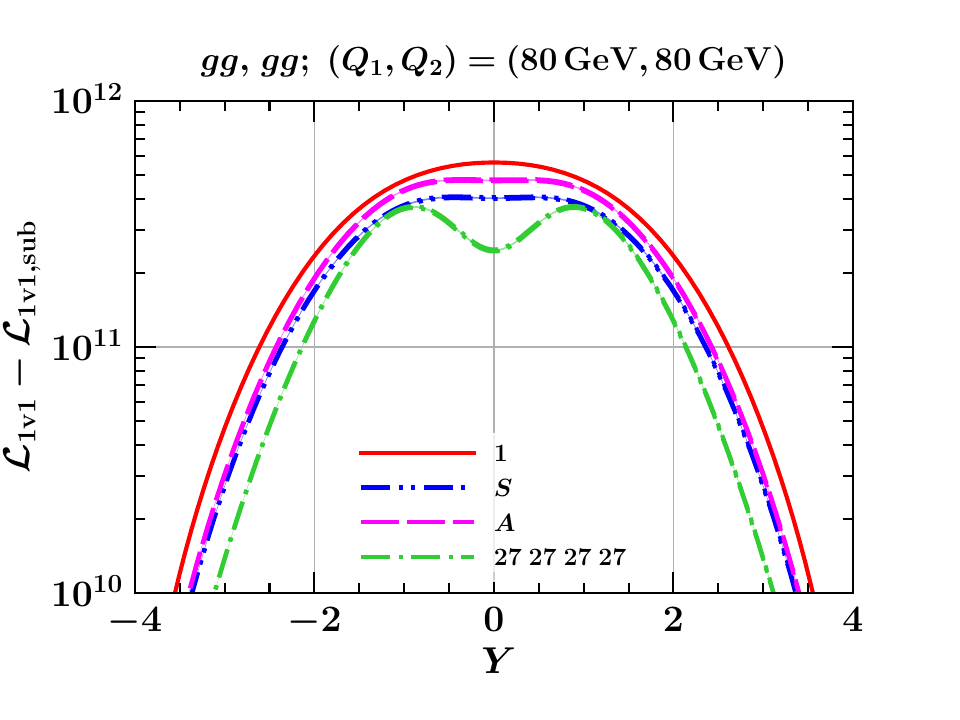}
}
\caption{\label{fig:lumis-4g-80-80} As \fig{\protect\ref{fig:lumis-4g-10-10}},
but for $(M_1, M_2) = (80 \gev, 80 \gev)$.}
\end{figure}

In \figs{\ref{fig:lumis-4g-10-10}} to \ref{fig:lumis-4g-80-80} we show the
luminosities for four gluons for different combinations of $10 \gev$ and $80
\gev$ for the invariant masses $M_1$ and $M_2$.  We remark that two jets with
$10 \gev$ invariant mass may not be identified as jets in the LHC experiments,
but they contribute to the underlying event in a region accessible to a
factorised description in terms of hard parton-level scattering.

\begin{figure}[p]
\centering
\subfloat[2v2 \, ($g g, g g$)]{
   \includegraphics[width=0.48\textwidth,trim=0 15 25 38,clip]{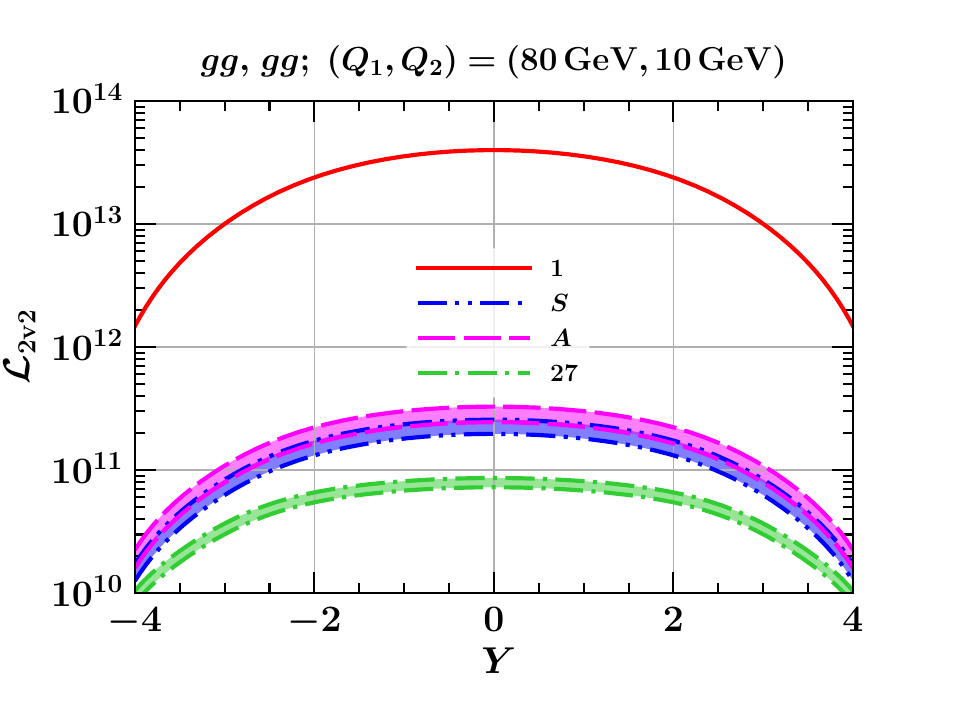}
}
\subfloat[1v2 + 2v1 \, ($g g, g g$)]{
   \includegraphics[width=0.48\textwidth,trim=0 15 25 38,clip]{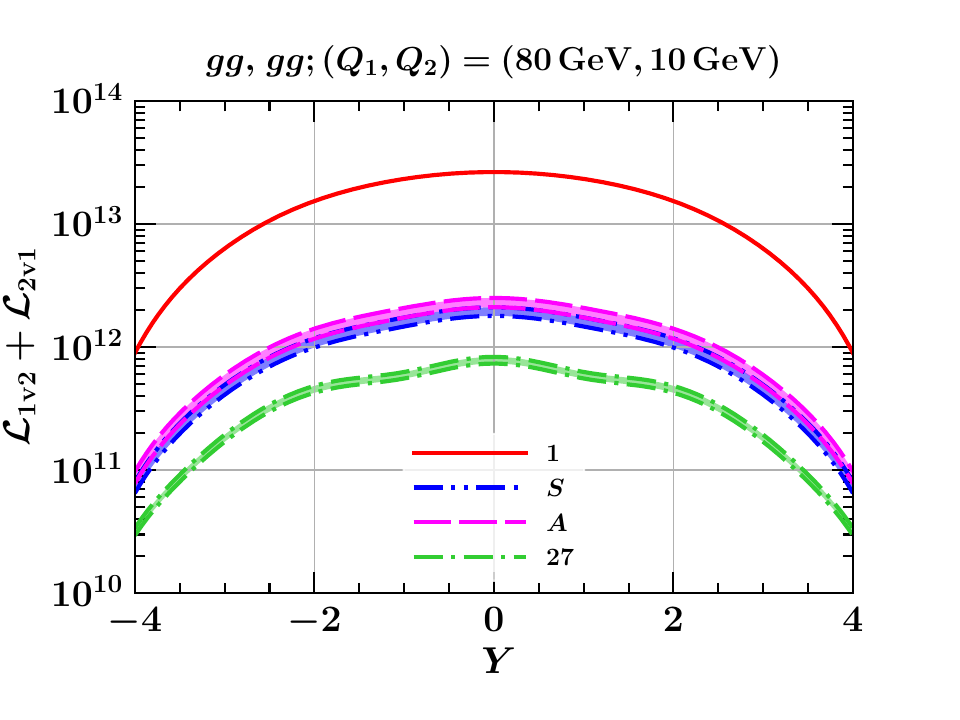}
}
\caption{\label{fig:lumis-4g-80-10} As \fig{\protect\ref{fig:lumis-4g-10-10}},
but for $(M_1, M_2) = (80 \gev, 10 \gev)$.}
\end{figure}
\begin{figure}[p]
\centering
\ContinuedFloat
\subfloat[1v1 \, ($g g, g g$)]{
   \includegraphics[width=0.48\textwidth,trim=0 15 25 38,clip]{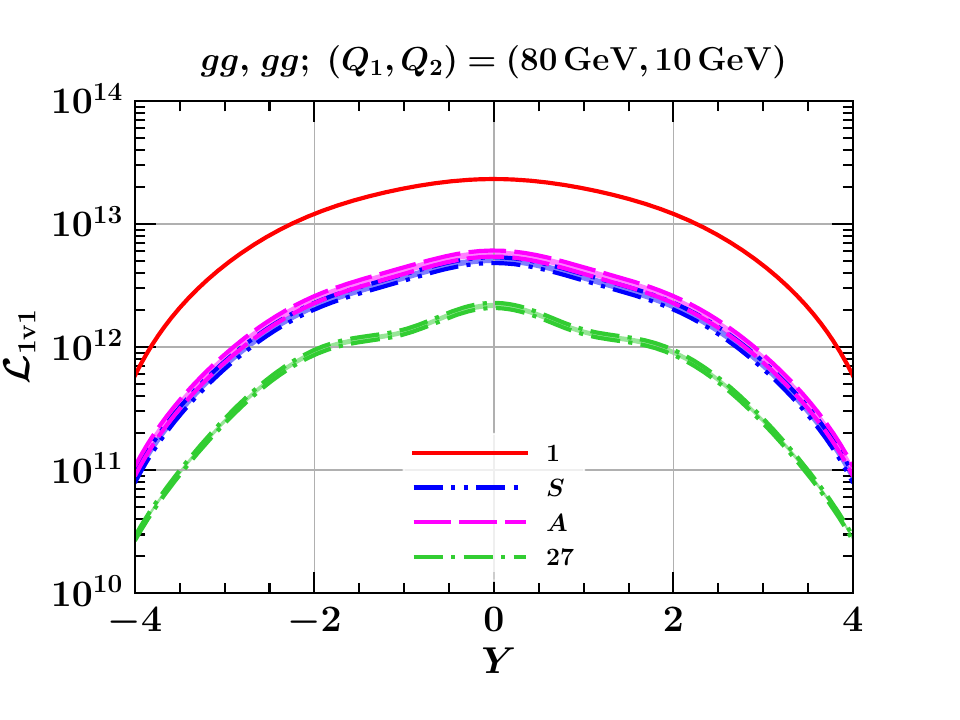}
}
\caption*{Figure~\ref{fig:lumis-4g-80-10} (continued)}
\end{figure}

In all cases, the 2v2 contributions for colour non-singlet channels are
suppressed compared with the colour singlet one by about a factor 30 or more, so
that they can be safely neglected for phenomenological purposes.  This is
because $\mathcal{L}_{\text{2v2}}$ for the colour singlet receives the bulk of
its contribution from large $y$, where the colour non-singlet distributions are
strongly suppressed as we saw in \fig{\ref{fig:F-vs-y}}.  Whilst the specific
values in our plots depend of course on our model for the input distribution, we
think that this is a rather robust conclusion, given what is known about the
size of the Collins-Soper kernel for quark TMDs at distances comparable with the
proton size.  It corresponds to the expectation in \cite{Mekhfi:1988kj} that
colour correlations would be negligible in double hard scattering.  The
suppression we see for the 2v2 term is much stronger than suggested by the
estimates of the Sudakov factor in \fig{10} of \cite{Manohar:2012jr}, where the
non-perturbative part of the Collins-Soper kernel was not included.

The 1v2 and 2v1 contributions receive important contributions from large $y$ if
at least one of the invariant masses is equal to $10 \gev$, as is indicated by
the size of the uncertainty bands for the choice of $\Delta J(y)$.  However,
they also receive important contributions from lower $y$, where the Sudakov
suppression is reduced because of the decreasing distance between the initial
and final scales for DPD evolution.  As a result, the colour non-singlet
channels in panels b of \figs{\ref{fig:lumis-4g-10-10}} and
\ref{fig:lumis-4g-80-10} are suppressed by about a factor 10 relative to
the singlet.  This confirms the observation of a reduced suppression of the 2v1
mechanism made in \cite{Blok:2022mtv}. For $(M_1, M_2) = (80 \gev, 80 \gev)$, the suppression is weaker than for $(M_1, M_2) = (80 \gev, 10 \gev)$, because the $y$ integral then extends down to $b_0 / (80 \gev) \approx 1.4 \times 10^{-2} \gev^{-1}$ and accumulates even more contributions from low $y$, where Sudakov suppression is not at work.

Not surprisingly, the suppression of colour non-singlet channels is even less for the 1v1 term in \figs{\ref{fig:lumis-4g-10-10}} and
\ref{fig:lumis-4g-80-10}.  For both invariant masses equal to $80 \gev$ these
channels are even enhanced over the colour singlet, as they are at the starting
scale $\mustar$ due to the colour factors $k_{a_1 a_2}(R_1 R_2)$ in the
perturbative splitting formula \eqref{splitting-DPD-pt}.

%
Comparing the unsubtracted and the subtracted partial luminosities in panels c
and d of \fig{\ref{fig:lumis-4g-10-10}}, we observe a very small difference in
the colour singlet and a moderate one for the colour octets.  By contrast, the
subtracted result for $R=27$ is much smaller than the unsubtracted one and has
the opposite sign.  In \fig{\ref{fig:lumis-4g-80-80}}, the effect of the
subtraction is noticeable in all colour channel and reverses the hierarchy
between the different colour channels (but not their sign).  For $Y$ around
zero, the subtracted 1v1 contributions turn out to be larger than the sum of 1v2 and 2v1 contributions, both in the colour singlet and non-singlet channels.

\begin{figure}
\centering
\subfloat[$M_1 = M_2 = 10 \gev$]{
   \includegraphics[width=0.48\textwidth,trim=0 0 25 28,clip]{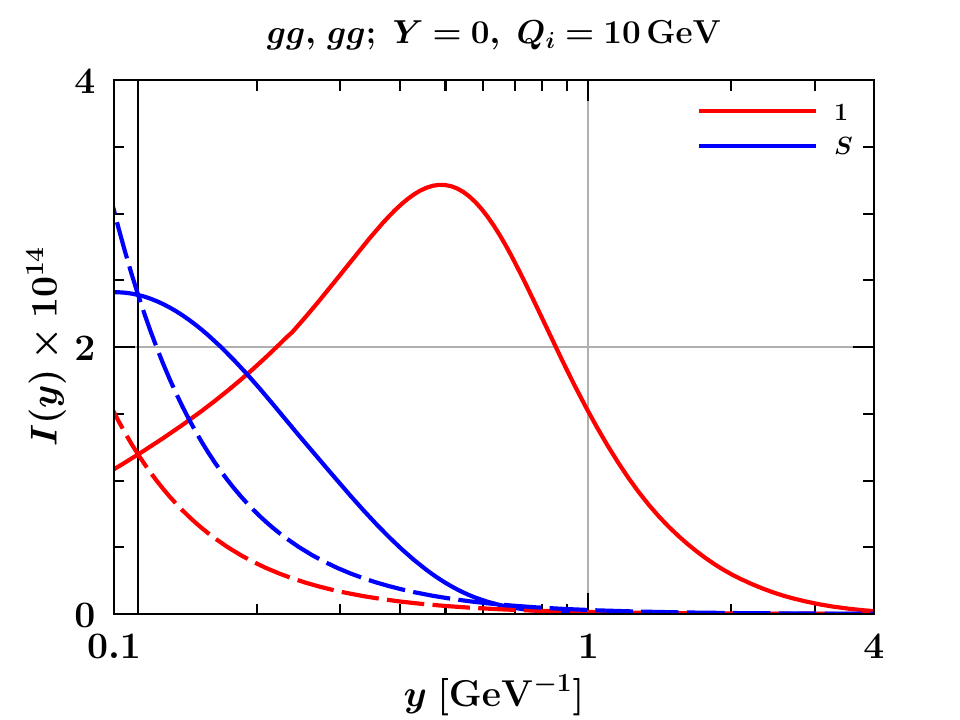}
   \includegraphics[width=0.48\textwidth,trim=0 0 25 28,clip]{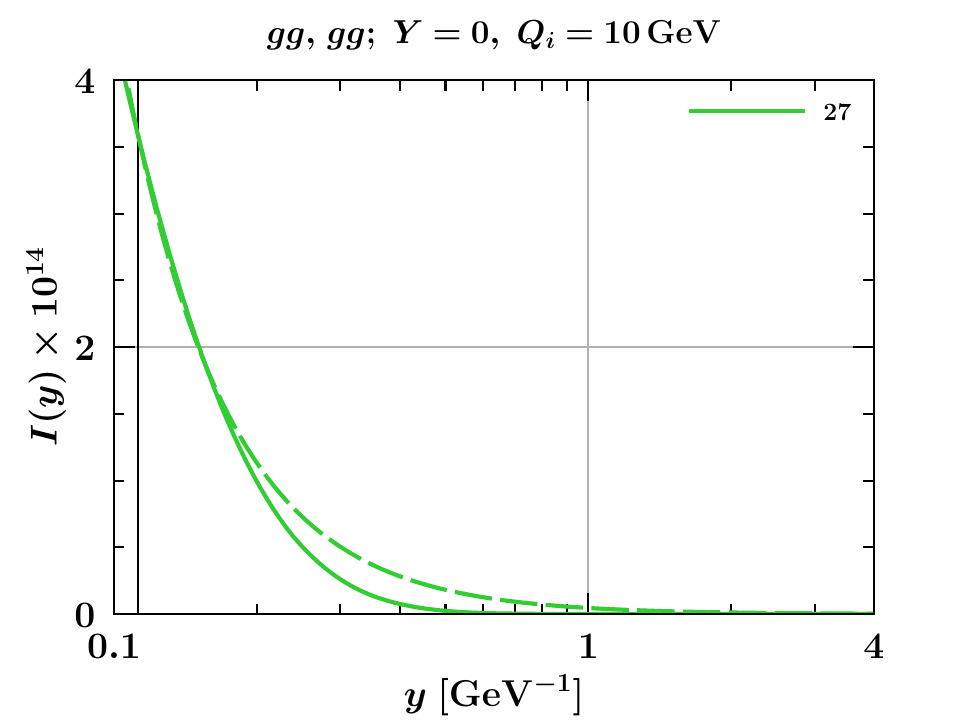}
}
\\[1.5em]
\subfloat[$M_1 = M_2 = 80 \gev$]{
   \includegraphics[width=0.48\textwidth,trim=0 0 25 28,clip]{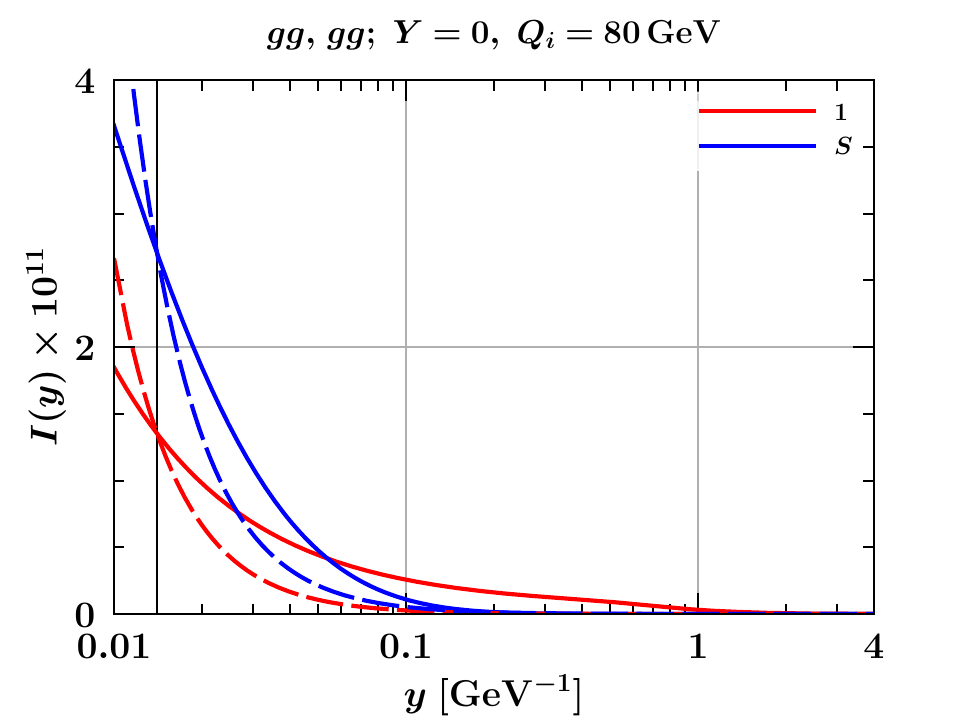}
   \includegraphics[width=0.48\textwidth,trim=0 0 25 28,clip]{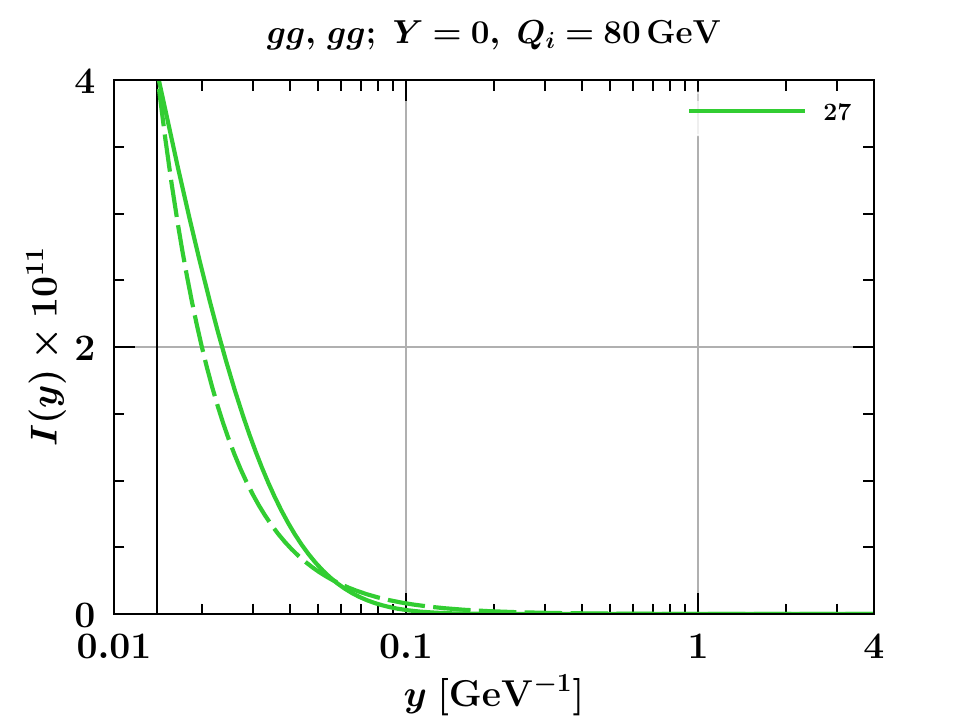}
}
\caption{\label{fig:lumi-integds} The integrand $I(y)$ of the double parton
luminosity for four gluons, scaled such that the integral is $2\pi$ times the
area under the corresponding curve.  All plots are for central rapidity $Y=0$.
Solid lines are for $\mathcal{L}_{\text{1v1}}$ and dashed ones for the
subtraction term $\mathcal{L}_{\text{1v1, sub}}$.  Vertical lines indicate the
lower integration bound $b_0 / \nu$.}
\end{figure}

To gain more insight, we show in \fig{\ref{fig:lumi-integds}} the function
\begin{align}
   \label{lumi-integrand}
   I(y)
   &=
   y^2 \, F(x_1, x_2, y; \ldots) \, F(\bar{x}_1, \bar{x}_2, y; \ldots)
   \,,
\end{align}
which appears as integrand for double parton luminosities if the integration is
performed over $\ln y$.  The curves for $R = A$ are very close to the ones for
$R = S$ and not shown for clarity.
Dashed curves are for the subtraction terms, in which the individual DPD factors
have a strict $1/y^2$ behaviour.  Solid curves are for the evolved DPDs in the
1v1 term, and we see that evolution has significantly altered their shape in $y$
away from a $1/y^2$ power law.  By construction, the two sets of curves
coincide at the lower integration limit $y = b_0/\nu$. Perhaps surprisingly,
the curves with evolved DPDs lie above their counterparts for the subtraction
term, not only for the colour singlet (where this trend was already observed in
\cite{Diehl:2017kgu}) but also for the colour octet in the $y$ region that
dominates the integral.  For $R=27$, the respective curves are very close to
each other, so that the subtracted result is very small compared with the
unsubtracted one.

Let us emphasise that, in contrast to the 1v2 and 2v1 terms, our results for the
unsubtracted and subtracted 1v1 terms are \emph{not} model dependent, since they
are dominated by the region of $y$ in which the splitting DPDs and their
evolution are under perturbative control.


\subsubsection{\texorpdfstring{$W$}{W} pair production}
\label{sec:like-sign-Ws}

The production of equal-sign $W$ pairs is a prominent channel for double parton
scattering.  It has been extensively studied in theory \cite{Kulesza:1999zh,
Gaunt:2010pi, Ceccopieri:2017oqe, Cotogno:2018mfv, Cotogno:2020iio}, and a cross
section measurement has been reported in \cite{CMS:2022pio}.

The parton combinations in the DPDs required for this channel are zero in the LO
splitting formula \eqref{splitting-DPD-pt}, which reflects that two $W$ bosons
with equal charge cannot be produced by the box graph in
\fig{\ref{fig:n-vs-m}}a.  As a result, the 1v1 subtraction terms for this
channel are all zero at the order we are working at.  However, splitting DPDs
with the relevant parton combinations are produced by quark-gluon mixing under
evolution, which gives rise to 1v2, 2v1, and 1v1 terms.

\begin{figure}[p]
\centering
\subfloat[2v2 \, ($u \bar{d}, \bar{d} u$)]{
   \includegraphics[width=0.48\textwidth,trim=0 20 25 38,clip]{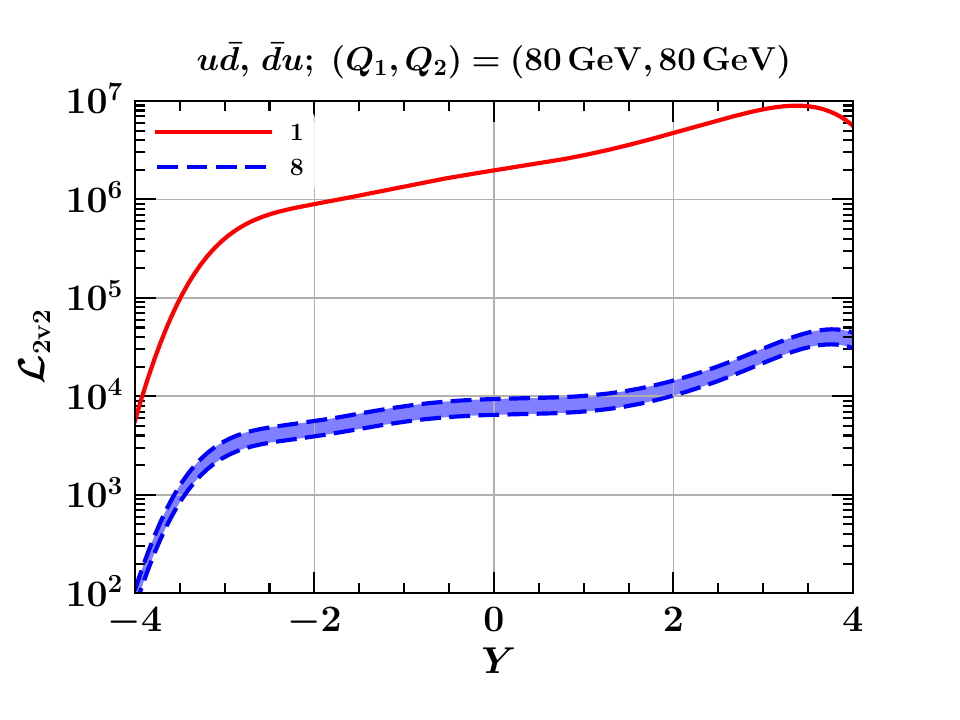}
}
\subfloat[1v2 + 2v1 \, ($u \bar{d}, \bar{d} u$)]{
   \includegraphics[width=0.48\textwidth,trim=0 20 25 38,clip]{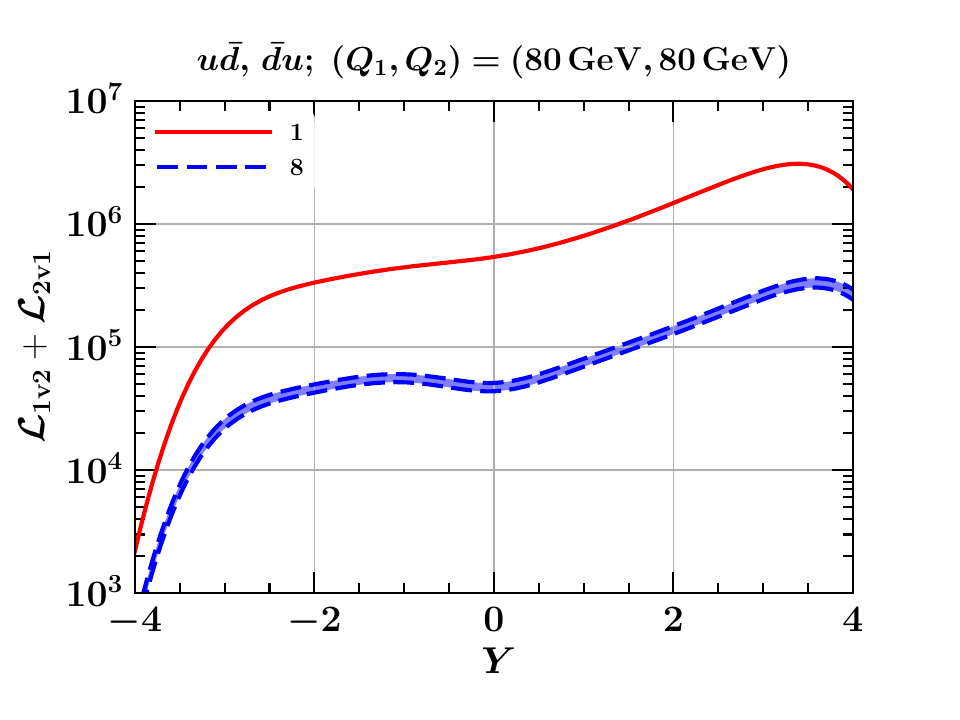}
}
\\[1.5em]
\subfloat[1v1 \, ($u \bar{d}, \bar{d} u$)]{
   \includegraphics[width=0.48\textwidth,trim=0 20 25 38,clip]{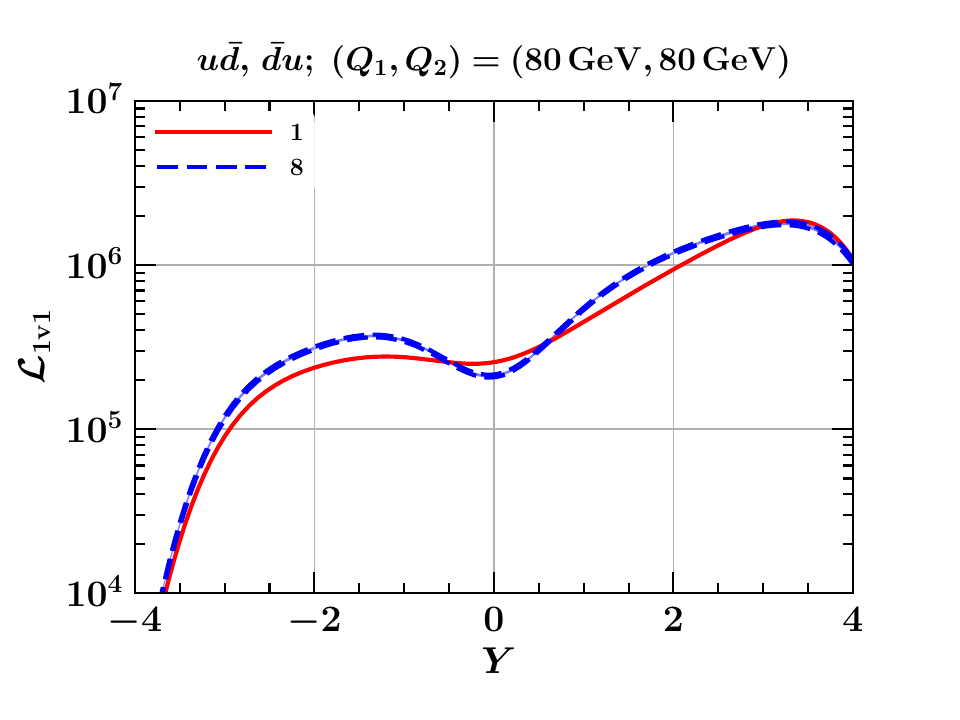}
}
\caption{\label{fig:lumis-WW-same-light} Double parton luminosities
$\mathcal{L}_{u \smash{\bar{d}, \bar{d}} u}$ for $M_1 = M_2 = 80 \gev$.  This
parton combination contributes to $W^+ W^+$ production.  The subtraction term
$\mathcal{L}_{\text{1v1, sub}}$ is zero in this case.  The bands have the same
meaning as in \fig{\protect\ref{fig:lumis-4g-10-10}}.}
\end{figure}

We see in \fig{\ref{fig:lumis-WW-same-light}} that in the colour singlet sector,
the 2v2 term is larger than the other ones, but these are not negligible.  The
relative size of the colour octet contribution follows the same general pattern
as we have seen for four gluons: it is strongly suppressed for 2v2, less
strongly for 1v2 and 2v1, and of the same size as the colour singlet for 1v1.

In \fig{\ref{fig:lumis-WW-opposite}} we show double parton luminosities that
contribute to $W^+ W^-$ production.  We observe several zero crossings of the
colour octet term for 1v2 and 2v1, as well as for the subtracted 1v1 term.  This
reflects the intricate pattern of zeros of the splitting DPDs we already
encountered in \fig{\ref{fig:F_uubar-x2}}.

\begin{figure}[p]
\centering
\subfloat[2v2 \, ($u \bar{u}, {\bar{d}} d$)]{
   \includegraphics[width=0.48\textwidth,trim=0 20 25 38,clip]{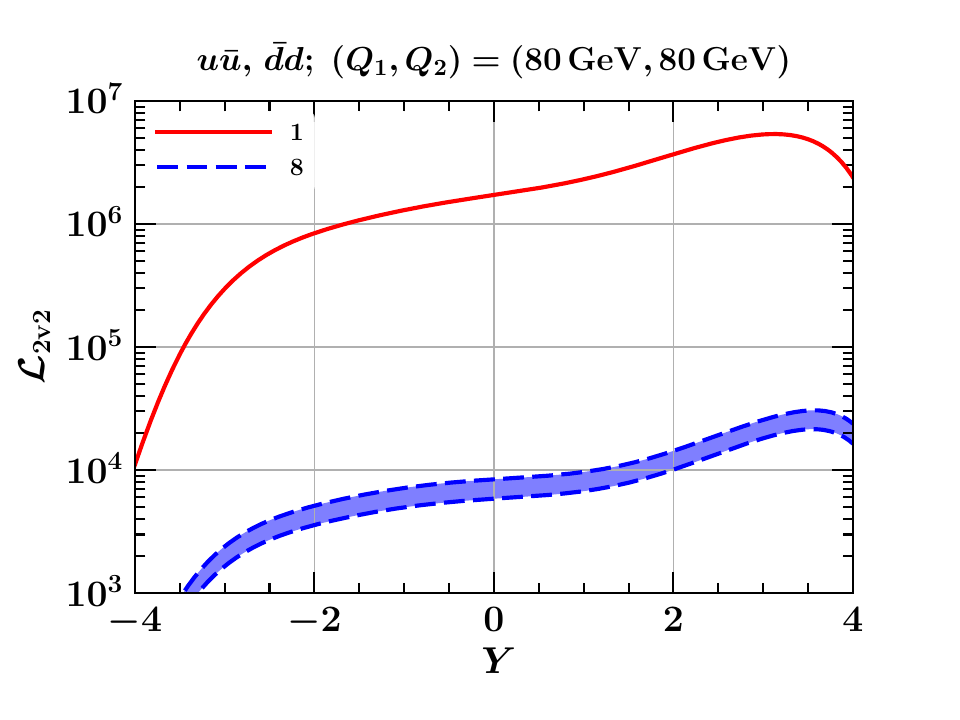}
}
\subfloat[1v2 + 2v1 \, ($u \bar{u}, {\bar{d}} d$)]{
   \includegraphics[width=0.48\textwidth,trim=0 20 25 38,clip]{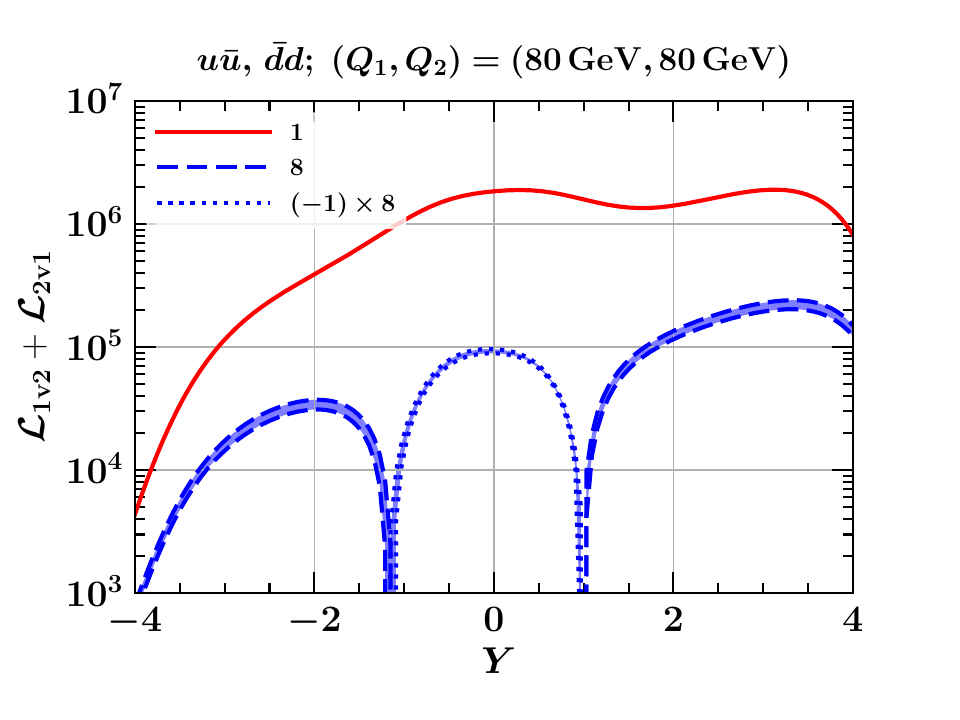}
}
\caption{\label{fig:lumis-WW-opposite} Double parton luminosities
$\mathcal{L}_{u \bar{u}, \smash{\bar{d}} d}$ for  $M_1 = M_2 = 80 \gev$.  This
parton combination contributes to $W^+ W^-$ production.}
\end{figure}

\begin{figure}[p]
\centering
\ContinuedFloat
\subfloat[1v1 \, ($u \bar{u}, {\bar{d}} d$)]{
   \includegraphics[width=0.46\textwidth,trim=0 20 25 38,clip]{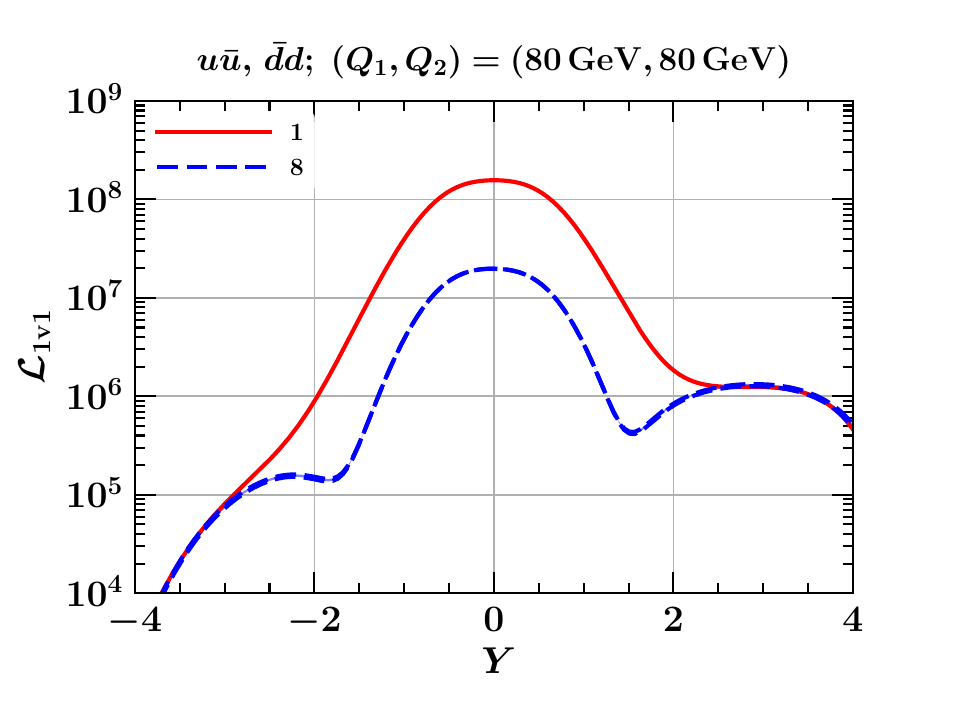}
}
\subfloat[subtracted 1v1 \, ($u \bar{u}, {\bar{d}} d$)]{
   \includegraphics[width=0.46\textwidth,trim=0 20 25 38,clip]{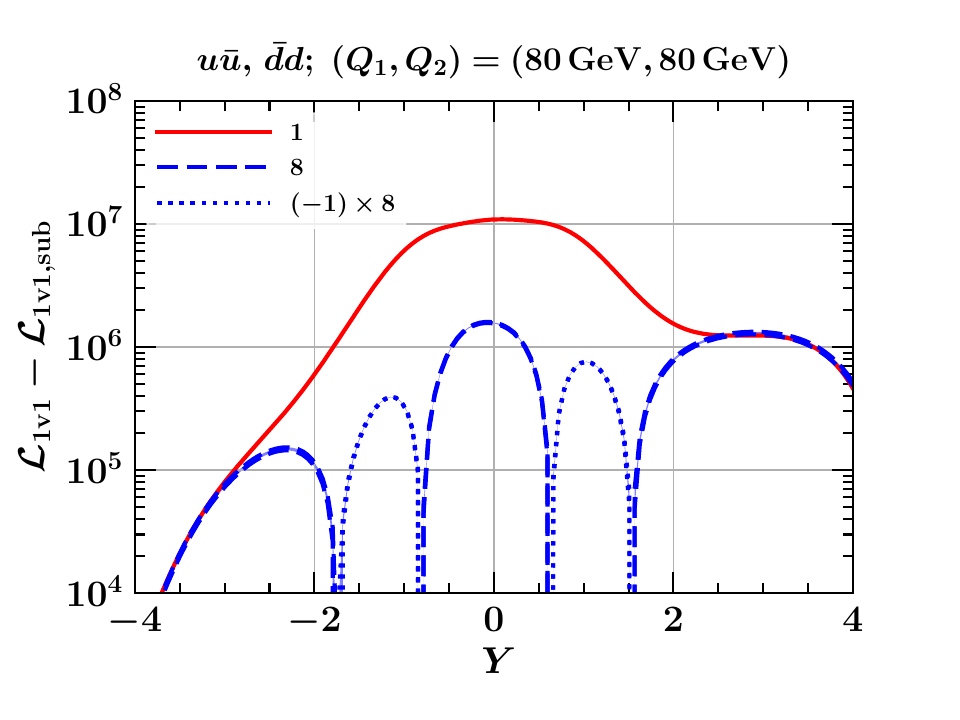}
}
\caption*{Figure~\protect\ref{fig:lumis-WW-opposite} (continued)}
\end{figure}


\subsection{Impact of higher orders}
\label{sec:lumis-higher-orders}

We saw in \sect{\ref{sec:dpds-higher-orders}} that the step from LL to \NLLp\
accuracy in DPD evolution can have significant quantitative effects.  Let us see
how this looks like at the level of double parton luminosities.  One should
however bear in mind that the change in these quantities is not directly visible
in the physical cross section, since it is accompanied by the change from LO to
NLO hard-scattering cross sections.  Nevertheless, one should
expect significant corrections in the overall cross sections when there are
important corrections to the parton luminosities.

\begin{figure}[p]
\centering
\subfloat[($g g, g g$) at $(M_1, M_2) = (10 \gev, 10 \gev)$]{
   \includegraphics[width=0.46\textwidth,trim=0 35 45 54,clip]{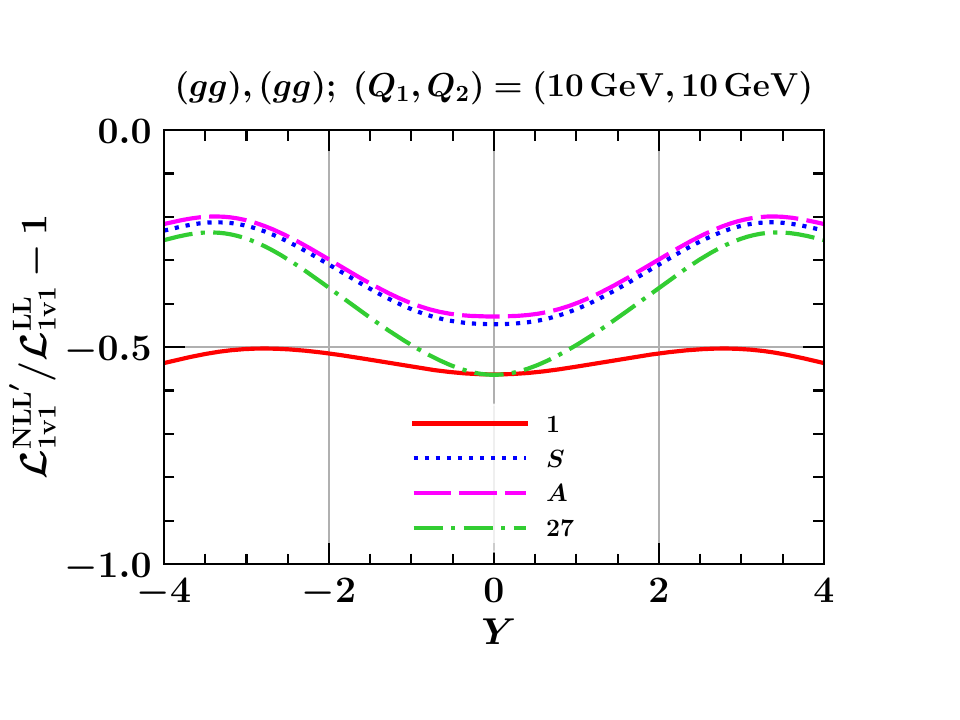}
}
\subfloat[($g g, g g$) at $(M_1, M_2) = (80 \gev, 80 \gev)$]{
   \includegraphics[width=0.46\textwidth,trim=0 35 45 54,clip]{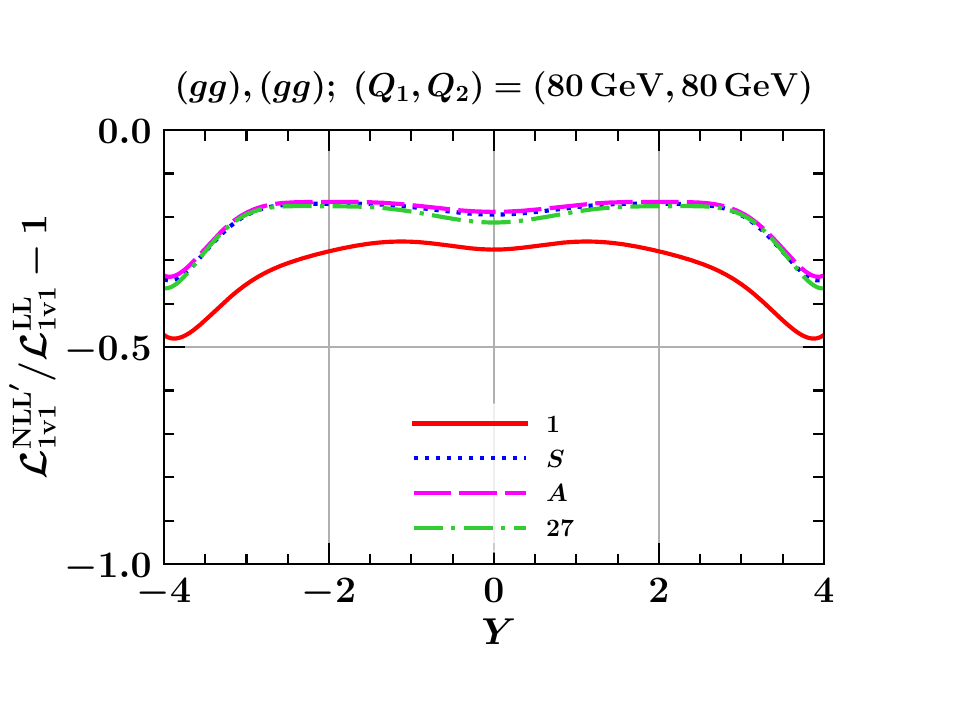}
}
\\[1em]
\subfloat[($g g, g g$) at $(M_1, M_2) = (80 \gev, 10 \gev)$]{
   \includegraphics[width=0.46\textwidth,trim=0 35 45 54,clip]{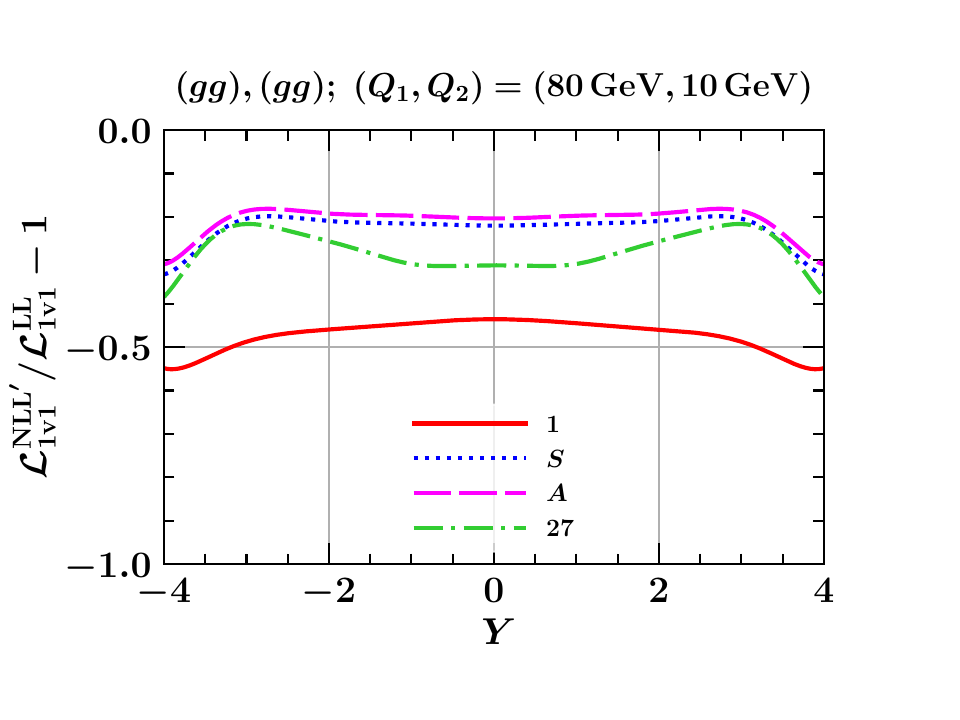}
}
\subfloat[($b g, g b$) at $(M_1, M_2) = (80 \gev, 10 \gev)$]{
   \includegraphics[width=0.46\textwidth,trim=0 35 45 54,clip]{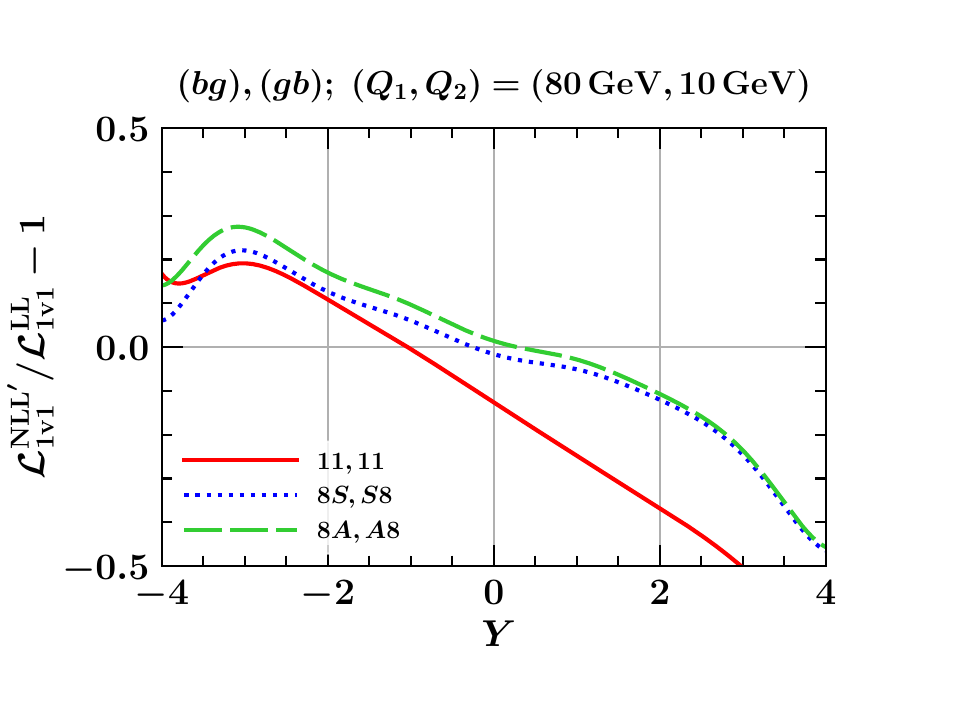}
}
\\[1em]
\subfloat[($u \bar{d}, \bar{d} u$) at $(M_1, M_2) = (80 \gev, 80 \gev)$]{
   \includegraphics[width=0.46\textwidth,trim=0 35 45 54,clip]{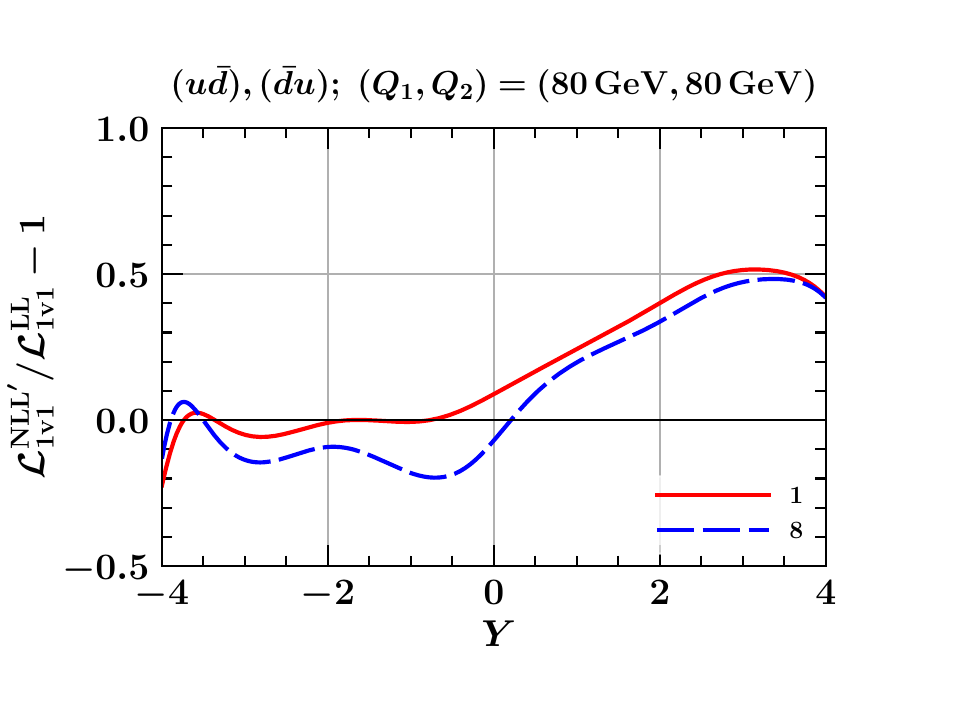}
}
\subfloat[($u \bar{u}, \bar{d} d$) at $(M_1, M_2) = (80 \gev, 80 \gev)$]{
   \includegraphics[width=0.46\textwidth,trim=0 35 45 54,clip]{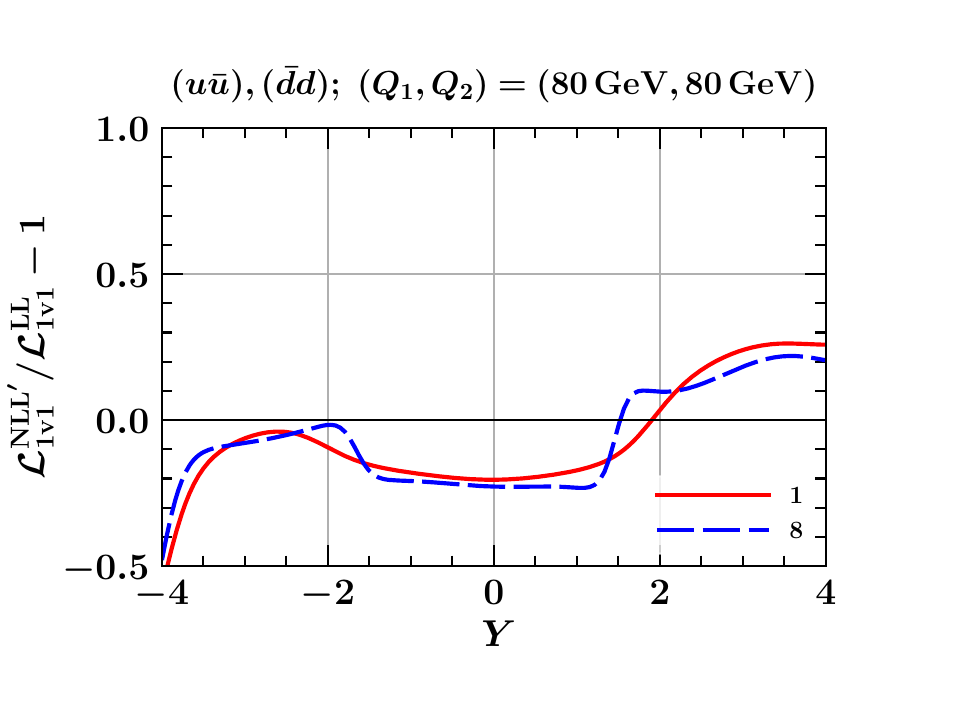}
}
\caption{\label{fig:lumis-NLL} Ratios of 1v1 luminosities computed
at \NLLp\ or LL accuracy.}
\end{figure}

In \fig{\ref{fig:lumis-NLL}} we show the ratio of 1v1 luminosities computed at
\NLLp\ or LL accuracy for representative parton channels.  We see corrections of
up to 50\%, without a universal pattern concerning their size in different
colour channels.  In some case, they have a pronounced dependence on the
rapidity $Y$, so that a simple $K$ factor would not be suitable to describe
them.

\section{Summary}
\label{sec:summary}

We have presented a detailed analysis of evolution for DPDs, with a focus on the
comparison between colour non-singlet and colour singlet channels.

The evolution of colour non-singlet DPDs shares features of PDF evolution, with
a convolution between splitting kernels $P(z)$ and the evolving distribution,
and of TMD evolution, with a Collins-Soper equation for the dependence on the
rapidity parameter $\zeta$.  At large transverse distances, the corresponding
Collins-Soper kernel is a non-perturbative quantity and currently needs to be
modelled.  For colour octet DPDs, it is equal to the Collins-Soper kernel for
gluon TMDs, and by assuming Casimir scaling we can relate it to the one for
quark TMDs, which is reasonably well known for distances up to $0.8 \fm$ thanks
to recent work in TMD fitting and in lattice QCD.  To model the kernels for DPDs
in the higher colour representations $R=10$ or $27$, we again assume Casimir
scaling.

For small transverse distances $y$ between the two partons, the dominant part of
the initial conditions for DPD evolution can be computed  for all colour
channels from the perturbative splitting of one parton into two partons. Beyond
this perturbative input, phenomenological and theoretical work provides us with
some guidance for modelling the DPDs, but such work is so far limited to the
colour singlet sector.  For the initial conditions of colour non-singlet DPDs we
make a simple ansatz by saturating the colour space bounds proposed in
\cite{Kasemets:2014yna}.

The evolution equations for DPDs are simplest when the rapidity parameter
$\zeta$ (defined with reference to the proton momentum) is kept fixed.
However, we find it more convenient to study the $x_1$ and $x_2$ dependence of
DPDs at $\zeta = \xi / (x_1 x_2)$ with fixed $\xi$.  This is natural for the
perturbative splitting contribution, and it yields a less steep behaviour in
the momentum fractions.  We have derived a representation \eqref{solution-2}
for this case in terms of a $\xi$ dependent exponential (which includes
resummed Sudakov logarithms) and distributions $\widetilde{F}$ that obey a DGLAP
type equation with effective kernels $\widetilde{P}(z)$ that depend on the
initial scales but not on a rapidity parameter.  As one evolves to higher
scales, these effective kernels are increasingly concentrated around the end
point $z=1$, so that the evolution of $\widetilde{F}$ becomes increasingly local
in the momentum fractions.

In our numerical studies, we find that for intermediate distances $y$ and
sufficiently high final scales, colour non-singlet DPDs are suppressed compared
with their colour singlet counterparts for the same parton combinations.
Exceptions to this rule are however seen for large momentum fractions.  Going to
small $y$, one has an increasingly small evolution path between the initial
scale of order $1/y$ and the final scale of the hard-scattering process, and
colour non-singlet configurations become prominent as a result of the colour
factors in their initial conditions.  Going to large $y$, one finds the colour
non-singlet distributions decrease very rapidly, driven by Collins-Soper
evolution with a large negative kernel.

In double parton luminosities, which are products of DPDs integrated over $y$
and directly appear in the cross section formula, the large $y$ region is
therefore much less important for colour non-singlet distributions than for
colour singlet ones.  In a sense, this is good for phenomenology, since it
limits the $y$ region where colour non-singlet distributions need to be known.
We find that for several kinematic settings and parton combinations, colour
non-singlet channels are suppressed by less than a factor $10$ if perturbative
splitting takes place in exactly one of the two DPDs.  For the contribution
with perturbative splitting in both DPDs, colour non-singlet contributions may
even be comparable to or larger than the colour singlet term.  This finding does
not depend on the non-perturbative model input of our study and hence is a firm
prediction.

Increasing the perturbative accuracy of evolution from leading logarithmic
accuracy to \NLLp\ as specified in \tab{\ref{tab:orders}}, we find that DPDs
change by up to 100\% in some cases.  For double parton luminosities, we see
corrections of up to 50\% in the kinematics we investigated.  Going to NNLL
accuracy by  increasing the order of the Collins-Soper kernel and its anomalous
dimension leads to changes below 20\% relative to \NLLp.

In summary, we have shown quantitatively that colour correlations in DPDs are
not generally negligible in kinematics relevant for double parton scattering
processes.  This is due to contributions from the region of small parton
distances $y$, where Sudakov suppression has only a limited effect.  Our
findings motivate further theoretical work.  One direction would be to obtain a
more comprehensive set of hard-scattering cross sections for colour non-singlet
channels than what is presently available.  It also seems important to explore
colour non-singlet DPDs  in a more comprehensive way than we have done here.  An
interesting possibility is presented by lattice calculations using the
quasi-distribution approach, which was recently proposed in
\cite{Jaarsma:2023woo} for DPDs including colour correlations.

\section*{Acknowledgements}

It is a pleasure to thank Tom Cridge, Jonathan Gaunt, Oskar Grocholski, and
Ignazio Scimemi for discussions about various aspects of this work.  The
numerical studies have been performed using the \textsc{ChiliPDF} library
\cite{Diehl:2021gvs, Diehl:2023cth}, which is under development.  We gratefully
acknowledge the contributions of our collaborators Oskar Grocholski, Riccardo
Nagar, and Frank Tackmann to this project.  The Feynman graphs in this
manuscript were produced with JaxoDraw \cite{Binosi:2003yf, Binosi:2008ig}.

This work is in part supported by the Deutsche Forschungsgemeinschaft (DFG,
German Research Foundation) -- grant number 409651613 (Research Unit FOR 2926)
and grant number 491245950.  M.D.\ thanks the Erwin-Schrödinger International
Institute for Mathematics and Physics at the University of Vienna for partial
support during the Programme ``Quantum Field Theory at the Frontiers of the
Strong Interactions'', July 31 to September 1, 2023.

\appendix

\section{Conventions for colour projections}
\label{sec:colour-proj}

In this appendix we give our conventions for colour projected DPDs and
parton-level cross sections.  Starting with quantities that have open colour
indices as shown in \fig{\ref{fig:xsect-graph}}, we define
\begin{align}
   \label{colour-DPD-def}
   \pr{\Rbar_{1} \Rbar_{2}}{F}_{a_1 a_2}
   &=
   \mathcal{N}_{a_1} \, \mathcal{N}_{a_2} \;
   \epsilon(R_1) \, \epsilon(R_2) \,
   \frac{1}{\sqrt{\dim(R_1)}} \;
   P^{r_1^{} r_1' \, r_2^{} r_2'}_{R_1 R_2} \,
   F_{a_1 a_2 \vphantom{R_1}}^{r_1^{} r_1' \, r_2^{} r_2'}
\end{align}
and
\begin{align}
   \label{colour-Xsec-def}
   \pr{\Rbar \Rpbar{}}{\hat{\sigma}}_{a b}
   &=
   \frac{1}{\mathcal{N}_a \, \mathcal{N}_b} \,
   \frac{1}{\epsilon(R) \, \epsilon(\Rp{})} \,
   \frac{1}{\dim(R)} \;
   P^{r r'  s s' \vphantom{r_1}}_{R \Rp{}} \,
   \hat{\sigma}^{r r'  s s'}_{a b \vphantom{R'}}
   \,,
\end{align}
where
\begin{align}
   \mathcal{N}_q &= \mathcal{N}_{\bar{q}} = \sqrt{3}
   \,,
   &
   \mathcal{N}_g &= \sqrt{8}
\end{align}
and
\begin{align}
   \epsilon(A) &= i
   \,,
   &
   \epsilon(R) &= 1 \text{ for } R \neq A
   \,.
\end{align}
$\dim(R)$ is the dimension of the representation $R$, and it is understood that
$\dim(R_1) = \dim(R_2)$ in \eqref{colour-DPD-def} and $\dim(R) = \dim(\Rp{})$ in
\eqref{colour-Xsec-def}.

The matrix $P^{r r'  s s'}_{R \Rp{}}$ projects the indices $(r r')$ onto
$\Rbar{}$ and $(s s')$ onto $\Rpbar{}$; explicit expressions are given in
\sect{4.1} and \app{I.1} of \cite{Buffing:2017mqm}.  As specified there, indices
for the fundamental representation may need to be permuted in
\eqref{colour-DPD-def} and \eqref{colour-Xsec-def} such that a contraction
always connects one index for the triplet with another for the anti-triplet
representation.

For the production of a colour singlet final state, one obviously has
\begin{align}
   \hat{\sigma}^{r r'  s s'}_{a b}
   & \propto
   \delta_{r s}\, \delta_{r' \bs s'}
   \,.
\end{align}
Inserting this into \eqref{colour-Xsec-def} and using the rules of computation
for the colour projectors, one readily obtains the relation
\eqref{Xsect-for-singlet}. The normalisation conventions \eqref{colour-DPD-def}
for DPDs have been chosen such as to obtain the simple form
\eqref{fact-for-singlet} in the factorisation formula for this case, and such
that $\pr{11}{F}$ corresponds to the sum over the colour states of each parton
$a_1$ and $a_2$.  The factors $\epsilon(R_i)$ ensure that DPDs are real valued
for all colour representations other than the decuplet and antidecuplet.

DPDs in the $s$-channel colour basis are defined as
\begin{align}
   \label{colour-DPD-def-s}
   F_{a_1 a_2}^{\Rbar \Rpbar{}}
   &=
   \frac{1}{\dim(R)} \;
   P^{r_1^{} r_2^{} \, r_1' r_2'}_{R \Rp{}} \,
   F_{a_1 a_2 \vphantom{\Rp{}}}^{r_1^{} r_1' \, r_2^{} r_2'}
\end{align}
with the same specification regarding the transposition of fundamental indices
as above.  The normalisation is such that $F^{R \Rbar{}}_{a_1 a_2}$ corresponds
to the probability that the two-parton system is in \emph{one} of the $\dim(R)$
colour states of the representation $R$.  Correspondingly, one has the sum rule
\begin{align}
   \sum_{R} \dim(R) \, F^{R \Rbar{}}_{a_1 a_2}
   &=
   \pr{11}{F}_{a_1 a_2}^{}
   \,.
\end{align}

The transformation between the distributions in \eqref{colour-DPD-def} and
\eqref{colour-DPD-def-s} can be found in \sect{2} of \cite{Diehl:2021wvd} with
the exception of two-gluon distributions, where we find
\begin{align}
   \label{gluon-colour-transform}
   \scriptstyle
   \begin{pmatrix}
      \prn{11}{F} \\[1.25ex] \prn{S S}{F} \\[1.25ex] \prn{A A}{F} \\[1.25ex]
      \prn{\twensev}{F} \\[1.05ex]
      \re \prn{\smalltentenbar}{F} \\[1.05ex]
      \im \prn{\smalltentenbar}{F} \\[1.25ex]
      \prn{\SA}{F} \\[1.3ex] \prn{\AS}{F} \\[0.3ex]
   \end{pmatrix}
   &=
   \; \scriptstyle
   \begin{pmatrix}
      1 & 1 & 1 & 1 & 1 & 1 & 0 & 0 \\[0.8ex]
      \sqrt{8} & -\frac{3 \sqrt{2}}{5} & \sqrt{2} &
         \frac{\sqrt{8}}{5} & -\frac{4 \sqrt{2}}{5} & -\frac{4 \sqrt{2}}{5} &
         0 & 0 \vphantom{\frac{\sqrt{1}}{\sqrt{1}}} \\[0.8ex]
      -\sqrt{8} & -\sqrt{2} & -\sqrt{2} &
         \frac{\sqrt{8}}{3} & 0 & 0 & 0 & 0 \\[0.8ex]
      \sqrt{27} & \frac{\sqrt{27}}{5} & - \sqrt{3} &
         \frac{7 \sqrt{3}}{45} & -\frac{\sqrt{3}}{5} & -\frac{\sqrt{3}}{5} &
         0 & 0 \\[0.8ex]
      \sqrt{10} & -\frac{\sqrt{8}}{\sqrt{5}} & 0
         & -\frac{\sqrt{2}}{3 \sqrt{5}} &
         \frac{\sqrt{2}}{\sqrt{5}} & \frac{\sqrt{2}}{\sqrt{5}} &
         0 & 0 \\[0.8ex]
      0 & 0 & 0 & 0 & 0 & 0 & \sqrt{8} & 0
         \vphantom{\frac{\sqrt{1}}{\sqrt{1}}} \\[0.8ex]
      0 & 0 & 0 & 0 &
         \frac{\sqrt{8}}{\sqrt{5}} & -\frac{\sqrt{8}}{\sqrt{5}} &
         0 & \sqrt{8} \\[0.8ex]
      0 & 0 & 0 & 0 &
         \frac{\sqrt{8}}{\sqrt{5}} & -\frac{\sqrt{8}}{\sqrt{5}} &
         0 & -\sqrt{8}
   \end{pmatrix}
   \,
   \begin{pmatrix}
      F^{11} \\[1.25ex] 8 \, F^{S S} \\[1.25ex] 8 \, F^{A A} \\[1.25ex]
      27 \, F^{\twensev} \\[1.05ex]
      10 \, F^{\smalltentenbar} \\[1.05ex] 10 \, F^{\smalltenbarten} \\[1.25ex]
      8 \, \re F^{\SA} \\[1.3ex] 8 \, \im F^{\SA} \\[0.3ex]
   \end{pmatrix}
   \,.
\end{align}
Here we have split the complex valued DPDs $\prn{\smalltentenbar}{F} = \bigl[\ms
\prn{\smalltenbarten}{F} \bigr]^{*}$ and $F^{\SA} = \bigl[ F^{\AS} \bigr]^{*}$
into their real and imaginary parts.

\section{Evolution of DPDs near the phase space boundary}
\label{sec:large-x}

It has long been known that the DGLAP equations admit an approximate solution in
closed form for PDFs at $x \approx 1$ (see e.g.~\sect{5} in
\cite{Peterman:1978tb} for a corresponding result for deep inelastic structure
functions).  In this section, we derive an analogous solution for DPDs near the
phase space boundary $x_1 + x_2 \approx 1$.  For simplicity, we restrict
ourselves LO evolution; the argument is however easy to generalise to NLO.

To be precise, we consider DPDs in a region of $x_1$ and $x_2$ in which
\begin{align}
   \label{eps-def}
   \varepsilon_1 &= \frac{1-x_1-x_2}{x_1}
   \,,
   &
   \varepsilon_2 &= \frac{1-x_1-x_2}{x_2}
\end{align}
are both much smaller than $1$.  It is required that for each point $(x_1, x_2)$
the region also includes the points $(x_1/z, x_2)$ and $(x_1, x_2/z)$ with
$z<1$, which appear in the integrals of the evolution equations.  We call this
the ``region of interest'' in the following.

We consider the distributions $\widetilde{F}$ introduced in
\sect{\ref{sec:second-solution}} and aim to solve their evolution equation
\begin{align}
   \label{DGLAP-large-x}
   &
   \frac{d}{d \ln\mu_1}\,
      \prn{R_1 R_2}{\widetilde{F}}_{a_1 a_2}(x_1,x_2, {y}; \mu_1,\mu_2)
   \notag \\
   & \qquad
   =
   2 \int_{x_1}^{1-x_2} \frac{d z}{z} \;
      \prn{R_1 \Rbar_1}{\widetilde{P}}_{a_1 a_1}\biggl(
         \frac{x_1}{z}; \mu_1^{} \biggr)
      \prn{R_1 R_2}{\widetilde{F}}_{a_1 a_2}(z, x_2^{}, {y}; \mu_1, \mu_2)
\end{align}
and the analogous equation for evolution in $\mu_2$, with a kernel
\begin{align}
   \label{large-z-kernel}
   \prn{R \Rbar}{\widetilde{P}}_{a a}(z; \mu)
   &= \frac{\alpha_s(\mu)}{2 \pi} \; \frac{1}{2} \,
      \biggl[
         \frac{\prn{R}{s}_a}{(1 - z)_+} + d_a\, \delta(1-z) + \mathcal{O}(1)
      \biggr]
   & \text{ for } z\to 1.
\end{align}
For brevity we omit the arguments $(\mu_{01},\mu_{02},\xi_0)$ in $\widetilde{F}$
here, as well as the arguments $\mu_{01}$ and $J$ in $\widetilde{P}$.
The coefficients of the plus-distribution term in \eqref{large-z-kernel} read
\begin{align}
   \label{colour-factors-plus}
   \prn{1}{s}_{q} &= 4 C_F = 16/3 \,,
   &
   \prn{8}{s}_{q} &= -2/3 \,,
   \notag \\
   \prn{1}{s}_{g} &= 4 C_A = 12 \,,
   &
   \prn{S}{s}_{g} &= \prn{A}{s}_{g} = 6 \,,
   &
   \prn{\smallten}{s}_{g} &= \prn{\smalltenbar}{s}_{g} = 0 \,,
   &
   \prn{27}{s}_{g} &= -4 \,,
\end{align}
where we have used \eqref{LO-scaling} to \eqref{P-factors-g}.

Having evolved $\widetilde{F}$, one obtains the full DPD $F$ by multiplying with
the exponential factors given in \eqref{solution-2}. The kernels in the
evolution equations \eqref{DGLAP-red-2} indeed have the form
\eqref{large-z-kernel}, as is easily seen by expanding the exponential in
\eqref{eff-kernel} around $z=1$.  Note however that we have dropped flavour
mixing terms in the evolution equation \eqref{DGLAP-large-x}; we will come back
to this point at the very end of this appendix.

Let us change variables from the scales $\mu_{i}$ to the evolution times
\begin{align}
   \tau_{i} &= - \ln \alpha_s(\mu_i)
   &
   \text{ for } i=1,2
\end{align}
and consider initial conditions at $\tau_{0\ms i} = - \ln \alpha_s(\mu_{0\ms
i})$ of the form
\begin{align}
   \label{large-x-init}
   \widetilde{F}_{a_1 a_2}(x_1,x_2; \tau_{01},\tau_{02})
   &=
   \frac{\exp\bigl[ - \gamma_E \, \beta_{a_1 a_2}(\tau_{01},\tau_{02}) \bigr]}{
   \Gamma\bigl( 1 + \beta_{a_1 a_2}(\tau_{01},\tau_{02}) \bigr)} \;
   \bigl( 1-x_1-x_2 \bigr)^{\beta_{a_1 a_2}(\tau_{01},\tau_{02})}
   \notag \\[0.2em]
   &\quad \times
   \exp\bigl[ \delta_{a_1 a_2}(x_1, x_2; \tau_{01},\tau_{02}) \bigr]
   \,,
\end{align}
where $\delta_{a_1 a_2}(x_1, x_2)$ is a regular function that in the $x_1, x_2$
region of interest has first derivatives satisfying
\begin{align}
   \label{deriv-condition}
   \biggl|\ms x_1\, \frac{\partial}{\partial x_1} \,
      \delta_{a_1 a_2}(x_1, x_2) \ms\biggr|
   &< C
   \,,
   &
   \biggl|\ms x_2\, \frac{\partial}{\partial x_2} \,
      \delta_{a_1 a_2}(x_1, x_2) \ms\biggr|
   &< C
\end{align}
with some constant $C$.  We suppress the dependence of the DPDs on ${y}$
and on the colour representations from now on, and it is understood that
$\beta_{a_1 a_2}$, $\delta_{a_1 a_2}$, and $C$ may depend on these parameters.

We will show that \eqref{DGLAP-large-x} is approximately solved by
\begin{align}
   \label{large-x-solution}
   \widetilde{F}_{a_1 a_2}(x_1,x_2; \tau_1,\tau_2)
   &=
   \frac{\exp\bigl[ - \gamma_E \, \beta_{a_1 a_2}(\tau_1, \tau_2) \bigr]}{
      \Gamma\bigl( 1 + \beta_{a_1 a_2}(\tau_1, \tau_2) \bigr)} \;
   (1-x_1-x_2)^{\beta_{a_1 a_2}(\tau_1, \tau_2)}
   \notag \\[0.2em]
   & \quad \times
   \exp\bigl[ \delta_{a_1 a_2}(x_1, x_2; \tau_1, \tau_2) \bigr]
\end{align}
with
\begin{align}
   \label{beta-delta-defs-2}
   \beta_{a_1 a_2}(\tau_1, \tau_2)
   &= \beta_{a_1 a_2}(\tau_{01}, \tau_{02})
      + \Delta\beta_{a_1}(\tau_1, \tau_{01})
      + \Delta\beta_{a_2}(\tau_2, \tau_{02})
   \,,
   \notag \\[0.3em]
   \delta_{a_1 a_2}(x_1, x_2; \tau_1, \tau_2)
   &= \delta_{a_1 a_2}(x_1, x_2; \tau_{01}, \tau_{02})
      + \Delta\delta_{a_1}(\tau_1, \tau_{01})
      + \Delta\delta_{a_2}(\tau_2, \tau_{02})
   \notag \\[0.15em]
   & \quad
   - \Delta\beta_{a_1}(\tau_1, \tau_{01}) \, \ln x_1
   - \Delta\beta_{a_2}(\tau_2, \tau_{02}) \, \ln x_2
   \,,
\end{align}
and
\begin{align}
   \label{beta-delta-defs-1}
   \Delta\beta_{a_i}(\tau_i, \tau_{0\ms i})
   &= \frac{s_{a_i}}{\beta_0}\, (\tau_i - \tau_{0\ms i})
   \,,
   &
   \Delta\delta_{a_i}(\tau_i, \tau_{0\ms i})
   &= \frac{d_{a_i}}{\beta_0}\, (\tau_i - \tau_{0\ms i})
   &
   \text{ for } i=1,2,
\end{align}
where $\beta_0$ is given in \eqref{beta-0}.  The function $\delta_{a_1 a_2}(x_1,
x_2; \tau_1, \tau_2)$ satisfies \eqref{deriv-condition} with a constant
depending on $\tau_1$ and $\tau_2$.
The solution \eqref{large-x-solution} is approximate in the sense that there are
correction terms of order $\varepsilon_1 \ms \alpha_s(\mu_1) \,
\widetilde{F}_{a_1 a_2}(x_1,x_2; \mu_1,\mu_2)$ on the r.h.s.\ of the evolution
equation \eqref{DGLAP-large-x} in $\mu_1$ and corresponding corrections of order
$\varepsilon_2 \, \alpha_s(\mu_2) \, \widetilde{F}_{a_1 a_2}(x_1,x_2;
\mu_1,\mu_2)$ in its analogue for evolution in~$\mu_2$.

As we will see below, the derivation of \eqref{large-x-solution} is valid only
for
\begin{align}
   \label{beta-condition}
   \beta_{a_1 a_2}(\tau_1, \tau_2) > -1
   \,.
\end{align}
If this power is smaller than $-1$, then the convolution integrals in the
evolution equations diverge.  If the power is exactly equal to $-1$, the
prefactor $1 / \Gamma\bigl( 1 + \beta_{a_1 a_2}(\tau_1, \tau_2) \bigr)$ goes to
zero.  What happens at this point depends on the order in which one takes the
limits $x_1 + x_2 \to 1$ and $\beta_{a_1 a_2}(\tau_1, \tau_2) \to -1$.  This is
a problem of mathematical interest, which we shall not pursue here.

Before proving the result \eqref{large-x-solution} let us explore its
consequences.  The power $\beta_{a_1 a_2}(\tau_1, \tau_2)$ describes how fast a
distribution vanishes at the phase space boundary.  It increases when going to
higher scales for the colour singlet representations, as well as for the two
octet representations for gluons.  By contrast, this power \emph{decreases} when
going to higher scales for quarks in the colour octet and for gluons in the
icosaheptaplet.  At sufficiently high scales, one will therefore find scales for
which $\beta_{a_1 a_2}(\tau_1, \tau_2)$ approaches the critical value
$-1$.

However, these scales turn out to be so large that they are not of practical
relevance.\footnote{A similar issue was discussed in
\protect\cite{Blok:2022mtv}, albeit in a different form and with different
conclusions.}
This can be seen in \tab{\ref{tab:alpha_s_etc}}, which shows the change of
$\beta_{g g}$ for $R=27$.  The highest scale $2.9 \times 10^6 \gev$ corresponds
to the c.m.\ energy of a proton with energy $320 \times 10^{9} \gev$ hitting a
nitrogen nucleus at rest.  This energy was reported for a cosmic ray event in
\cite{Bird:1994uy}.  We see in the table that even for such extreme scales,
$\beta_{g g}$ decreases by only 2 units if one evolves in both $\mu_1$ and
$\mu_2$.  According to \eqref{colour-factors-plus}, the corresponding decrease
of $\beta_{q \bar{q}}$ for $R=8$ is smaller by a factor $1/6$.

The value of $\beta_{a_1 a_2}$ at low scales is determined by the large $x$
behaviour of the PDF $f_{a_0}(x)$ for the splitting form
\eqref{splitting-DPD-pt}, whilst it is equal to $2$ for the product ansatz given
by \eqref{int-DPD-singlet} and \eqref{int-DPD-colour}.  In all cases,
$\beta_{a_1 a_2}$ remains well above $-1$ after evolution to the highest scale
in \tab{\ref{tab:alpha_s_etc}}.

\begin{table}
\renewcommand{\arraystretch}{1.2}
\begin{center}
\begin{tabular}{c c c c c c}
\hline
$n_f$ & $\mu_{0}$ & $\mu$ &
$\alpha_s(\mu_0)$ & $\alpha_s(\mu)$ &
$\pr{27}{\Delta\beta}_{g}(\tau, \tau_0)$ \\
\hline
$4$ & $m_c$ & $m_b$                  & $0.405$ & $0.245$  & $-0.242$ \\
$5$ & $m_b$ & $m_t$                  & $0.245$ & $0.118$  & $-0.380$ \\
$6$ & $m_t$ & $2.9 \times 10^6 \gev$ & $0.118$ & $0.0518$ & $-0.471$ \\
\hline
    &       &                   &         &          & $-1.093$ \\
\hline
\end{tabular}
\caption{\label{tab:alpha_s_etc} The function $\Delta \beta_{g}$ for $R=27$ and
different initial and final scales.  Values are computed for the masses $m_c =
1.4 \gev$, $m_b = 4.75 \gev$, $m_t = 172.5 \gev$ and the coupling $\alpha_s(m_Z)
= 0.13$ used in the default MSHT20 PDF set at LO.  The last row shows the sum
of the three values for $n_f = 4, 5, 6$.}
\end{center}
\end{table}



\paragraph{Proof of \eqn{\eqref{large-x-solution}}.}

In the following, we suppress not only the argument ${y}$ and the colour labels,
but also the labels $a_1, a_2$ designating the partons.
After changing variables from $\mu_i$ to $\tau_i$ and inserting the form
\eqref{large-z-kernel} of the kernel, the evolution equation
\eqref{DGLAP-large-x} reads
\begin{align}
   \label{large-x-evol-1}
   \frac{d}{d \tau_1}\,
      \widetilde{F}(x_1,x_2; \tau_1,\tau_2)
   &=
   \frac{4 \pi}{\beta_0 \, \alpha_s(\mu_1)} \int_{x_1}^{1-x_2} \frac{d z}{z} \,
      \widetilde{P}\biggl(
         \frac{x_1}{z}; \mu_1^{} \biggr)
      \widetilde{F}(z,x_2^{}; \tau_1,\tau_2)
   \notag \\[0.4em]
   &=
   \frac{1}{\beta_0} \int_{z_0}^1 \frac{d z}{z} \,
   \biggl[ \frac{s}{(1 - z)_+}
      + d \, \delta(1-z) + \mathcal{O}(1) \biggr] \,
   \widetilde{F}\Bigl( \frac{x_1}{z}, x_2; \tau_1,\tau_2 \Bigr)
   \notag \\[0.4em]
   &=
   \frac{1}{\beta_0} \,
   \biggl\{
      s \int_{z_0}^1 \frac{d z}{z} \,
      \frac{1}{1-z} \, \biggl[
         \widetilde{F}\Bigl( \frac{x_1}{z},x_2; \tau_1,\tau_2 \Bigr)
         - \widetilde{F}(x_1, x_2; \tau_1,\tau_2) \biggr]
   \notag \\[0.2em]
   &\qquad
   + \biggl[
      s \, \ln \frac{1 - z_0}{z_0}
      + d + \mathcal{O}(1 - z_0)
     \biggr] \,
     \widetilde{F}(x_1, x_2; \tau_1,\tau_2)
   \biggr\}
\end{align}
with
\begin{align}
   z_0
   &= \frac{x_1}{1 - x_2}
   = 1 - \frac{x_1}{1 - x_2} \, \varepsilon_1
   \,,
\end{align}
where in the last step of \eqref{large-x-evol-1} we used the definition of the
plus-distribution.  We have
\begin{align}
   \label{approx-difference}
   &
   \widetilde{F}\Bigl( \frac{x_1}{z},x_2; \tau_1,\tau_2 \Bigr)
   \Big/ \widetilde{F}(x_1, x_2; \tau_1,\tau_2)
   \notag \\
   & \quad =
   \biggl(
      \frac{1 - x_1/z - x_2}{1 - x_1 - x_2}
   \biggr)^{\beta(\tau_1, \tau_2)}
   \exp\biggl[
      \delta\Bigl( \frac{x_1}{z}, x_2; \tau_1, \tau_2 \Bigr)
      - \delta(x_1, x_2; \tau_1, \tau_2)
   \biggr]
   \notag \\
   & \quad =
   \biggl(
      \frac{1 - x_1/z - x_2}{1 - x_1 - x_2}
   \biggr)^{\beta(\tau_1, \tau_2)}
   \exp\biggl[
      \Bigl( \frac{1}{z} - 1 \Bigr) \,
         x_1\, \frac{\partial}{\partial x_1} \,
         \delta(\xi_1, x_2; \tau_1, \tau_2)
   \biggr]
   \,,
\end{align}
where in the last expression we used the fundamental theorem of calculus, with
$\xi_1$ being some value between $x_1/z$ and $x_1$.  Changing variables from $z
\in [z_0, 1]$ to $u \in [0,1]$, with
\begin{align}
   u &= \frac{1}{\varepsilon_1} \, \frac{1-z}{z}
   \,,
   &
   z &= \frac{1}{1 + u \varepsilon_1}
   \,,
   &
   \frac{d z}{z (1-z)} &= - \frac{d u}{u}
   \,,
\end{align}
we obtain
\begin{align}
   \widetilde{F}\Bigl( \frac{x_1}{z},x_2; \tau_1,\tau_2 \Bigr)
   \Big/ \widetilde{F}(x_1, x_2; \tau_1,\tau_2)
   &=
   (1 - u)^{\beta(\tau_1, \tau_2)} \,
   \bigl[ 1 + \mathcal{O}(\varepsilon_1 u) \bigr]
   \,,
\end{align}
where for the approximation of the exponential in \eqref{approx-difference} we
used that \eqref{deriv-condition} holds in the $x_1, x_2$ region of interest.
Inserting this into \eqref{large-x-evol-1} and dividing both sides by $F(x_1,
x_2)$, we get
\begin{align}
   \label{large-x-evol-2}
   &
   \frac{d}{d \tau_1}\,
      \ln \widetilde{F}(x_1,x_2; \tau_1,\tau_2)
   \notag \\
   & \quad =
   \frac{1}{\beta_0} \,
   \biggl\{ \,
      s \int_0^1 \frac{d u}{u} \,
         \bigl[ (1 - u)^{\beta(\tau_1, \tau_2)} - 1 \bigr]
      + \mathcal{O}(\varepsilon_1) \,
         s \int_0^1 d u\; (1 - u)^{\beta(\tau_1, \tau_2)}
   \notag \\
   & \qquad\qquad
      + s \ms \ln \frac{1 - x_1 - x_2}{x_1} + d + \mathcal{O}(\varepsilon_1)
   \biggr\}
   \notag \\
   & \quad =
   \frac{1}{\beta_0} \,
   \biggl\{
      - s \ms \Bigl[ \gamma_E + \Psi\bigl( 1 + \beta(\tau_1, \tau_2) \bigr) \Bigr]
      + s \ms \ln (1 - x_1 - x_2) + d - s \ms \ln x_1
   \biggr\}
   + \mathcal{O}(\varepsilon_1)
   \,,
\end{align}
where $\Psi(z) = d \ln \Gamma(z) / d z$.  Note that to perform the integral over
$u$ we used the condition in \eqref{beta-condition}.  For $\beta(\tau_1, \tau_2)
\le -1$ the integral diverges, and the evolution equations become ill defined.
Using \eqref{beta-delta-defs-2} and \eqref{beta-delta-defs-1}, one readily sees
that \eqref{large-x-solution} satisfies \eqref{large-x-evol-2}.


\paragraph{Quark-gluon mixing.}  If one restores the flavour mixing terms in
\eqref{DGLAP-large-x}, one finds that the r.h.s.\ of the evolution equation
includes terms of order $\varepsilon_1 \ms \alpha_s(\mu_1) \, F_{b_1 a_2}$ for
evolution in $\mu_1$. These can be comparable to or larger than the terms
retained in \eqref{DGLAP-large-x} if $\beta_{b_1 a_2}(\tau_1, \tau_2)$ is
significantly smaller than $\beta_{a_1 a_2}(\tau_1, \tau_2)$. Whether this
happens for the evolution of $\pr{88}{F}_{q \bar{q}}$ depends on the initial
conditions.  On the other hand, the distribution $\pr{\twensev}{F}_{g g}$ does
not mix with quarks under evolution, so that the analysis of this appendix is
always valid for this case.

\section{Numerical setup and accuracy in ChiliPDF}
\label{sec:chili}

In this appendix, we specify the accuracy settings used for the discretisation
and evolution of DPDs in the present work.  The settings for colour singlet DPDs
are identical to the ones documented in the detailed study \cite{Diehl:2023cth},
except for the grids in $y$.

The evolution equations for DPDs are solved using the eighth-order Runge-Kutta
algorithm DOPRI8 \cite{Dormand:1980aa}, with step sizes
\begin{align}
   \label{RK-steps}
   h &=
   \begin{cases}
      0.22 & \text{ for colour singlet DPDs,} \\
      0.15 & \text{ for $\dim(R) = 8$,} \\
      0.07 & \text{ for $\dim(R) = 10$,} \\
      0.02 & \text{ for $\dim(R) = 27$}
   \end{cases}
\end{align}
in the evolution times $\tau_i = - \ln \alpha_s(\mu_i)$.

The momentum fractions $x_1$ and $x_2$ of DPDs are discretised on Chebyshev
grids in the variables $u_i = \ln x_i$.  The grids are taken as
\begin{align}
   \label{x-grids}
   & \bigl[ 10^{-5}, 5 \times 10^{-3}, 0.5, 1 \bigr]_{(16,16,24)}
   && \text{for colour singlet DPDs,}
   \notag \\
   & \bigl[ 10^{-5}, 3 \times 10^{-4}, 10^{-2}, 0.5, 1 \bigr]_{(16,16,24,24)}
   && \text{for colour non-singlet DPDs,}
\end{align}
where $[ x_0, x_1, \ldots x_k ]_{( n_1, \ldots n_k )}$ denotes a set of $k$
subgrids, with $n_{\ell}$ grid points in the subinterval $[x_{\ell - 1},
x_{\ell}]$.  Using several subintervals instead of a single one improves the
interpolation accuracy for functions that assume a wide range of values, as
explained in \cite{Diehl:2021gvs, Diehl:2023cth}.

As discussed below \eqref{internal-convol}, large negative values of the
Collins-Soper kernel give rise to very steep functions of $x_1$ and $x_2$ in the
intermediate steps of evolution, which degrades the accuracy of Chebyshev
interpolation.  To avoid associated numerical instabilities, we freeze the
functions $\pr{R}{\Delta J}(y)$ when they reach the value $-10$. This has no
impact on physics results, because at such large values of the Collins-Soper
kernel, evolved DPDs are suppressed so strongly that their contribution to
double parton luminosities is negligible.

\begin{figure}
\centering
\subfloat[$F_{g g}$ at $y = 0.5 \gev^{-1}$]{
   \includegraphics[width=0.48\textwidth,trim=0 20 35 46,clip]{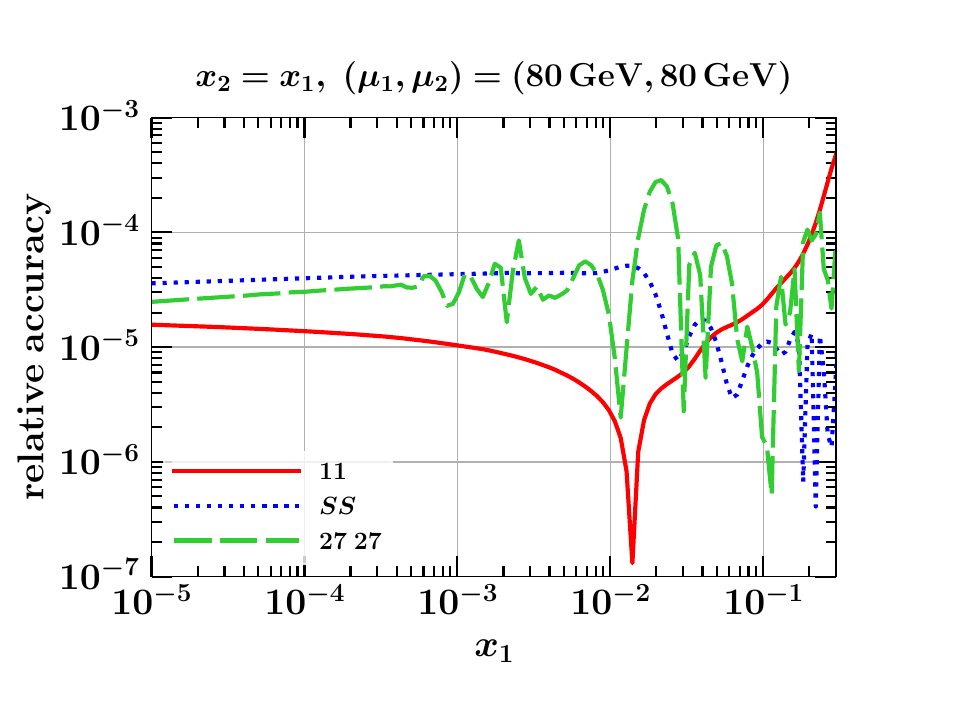}
}
\subfloat[$F_{g g}$ at $y = 1.5 \gev^{-1}$]{
   \includegraphics[width=0.48\textwidth,trim=0 20 35 46,clip]{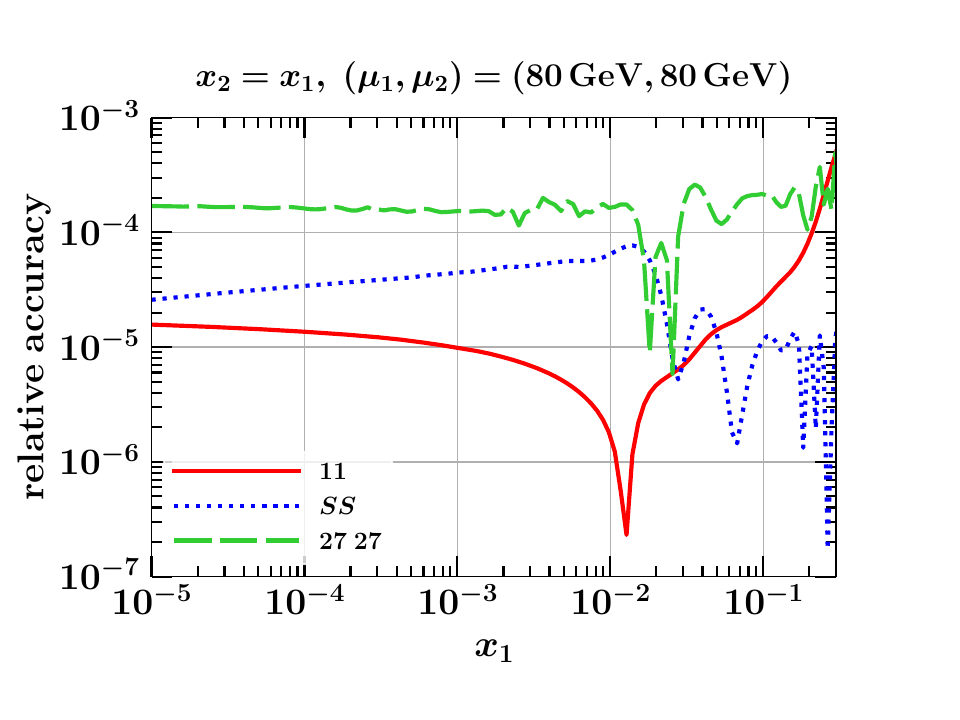}
}
\caption{\label{fig:acc-xgrids} Relative Runge-Kutta and discretisation errors
for DPD evolution in different colour channels.  The DPDs are initialised with
the splitting form at $2.54 \gev$ and evolved to $80 \gev$ for both partons,
and the accuracy is evaluated for $x_1 = x_2$ and two values of $y$.  More
details are given in the text.}
\end{figure}

To assess the accuracy of our settings, we evolve the splitting form
\eqref{splitting-DPD} at LO accuracy from $\mu_1 = \mu_2 = 2.54 \gev$ (the
initial scale for $y = 0.5 \gev^{-1}$) to $\mu_1 = \mu_2 = 80 \gev$, with
matching from $n_f = 4$ to $5$ flavours at $\mu = m_b$.  The joint Runge-Kutta
and discretisation error is estimated by repeating the calculation with more
precise settings in which $(i)$ the step size in \eqref{RK-steps} is multiplied
by $0.1$ in each colour channel and $(ii)$ the number of points is raised to
$40$ in each subinterval of the grids in \eqref{x-grids}.  The relative
difference between the two-gluon DPDs obtained with the two settings is shown in
\fig{\ref{fig:acc-xgrids}} for different colour channels.  As anticipated in
\sect{\ref{sec:solutions}}, one needs denser grids in the colour non-singlet
sector in order to obtain a competitive accuracy, and the accuracy of colour
non-singlet evolution degrades as one goes to higher $y$.  The latter happens,
however, in a region in which the corresponding DPDs are very small and hence
play a minor role in cross sections, as one can see in \fig{\ref{fig:F-vs-y}}.

\begin{figure}
\centering
\subfloat[grids \eqref{low-y-grids} and \eqref{high-y-grid}]{
   \includegraphics[width=0.48\textwidth,trim=0 25 50 57,clip]{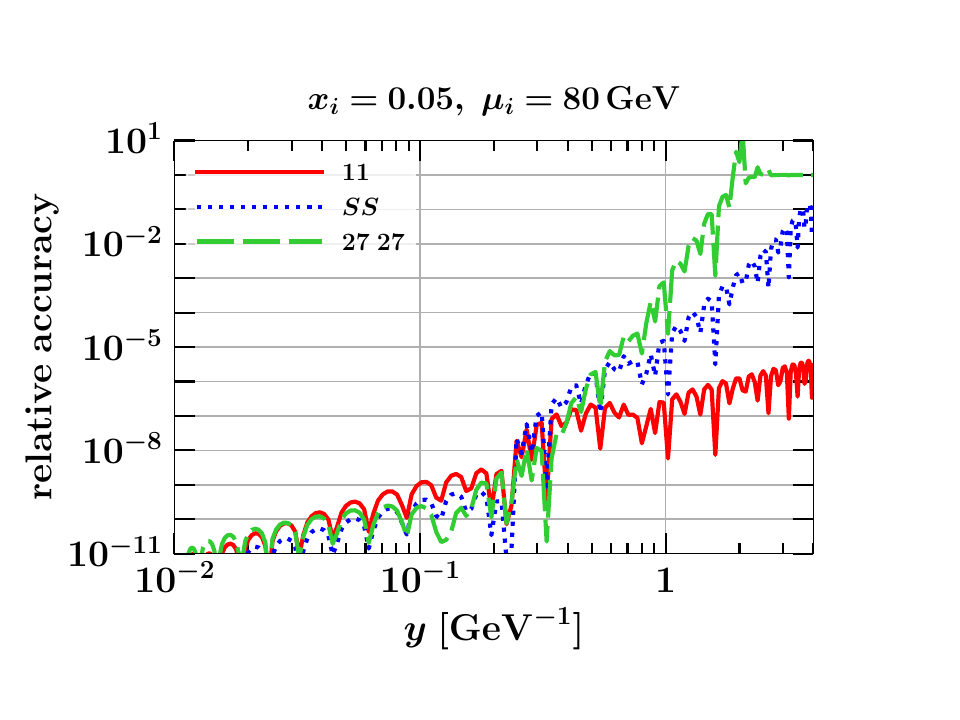}
}
\subfloat[grids \eqref{low-y-grids} and \eqref{prec-y-grid}]{
   \includegraphics[width=0.48\textwidth,trim=0 25 50 57,clip]{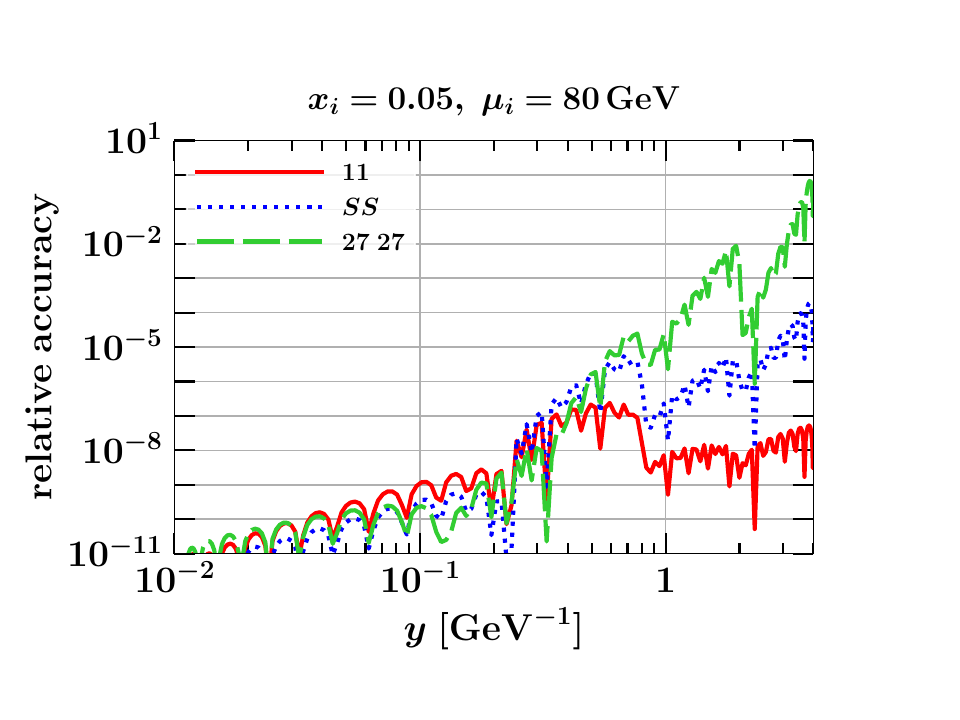}
}
\caption{\label{fig:acc-ygrids} Relative interpolation accuracy of $F_{g
g}^{\smash{\text{spl}}}$ at $x_1 = x_2 = 0.05$ and $\mu_1 = \mu_2 = 80 \gev$.
The $y$ dependence of $F_{g g}^{\smash{\text{spl}}}$ in the different colour
channels is shown in \fig{\protect\ref{fig:F-vs-y}}b.  Panel (a) is for our
default grids, and panel (b) for the improved grid in
\protect\eqref{prec-y-grid}.  The accuracy is computed by comparing the
specified grids with grids that have twice as many points in each subinterval.}
\end{figure}

For the $y$ dependence of DPDs, we take the grids
\begin{align}
   \label{low-y-grids}
   & \bigl[ 0.01 \gev^{-1},\, b_0 / m_b \bigr]_{(16)}
   &
   \text{ and }
   &
   & \bigl[ b_0 / m_b,\, b_0, / m_c \bigr]_{(12)}
\end{align}
with $u(y) = - y^{-0.2}$, and
\begin{align}
   \label{high-y-grid}
   & \bigl[ b_0 / m_c,\, 5 \gev^{-1}, \infty \bigr]_{(24, 12)}
\end{align}
with $u(y) = - \exp\bigl[\ms - (m^2 y^2 + m y ) \big/ 4 \ms\bigr]$ and $m =
0.232 \gev$, where $u(y)$ is the variable used for Chebyshev interpolation.
DPDs are initialised with $n_f = 5$ flavours on the first grid in
\eqref{low-y-grids} and with $n_f = 4$ flavours on the other grids.
As seen in \fig{\ref{fig:acc-ygrids}}a, this setup yields very accurate
interpolation for $y$ up to $b_0/m_c \approx 0.8 \gev^{-1}$.  The grids are also
sufficient for computing the integrals over $y$ appearing in double parton
luminosities.
However, they become insufficient for the interpolation of colour non-singlet
DPDs at large $y$, where these functions decrease very fast. For the DPD plots
in \fig{\ref{fig:F-vs-y}}, we have therefore replaced the grid in
\eqref{high-y-grid} with
\begin{align}
   \label{prec-y-grid}
   & \bigl[ b_0 / m_c,\,  2 \gev^{-1},\ms  5 \gev^{-1},
      \infty \bigr]_{(16, 16, 12)}
   \,.
\end{align}
Figure~\ref{fig:acc-ygrids}b shows that this significantly improves the accuracy
in the relevant $y$ range.

\section{Additional plots of double parton luminosities}
\label{sec:more-lumis}

This appendix contains results for double parton luminosities for parton
combinations not discussed in the main part of the paper.  The settings for all
plots are as in \sect{\ref{sec:lumis-LL}}, to which we refer for further
explanation.


\subsection{\texorpdfstring{$W$}{W} pairs and \texorpdfstring{$W$}{W} plus jets}
\label{sec:lumis-Ws-jets}

\begin{figure}
\centering
\subfloat[2v2 \, $(c \bar{b}, \bar{s} c) + (c \bar{s}, \bar{b} c)$]{
   \includegraphics[width=0.48\textwidth,trim=0 20 30 39,clip]{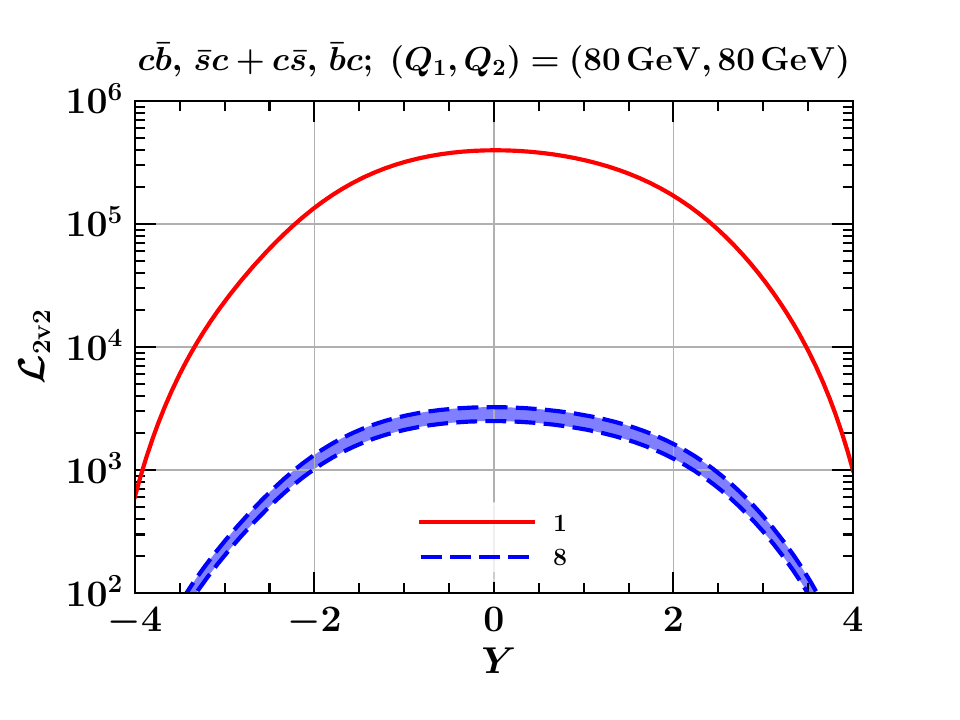}
}
\subfloat[1v2 + 2v1 \, $(c \bar{b}, \bar{s} c) + (c \bar{s}, \bar{b} c)$]{
   \includegraphics[width=0.48\textwidth,trim=0 20 30 39,clip]{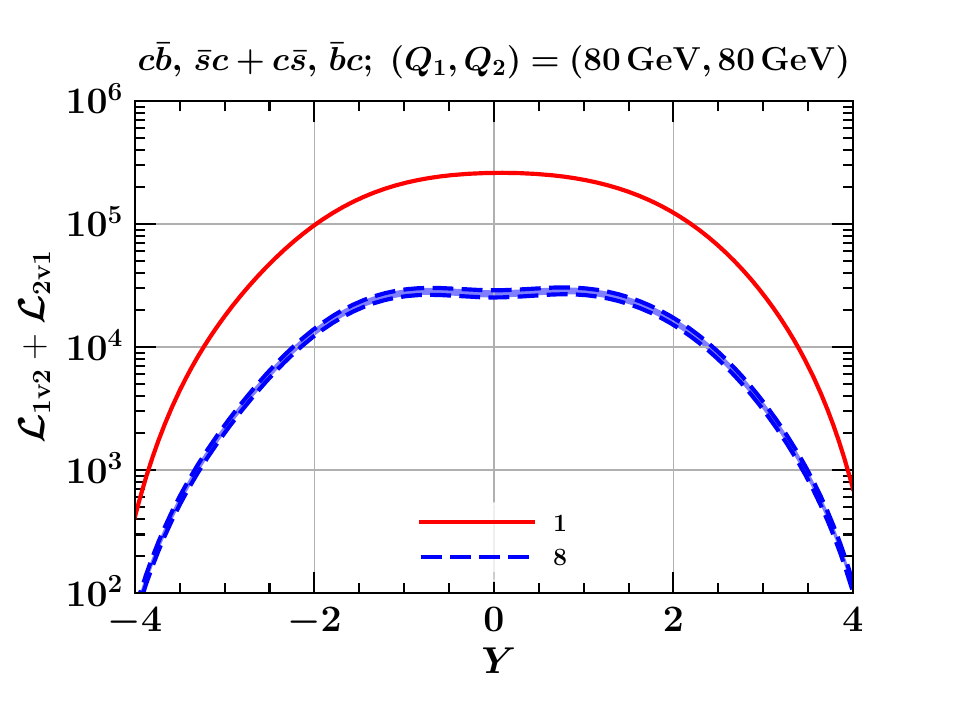}
}
\\[1.5em]
\subfloat[1v1 \, $(c \bar{b}, \bar{s} c) + (c \bar{s}, \bar{b} c)$]{
   \includegraphics[width=0.48\textwidth,trim=0 20 30 39,clip]{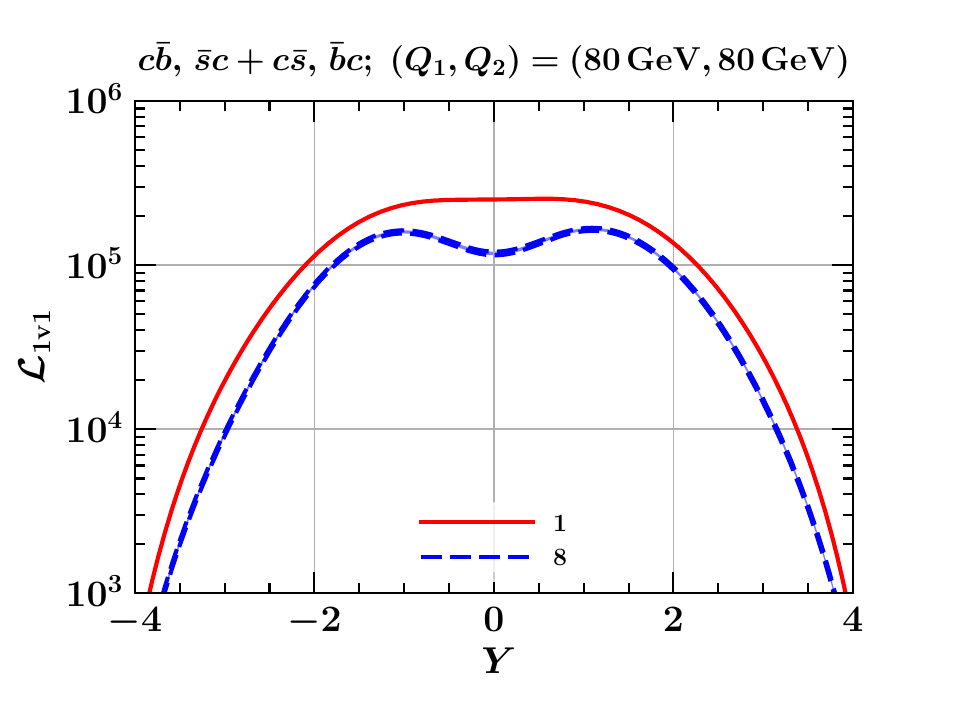}
}
\caption{\label{fig:lumis-WW-same-heavy} As
\fig{\protect\ref{fig:lumis-WW-same-light}}, but for $\mathcal{L}_{c
\smash{\bar{b}}, \bar{s} c} + \mathcal{L}_{c \bar{s}, \smash{\bar{b}} c}$.  Here
we sum over two parton combinations that have the same product of
hard-scattering cross sections.}
\end{figure}

Figure~\ref{fig:lumis-WW-same-heavy} shows luminosities that contribute to $W^+
W^+$ production and contain heavy quarks.  The relative size of colour singlet
and octet contributions is similar to the one for light quarks in
\fig{\ref{fig:lumis-WW-same-light}} but has a different dependence on the
rapidity $Y$.

\begin{figure}
\centering
\subfloat[2v2 \, ($u g, \bar{d} g$)]{
   \includegraphics[width=0.48\textwidth,trim=0 20 30 39,clip]{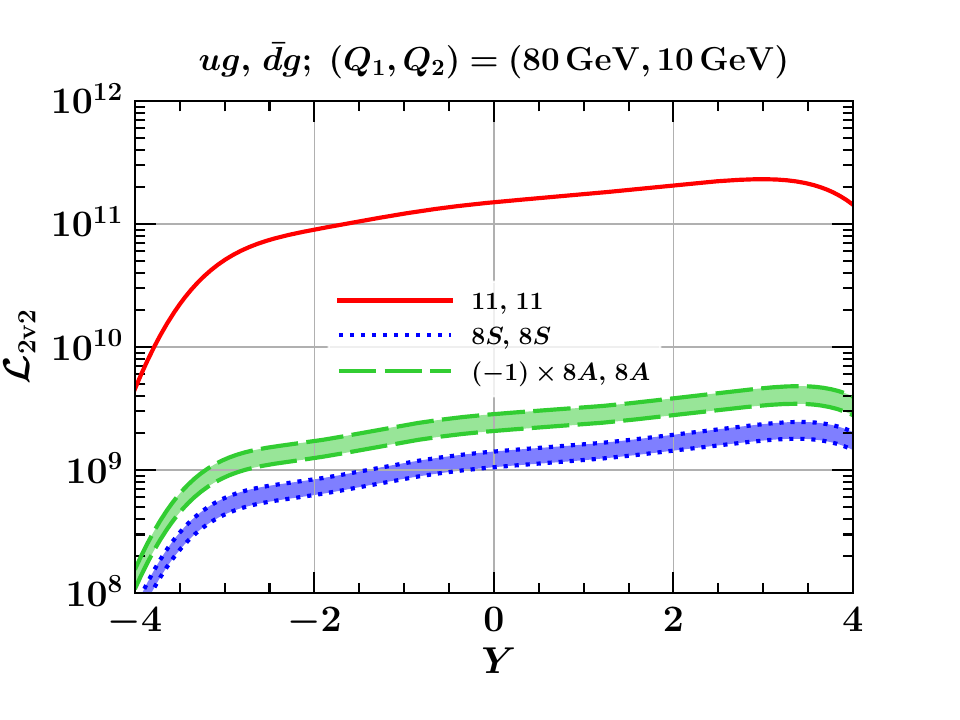}
}
\subfloat[1v2 + 2v1 \, ($u g, \bar{d} g$)]{
   \includegraphics[width=0.48\textwidth,trim=0 20 30 39,clip]{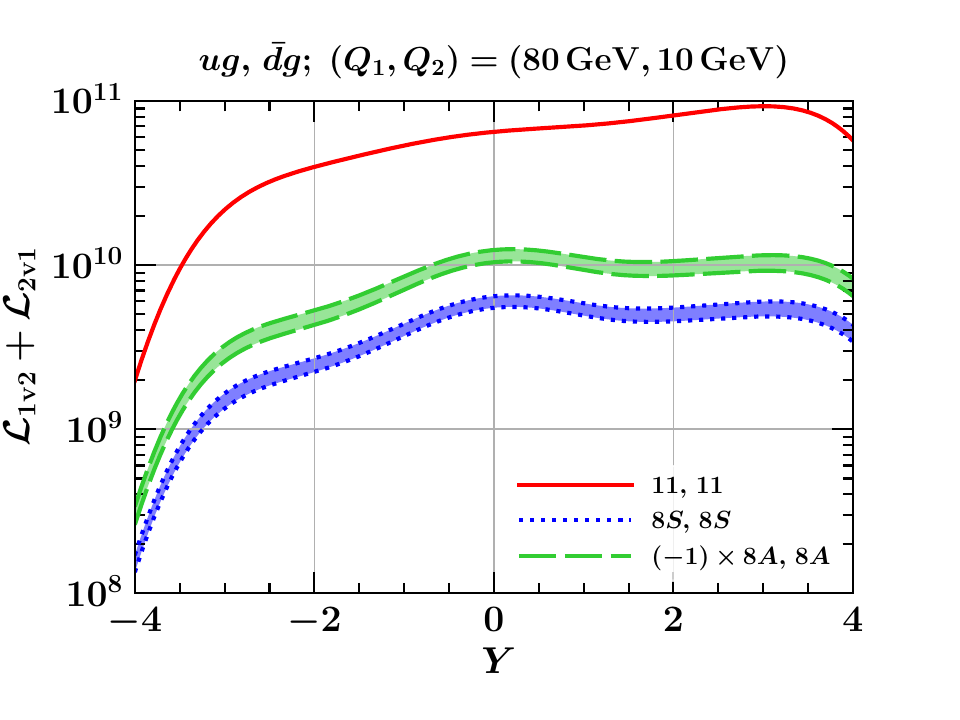}
}
\\[1.5em]
\subfloat[1v1 \, ($u g, \bar{d} g$)]{
   \includegraphics[width=0.48\textwidth,trim=0 20 30 39,clip]{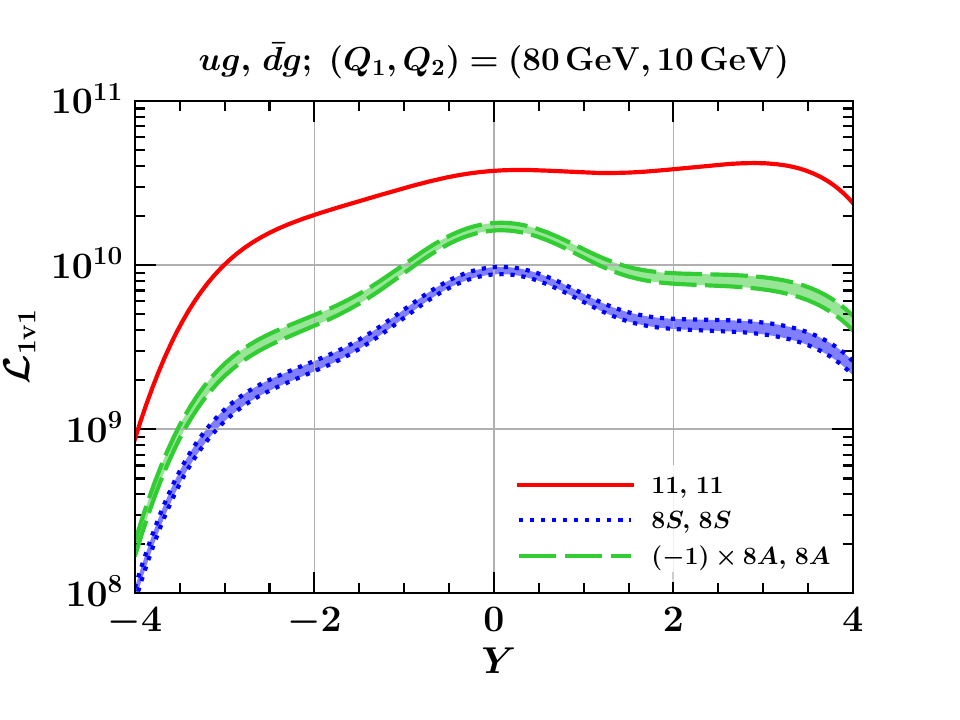}
}
\caption{\label{fig:lumis-W-plus-jets-10} Double parton luminosities
$\mathcal{L}_{u g, \smash{\bar{d}} g}$ for  $M_1 = 80 \gev$ and $M_2 = 10 \gev$.
This parton combination contributes to the production of a $W^+$ and a dijet.}
\end{figure}

\begin{figure}
\centering
\subfloat[2v2 \, ($u g, \bar{d} g$)]{
   \includegraphics[width=0.48\textwidth,trim=0 20 30 39,clip]{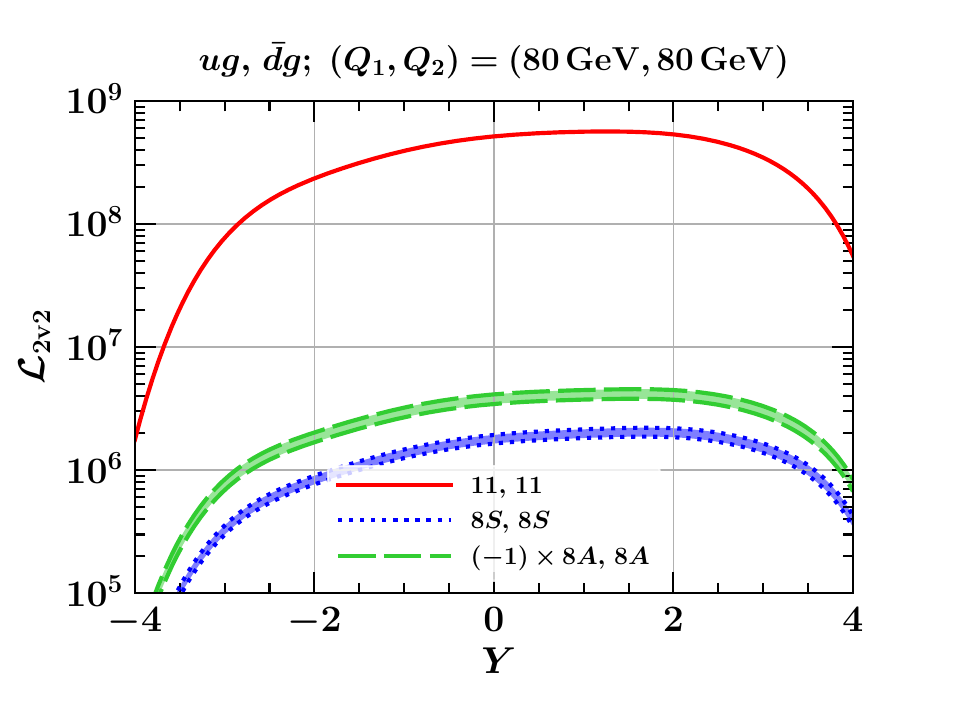}
}
\subfloat[1v2 + 2v1 \, ($u g, \bar{d} g$)]{
   \includegraphics[width=0.48\textwidth,trim=0 20 30 39,clip]{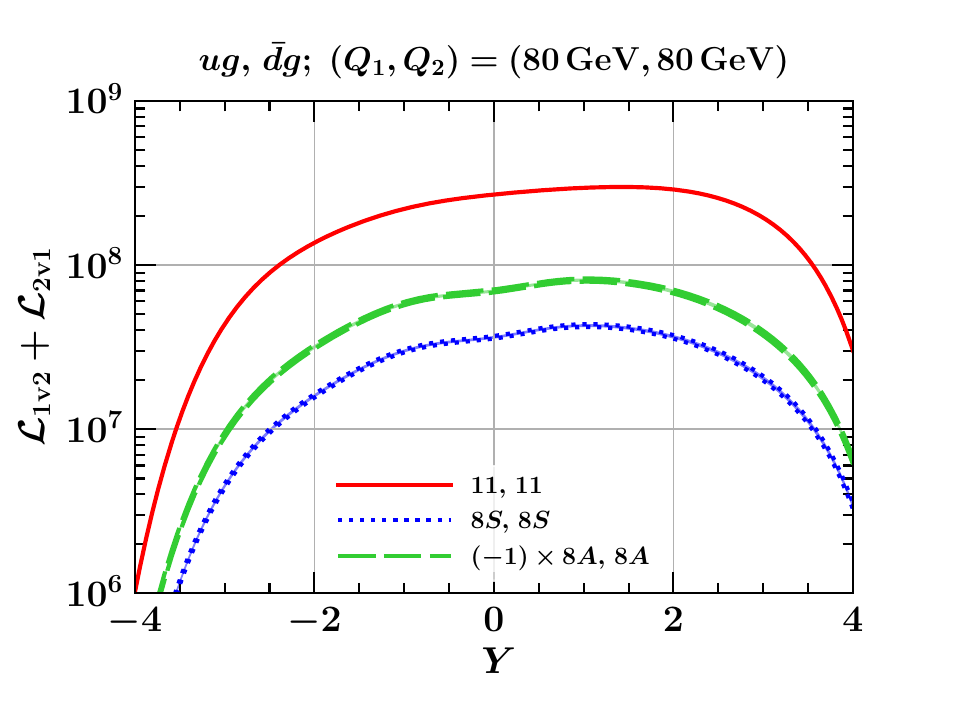}
}
\\[1.5em]
\subfloat[1v1 \, ($u g, \bar{d} g$)]{
   \includegraphics[width=0.48\textwidth,trim=0 20 30 39,clip]{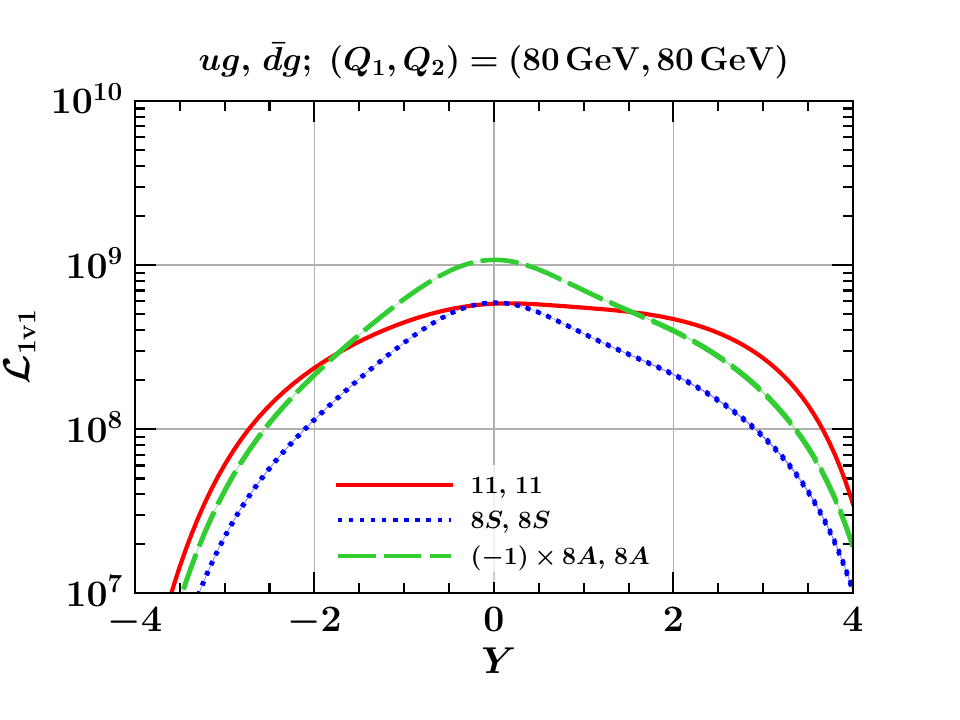}
}
\subfloat[1v1 subtracted \, ($u g, \bar{d} g$)]{
   \includegraphics[width=0.48\textwidth,trim=0 20 30 39,clip]{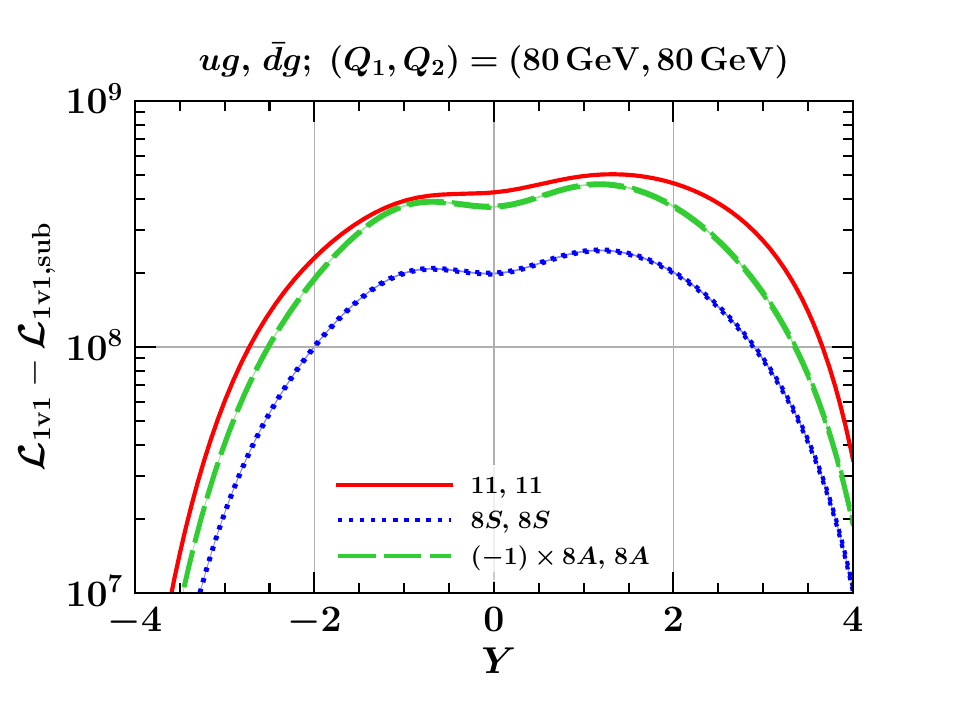}
}
\caption{\label{fig:lumis-W-plus-jets-80} As \fig{\protect\ref{fig:lumis-W-plus-jets-10}}, but for $M_2 = 80 \gev$.}
\end{figure}

The production of a $W$ and a dijet by double parton scattering has been
observed already in run 1 at the LHC \cite{ATLAS:2013aph, CMS:2013huw}.  In
\figs{\ref{fig:lumis-W-plus-jets-10}} and \ref{fig:lumis-W-plus-jets-80} we show
double parton luminosities that contribute to this channel.  The first figure is
for a low invariant mass $M_2 = 10 \gev$ of the dijet pair, whereas the second
one if for $M_2 = 80 \gev$.  The patterns of the relative size between colour
singlet and octet contributions is very similar to the one we observed for four
gluons in \sect{\ref{sec:double-dijets}}.

\subsection{Heavy-quark jets}
\label{sec:lumis-b-quarks}

\begin{figure}[p]
\centering
\subfloat[2v2 \, ($b g, g b$)]{
   \includegraphics[width=0.48\textwidth,trim=0 20 30 39,clip]{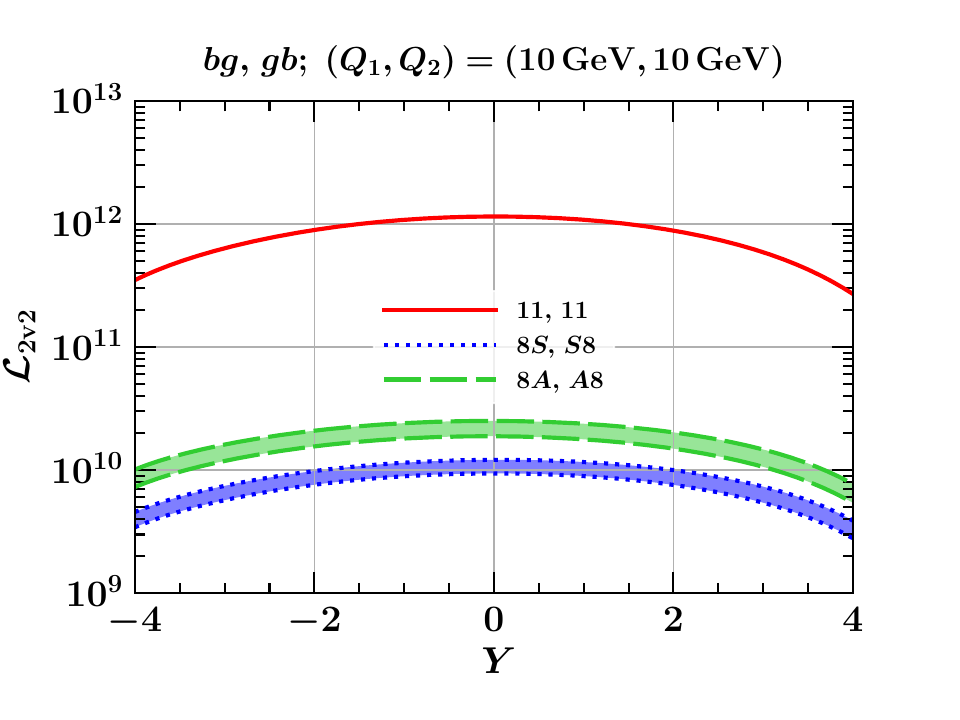}
}
\subfloat[1v2 + 2v1 \, ($b g, g b$)]{
   \includegraphics[width=0.48\textwidth,trim=0 20 30 39,clip]{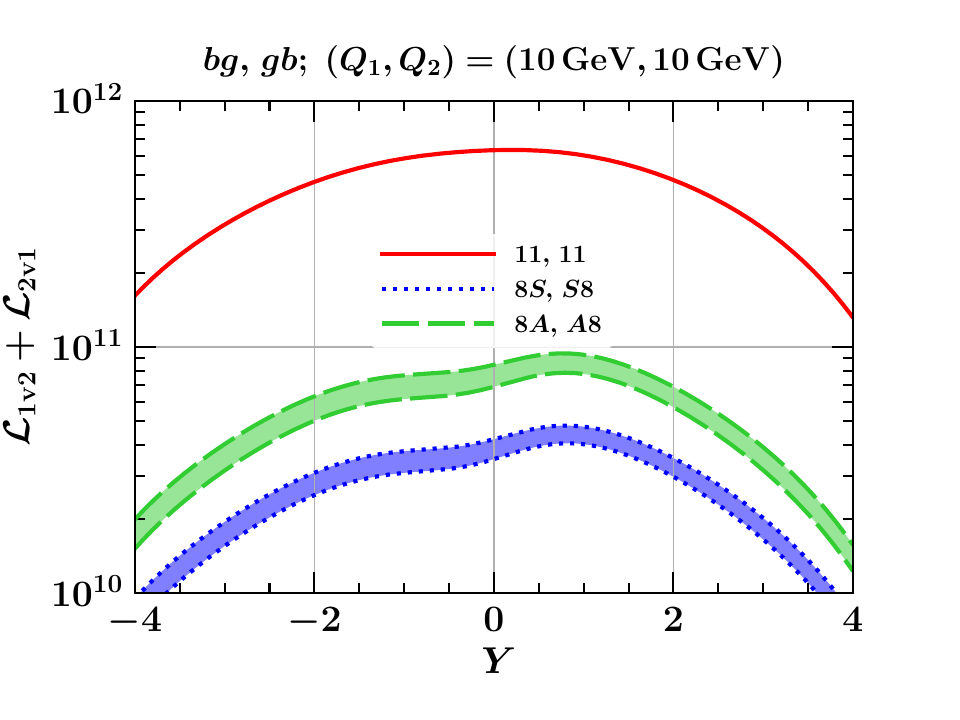}
}
\\[1.5em]
\subfloat[1v1 \, ($b g, g b$)]{
   \includegraphics[width=0.48\textwidth,trim=0 20 30 39,clip]{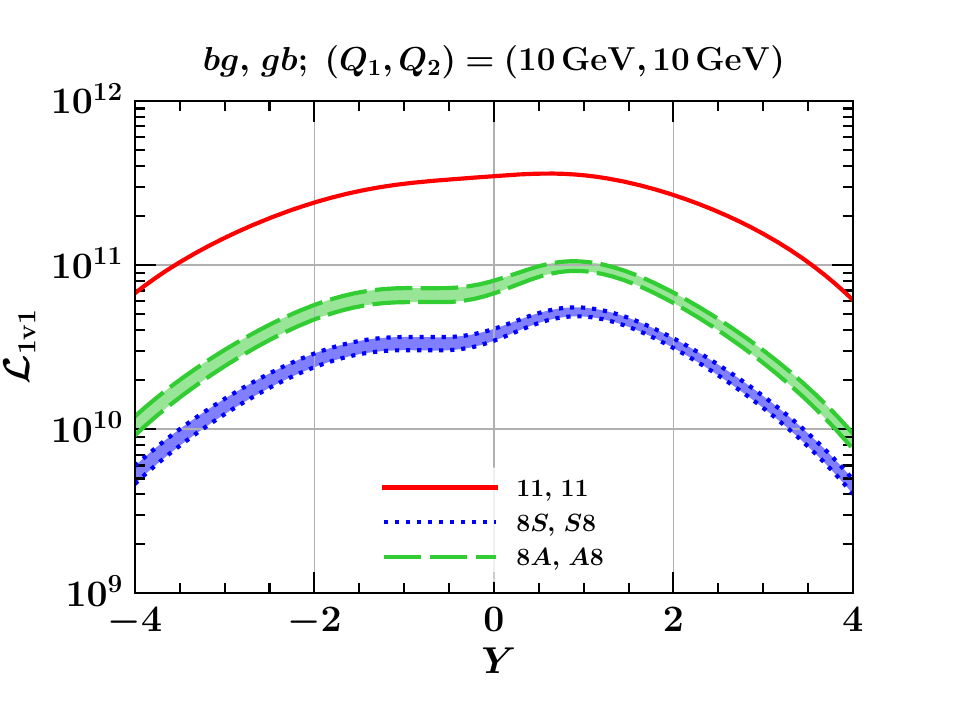}
}
\subfloat[1v1 subtracted \, ($b g, g b$)]{
   \includegraphics[width=0.48\textwidth,trim=0 20 30 39,clip]{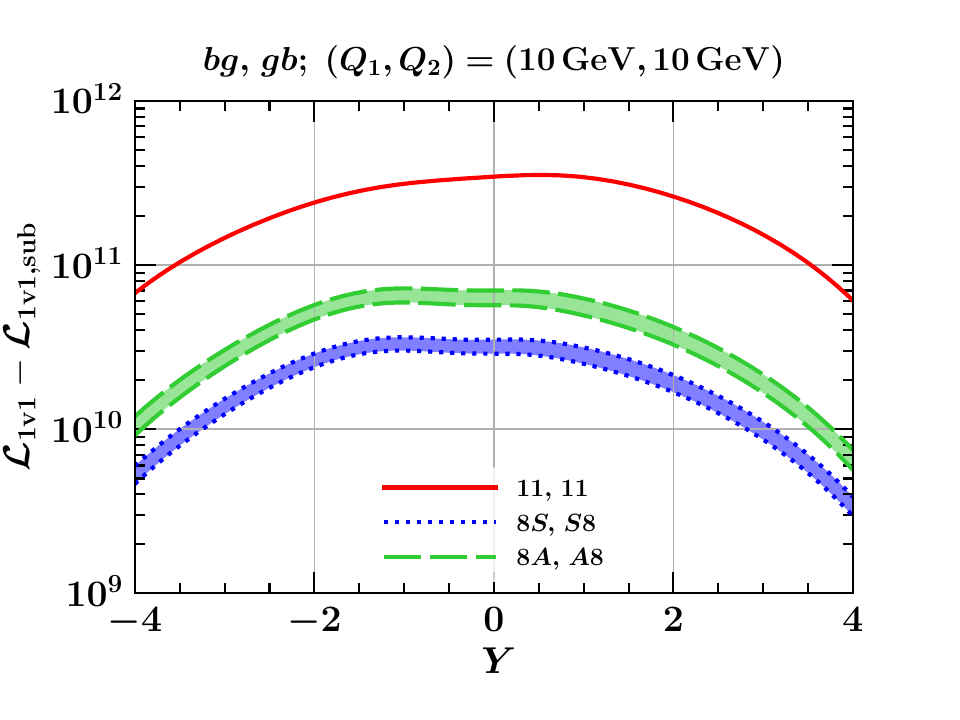}
}
\caption{\label{fig:lumis-b-quarks-10-10} Double parton luminosities
$\mathcal{L}_{\smash{b g, g b}}$ for $M_1 = M_2 = 10 \gev$.}
\end{figure}
\begin{figure}[p]
\centering
\subfloat[2v2 \, ($b g, g b$)]{
   \includegraphics[width=0.48\textwidth,trim=0 20 30 39,clip]{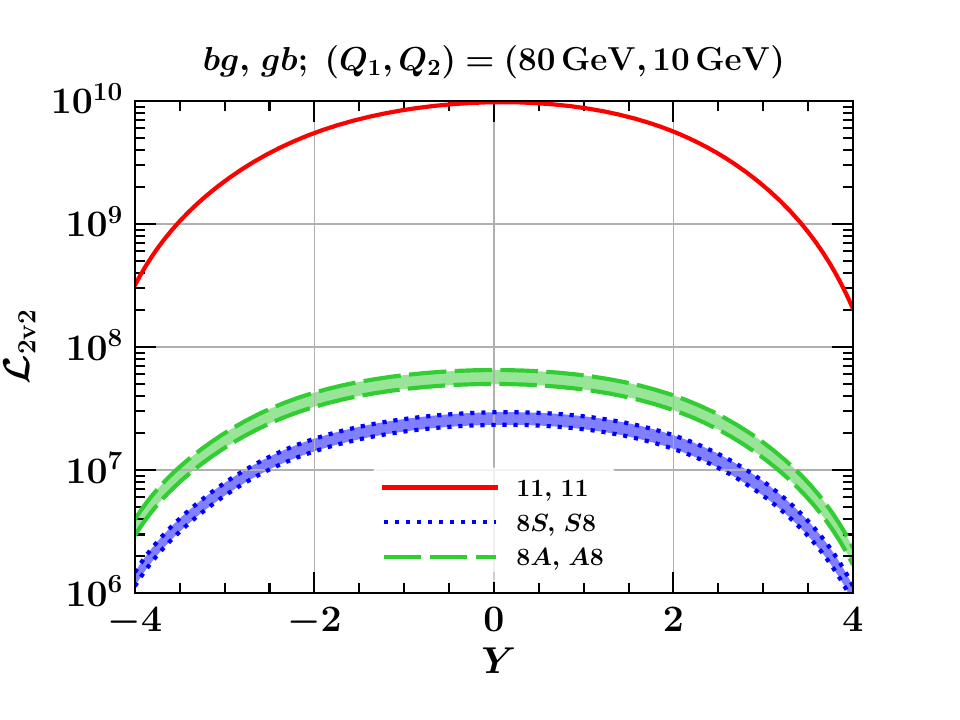}
}
\subfloat[1v2 + 2v1 \, ($b g, g b$)]{
   \includegraphics[width=0.48\textwidth,trim=0 20 30 39,clip]{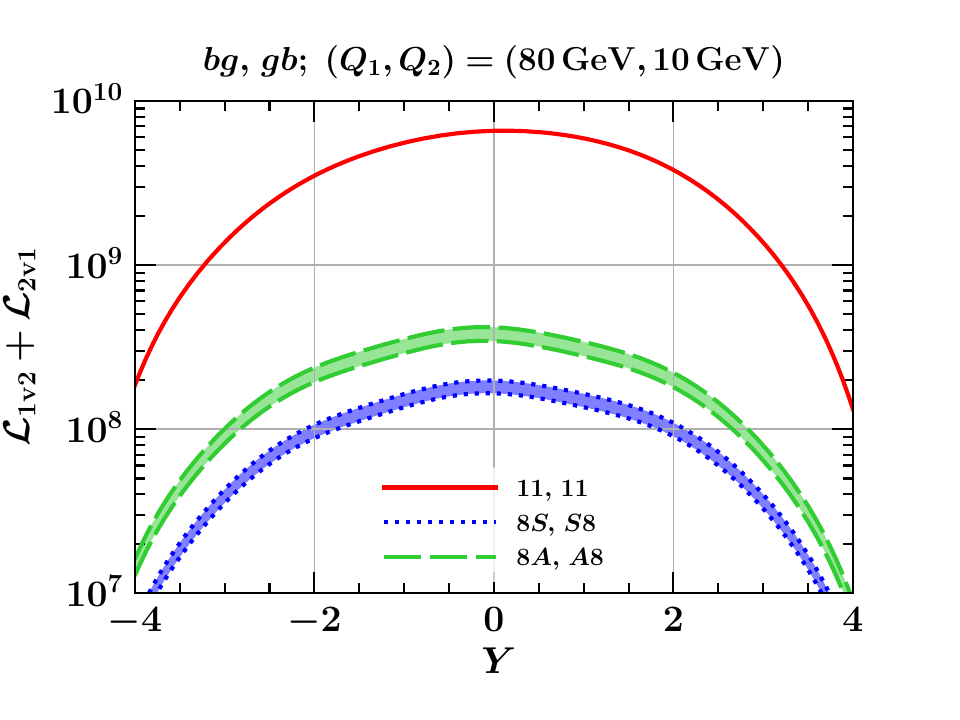}
}
\caption{\label{fig:lumis-b-quarks-80-10} As
\fig{\protect\ref{fig:lumis-b-quarks-10-10}}, but for $M_1 = 80 \gev$ and $M_2 =
10 \gev$.}
\end{figure}

\begin{figure}[p]
\ContinuedFloat
\centering
\subfloat[1v1 \, ($b g, g b$)]{
   \includegraphics[width=0.48\textwidth,trim=0 20 30 39,clip]{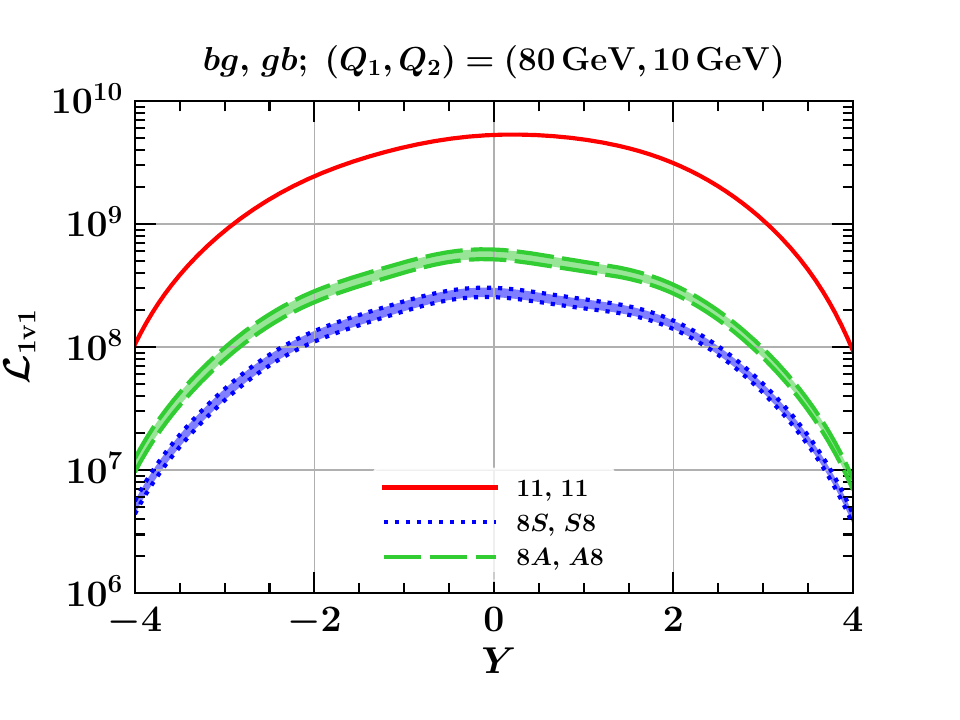}
}
\caption*{Figure~\protect\ref{fig:lumis-b-quarks-80-10} (continued)}
\end{figure}
\begin{figure}[p]
\centering
\subfloat[2v2 \, ($b g, g b$)]{
   \includegraphics[width=0.48\textwidth,trim=0 20 30 39,clip]{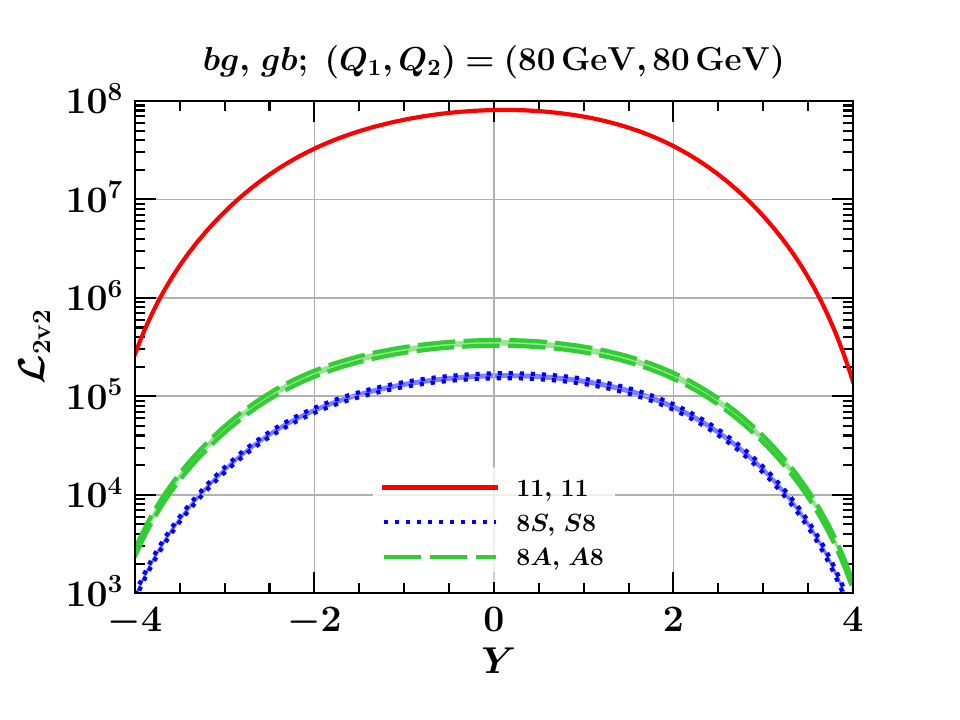}
}
\subfloat[1v2 + 2v1 \, ($b g, g b$)]{
   \includegraphics[width=0.48\textwidth,trim=0 20 30 39,clip]{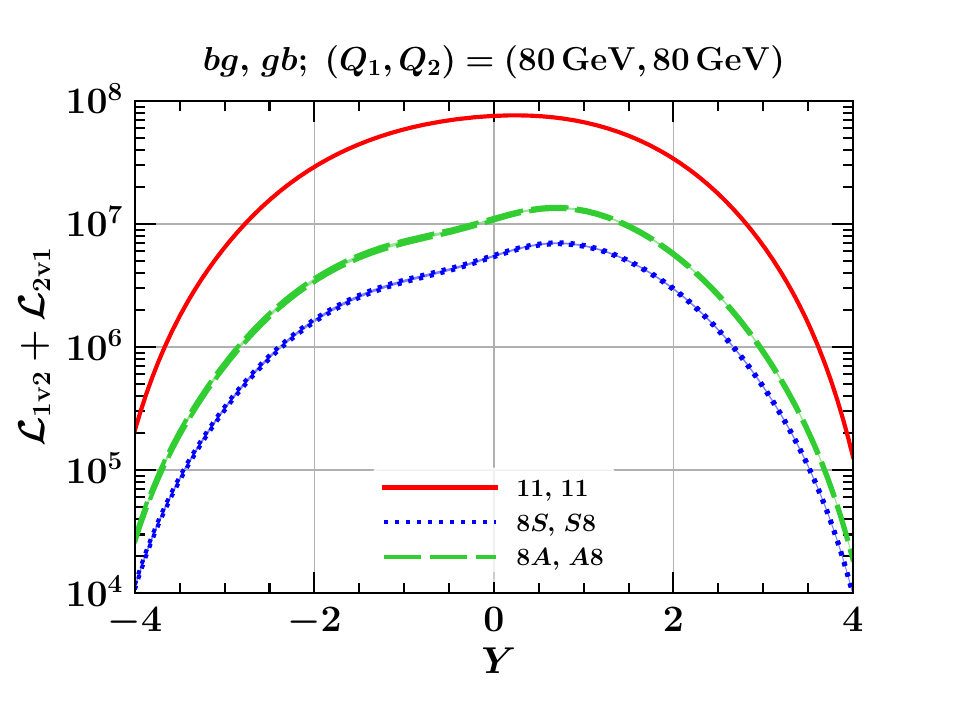}
}
\\[1.5em]
\subfloat[1v1 \, ($b g, g b$)]{
   \includegraphics[width=0.48\textwidth,trim=0 20 30 39,clip]{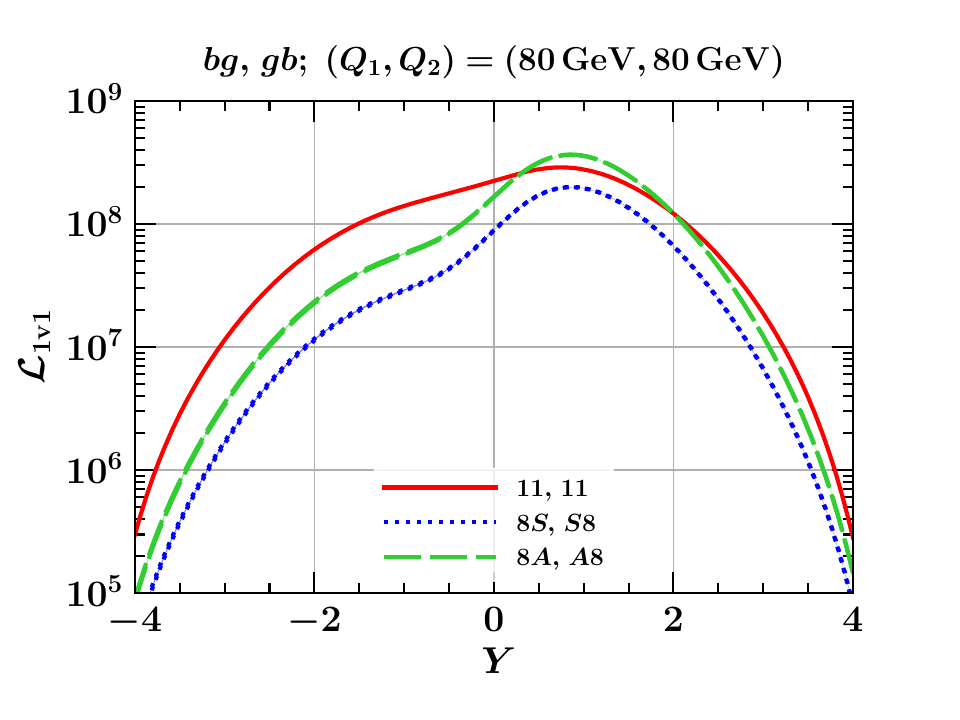}
}
\subfloat[1v1 subtracted \, ($b g, g b$)]{
   \includegraphics[width=0.48\textwidth,trim=0 20 30 39,clip]{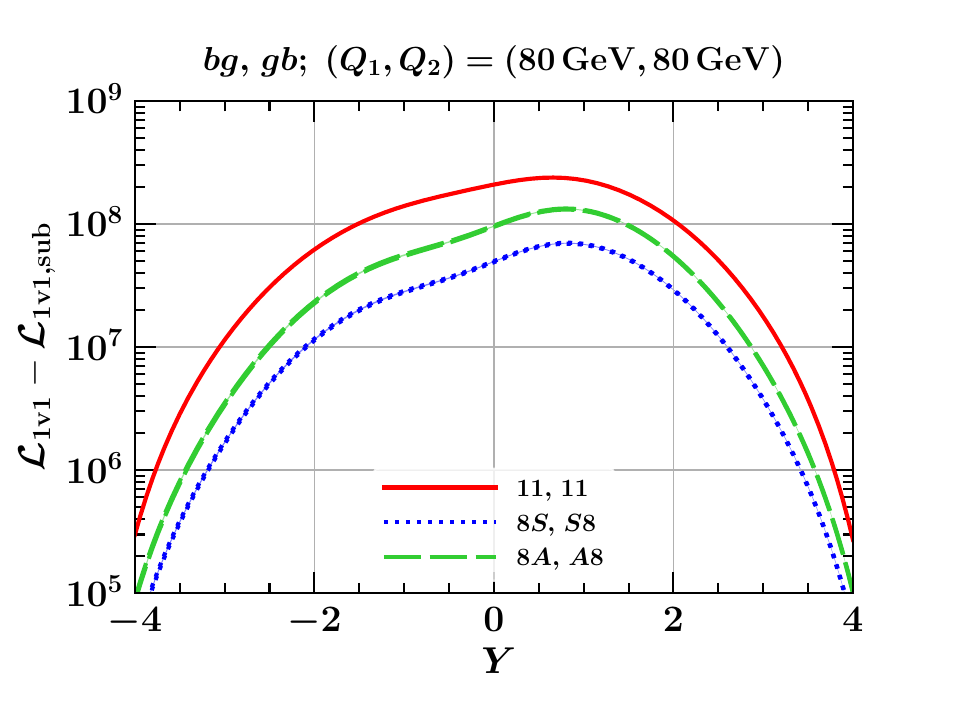}
}
\caption{\label{fig:lumis-b-quarks-80-80} As
\fig{\protect\ref{fig:lumis-b-quarks-10-10}}, but for $M_1 = M_2 = 80 \gev$.}
\end{figure}

Double parton luminosities $\mathcal{L}_{\smash{b g, g b}}$ are shown in
\figs{\ref{fig:lumis-b-quarks-10-10}} to \ref{fig:lumis-b-quarks-80-80} for
different combinations of $M_1$ and $M_2$.  Such a parton combination gives rise
to the production of two dijets with exactly one $b$ quark in each dijet.  We
find that the relative size of colour singlet and octet contributions is again
similar to what we saw for four gluons in \sect{\ref{sec:double-dijets}}.  In
particular, there is barely a suppression of the octet channels in the
subtracted 1v1 contribution at high invariant masses, as shown in
\fig{\ref{fig:lumis-b-quarks-80-80}}d.

\FloatBarrier


\phantomsection
\addcontentsline{toc}{section}{References}

\bibliographystyle{JHEP}
\bibliography{evo-nlo.bib}

\end{document}